\documentclass[useAMS,usenatbib]{mn2e}

\usepackage[latin1]{inputenc}
\usepackage{graphicx}
\usepackage{multicol}
\usepackage{float}
\usepackage{footnote}
\usepackage{amssymb}
\usepackage[figuresright]{rotating}
\usepackage{amsmath}
\usepackage{url}
\usepackage{rotating}
\usepackage{caption}
\usepackage[caption=false]{subfig} 
\newcommand{\Mpc}{\rm\; Mpc}

\newcommand{\km}{\rm\; km}

\newcommand{\mum}{\hbox{$\rm\; \mu m\,$}}

\newcommand {\apgt} {\ {\raise-.5ex\hbox{$\buildrel>\over\sim$}}\ }
\newcommand {\aplt} {\ {\raise-.5ex\hbox{$\buildrel<\over\sim$}}\ } 
\newcommand{\s}{\rm\; s}

\newcommand{\rcool}{\hbox{$\thinspace r_\mathrm{cool}$}}

\newcommand{\keV}{\rm\; keV}

\newcommand{\erg}{\rm\; erg}

\newcommand{\ergps}{\hbox{$\erg\s^{-1}\,$}}

\newcommand{\kmps}{\hbox{$\km\s^{-1}\,$}}

\newcommand{\kmpspMpc}{\hbox{$\kmps\Mpc^{-1}\,$}}
\newcommand{\Lx}{\hbox{$\thinspace L_\mathrm{X}$}}

\newcommand{\Omm}{\hbox{$\rm\thinspace \Omega_{m}$}}
\newcommand{\OmL}{\hbox{$\rm\thinspace \Omega_{\Lambda}$}}
\title[X-ray cavities in the MACS cluster sample]{Extreme AGN feedback in the MAssive Cluster Survey (MACS): a detailed study of X-ray cavities at $z>0.3$}
\author[J. Hlavacek-Larrondo, et al.]{J. Hlavacek-Larrondo$^{1}$\thanks{E-mail: juliehl@ast.cam.ac.uk}\thanks{A pdf version of the paper with high-resolution images can be found at http://www-xray.ast.cam.ac.uk/$\sim$juliehl/MACSpaper/}, A.C. Fabian$^{1}$, A. C. Edge$^{2}$, H. Ebeling$^{3}$, J.S. Sanders$^{1}$, \newauthor{M. T. Hogan$^{2}$ and G.B. Taylor$^{4,5}$}\\
$^{1}$Institute of Astronomy, University of Cambridge, Madingley Road, Cambridge CB3 0HA\\ $^{2}$Institute of Computational Cosmology, Department of Physics, Durham University, Durham, DH1 3LE\\ 
$^{3}$Institute of Astronomy, University of Hawaii, 2680 Woodlawn Drive, Honolulu, HI 96822, USA\\
$^{4}$Department of Physics and Astronomy, University of New-Mexico, Albuquerque, NM 87131, USA\\
$^{5}$National Radio Astronomy Observatory, Socorro, NM 87801, USA}
\begin{document}

\date{Accepted 2011 December 16.  Received 2011 December 1; in original form 2011 September 27}

\pagerange{\pageref{firstpage}--\pageref{lastpage}} \pubyear{2011}

\maketitle

\begin{abstract}
We present the first statistical study of X-ray cavities in distant clusters of galaxies ($z>{0.3}$). With the aim of providing further insight into how AGN feedback operates at higher redshift, we have analysed the $Chandra$ X-ray observations of the Massive Cluster Survey (MACS) and searched for surface-brightness depressions associated with the Brightest Cluster Galaxy (BCG). The MACS sample consists of the most X-ray luminous clusters within $0.3\leq{z}\leq{0.7}$ (median $L_{\rm X, RASS}=7\times10^{44}\ergps$), and out of 76 clusters, we find 13 with ``clear" cavities and 7 with ``potential" cavities (detection rate $\sim25$ per cent). Most of the clusters in which we find cavities have a short central cooling time below $3-5$ Gyrs, consistent with the idea that cavities sit predominantly in cool core clusters. We also find no evidence for evolution in any of the cavity properties with redshift, up to $z\sim0.6$. The cavities of powerful outbursts are not larger (or smaller) at higher redshift, and are not able to rise to further (or lesser) distances from the nucleus. The energetics of these outbursts also remain the same. This suggests that extreme ``radio mode" feedback ($L_{\rm mech}>10^{44}\ergps$) starts to operate as early as $7-8$ Gyrs after the Big Bang and shows no sign of evolution since then. In other words, AGNs lying at the centre of clusters are able to operate at early times with extreme mechanical powers, and have been operating in such a way for at least the past 5 Gyrs.
\end{abstract}


\section{Introduction}
Active galactic nucleus (AGN) feedback plays a major role in quenching star formation, enriching the surrounding medium with metals, and fuelling the supermassive black hole (SMBH) of the host galaxy. Some of the most extreme examples of AGN feedback are seen in clusters of galaxies, where the central SMBH inflates large cavities filled with radio emitting particles through jets. These cavities appear as depressions in the X-ray image and provide a direct measurement of the energy being injected into the surrounding medium by the central SMBH (see reviews on the topic by Peterson \& Fabian 2006 and McNamara \& Nulsen 2007) \nocite{Mcn200745,Pet2006427}. 

Detailed studies of individual nearby clusters such as the Perseus Cluster \citep[see][ and references therein]{Fab2006366} have shown that the AGN lying at the centre can energetically offset cooling of the intracluster medium (ICM) not only by inflating these large cavities, but also by inducing weak shocks and propagating energy through sound/pressure waves \citep{Bir2004607,Raf2006652,Dun2006373,Dun2008385,Mcn200745,Fab2003344,Fab2006366,For2005635,San2007381}. Statistical studies of X-ray cavities have also provided a wealth of information on how AGN feedback operates \citep{Bir2004607,Bir2008686,Dun2005364,Dun2006373,Dun2008385,Nul2007,Dun2010404,Cav2010720,Don2010712,OSu2011735}. Essentially, these studies find that the power output of cavities is substantial, in most cases sufficient to prevent cooling of the ICM, and that cavities are common, especially in systems which require some form of heating to prevent the gas from cooling. However, all the statistical compilations of cavities have until now focussed on the nearby Universe (only 4 objects are at $z>0.3$). The cavity properties at the higher redshift end therefore remain unexplored. It is also not clear if and how AGN heating evolves across time. According to cosmological simulations \citep{Cro2006365,Sij2006366}, present-day black holes are in a ``radio mode" phase, which involves a black hole accreting at sub-Eddington rates and driving powerful outflows, whereas at earlier times ($z\apgt1$), ``quasar mode" feedback dominates and consists of a merger-driven phase where the black hole grows rapidly. When and how this transition occurs remains unclear. 

\begin{figure}
\centering
\begin{minipage}[c]{0.99\linewidth}
\centering \includegraphics[width=\linewidth]{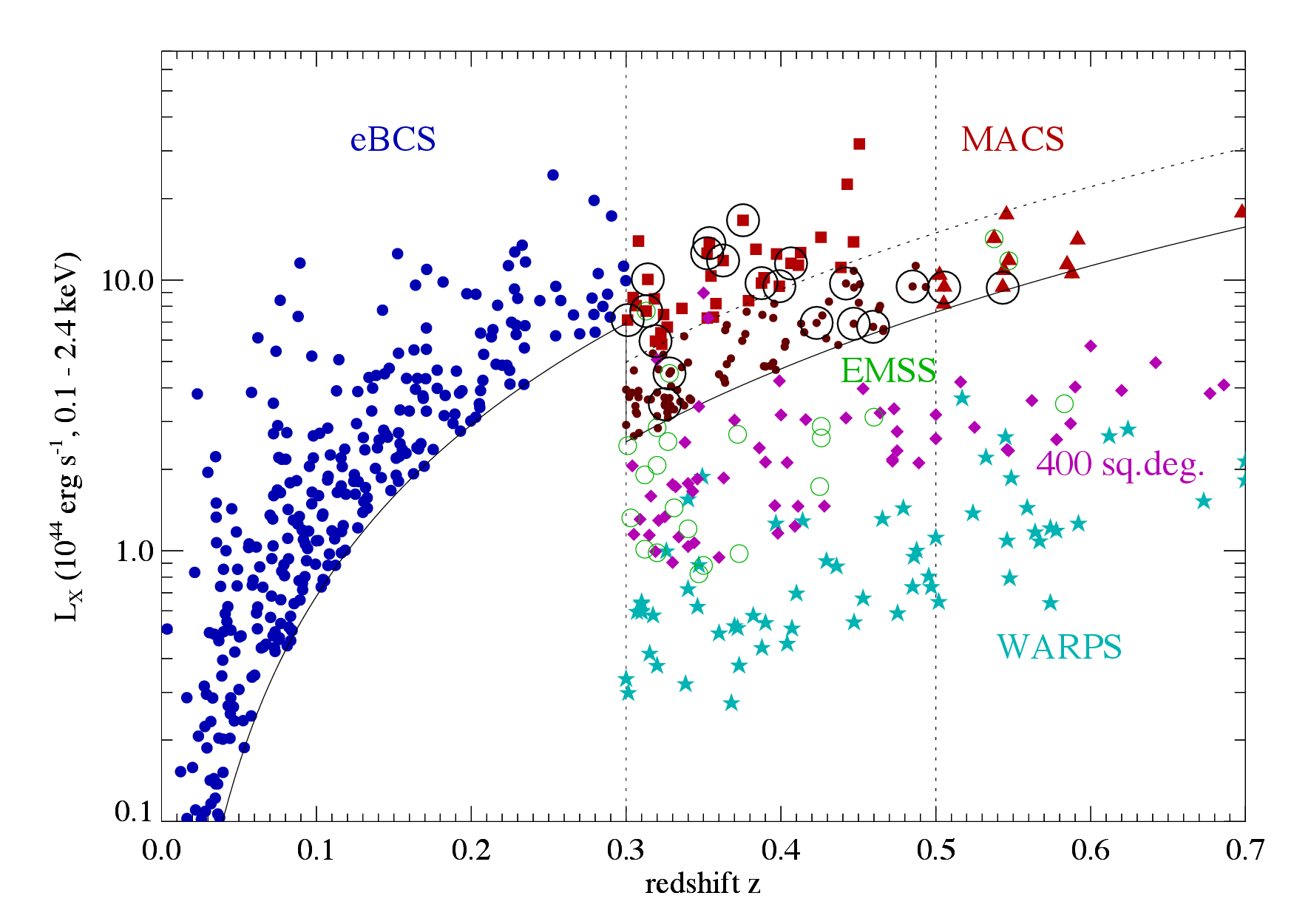}
\end{minipage}
\caption{The X-ray luminosity versus redshift distribution of different cluster samples. There are a total of 124 MACS clusters, all with $z>0.3$. The MACS clusters outlined with black circles are those in which we identified cavities. }
\label{fig1}
\end{figure}

We can start investigating how AGN feedback evolves with redshift by using targeted samples of $z>0.3$ clusters that have extensive follow-up X-ray observations, and in this case, the Massive Cluster Survey (MACS) provides an ideal example. MACS was launched in 1999 and compiled the first large sample of very X-ray luminous, and therefore very massive, clusters in the distant Universe \citep[][]{Ebe2001553,Ebe2007661,Ebe2010407}. The MACS sample was based on the $ROSAT$ All-Sky Survey \citep[RASS; ][ and references therein]{Tru1983} Bright Source Catalogue (BSC; Voges et al. 1999)\nocite{Vog1999}. It consists of clusters above a redshift of $z>0.3$ and covers a solid angle over 22000 deg$^2$ in the extragalactic sky.  In Fig. \ref{fig1}, we show the $L_{\rm X}-z$ distribution for various samples, including the Brightest Cluster Survey \citep{Ebe1998301,Ebe2000318}, the $Einstein$ Medium Sensitivity Survey \citep[EMSS;][]{Gio199494}, the Wide Angle Rosat Pointed Survey \citep[WARPS;][]{Per2002140}, the 400 deg$^2$ cluster samples \citep[][]{Bur2007172}, and the MACS sample.  This figure shows that MACS selects on average clusters which are $10-20$ times more luminous than the other samples in the same redshift range. The MACS sample now contains 124 spectroscopically confirmed clusters within $0.3 \leq{z}\leq{0.7}$, for which more than two thirds are new discoveries. 

Since MACS contains only very X-ray luminous clusters (median $L_{\rm X, RASS}=7\times10^{44}\ergps$), those harbouring a cool core should require extreme feedback from their central AGN to prevent the ICM from cooling ($L_{\rm mech}\sim10^{44-45}\ergps$). We therefore expect to see some of the most extreme examples of AGN outbursts in this sample, which will allow us to extend our knowledge of X-ray cavities not only in the distant Universe ($z>0.3$), but also at the higher end of mechanical power arising from the AGN. To date, more than two thirds of all MACS clusters have been targeted in $Chandra$ follow-up observations, and the unique and exquisite spatial resolution of $Chandra$ is essential to identify cavities at such redshifts. The MACS sample therefore provides a unique opportunity to study the cavity properties in clusters at $z>0.3$,  and determine if AGN feedback evolves with time.

We first present the data in Section 2, and then discuss how we proceeded in identifying systems with cavities in Section 3. Section 4 and Section 5 outline the techniques used to calculate the energetics of the cavities, as well as their associated time-scales. In Section 6, we derive the properties of the cores and plot scaling relations in Section 7. Finally, we present the results in Section 8 and discuss them in Section 9. We adopt $H_\mathrm{0}=70\kmpspMpc$ with $\Omm=0.30$ and $\OmL=0.7$ throughout this paper. All errors are $2\sigma$, and the abundance ratios of \citet{And198953} were used throughout this paper.

\section{Observations and data reduction}
\subsection{X-ray observations}

The X-ray data were obtained through the $Chandra$ data archive. We found observations with the Advanced CCD Imaging Spectrometer (ACIS) for 83 of the MACS clusters, with exposure times varying between $10$~ks and $100$~ks. Some objects were also observed multiple times. For each cluster, we selected the deepest observational data set available. These data were then processed, cleaned and calibrated using the latest version of the {\sc ciao} software ({\sc ciaov4.3}, {\sc caldb4.4.1}), and starting from the level 1 event file. We applied both CTI (charge time interval) and time-dependent gain corrections, as well as removed flares using the {\sc {lc$\_$clean}} script, with a $3\sigma$ threshold. When a cluster was observed multiple times with the target centred on the same detector and in the same observing mode (FAINT or VFAINT), we combined the different observations only if this would improve the image quality significantly. We then exposure-corrected the images, using an exposure map generated with a monoenergetic distribution of source photons at 1.5\keV  (which is almost the peak energy of the clusters). These images were used to identify systems with cavities (see Section 3). Out of the 83 clusters for which we have X-ray observations, 2 are clear mergers with multiple components merging together and 5 have a bright point source with no clear extended emission associated with it. We therefore only consider the remaining 76 in the analysis of X-ray cavities.
\begin{table*}
\caption{$Chandra$ observations of the MACS clusters with cavities - (1) Name; (2) Alternate name; (3) Redshift $z$; (4) Observation identification number; (5) Exposure time; (6) Chip name where the target was centred on; (7) References where the redshift was taken from: (i) \citet{Ebe2007661}; (ii) \citet{Ebe2010407}; (iii) \citet{Sto199176}; (iv) \citet{Wri1983205}; (v) \citet{Man2011}; (vi) \citet{Kle1988328}; (vii) \citet{Dre199278}; (viii) Ebeling et al. (in preparation); (ix) Sloan Digital Sky Survey; (x) \citet{Sma2007654}. }
\centering
\resizebox{14cm}{!} {
\begin{tabular}{lcccccc}
\hline
\hline
(1) & (2) & (3) & (4) & (5) & (6) & (7) \\
Cluster & Alternate & $z$  & Obs. ID & Exposure & Chip & Ref. \\
name & name & & & \rm{ks} \\
\\
\hline
MACS~J0159.8-0849 & ... & 0.404 & 3265 & 14.5 & ACIS-I3 &  (ii)\\
& & &   6106 & 28.9 & ACIS-I3 &\\
& & &  9376 & 18.0 & ACIS-I3 &\\
MACS~J0242.5-2132 & ... & 0.314 &  3266 & 8.6 & ACIS-I3 &  (iv)\\
MACS~J0429.6-0253 & ... & 0.397 &  3271 & 21.4 & ACIS-I3 & (ii) \\
MACS~J0547.0-3904 & ... & 0.319 &  3273 & 19.2 & ACIS-I3 & (ii) \\
MACS~J0913.7+4056 & IRAS 09104+4109 & 0.442 &  10445 & 70.4 & ACIS-I3 & (vi) \\
MACS~J0947.2+7623 & RBS 0797 & 0.354 & 7902 & 38.8 & ACIS-S3 & (ii) \\
MACS~J1411.3+5212 & 3C295 & 0.460 & 2254 & 76.4 & ACIS-I3 & (vii) \\
MACS~J1423.8+2404 & ... & 0.5449 & 4195 & 106.9 & ACIS-S3 & (i) \\
MACS~J1532.8+3021 & RX J1532.9+3021 & 0.3613 &  1649 & 9.5 & ACIS-S3 & (ii) \\
MACS~J1720.2+3536 & Z8201 & 0.3913 &  3280 & 17.6 & ACIS-I3 & (ii) \\
& & &  6107 & 27.3 & ACIS-I3 &\\
& & &  7718 & 6.8 & ACIS-I3 &\\
MACS~J1931.8-2634 & ... & 0.352 & 9382 & 95.0 & ACIS-I3 & (ii) \\
MACS~J2046.0-3430 & ... & 0.423 & 9377 & 35.9 & ACIS-I3 & (viii) \\
MACS~J2140.2-2339 & MS 2137.3-2353 & 0.313 &  4974 & 7.3 & ACIS-S3 & (iii) \\
				  & & & 5250 & 34.7 & ACIS-S3 &\\
				  & & & 928 & 34.8  & ACIS-S3 &\\
\hline
MACS~J0111.5+0855 & ... & 0.485 &  3256 & 15.2 & ACIS-I3 & (v) \\
MACS~J0257.1-2325 & ... & 0.5039 &  1654 & 18.0 & ACIS-I3 & (i) \\
				  & &  &  3581 & 16.1 & ACIS-I3 &\\
MACS~J1359.1-1929 & ... & 0.447 &  9378 & 46.1 & ACIS-I3 & (v) \\
MACS~J1359.8+6231 & MS 1358.4+6245 & 0.330 &  516 & 48.5 & ACIS-S3 & (iii) \\
MACS~J1447.4+0827 & RBS 1429 & 0.3755 & 10481 & 11.5 & ACIS-S3 & (ix) \\
MACS~J2135.2-0102 & 1RXS J213515.7-010208 & 0.325 &  11710 & 24.7 & ACIS-I3  & (x) \\
MACS~J2245.0+2637 & ... & 0.301 &  3287 & 11.5 & ACIS-I3 & (ii) \\
\hline
\end{tabular}}
\label{tab1}
%\end{table}
\end{table*}

For the clusters in which we identified cavities (see Table \ref{tab1}), the spectra were analysed using {\sc xspec} \citep[{v12.6.0 d,f, \rm e.g.}][]{Arn1996101}. For each object, the Galactic absorption was kept frozen at the \citet{Kal2005440} value. Varying this parameter did not improve significantly the fit in any of our clusters. For targets observed multiple times, we fitted simultaneously the spectra. Finally, although some of the objects in our sample with cavities have a bright X-ray point source at their centres, we find no significant pileup for any of these point sources. Pileup occurs when two or more photons are detected as one event \citep[see for more details][]{Dav2001562,Rus2009402}, and the amount of pileup can be estimated by comparing the fraction of good (grades 0,2,3,4,6) to bad grades (grades 1,5,7) for each point source. Typically, pileup becomes problematic when the fraction of bad grades exceeds 10 per cent of the good grades, but we find a ratio less then $6$ per cent for all of our central point sources.

\subsection{Radio observations}

For each cluster with cavities, we searched for radio emission associated with the central AGN by looking through different radio surveys available to the public. We first looked for radio emission associated with the AGN at 1.4 GHz by using the 1.4 GHz VLA Faint Images of the Radio Sky at Twenty-Centimeters survey \citep[FIRST, average resolution of $5''$;][]{Bec199461}, or the 1.4 GHz NRAO VLA Sky Survey catalogue \citep[NVSS, average resolution of $45''$;][]{Con1998115}. We then looked for radio emission associated with the AGN at lower frequency, either with the 326 MHz Westerbork Northern Sky Survey\footnote[1]{http://www.astron.nl/wow/testcode.php?survey=1} \citep[WENSS, resolution of $\sim50''$;][]{Ren1997124}, the 74 MHz VLA Low-frequency Sky Survey \citep[VLSS, resolution of $\sim80''$;][]{Coh2007134}, the 150 MHz TIFR GMRT Sky Survey (TGSS\footnote[2]{http://tgss.ncra.tifr.res.in}, resolution of $\sim20''$) or the 843 MHz Sydney University Molonglo Sky Survey \citep[SUMSS, resolution of $\sim40''$;][]{Boc1999117,Mau2003342}. Finally, we searched for high-frequency radio emission associated with the central AGN at 5 GHz with the Parkes-MIT-NRAO (PMN, resolution of $\sim5'$) radio survey \citep{Gri1993105} or the EINSTEIN Observatory Extended Medium-Sensitivity Survey \citep[EMSS, where the radio resolution of the survey lies between $\sim4-14''$;][]{Gio199072,Sto199176}, and then at 28.5 GHz with the Berkeley-Illinois-Maryland Association \citep[BIMA, sensitive to angular scales around $1.5'$;][]{Cob2008134}. If no point source was seen within the central regions ($r\aplt100$ kpc), we used the 2$\sigma_{\rm rms}$ value within the beam area as an upper limit for the flux. The values we obtain are shown in Table \ref{tab2}. Some sources also have flux densities available from the literature, and we include these in Table \ref{tab2}. 

Errors are derived as the quadratic sum of the rms noise level in the map and the systematic uncertainty associated with the value. For some of the surveys used, the error quoted in the catalogue did not account for systematic uncertainties. This type of error varies with frequency, but is on the order of 5 per cent \citep[see][]{Car1991383}. For simplicity, we therefore choose to compute the total uncertainty assuming a 5 per cent systematic error and a 2$\sigma_{\rm rms}$ noise level. MACS~J1931.8-2634 has a complicated morphology at radio wavelengths, with a Narrow Angle Tail (NAT) source located 45$''$ to the south that is contaminating the flux measurements of the central galaxy. The 1.4 GHz flux quoted in Table \ref{tab2} is taken from \citet{Ehl2011411} and only includes the contribution of the central galaxy. However, for the 5 GHz and 150 MHz data, we had to remove the contribution of the NAT source. In this case, we used the peak flux value of the galaxy and subtracted the peak contribution of the NAT source. 

Finally, by using the flux densities quoted in Table \ref{tab2}, we calculated a rough estimate of the total radio luminosity ($L_{\rm radio}$) with Eq. \ref{eq1} and integrating between $\nu_1=10$ MHz and $\nu_2=10000$ MHz.  
\begin{eqnarray}
L_{\rm radio}=4{\pi}D^2_{\rm L}\int_{\nu_1}^{\nu_2}(S_\nu)d\nu
\label{eq1}
\end{eqnarray}
Here, $D_{\rm L}$ is the luminosity distance to the source and $S_\nu$ is the flux density ($S_\nu\propto\nu^{-\alpha}$, where $\alpha$ is the spectral index). For the central galaxies of MACS~J0242.5-2132 and MACS~J1411.3+5212 (3C 295), we also use the extensive follow-up radio observations available in the NASA/IPAC Extragalactic Database (NED) to determine $L_{\rm radio}$. Only MACS~J1411.3+5212 has a complete coverage from $\sim10$ MHz to 10 GHz. For the remaining objects, we extrapolate the values at 10 MHz and 10 GHz based on the assumption that the flux density scales as $S_\nu\propto\nu^{-\alpha}$, and for each extremity (10 MHz and 10 GHz), we use the local spectral index as determined from the two nearest flux density data points. Using a simple trapezoid rule, we then integrate over the range 10 MHz to 10 GHz using the extrapolated values at 10 MHz and 10 GHz, as well as the other data points available for each source. Our results are shown in Table \ref{tab2}.

\subsection{Optical observations}

We also searched for optical observations of each cluster harbouring a cavity. These images were used to verify that the cavities identified were associated with the central dominant galaxy. For each cluster, we searched for Hubble Space Telescope ($HST$) archival images \citep[many of which were acquired through snapshot campaigns, see][]{Ebe2009}, and if not available, we used the R passband images from \citet{Ebe2007661,Ebe2010407} taken with the University of Hawaii (UH) 2.2-m telescope (MACS~J2046.0-3430 and MACS~J0111.5+0855) or the red images from the Second Palomar Observatory Sky Survey \citep[$POSS II$;][]{Rei1991103} for MACS~J1447.4+0827. In Fig. \ref{fig2}, we show these optical images along with the $0.5-7.0$ keV $Chandra$ X-ray images.

\section{Identifying systems with cavities}

For each cluster, we visually searched for circular or ellipsoidal surface brightness depressions in the X-ray images associated with the central regions ($r<100$ kpc). Beyond 100 kpc, the count rate in most observations is not sufficient to identify cavities. As a first indicator, we computed unsharp-masked images for each source. This technique enhances deviations in the original image and consists of subtracting a strongly smoothed image from a lightly smoothed image, and is similar to subtracting an elliptical model of the cluster emission. Both the unsharp-masked and ellipse-subtracted images are shown in Fig. \ref{figA2}. However, although these figures can be very useful, they can also be misleading and cause false cavities to appear. Therefore, we only considered a cavity as being truly there if we could also see a hint of a depression in the original image. Once the systems with cavities were identified, we proceeded in classifying each cavity (see Column 2 of Table \ref{tab3}). First, if the cavity had a clear contrast in the raw image, as well as in the unsharp-masked and ellipse-subtracted image, we classified it as being ``clear". Out of these ``clear" cavities, some had surrounding bright rims in the raw image which we identified with the annotation $1$ in Table \ref{tab3}. Those without the bright rims were given the annotation $2$. If there was just a hint of a depression in the raw image, but the cavity could clearly be seen in the unsharp-masked or ellipse-subtracted image, then we classified it as being a ``potential" one. 

In total, we find that 13 clusters have at least one ``clear" cavity and 7 have ``potential" cavities. Therefore, out of the 76 MACS clusters with $Chandra$ data, we find that 20 have X-ray cavities. If we calculate the fractional difference between the counts within the cavity and the surrounding region, then we find that the decrements are on the order of $20-30$ per cent for the ``clear" cavities and between $10-20$ per cent for the ``potential" cavities. These fractions are roughly consistent with the decrements seen in nearby clusters of galaxies. In the calculation, we use the two immediate regions opposite the cavity, at the same radius and same size as the cavity, as the surrounding region. Note also that many of the MACS clusters have only been observed for $10-20$~ks and some have as little as $500-1000$ counts within 200 kpc. We could therefore be missing many cavities and our detection rate should only be considered as a lower limit to the number of MACS clusters with cavities. 

Our sample of objects with cavities includes MACS 1411.3+5212 (3C 295). This is a particularly interesting cluster, which has a very powerful radio source at its centre and shows strong evidence for Inverse Compton scattered Cosmic Microwave Background (ICCMB) photons along the central jet axis. Out of the remaining clusters, MACS~J2135.2-0102 has a central galaxy harbouring a double or triple nucleus (see optical image in Fig. \ref{fig2}). Initially, we identified two cavities in this system, one to the north-west and another to the south-east. However, the BCG is located almost within the south-eastern cavity. Although this could be due to projection effects, we choose to discard the southern cavity and just consider the northern one as a ``potential" cavity. MACS~J2135.2-0102 is more well known for its lensing arcs in the shape of a ``Cosmic Eye" and ``Cosmic Eyelash" \citep[][]{Sma2007654,Sia2009698,Swi2010NAT}.

\begin{table*}
\caption{Radio properties of the MACS clusters with cavities - (1) Name; (2) 74 MHz radio flux density from VLSS; (3) 326 MHz radio flux density from WENSS; (4) 1400 MHz radio flux density; (5) 5 GHz radio flux density; (6) 28.5 GHz radio flux density from Berkeley-Illinois-Maryland Association \citep[BIMA;][]{Cob2008134}; (7) Other radio flux density; (8) Integrated radio luminosity from 10 MHz to 10 GHz, see Section 2.2 for details; (9) References: (i) 1.4 GHz First; (ii) 1.4 GHz NVSS; (iii) Parkes MIT-NRAO 4.85 GHz survey \citep{Gri1993105}; (iv) 5 GHz private communication (M. T. Hogan); (v) 5 GHz from \citet{Hin1993415}; (vi) 5 GHz from \citet{Cav2011732}; (vii) 5 GHz from \citet{Gre199175}; (viii) EMSS 5 GHz \citep{Gio199072,Sto199176}; (ix) 843 MHz SUMSS \citep[][]{Boc1999117,Mau2003342}; (x) 15 GHz Arcminute Microkelvin Imager (AMI), private communication (K. Grainge); (xi) 1.4 GHz from \citet{Ehl2011411}; (xii) 150 MHz TIFR GMRT Sky Survey. $^{a}$For the central galaxies of MACS~J0242.5-2132 and MACS~J1411.3+5212 (3C 295), we used the extensive follow-up radio observations available in the NASA/IPAC Extragalactic Database (NED) to determine $L_{\rm radio}$. $^{b}$MACS~J1931.8-2634 has a complicated morphology at radio wavelengths, see Section 2.2 for details on how we obtained the flux densities.   }
\centering
\resizebox{17.5cm}{!} {
\begin{tabular}{lccccccccc}
\hline
\hline
(1) & (2) & (3) & (4) & (5) & (6) & (7) & (8) & (9) \\
Cluster & $S_{74{\rm MHz}}$ & $S_{326{\rm MHz}}$ & $S_{1.4{\rm GHz}}$ & $S_{5{\rm GHz}}$ & $S_{28.5{\rm GHz}}$ & $S_{\rm other}$ & $L_{\rm radio}$ & Ref. \\
name & mJy & mJy  & mJy  & mJy  & mJy  & mJy & ($10^{42}\ergps$) & \\
\hline
MACS~J0159.8-0849 & $<112$ & ...              & $31.4\pm1.6$ & $58\pm11$  & ... & ... & $3.56$ &   (i,iii) \\
MACS~J0242.5-2132$^{a}$ & $890\pm145$     & ... & $1255\pm73$ & $795\pm43$ & ... & ... & 27.1 &   (ii,iii)\\
MACS~J0429.6-0253 & $<214$ & ...              & $138.8\pm8.1$ & ... & ... & ... &  $7.2$ &  (ii)\\
MACS~J0547.0-3904 & ... & ...                 & $31.4\pm1.9$ & $15.4\pm0.8$ & ... & $19.6\pm1.3$[@843MHz] & 0.59 &   (ii,iv,ix)\\
MACS~J0913.7+4056 & $<272$ & $54.0\pm7.1$     &$8.3\pm0.5$ & $1.6\pm0.1$ & $0.69\pm0.12$ & $0.80\pm0.04$[@15GHz] & $1.33$  & (i,v,x)\\
MACS~J0947.2+7623 & $<329$ & $91.0\pm6.1$     &$21.7\pm1.3$ & $4.0\pm0.2$ & ... & ... & $1.02$ &   (ii,vi)\\
MACS~J1411.3+5212$^{a}$ & $120270\pm6022$ & $61647\pm3082$ &$22171\pm1109$ & $7401\pm808$ & ... & ... & 1025.3 &  (i,vii)\\
MACS~J1423.8+2404 & $<232$ & ... &$5.2\pm0.4$ & ... & $1.49\pm0.12$ & ... & $3.58$ &   (i)\\
MACS~J1532.8+3021 & $<222$ & $71.0\pm8.2$ &$17.1\pm0.9$ & $8.8\pm0.5$ & $3.25\pm0.18$ & ... & $0.94$ &  (i,iv)\\
MACS~J1720.2+3536 & $<266$ & $103.0\pm7.4$ &$16.8\pm1.0$ & ... & ... & ... & $1.23$ &   (i)\\
MACS~J1931.8-2634$^{b}$ & ... & ...& $70\pm4$ & $6.0\pm1.3$ & ... & $2799\pm161$[@150MHz] & 80.4 &   (iv,xi,xii)\\
MACS~J2046.0-3430 & ... & ...& $8.1\pm0.6$ & ... & ... & $13\pm1.3$[@843MHz] & 2.49 &   (ii,ix)\\
MACS~J2140.2-2339 & $<116$ & ...& $3.8\pm0.5$ & $1.0\pm0.1$ & ... & ... & $0.42$ &   (ii,viii)\\
\hline
MACS~J0111.5+0855 &$<124$ &... & $<0.25$  & ... & ... & ... & $<3.20$ &   (i)\\
MACS~J0257.1-2325 & $<100$& ...& $<0.89$ & ... & ... & ... & $<1.50$ &   (ii)\\
MACS~J1359.1-1929 & $<266$& ...& $9.5\pm0.8$ & ... & ... & ... & $2.34$ &   (ii)\\
MACS~J1359.8+6231 & ... & $<7$ & $2.8\pm0.6$ & $3.8\pm0.2$ & $1.67\pm0.08$ & ... & $0.16$ &   (i,viii)\\
MACS~J1447.4+0827 & $<203$& ...& $39.0\pm2.0$ & ... & ... & ... & $2.00$ &   (i)\\
MACS~J2135.2-0102 & $<192$&...& $<0.40$  & ... &  ...& ... & $<1.90$ &   (i) \\
MACS~J2245.0+2637 & $<208$& ...& $5.8\pm0.6$ & ... &  ...& ... & $0.72$ &   (ii) \\
\hline
\end{tabular}}
\label{tab2}
%\end{table}
\end{table*}

\begin{table*}
\caption{Cavity energetics. The annotations in Column 2 indicate the quality of the cavities detected. Those indicated with a $^c$ are the cavities with clear surface brightness depressions. The second annotation $^{1,2}$ indicates whether a cavity is surrounding by bright rims ($^1$) or not ($^2$). Column 3 shows the position angle of the major axis along the cavity, and is measured counter-clockwise from the north.}
\centering
\resizebox{17.5cm}{!} {
\begin{tabular}{lcccccccccccc}
\hline
\hline
Cluster & Lobe & PA & $R_{\rm l}$ & $R_{\rm w}$ & $R$ & $E_{\rm bubble}=PV$ & $t_{\rm c_s}$ & $t_{\rm buoy}$ & $t_{\rm refill}$ & $P_{\rm c_s}$ & $P_{\rm buoy}$ & $P_{\rm refill}$\\
& & (degrees) & kpc & kpc & kpc & $10^{58}\erg$  &  $10^7$ yrs & $10^7$ yrs & $10^7$ yrs & $10^{44}\ergps$ & $10^{44}\ergps$ & $10^{44}\ergps$\\
\\
\hline
MACS~J0159.8-0849 & NE$^{c,2}$ & 29 & 16.4 & 8.8 & 25.6 & 7.81 & 2.1 & 3.3 & 6.8 & 4.73 & 2.99 & 1.45\\
				  & SW$^{c,2}$ & 217 & 14.9 & 7.7 & 28.6 & 5.45 & 2.3 & 3.9 & 6.5 & 2.95 & 1.78 & 1.07 \\
MACS~J0242.5-2132 & N$^{c,2}$ & 21 & 6.1 & 5.5 & 12.9 & 1.88 &  1.3 & 3.0 & 5.2 & 1.85 & 0.81 & 0.46\\
				  & SW$^{c,1}$ & 215 & 10.9 & 8.0 & 19.2 & 7.08 & 1.9 & 3.3 & 6.6 & 4.66 & 2.72 & 1.37 \\
MACS~J0429.6-0253 & NE$^{c,2}$ & 40 & 5.9 & 5.8 & 15.7 & 1.19 & 1.5 & 3.3 & 4.7 & 1.02 & 0.45 & 0.32\\
				  & W$^{c,2}$ & 255 & 6.0 & 5.2 & 15.5 & 0.98 & 1.5 & 3.3 & 4.6 & 0.85 & 0.38 & 0.27\\
MACS~J0547.0-3904 & NE$^{c,1}$ & 65 & 8.5 & 11.5 & 15.0 & 3.22 & 1.7 & 2.5 & 5.8 & 2.35 & 1.63 & 0.71\\
MACS~J0913.7+4056 & NW$^{c,2}$ & 335 & 25.2 & 25.6 & 30.1 & 120.0 & 3.0 & 2.2 & 7.0 & 50.1 & 68.5 & 21.6\\
				  & SE$^{c,2}$ & 155 & 23.7 & 17.2 & 46.1 & 30.0 & 4.2 & 5.4 & 9.6 & 9.01 & 7.06 & 3.94\\
MACS~J0947.2+7623 & NE$^{c,1}$ & 60 & 20.9 & 16.3 & 24.0 & 89.1 & 2.1 & 1.7 & 5.1 & 53.0 & 68.0 & 22.0\\
				  & SW$^{c,1}$ & 250 & 18.8 & 13.7 & 24.6 & 56.2 & 2.2 & 1.8 & 4.8 & 32.6 & 39.6 & 14.9\\
MACS~J1411.3+5212 & SE$^{c,2}$ & 135 & 28.9 & 22.7 & 36.6 & 49.1 & 3.3 & 1.8 & 4.9 & 19.1 & 35.6 & 12.7\\
MACS~J1423.8+2404 & E$^{c,2}$ & 90 & 12.6 & 16.5 & 23.7 & 38.7 & 2.3 & 4.2 & 9.0 & 21.3 & 11.7 & 5.49 \\
				  & W$^{c,1}$ & 280 & 14.0 & 11.9 & 19.3 & 22.5 & 1.9 & 3.2 & 8.5 & 15.2 & 8.83 & 3.36 \\
MACS~J1532.8+3021 & W$^{c,1}$ & 310 & 17.4 & 15.4 & 37.6 & 31.9 & 3.5 & 4.7 & 8.0 & 11.5 & 8.59 & 5.07\\
MACS~J1720.2+3536 & SE$^{c,1}$ & 150 & 4.8 & 5.3 & 6.7 & 0.84 & 0.58 & 3.4 & 9.4 & 1.82 & 0.31 & 0.11\\
				  & N$^{2}$ & 355 & 12.1 & 15.0 & 19.0 & 16.8 & 1.6 & 6.1 & 15.4 & 13.0 & 3.50 & 1.39\\
MACS~J1931.8-2634 & E$^{c,2}$ & 90 & 15.6 & 12.8 & 29.5 & 41.7 & 2.6 & 1.5 & 2.8 & 20.7 & 36.0 & 19.0   \\
				  & W$^{c,2}$ & 270 & 15.6 & 12.8 & 20.6 & 41.7 & 1.8 & 1.0 & 2.8 & 29.6 & 51.6 & 19.0 \\
MACS~J2046.0-3430 & S$^{c,1}$ & 178 & 6.8 & 13.7 & 11.6 & 7.46 & 1.2 & 4.1 & 10.9 & 7.73 & 2.30 & 0.87 \\
				  & N$^{2}$ & 358 & 17.4 & 12.3 & 23.6 & 15.4 & 2.4 & 6.3 & 16.2 & 7.88 & 3.75 & 1.46\\
MACS~J2140.2-2339 & S$^{c,2}$ & 180 & 7.1 & 6.8 & 14.4 & 3.43 & 1.4 & 2.4 & 4.5 & 3.06 & 1.79 & 0.98\\
\hline
MACS~J0111.5+0855 & N$^{2}$ & 350 & 32.8 & 33.2 & 77.5 & 56.5 & 6.5 & 7.2 & 11.5 & 11.0 & 9.97 & 6.22 \\
				  & SE$^{2}$ & 90 & 28.0 & 23.7 & 60.7 & 24.5 & 5.1 & 6.1 & 10.2 & 6.12 & 5.10 & 3.05 \\
MACS~J0257.1-2325 & NE$^{2}$ & 58 & 17.5 & 15.3 & 39.5 & 16.2 & 2.5 & 3.6 & 5.9 & 8.08 & 5.68 & 3.51 \\
MACS~J1359.1-1929 & N$^{2}$ & 345 & 23.9 & 22.4 & 53.9 & 35.3 & 5.2 & 16.2 & 26.6 & 8.56 & 2.77 & 1.68\\
				  & S$^{2}$ & 175 & 11.5 & 10.6 & 44.1 & 3.75 & 4.3 & 19.1 & 18.3 & 1.11 & 0.25 & 0.26\\
MACS~J1359.8+6231 & NE$^{2}$ & 70 & 4.2 & 3.2 & 9.3 & 0.14 & 0.85 & 3.4 & 5.4 & 0.20 & 0.052 & 0.032\\
MACS~J1447.4+0827 & N$^{2}$ & 5 & 21.0 & 20.4 & 26.9 & 123.4 & 2.4 & 3.5 & 10.3 & 64.0 & 44.2 & 15.1 \\
MACS~J2135.2-0102 & NW$^{2}$ & 335 & 27.2 & 28.2 & 75.3 & 42.5 & 5.1 & 12.9 & 17.7 & 10.6 & 4.18 & 3.04 \\
MACS~J2245.0+2637 & NW$^{2}$ & 325 & 8.5 & 7.8 & 15.5 & 1.8 & 1.3 & 2.9 & 5.9 & 1.76 & 0.79 & 0.39\\
\hline
\end{tabular}}
\label{tab3}
\end{table*}

\section{Cavity energetics}

We estimate the energy stored within each of the cavities using Eq. \ref{eq2} \citep[{\rm e.g.} ][]{Bir2004607,Dun2005364,Raf2006652,Dun2006373,Dun2008385}. Here, $p$ is the thermal pressure of the ICM at the radius of the bubble and estimated from X-ray data, $V$ is the volume of the cavity and for a relativistic fluid $\gamma_1$ = 4/3, therefore $E_{\rm bubble}=4pV$ \citep[this is also supported observationally, see][]{Gra2009386}. 
\begin{equation}
E_{\rm bubble}=\frac{\gamma_1}{\gamma_1-1}pV
\label{eq2}
\end{equation}

We assume that the cavities have a prolate shape. The volume is then given by $V=4{\pi}R^2_{\rm w}R_{\rm l}/3$, where $R_{\rm l}$ is the semi-major axis along the direction of the jet, and $R_{\rm w}$ is the semi-major axis perpendicular to the direction of the jet. We define ``jet" as the line that connects the nuclear X-ray source to the centre of the cavity. The dimensions of the cavities we derive are given in Table \ref{tab3}. 

We determine the local thermal pressure by selecting a set of annuli containing roughly the same signal-to-noise and centred on the X-ray peak. For some clusters, the data quality was sufficient to allow $\sim4000$ counts per annuli, but for others we could only consider annuli with $\sim900$ counts each. Once the regions were selected, we chose a background region for each cluster within the same chip but far from any cluster emission. Since all of our clusters are above a redshift of $z>0.3$, they only occupy a small portion of the chip, allowing ample area for selecting a background within the same chip. We then proceeded in deprojecting the data using the Direct Spectral Deprojection method of \citet{San2007381}. For each spectrum, we fitted an absorbed {\sc mekal} model using $\chi^2$ statistics and left the temperature, abundance and normalization parameter (related to the electron density) free to vary. Here, the absorption accounts for Galactic absorption and we keep it frozen at the \citet{Kal2005440} value. In some annuli, it was difficult to constrain the abundance. In this case, we kept the abundance value frozen at the cluster average which we determined by selecting a region within the entire cluster ($r<200$ kpc).

For MACS~J0111.5+0855 and MACS~J2135.2-0102, the count rate was so low that we could only extract regions containing $\sim400$ counts each, and the data quality was too poor to correct for deprojection effects. Here, we proceeded in the same way, but did not deproject the spectra and used C-statistics instead. For MACS~1411.3+5212 (3C 295), the central jet axis ($r<25$ kpc) and nucleus are dominated by non-thermal emission. During the spectral analysis, we therefore remove the regions along the jet axis, so that we are only left with the thermal emitting gas. In all systems, we start the inner-most annulus at $r=1''$ (which corresponds to the $Chandra$ Point Spread Function, PSF), so that the central point source is not included.  

Following this technique, we were able to derive the temperature ($kT$) and electron density ($n_{\rm e}$) as a function of radius for each cluster. We then calculate the deprojected electron pressure ($p_{\rm e}=n_{\rm e}kT$) and entropy ($S=kTn_{\rm e}^{-2/3}$) profiles, but only projected quantities for the 4 clusters mentioned in the previous paragraph. Our results are shown in Fig. \ref{figA1} of Appendix B. When deriving the cavity energetics, we use the local thermal pressure of the gas as an estimate of $p$ in Eq. \ref{eq1} ($p=nkT\sim1.92\times{n_{\rm e}}kT$). 

Note that deprojection relies on the cluster having a ``relaxed" morphology, i.e. that it can be approximated as being ``spherical". Ten of the systems in which we find surface-brightness depressions are part of the X-ray brightest MACS clusters and have been regarded as having a ``relaxed" morphology since they either have a pronounced cool core and very good alignment with a single central dominant cD galaxy or a good optical/X-ray alignment and concentric contours \citep[][ MACS~J0159.8-0849, MACS~J0242.5-2132, MACS~J0429.6-0253, MACS~J0547.0-3904, MACS~J0947.2+7623, MACS~J1532.8+3021, MACS~J1720.2+3536, MACS~J1931.8-2634, MACS~J2140.2-2339 and MACS~J2245.0+2637]{Ebe2010407}. Two others are part of the 12 most distant MACS clusters \citep[][ MACS~J0257.1-2325 and MACS~J1423.8+2404]{Ebe2007661}, and have also been considered as having a ``relaxed" morphology. We can therefore approximate these 12 systems as being spherical and deproject the spectra. The remaining 8 systems are not part of the \citet{Ebe2007661,Ebe2010407} studies and have not yet been classified. MACS~J2046.0-3430 shows a slight offset between the central X-ray peak and galaxy, but the cluster still has a pronounced cool core and concentric X-ray contours. We therefore choose to pursue with spectral deprojection for this cluster. All of the remaining clusters have concentric X-ray contours and good optical/X-ray alignment, apart for MACS~J2135.2-0102. However, as mentioned in the previous paragraphs, the count rate was so low in this cluster that we could not deproject the spectra and used the projected quantities to calculate the cavity energetics.

 \begin{table}
\caption{Cooling flow properties of the MACS clusters with cavities - (1) Name; (2) Bolometric X-ray luminosity ($0.01-50$ keV) estimated from the $Chandra$ data; (3) Cooling time at $r=50$ kpc; (4) Cooling radius estimated as the radius at which $t_{\rm cool}=7.7$ Gyrs; (5) Bolometric X-ray luminosity ($0.01-100$ keV) of the clusters within the cooling radius. $^\star$The $Chandra$ data of these clusters were too poor to deproject the spectra. In this case, we deprojected the surface brightness profiles to obtain an estimate of the deprojected $n_{\rm e}$ and $L_{\rm X}$. Using these values, combined with the average cluster temperature within $r<200$, we then determined $t_{\rm cool}$. }
%\centering
\resizebox{8.5cm}{!} {
\begin{tabular}{lcccc}
%\small
\hline
\hline
(1) & (2) & (3) & (4) & (5) \\
Cluster & $\Lx$ & $t_{\rm cool}$ (50kpc) &  $r_{\rm cool}$  & $\Lx(<\rcool)$\\
name & ($0.01-100\keV$) & \rm{Gyr} & kpc & ($0.01-100\keV$) \\
 & $10^{44}\ergps$ &  &  & $10^{44}\ergps$ \\
\\
\hline
MACS~J0159.8-0849  & 23.1$^{+0.7}_{-0.6}$ & 3.0 & 86 & 11.8$^{+0.4}_{-0.3}$\\
MACS~J0242.5-2132 & 21.1$^{+0.8}_{-0.8}$ & 2.3 & 110 & 16.5$^{+0.6}_{-0.6}$\\
MACS~J0429.6-0253  & 16.8$^{+0.7}_{-0.6}$ & 3.1 & 105 & 11.6$^{+0.5}_{-0.5}$\\
MACS~J0547.0-3904 & 6.9$^{+0.3}_{-0.3}$ & 2.7 & 100 & 4.8$^{+0.2}_{-0.2}$\\
MACS~J0913.7+4056 & 19.3$^{+0.4}_{-0.4}$ & 2.1 & 107 & 15.2$^{+0.3}_{-0.3}$\\
MACS~J0947.2+7623  & 37.6$^{+0.7}_{-0.7}$ & 1.6 & 120 & 29.9$^{+0.6}_{-0.6}$\\
MACS~J1411.3+5212 & 8.9$^{+0.3}_{-0.3}$ & 4.8 & 60 & 4.0$^{+0.2}_{-0.2}$ \\
MACS~J1423.8+2404 & 23.3$^{+0.5}_{-0.4}$ & 2.1 & 95 & 16.3$^{+0.3}_{-0.3}$\\
MACS~J1532.8+3021 & 33.2$^{+-1.1}_{-1.1}$ & 1.5 & 115 & 24.9$^{+0.9}_{-0.8}$\\
MACS~J1720.2+3536  & 15.1$^{+0.4}_{-0.4}$ & 3.8 & 100 & 8.9$^{+0.3}_{-0.3}$\\
MACS~J1931.8-2634 & 9.0$^{+0.1}_{-0.1}$ & 1.8 & 112 & 6.50$^{+0.09}_{-0.09}$\\
MACS~J2046.0-3430  & 13.4$^{+0.4}_{-0.4}$ & 3.1 & 81 & 8.3$^{+0.3}_{-0.3}$\\
MACS~J2140.2-2339  & 18.6$^{+0.2}_{-0.2}$ & 1.9 & 107 & 14.2$^{+0.2}_{-0.2}$\\
\hline
$^\star$MACS~J0111.5+0855 & 5.3$^{+0.6}_{-0.5}$ & 5.7 & 78 & 2.3$^{+0.5}_{-0.3}$ \\
MACS~J0257.1-2325 & 17.4$^{+1.6}_{-1.2}$ & 8.0 & 54 & 3.4$^{+0.6}_{-0.4}$\\
MACS~J1359.1-1929 & 7.5$^{+0.3}_{-0.3}$ & 3.9 & 76 & 4.2$^{+0.2}_{-0.2}$\\
MACS~J1359.8+6231 & 8.8$^{+0.4}_{-0.3}$ & 5.9 & 67 & 2.9$^{+0.1}_{-0.1}$\\
MACS~J1447.4+0827 & 44.0 $^{+1.4}_{-1.3}$ & 1.3 & 125 & 36.2$^{+1.2}_{-1.1}$\\
$^\star$MACS~J2135.2-0102 & 4.1$^{+0.4}_{-0.3}$ & 11.9 & ... & ...\\
MACS~J2245.0+2637 & 11.9$^{+0.6}_{-0.6}$ & 4.0 & 89 & 6.3$^{+0.4}_{-0.4}$\\
\hline
\end{tabular}}
\label{tab4}
%\end{table}
\end{table}

\begin{figure*}
\centering
\begin{minipage}[c]{0.49\linewidth}
\centering \includegraphics[width=\linewidth]{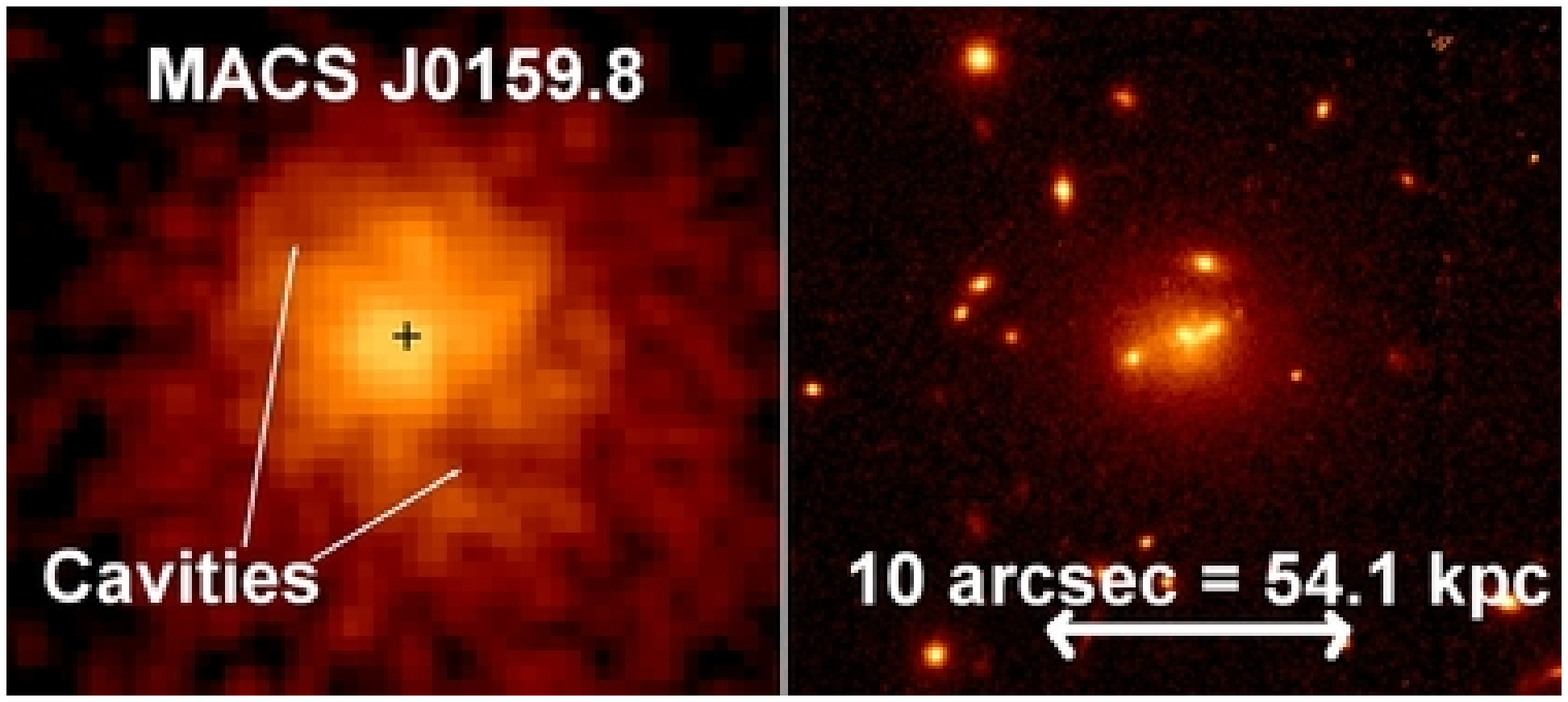}
\end{minipage}
\begin{minipage}[c]{0.49\linewidth}
\centering \includegraphics[width=\linewidth]{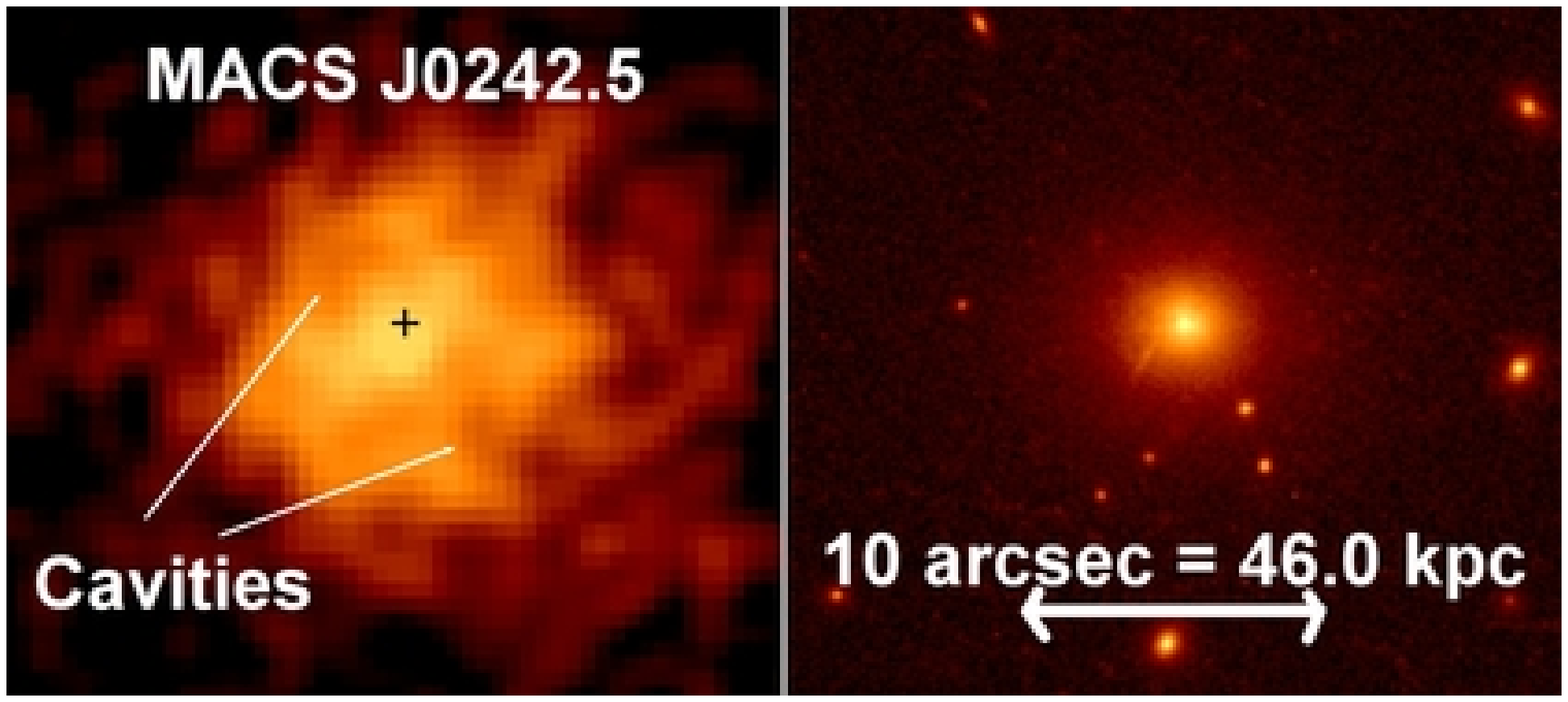}
\end{minipage}
\begin{minipage}[c]{0.49\linewidth}
\centering \includegraphics[width=\linewidth]{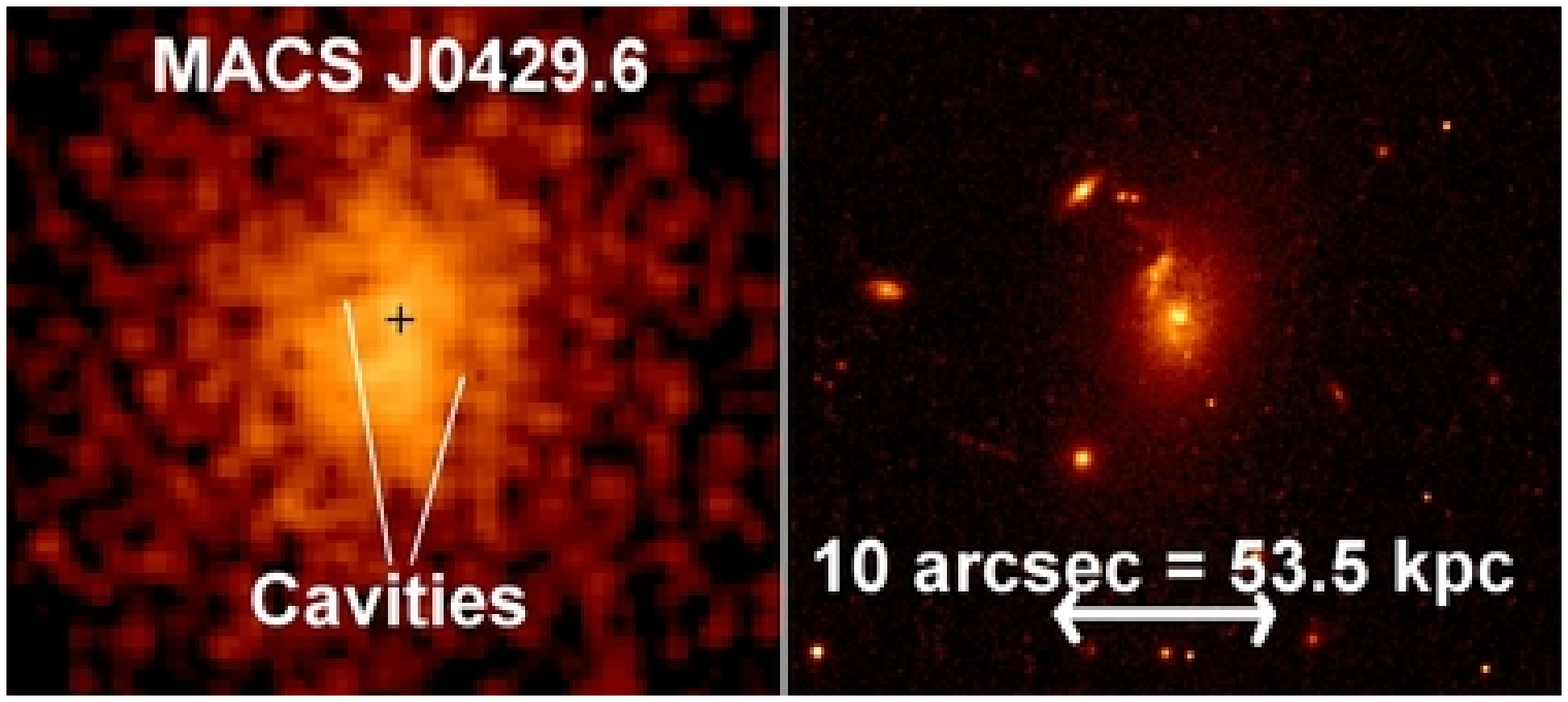}
\end{minipage}
\begin{minipage}[c]{0.49\linewidth}
\centering \includegraphics[width=\linewidth]{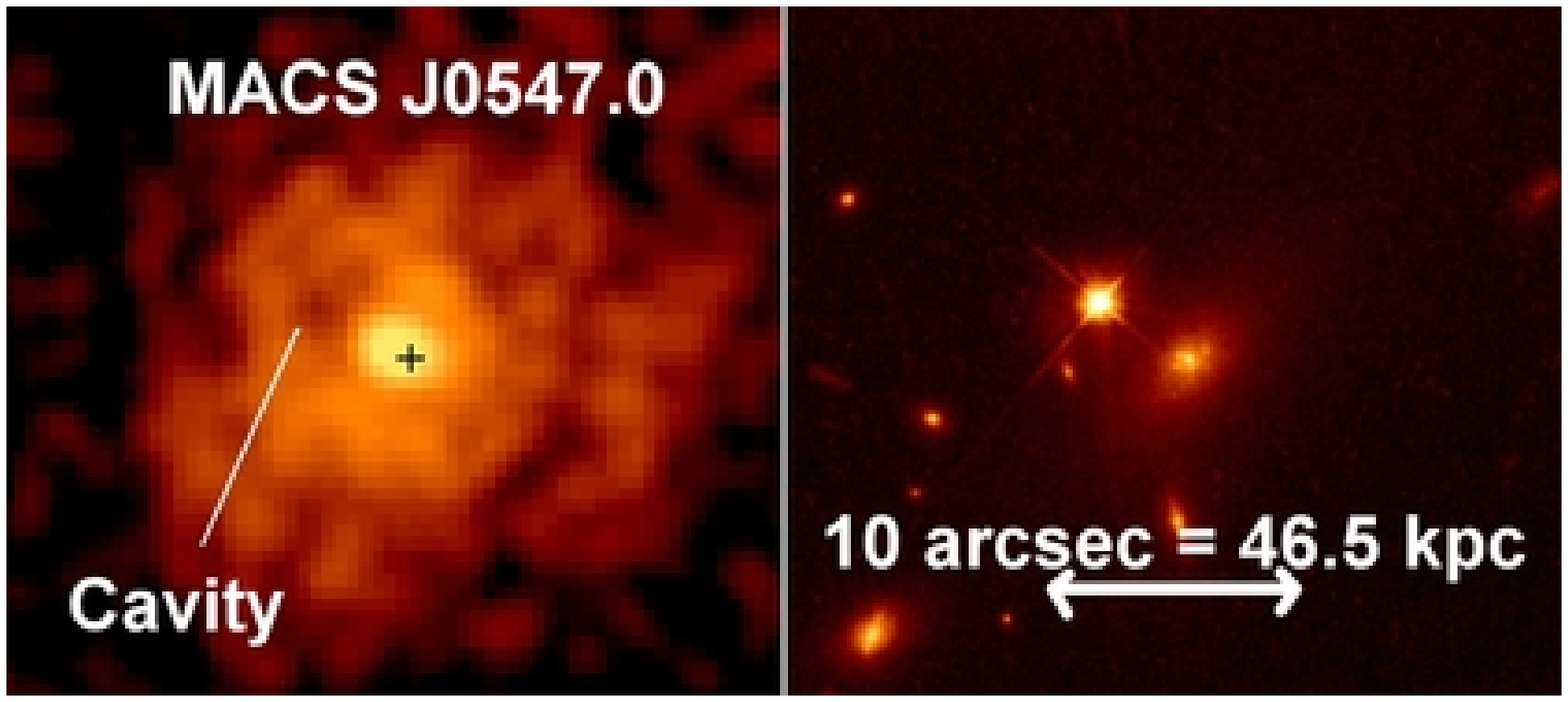}
\end{minipage}
\begin{minipage}[c]{0.49\linewidth}
\centering \includegraphics[width=\linewidth]{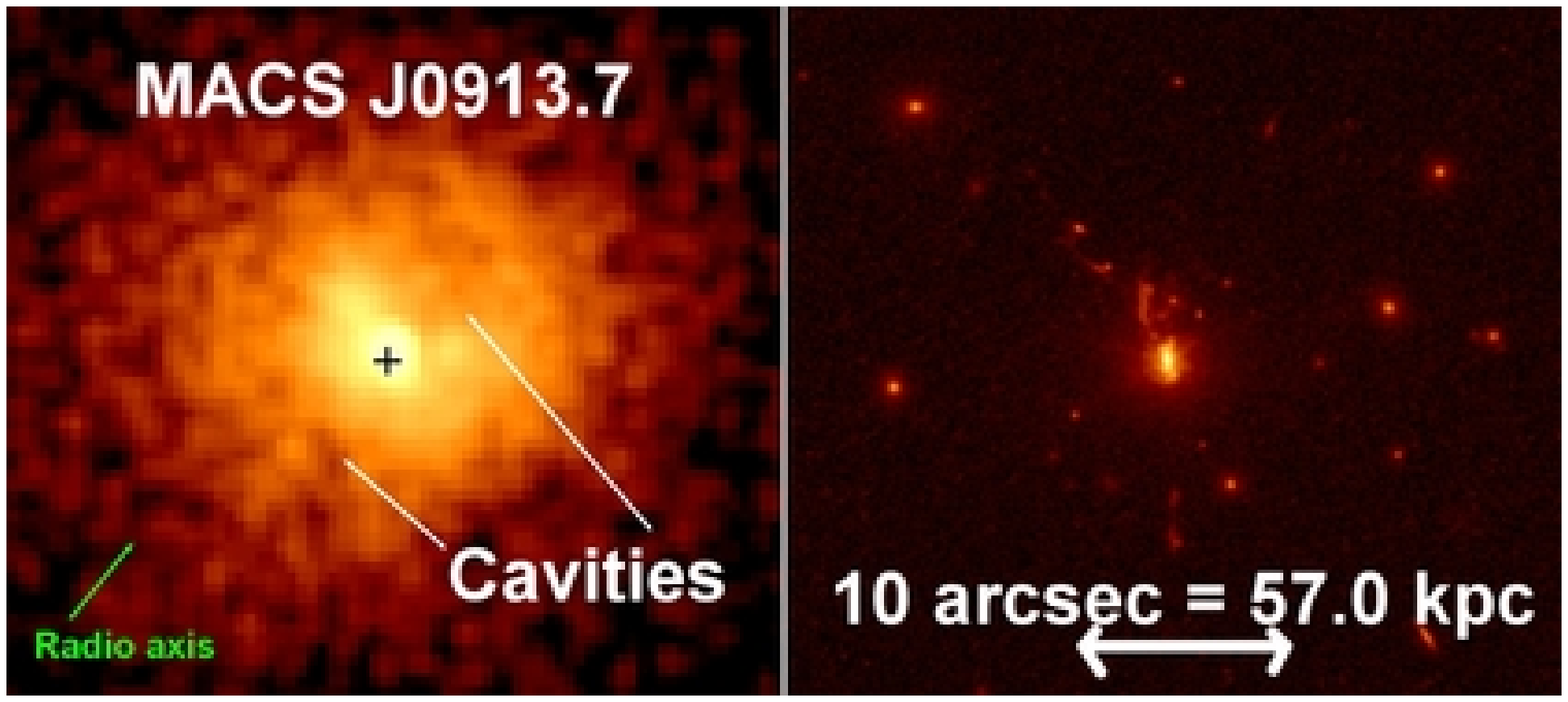}
\end{minipage}
\begin{minipage}[c]{0.49\linewidth}
\centering \includegraphics[width=\linewidth]{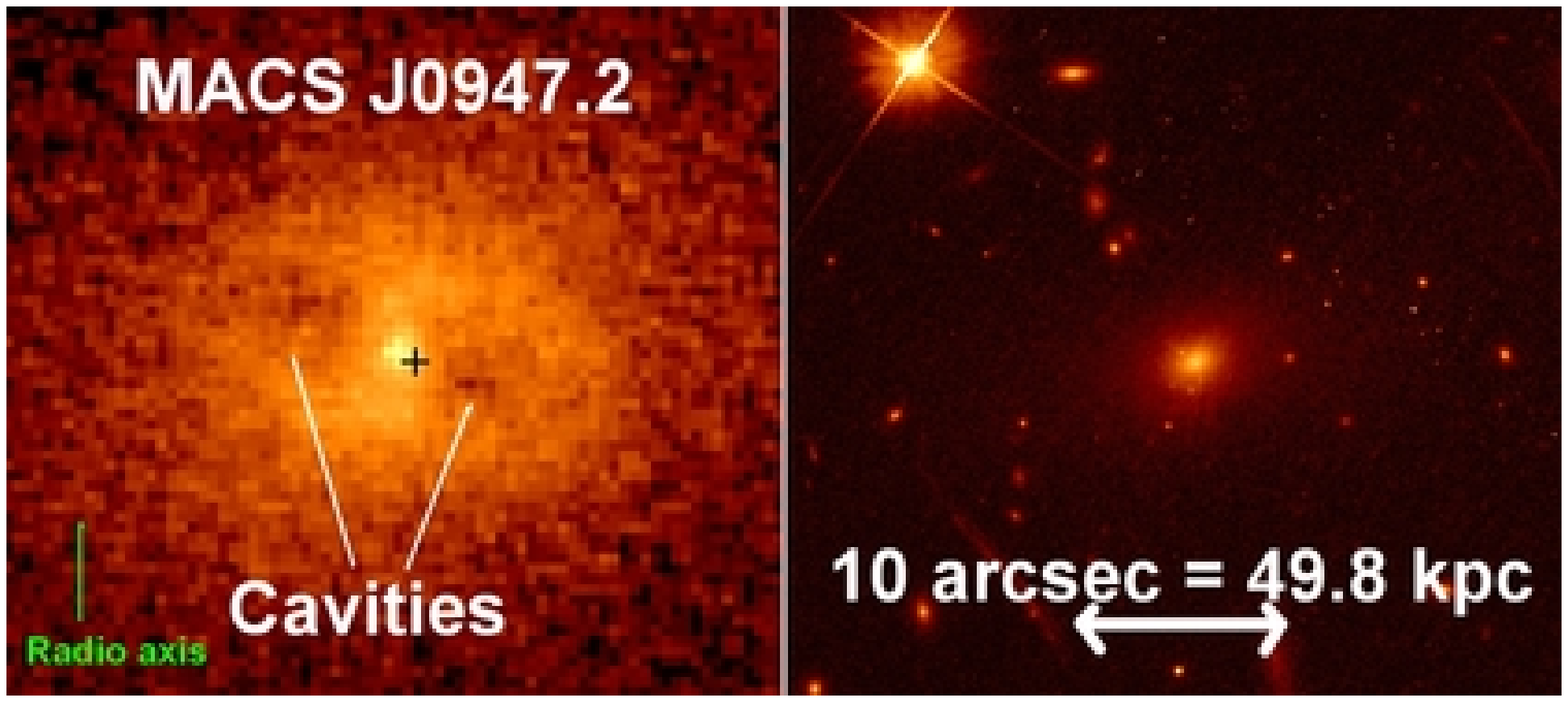}
\end{minipage}
\begin{minipage}[c]{0.49\linewidth}
\centering \includegraphics[width=\linewidth]{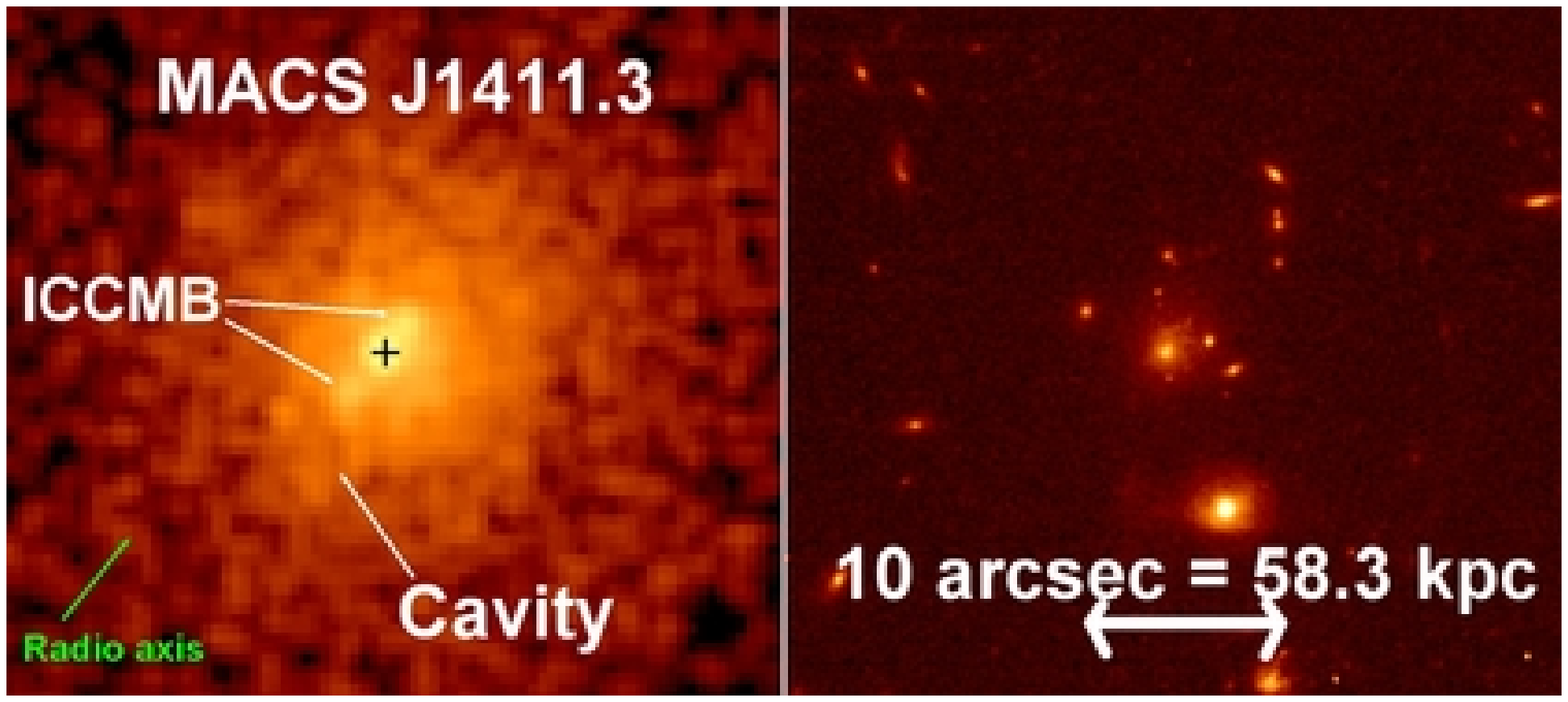}
\end{minipage}
\begin{minipage}[c]{0.49\linewidth}
\centering \includegraphics[width=\linewidth]{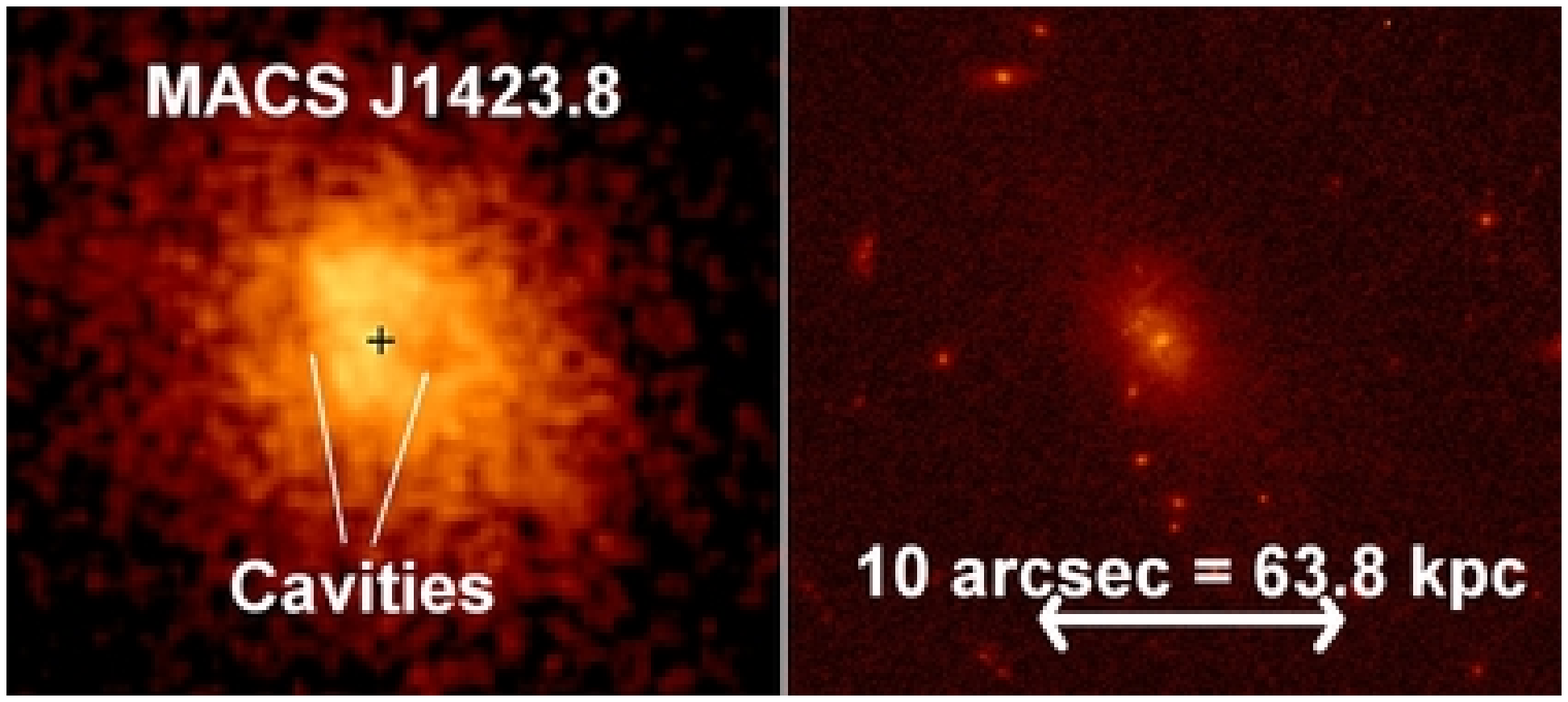}
\end{minipage}
\begin{minipage}[c]{0.49\linewidth}
\centering \includegraphics[width=\linewidth]{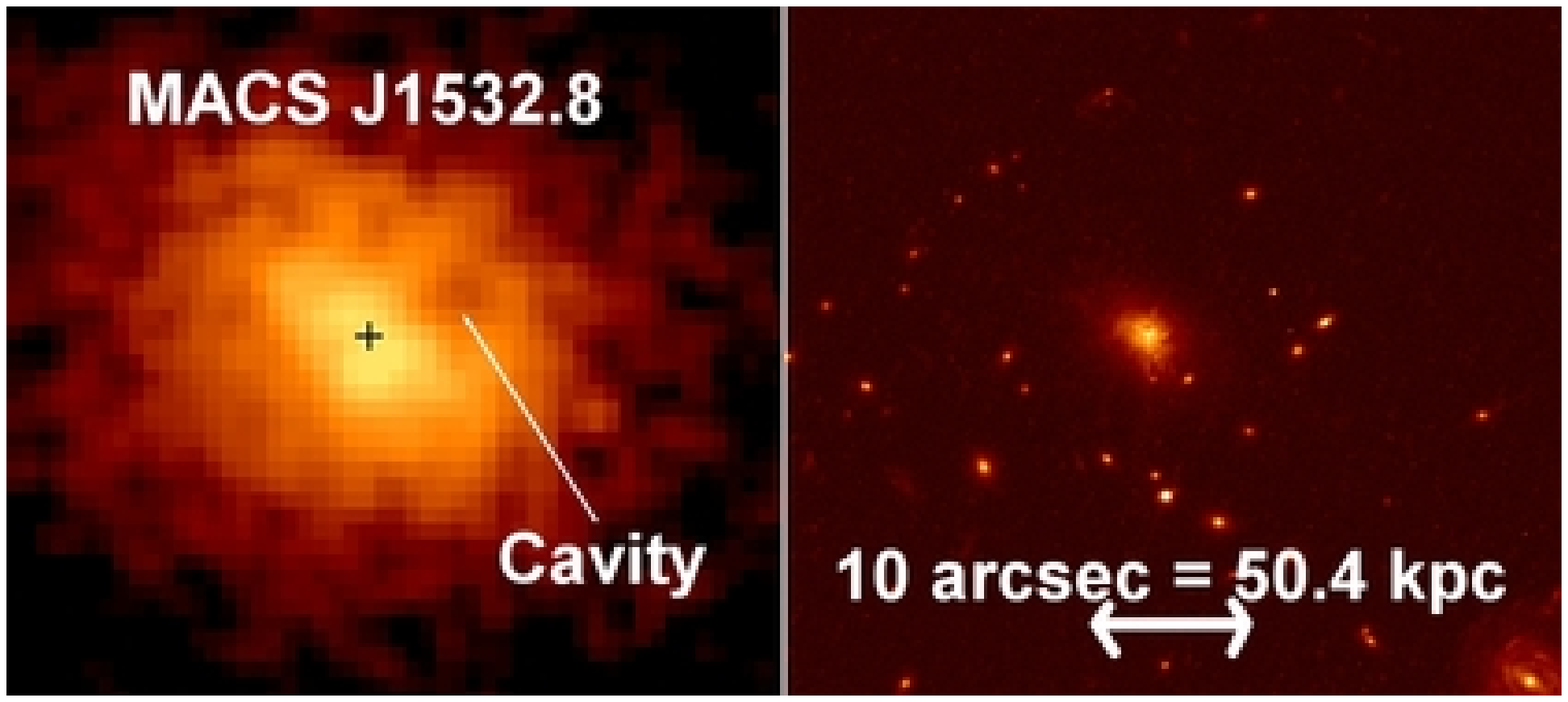}
\end{minipage}
\begin{minipage}[c]{0.49\linewidth}
\centering \includegraphics[width=\linewidth]{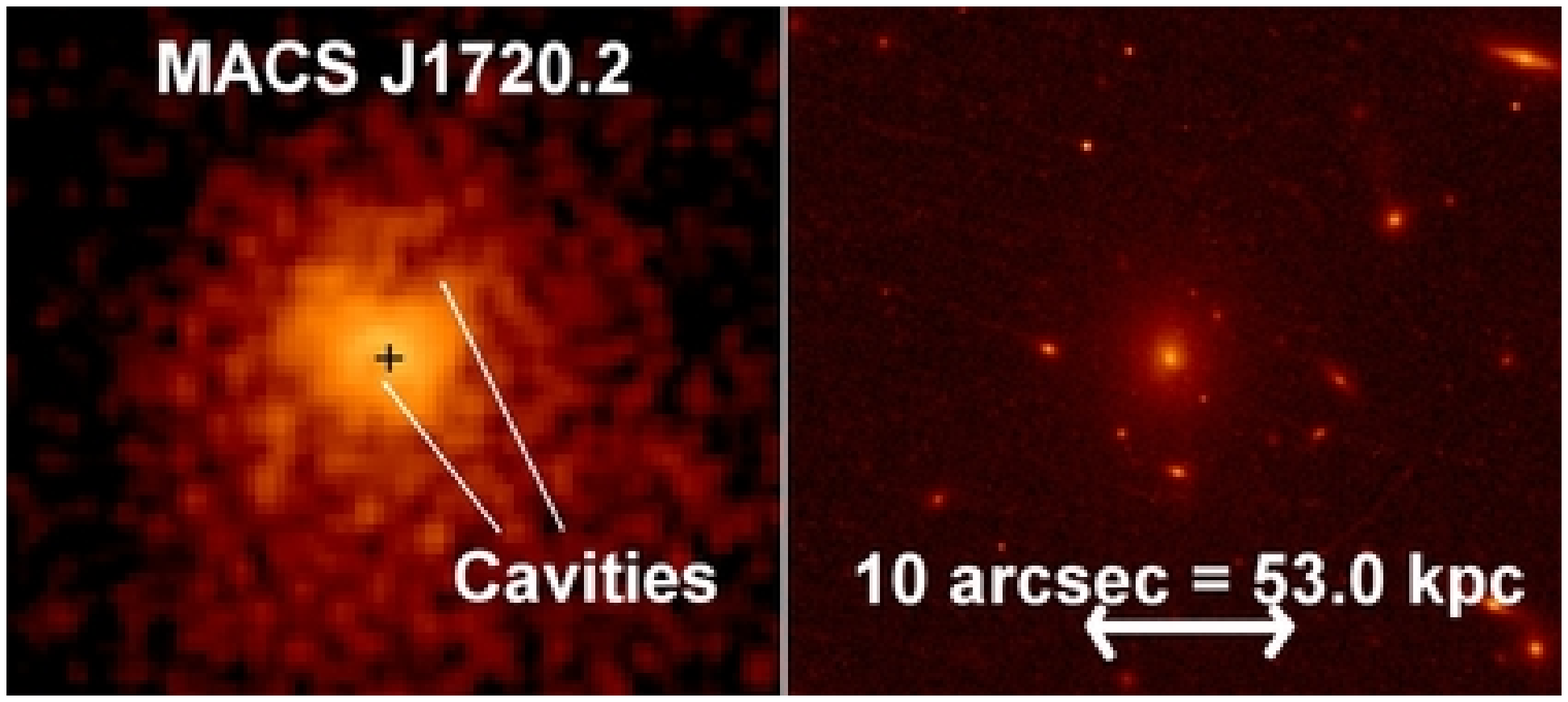}
\end{minipage}
\begin{minipage}[c]{0.49\linewidth}
\centering \includegraphics[width=\linewidth]{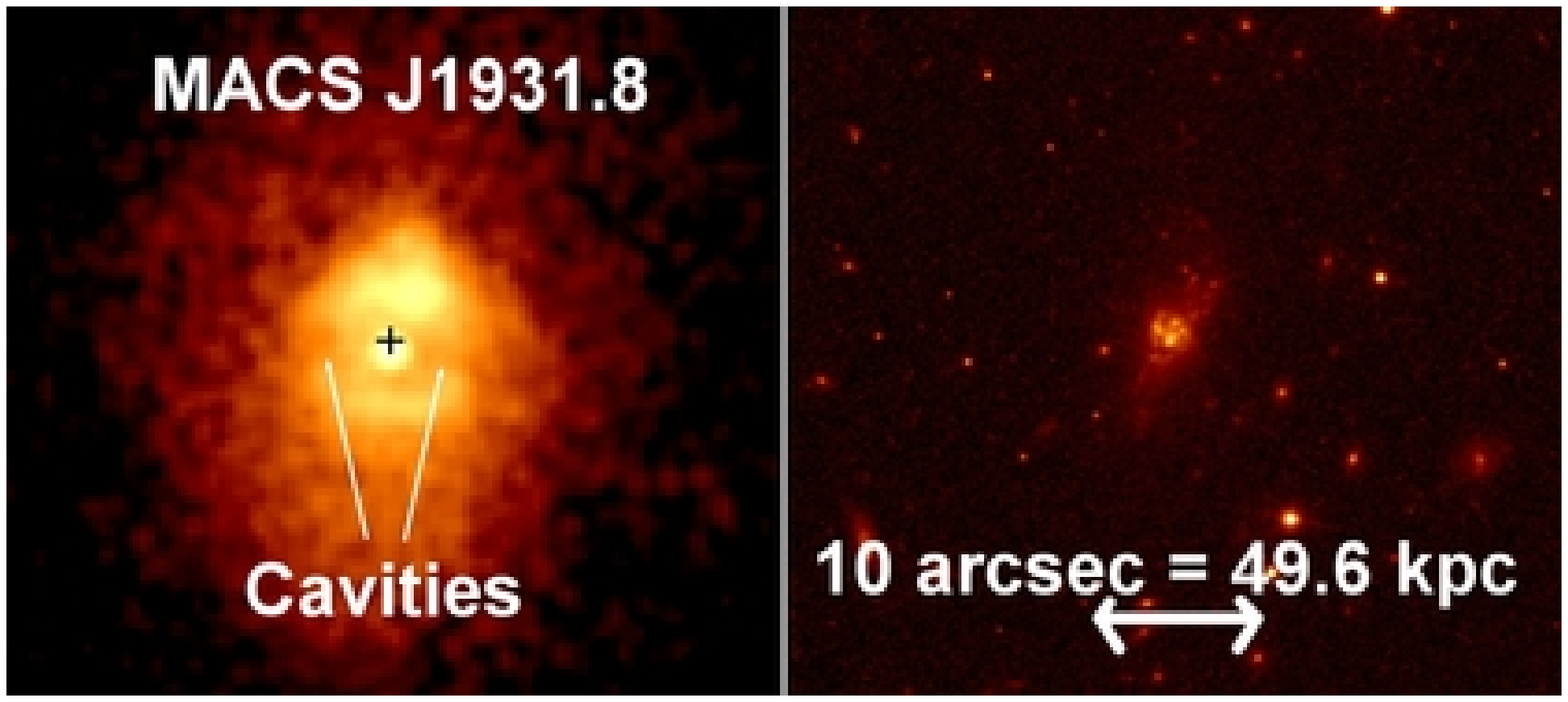}
\end{minipage}
\begin{minipage}[c]{0.49\linewidth}
\centering \includegraphics[width=\linewidth]{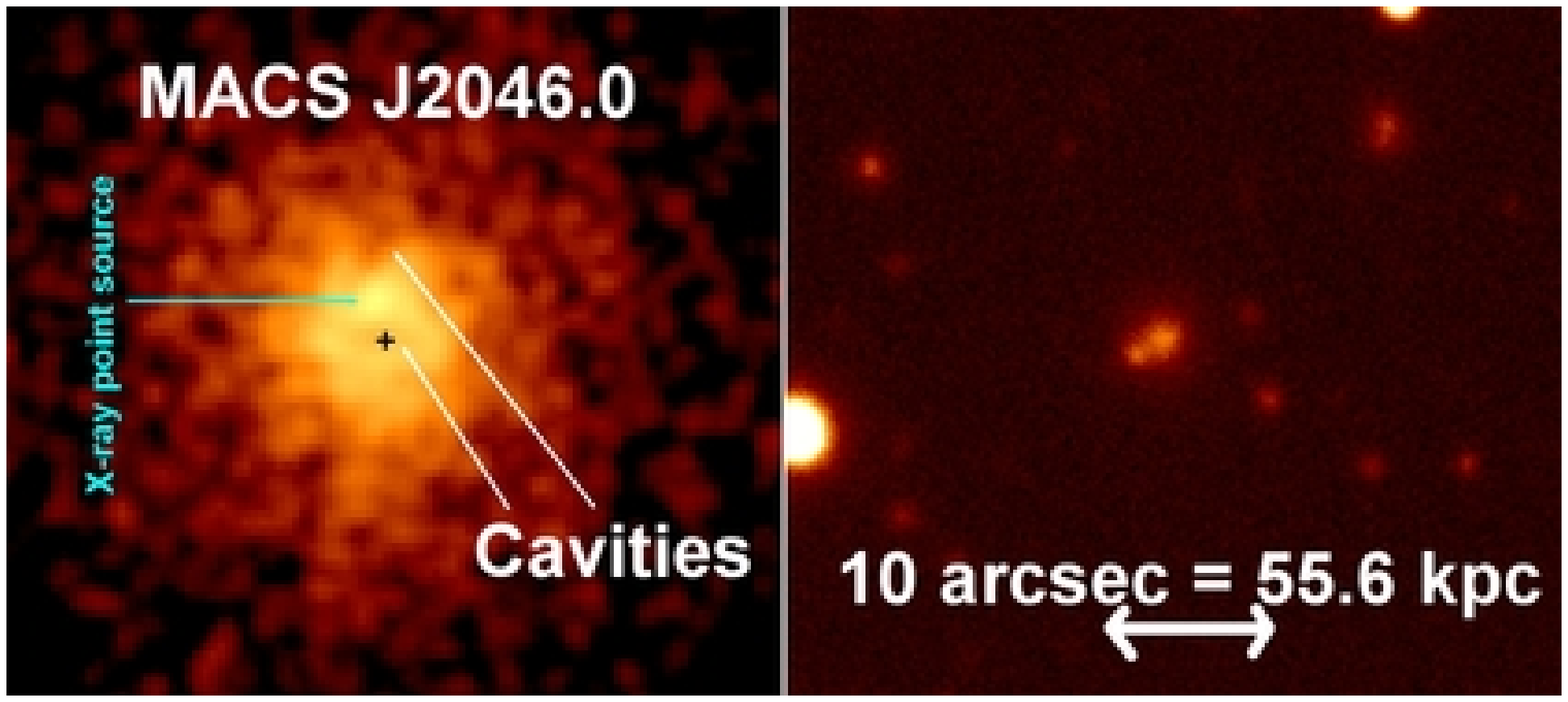}
\end{minipage}
\caption{See caption on next page.}
\end{figure*}
\begin{figure*}
\ContinuedFloat
\centering
\begin{minipage}[c]{0.49\linewidth}
\centering \includegraphics[width=\linewidth]{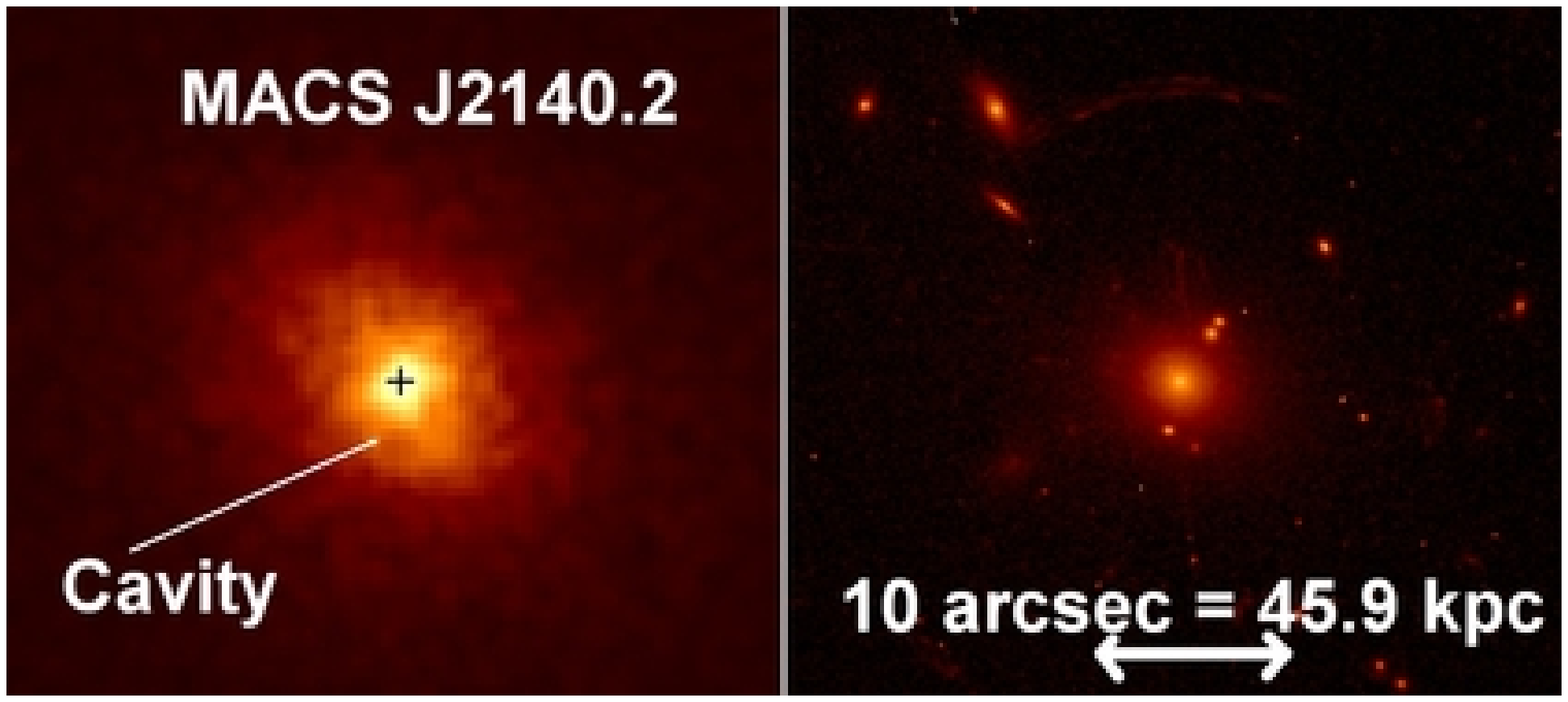}
\end{minipage}
\begin{minipage}[c]{0.49\linewidth}
\centering \includegraphics[width=\linewidth]{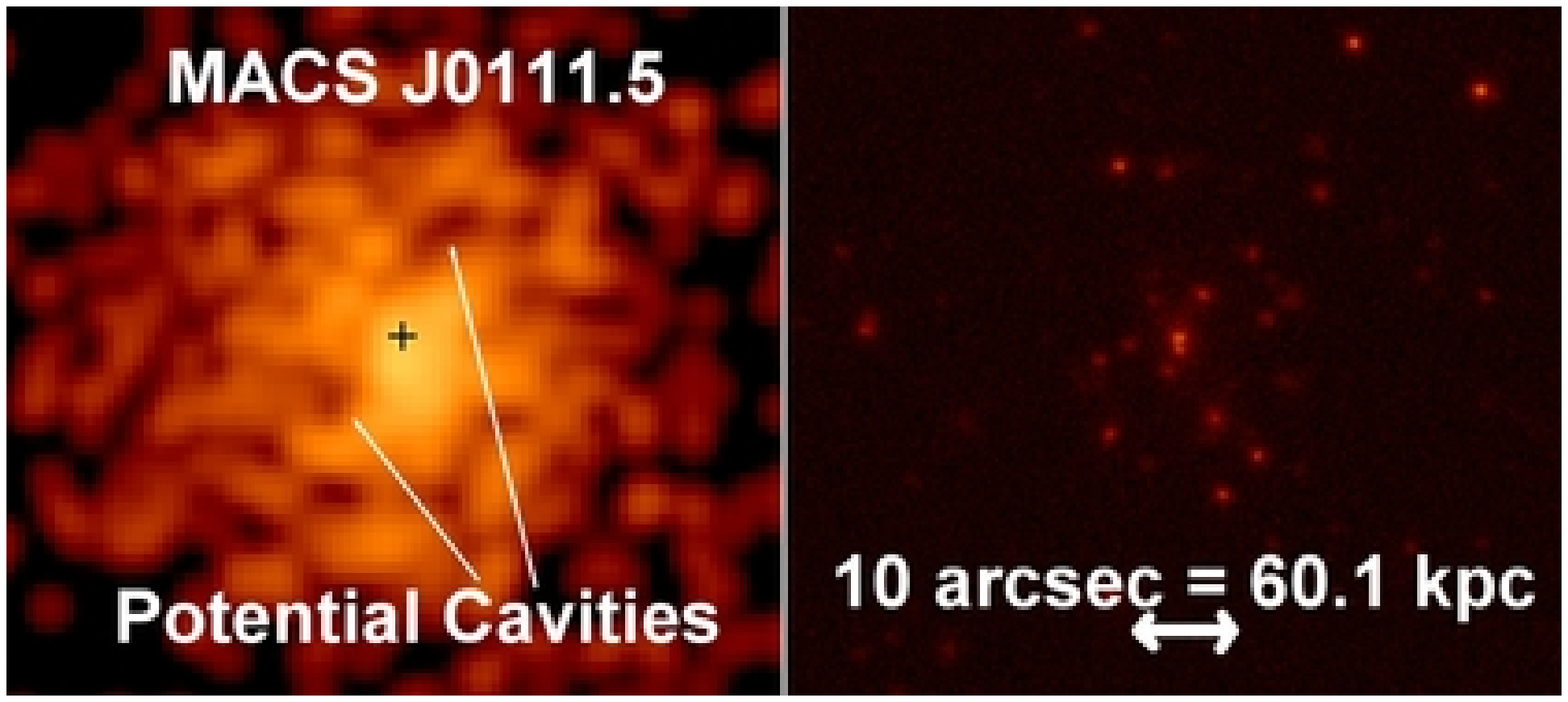}
\end{minipage}
\begin{minipage}[c]{0.49\linewidth}
\centering \includegraphics[width=\linewidth]{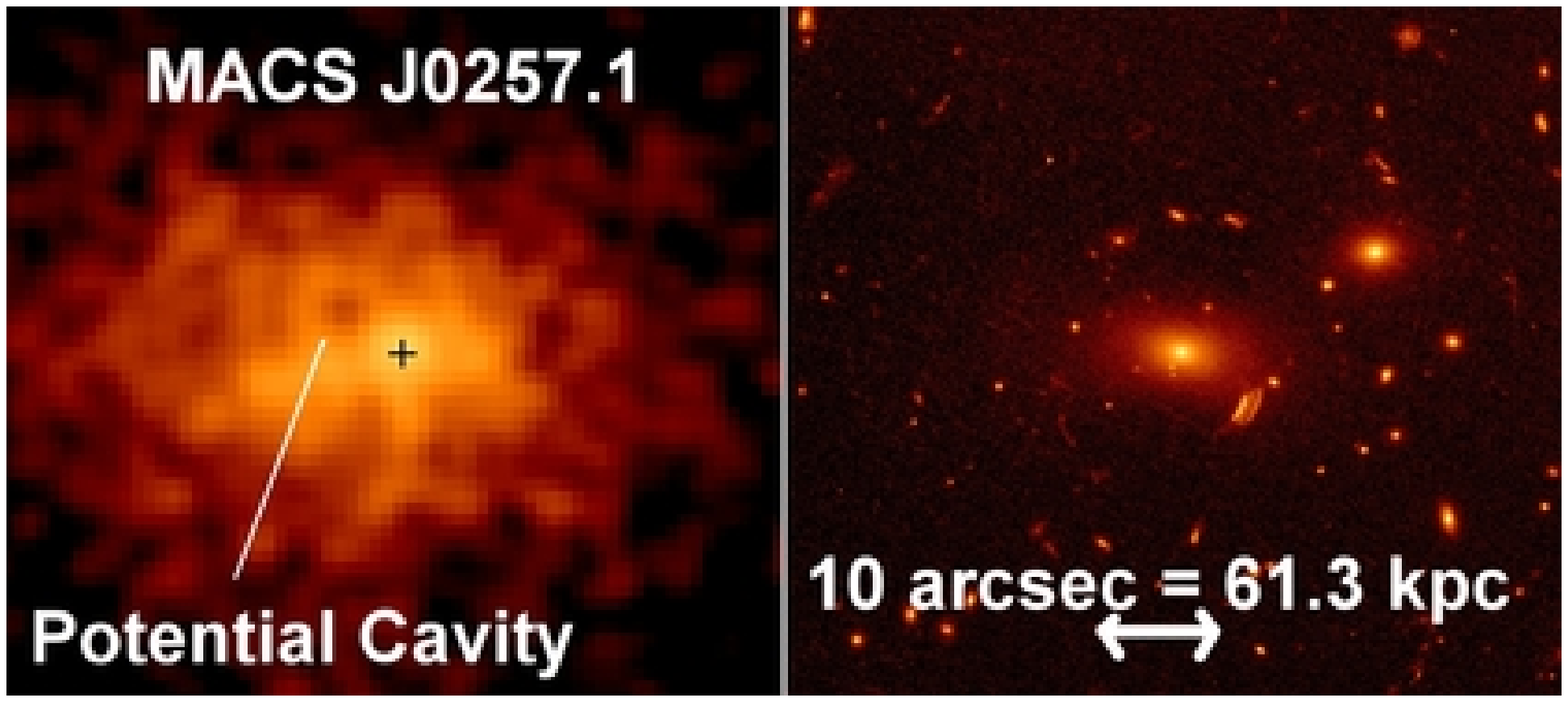}
\end{minipage}
\begin{minipage}[c]{0.49\linewidth}
\centering \includegraphics[width=\linewidth]{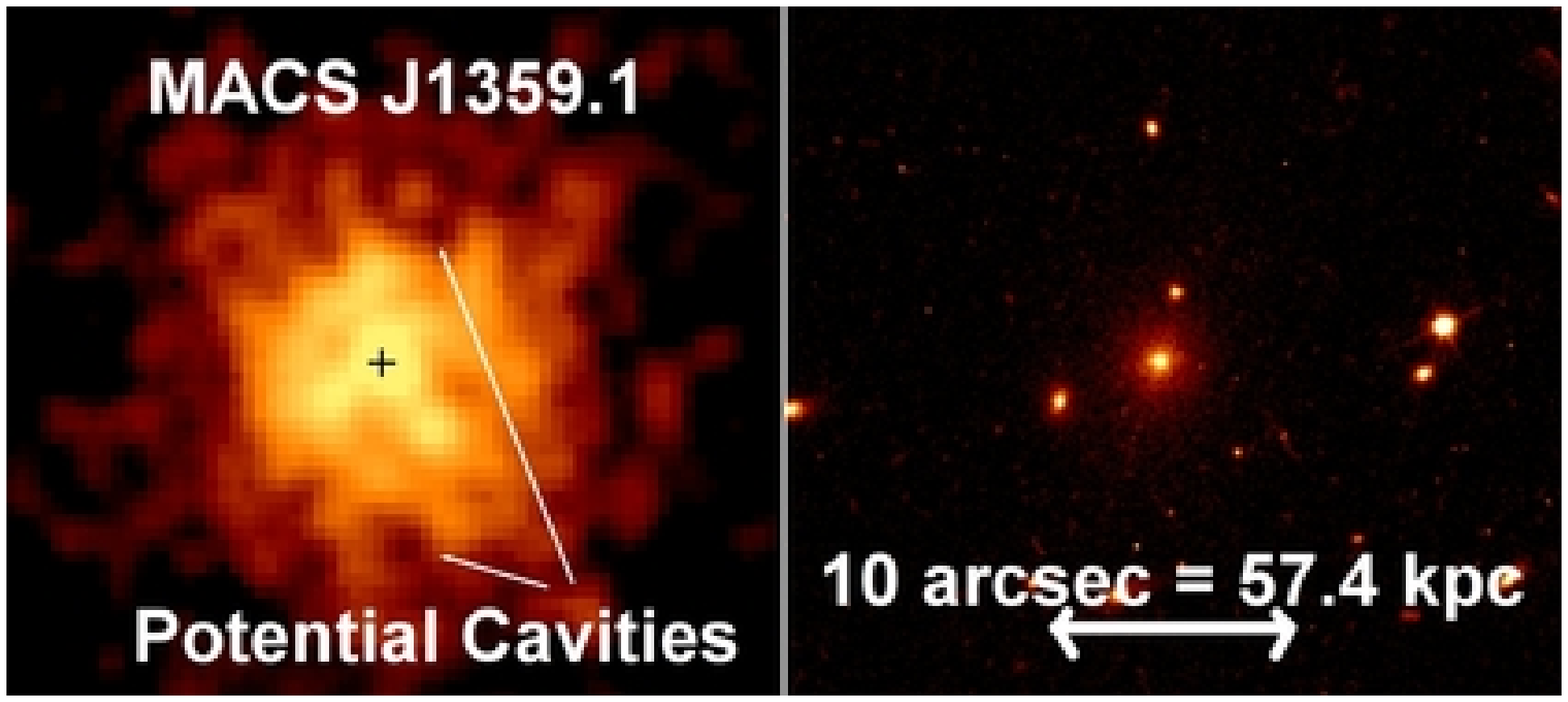}
\end{minipage}
\begin{minipage}[c]{0.49\linewidth}
\centering \includegraphics[width=\linewidth]{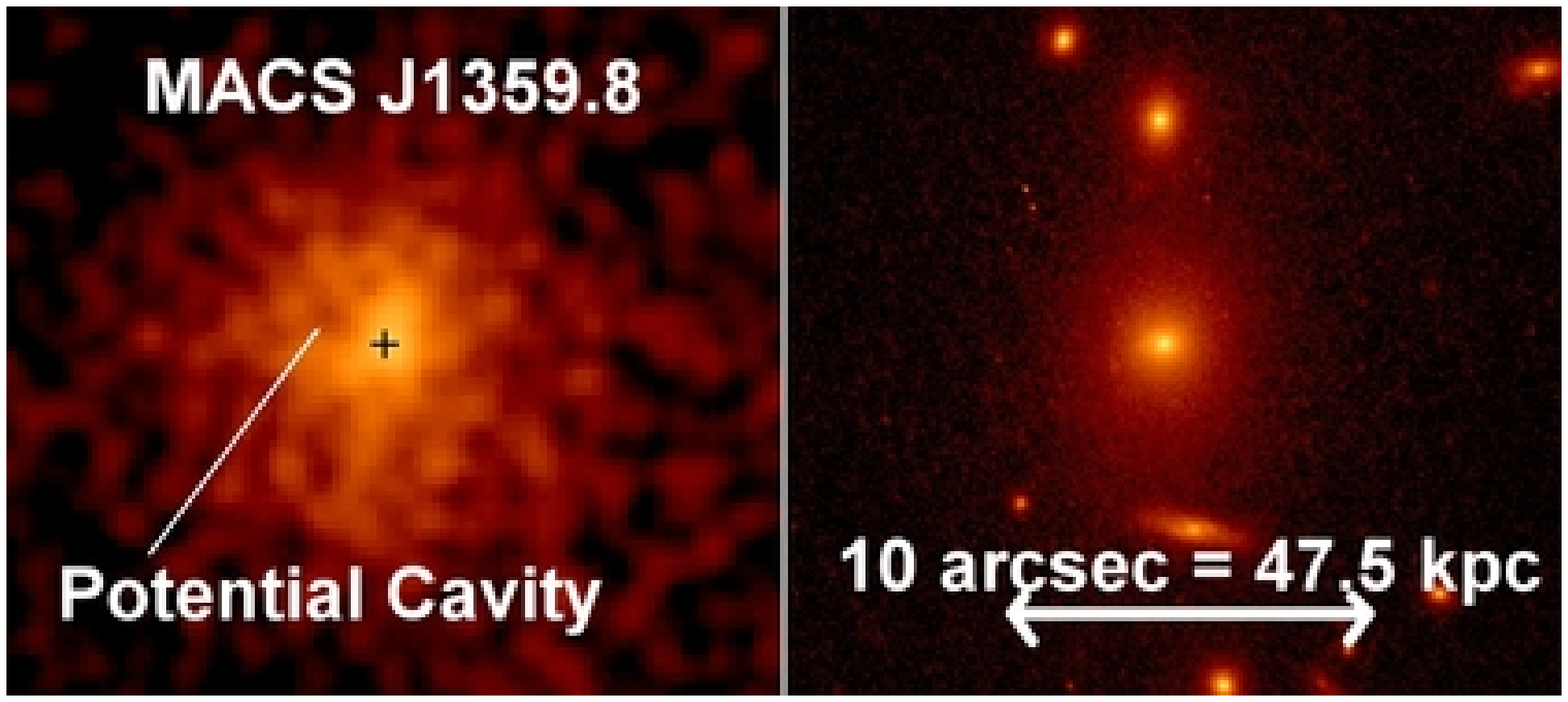}
\end{minipage}
\begin{minipage}[c]{0.49\linewidth}
\centering \includegraphics[width=\linewidth]{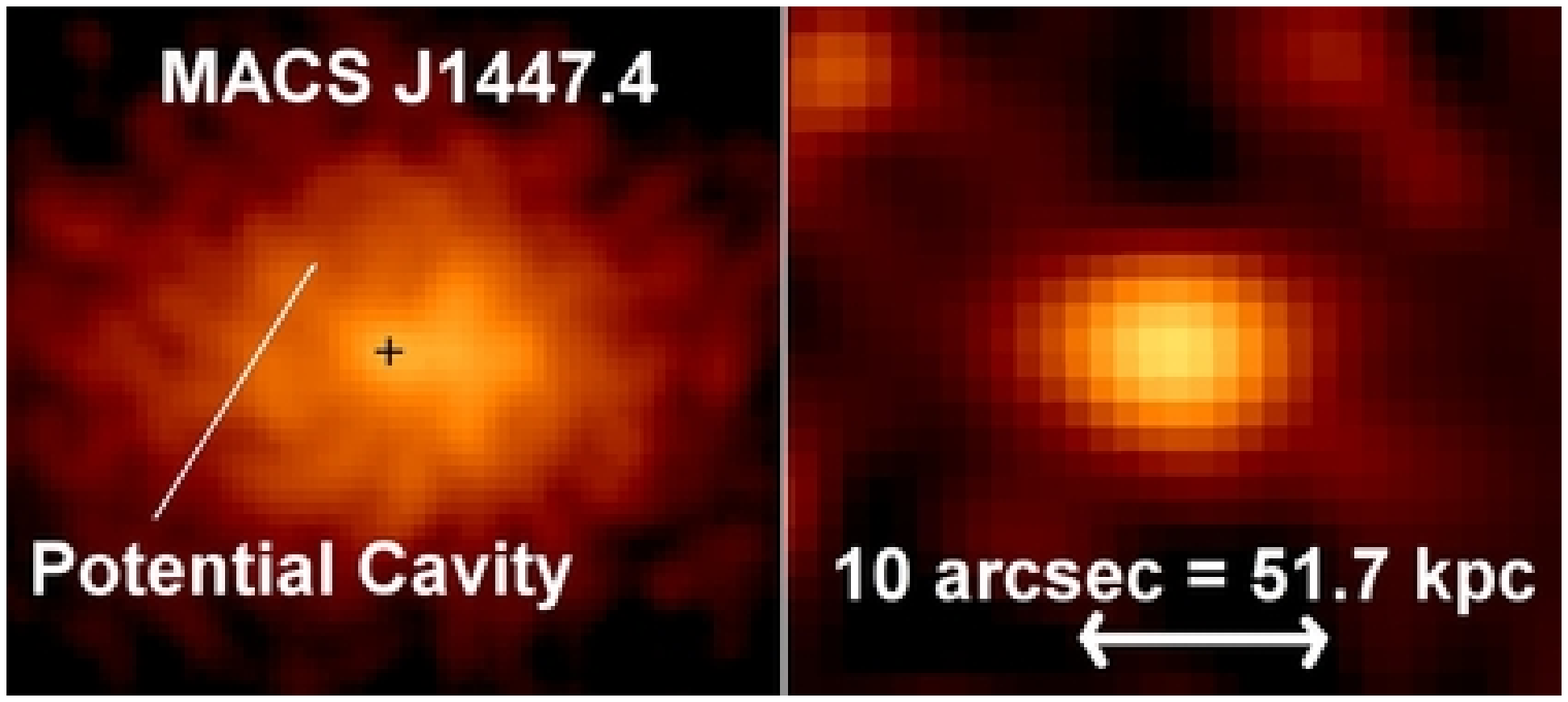}
\end{minipage}
\begin{minipage}[c]{0.49\linewidth}
\centering \includegraphics[width=\linewidth]{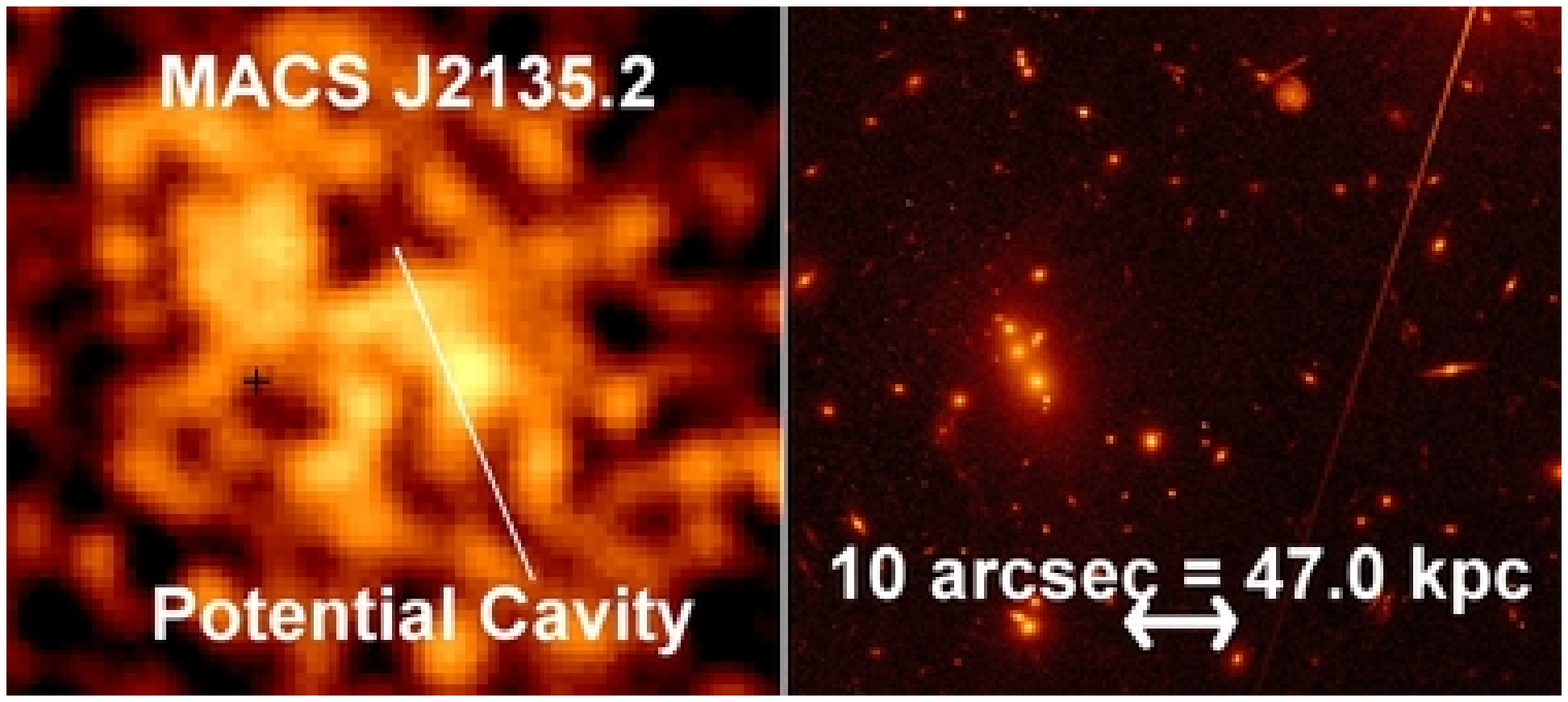}
\end{minipage}
\begin{minipage}[c]{0.49\linewidth}
\centering \includegraphics[width=\linewidth]{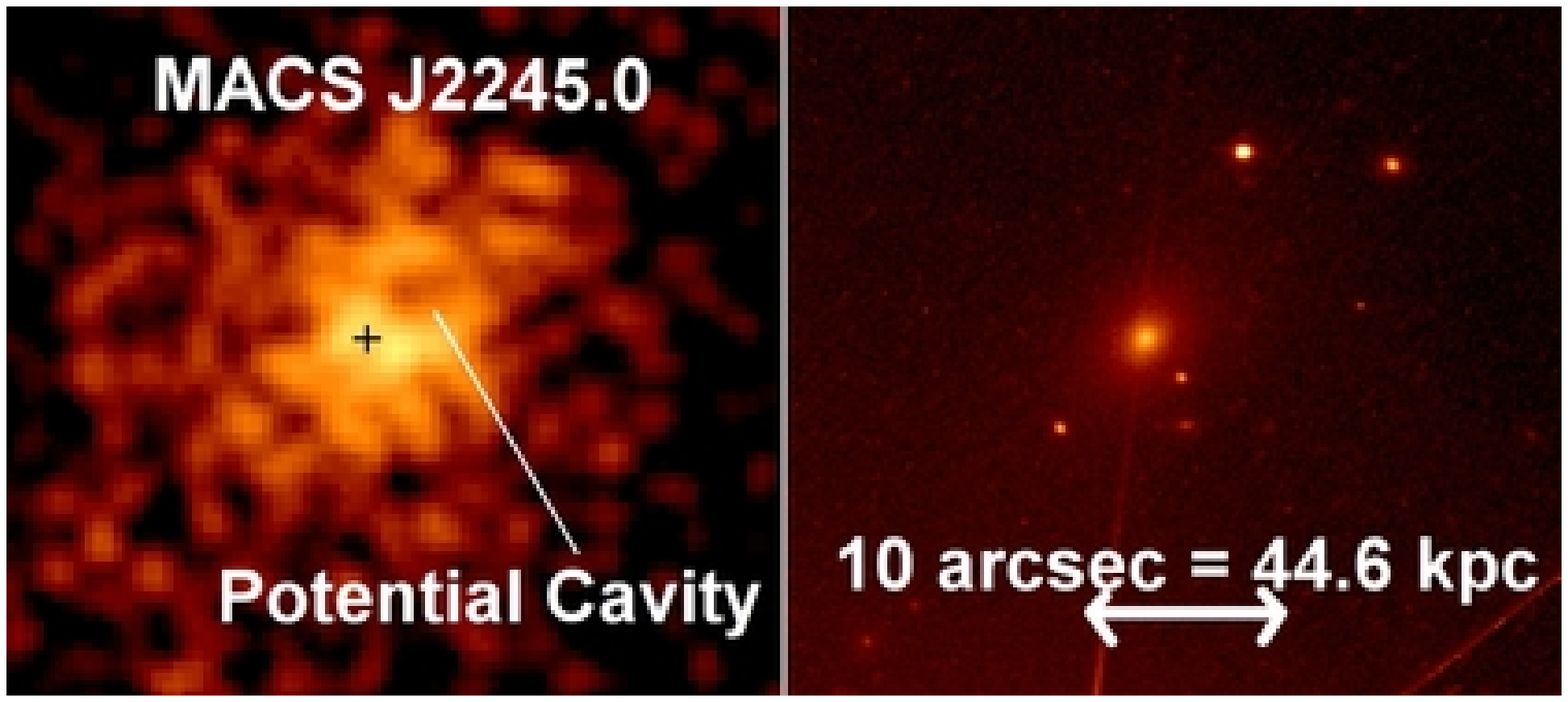}
\end{minipage}
\caption{Images of the MACS clusters with cavities. There are two clusters in each horizontal line, and for each cluster we show the $0.5-7$ keV X-ray image along with a $HST$ image, and if not available, the R passband images from \citet{Ebe2007661,Ebe2010407} taken with the UH 2.2-m telescope (MACS~J2046.0-3430 and MACS~J0111.5+0855) or the Second Palomar Observatory Sky Survey image, $POSS II$, for MACS~J1447.4+0827. The 10$''$ scale for each object is shown with the vector in the optical images, and the black cross symbol shows the location of the BCG. The first 13 clusters have ``clear" cavities, while the remaining 7 only have ``potential" cavities (see ``potential cavities" indicated in the figure). We emphasize that deeper observations are needed to confirm if the ``potential" cavities are real. For the clusters with published radio maps, we plot the radio axis with a green arrow. }
\label{fig2}
\end{figure*}

\section{Cavity ages}

The power capabilities of the cavities are determined by dividing the energy in the bubble with its age. The latter is given by the buoyancy rise time, the refill time or the sound crossing time. See respectively Eq. \ref{eq3} \citep[][]{Chu2001554}, Eq. \ref{eq4} \citep[][]{McN2000534} and Eq. \ref{eq5} \citep[][]{Mcn200745}. Here, $R$ is the distance from the central BCG to the middle of the cavity (projected), $S$ is the cross-sectional area of the bubble ($S={\pi}R_{\rm w}^2$), $C_{\rm D}$=0.75 is the drag coefficient \citep{Chu2001554}, $g$ is the local gravitational acceleration ($g=GM(<R)/R^2$), $r$ is the bubble radius ($r=(R_{\rm l}R_{\rm w})^{1/2}$ for an ellipsoidal bubble), and $c_\mathrm{s}$ is the sound crossing time ($c_\mathrm{s}=\sqrt{{\gamma_2}kT/(\mum_\mathrm{H})}$, where $kT$ is the plasma temperature at the radius of the bubble, $\gamma_2=5/3$ and $\mu=0.62$). 

\begin{equation}
t_\mathrm{buoyancy}= R\sqrt{\frac{SC_{\rm D}}{2gV}}
\label{eq3}
\end{equation}
\begin{equation}
t_\mathrm{refill}= 2\sqrt{\frac{r}{g}}
\label{eq4}
\end{equation}
\begin{equation}
t_{c\mathrm{s}}=\frac{R}{c_\mathrm{s}}
\label{eq5}
\end{equation}

The buoyancy rise time is the time it takes a bubble to reach its terminal buoyancy velocity and depends on drag forces. This is a good estimate of a bubble's age if it has clearly detached from its AGN and has risen, which is the case for most of our cavities. The refill time is the time it takes a bubble to rise buoyantly through its own diameter starting from rest. The sound crossing time is the time it takes a bubble to travel the projected distance from the central AGN to its current location at the sound speed. The latter is used under the assumption that the bubbles travel at subsonic speeds. Between the three time estimates, it is still not clear which is the best to use, although they should not vary significantly.

The values we derive in the spectral deprojecting technique are used to estimate the thermal properties of the gas (see Fig. \ref{figA1} in Appendix B). To determine $M(<R)$, we assume hydrostatic equilibrium (Eq. \ref{eq6}), where $G$ is the gravitational constant, $\rho_{\rm gas}\sim1.92\mum_{\rm H}n_{\rm e}$ and $dp/dr$ is the pressure gradient.   
\begin{equation}
M(<r)=-1\frac{1}{G}\frac{r^2}{\rho_{\rm gas}}\frac{dp}{dr}
\label{eq6}
\end{equation}
Table \ref{tab3} shows the values we obtain for the power stored in each cavity using $t_{\rm cs}$, $t_{\rm buoy}$ and $t_{\rm refill}$.  

\section{Cool cores}
\subsection{Clusters in the MACS sample with cavities}
\begin{figure}
\centering
\begin{minipage}[c]{0.99\linewidth}
\centering \includegraphics[width=\linewidth]{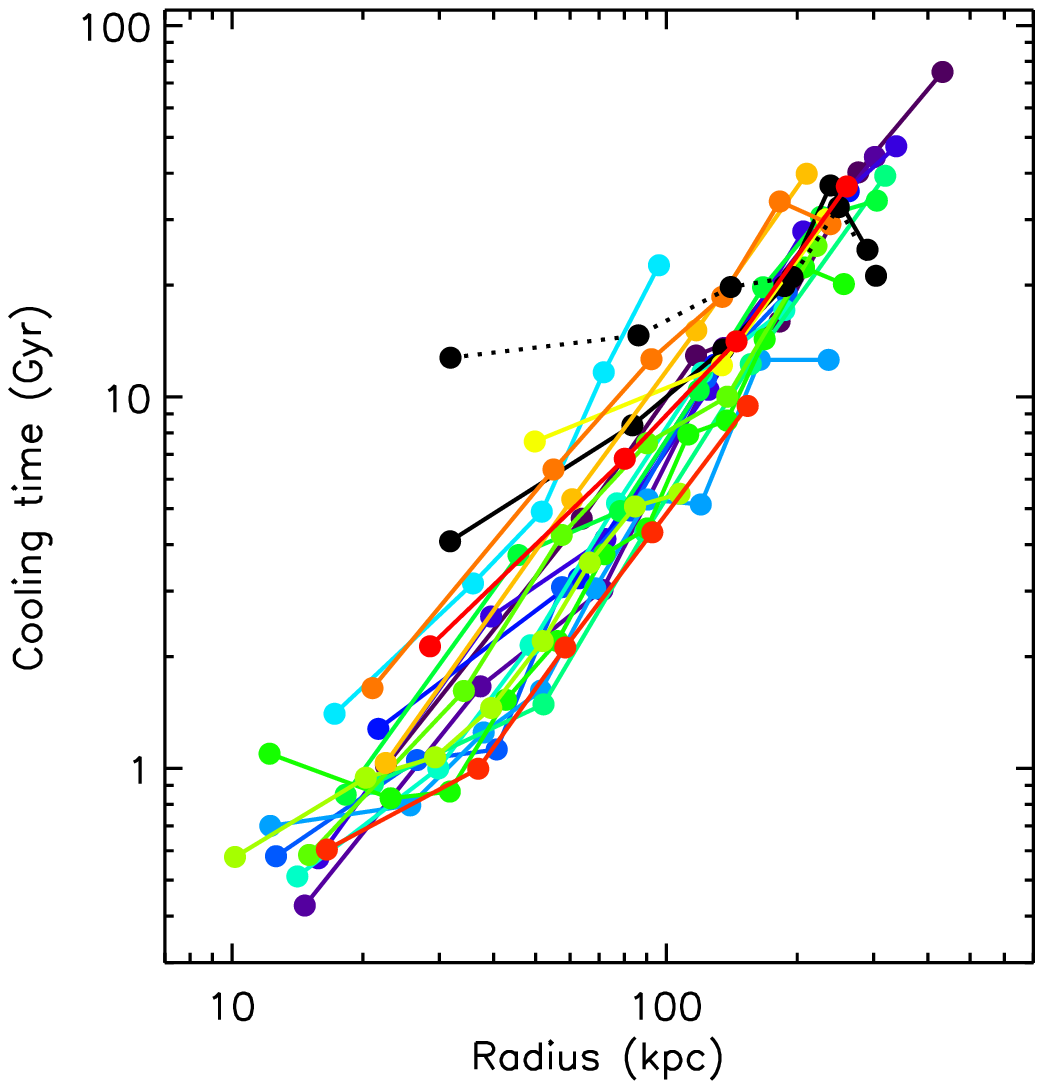}
\end{minipage}
\caption{Cooling time profiles for all clusters in our sample for which we were able to identify cavities. Each colour illustrates a different cluster. The clusters shown with the black lines are those in which the $Chandra$ data were too poor to deproject the spectra. In this case, we deprojected the surface brightness profiles to obtain an estimate of the deprojected $n_{\rm e}$ and $L_{\rm X}$. Using these values, combined with the average cluster temperature within $r<200$, we then determined $t_{\rm cool}$}. MACS~J2135.2-0102 is the only object in our sample with cavities that has a central cooling time larger than 10 Gyrs (shown with the black dotted line). 
\label{fig3}
\end{figure}

Statistical studies of cavities in clusters of galaxies have shown that cavities sit predominately in cool core clusters \citep{Dun2005364,Dun2006373}. In order to compare our results with theirs, we have computed detailed cooling time profiles for all of our clusters with cavities. More precisely, we compute the cooling time ($t_{\rm cool}$) by using Eq. \ref{eq7} and the thermal gas properties derived in Section 4 (shown in Fig. \ref{figA1}).
\begin{equation}
t_{\rm cool} = \frac{5}{2}\frac{1.92~n_{\rm e}~kT~V}{L_{\rm X}}
\label{eq7}
\end{equation}
Here $n_{\rm e}$ is the electron density, $kT$ is the gas temperature, $L_{\rm X}$ is the gas X-ray luminosity and $V$ is the gas volume contained within each annulus. We use the deprojected spectral quantities as an estimate of the plasma parameters and show our results in Fig. \ref{fig3}. However, we show in black the two clusters for which we were not able to deproject the spectra (MACS~J0111.5+0855 and MACS~J2135.2-0102). In this case, we deprojected the surface brightness distribution, and then converted the counts per second of each annulus into the predicted electron density and flux of a {\sc mekal} model for which the temperature and abundance were frozen at the average value within $r<200$ kpc. We then used these values of $n_{\rm e}$ and $L_{\rm X}$ combined with the average cluster temperature within $r<200$ kpc to determine $t_{\rm cool}$ using Eq. \ref{eq7}. This is possible since X-ray observations in clusters give very precise estimates of the surface brightness profiles. It is therefore much easier to deproject the surface brightness profiles than the spectra, allowing us to obtain an estimate of the ``deprojected" cooling time profiles for these two objects. 

Fig. \ref{fig3} shows that many of the clusters in which we detected cavities are strong cool cores, with cooling times reaching down to $\sim0.5$ Gyr in the central regions. The estimates we obtain for the central cooling times are shown in Table \ref{tab4}. We define the central cooling time as the cooling time we derive at a radius of 50 kpc and not the cooling time of the inner most bin. This is because the radius of the inner most bin depends not only on the cluster distance, but more importantly on the data quality, and is therefore not necessarily the best estimate of the central cooling time \citep[see also][]{Bau2005359}. For almost all of the clusters in our sample we have cooling time estimates below 50 kpc, allowing us to obtain an accurate central cooling time at 50 kpc. For the remaining clusters, we determine a rough estimate by extrapolating the values down to 50 kpc. 

Adopting this definition of the central cooling time, we find that 19 of the 20 clusters in which we identified cavities are mild cool cores with $t_{\rm cool}<8$ Gyrs and that all of the systems with ``clear" cavities are strong cool cores with $t_{\rm cool}<3-5$ Gyrs. We also illustrate this in the top panel of Fig. \ref{fig4}, where we plot the distribution of cooling times in our sample of clusters with cavities. 

Since many clusters at $z=1$ appear to have similar properties to present-day ones, they can be regarded as relaxed systems and more importantly, we can assume that a cooling flow should have had the time to establish itself since then. We therefore choose to define the cooling radius of each cluster as the radius at which the cooling time is equal to the $z=1$ look-back time. For our cosmology, this corresponds to $7.7$ Gyrs. This definition of the cooling radius was also adopted by \citet{Raf2006652}, allowing us to directly compare our results with theirs. Although there are other statistical studies of cavities in the literature \citep[e.g. ][]{Bir2004607,Dun2005364,Dun2006373,Bir2008686,Dun2008385}, many of the objects have already been included in \citet{Raf2006652}. 

Finally, we derive the bolometric X-ray luminosity of each cluster ($L_{\rm X}$(0.01-100 keV)), and the cooling luminosity which we define as the bolometric X-ray luminosity ($0.01-100$ keV) within the cooling radius, $L_{\rm X}$($<r_{\rm cool}$). When deriving these quantities, we exclude the central point source ($r<1''$) for clusters with point-like nuclei at their centres, since some have important non-thermal emission associated with them. These include MACS~J0547.0-3904, MACS~J0913.7+4056, MACS~J0947.2+7623, MACS~J1411.3+5212, MACS~J1423.8+2404, MACS~J1931.8-2634 and MACS~J2046.0-3430.

\subsection{MACS cluster sample}
In order to derive the fraction of cool core clusters with cavities in the MACS sample, we computed the central cooling time within $50$ kpc for all 76 MACS clusters with $Chandra$ archival data. Here, we extracted a spectrum for each cluster within $r<50$ kpc and fitted a {\sc mekal} model to the spectrum while letting the temperature, abundance and normalisation parameter free to vary. For some clusters, the counts were not sufficient to constrain the fit. In this case, we kept the abundance and temperature frozen at the value derived for the entire cluster ($r<200$ kpc). 

\begin{figure}
%\centering
\begin{minipage}[c]{0.99\linewidth}
\includegraphics[width=\linewidth]{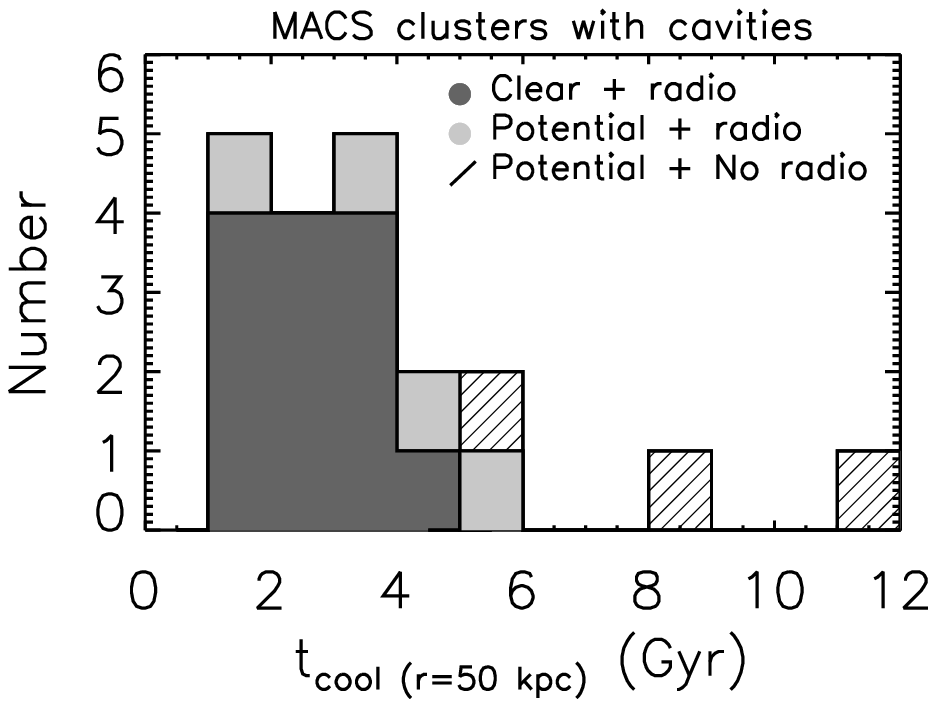}
\end{minipage}
\begin{minipage}[c]{0.99\linewidth}
\includegraphics[width=\linewidth]{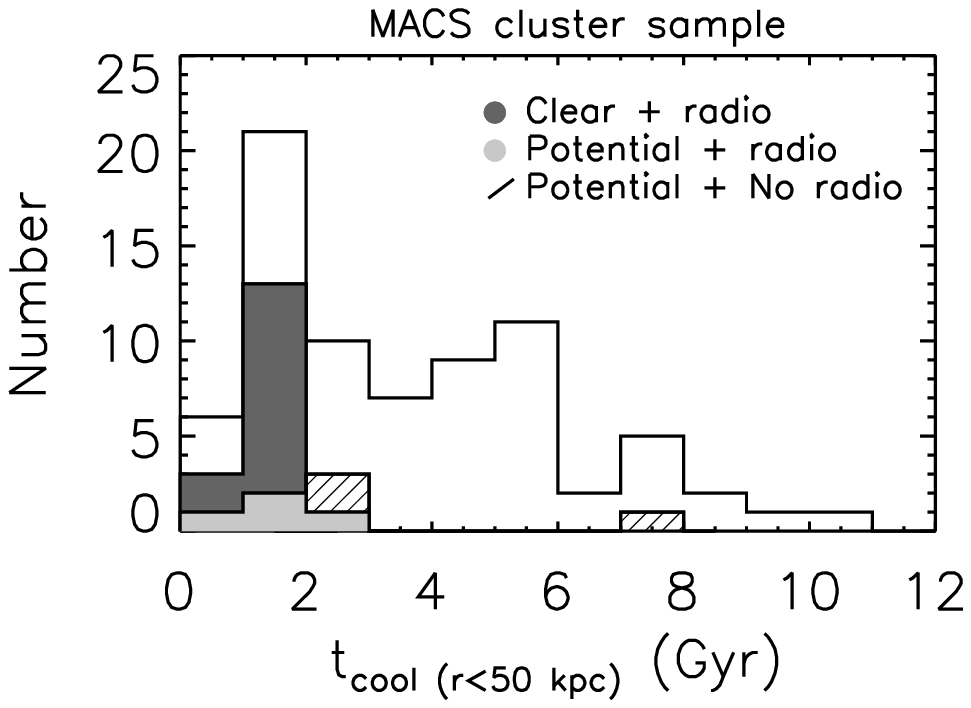}
\end{minipage}
\caption{{Top panel: Distribution of $t_{\rm cool}$ at 50 kpc for the MACS clusters in which we identified cavities. Here, we derived the central cooling time by extracting a series of annuli and determining the cooling time at $r=50$ kpc. Bottom panel: Distribution of $t_{\rm cool}$ within 50 kpc ($r<50$ kpc) for all 76 MACS clusters with $Chandra$ observations. Among these, we outline those with clear bubbles in black (all of these have a radio source associated with the central galaxy) and those with potential cavities and a radio source in grey, as well as those with potential cavities but no radio source in diagonal lines.} }
\label{fig4}
\end{figure}

We could not compute detailed cooling time profiles for all the MACS clusters since the data quality in many is not sufficient to derive deprojected or even projected profiles. We therefore choose to use the cooling time within the central 50 kpc ($t_{\rm cool}(r<50\rm{kpc})$) as an estimate of the central cooling time. This definition of the central cooling time is not the same as the one used previously and defined as the cooling time derived at a radius of 50 kpc ($t_{\rm cool}(r=50\rm{kpc})$). To determine the fraction of cool core clusters with cavities in the MACS sample we therefore recalculated the central cooling time of clusters with cavities adopting the new definition ($t_{\rm cool}(r<50\rm{kpc})$). Our results are shown in the bottom panel of Fig. \ref{fig4}. 

\begin{figure}
%\centering
\begin{minipage}[c]{0.99\linewidth}
\includegraphics[width=\linewidth]{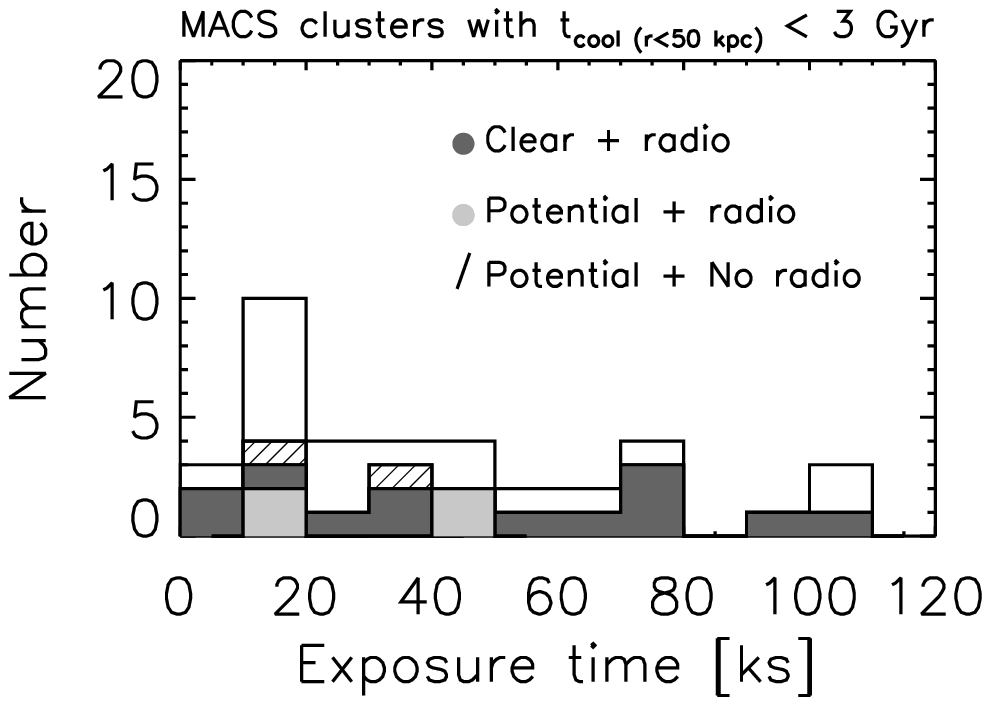}
\end{minipage}
\begin{minipage}[c]{0.99\linewidth}
\includegraphics[width=\linewidth]{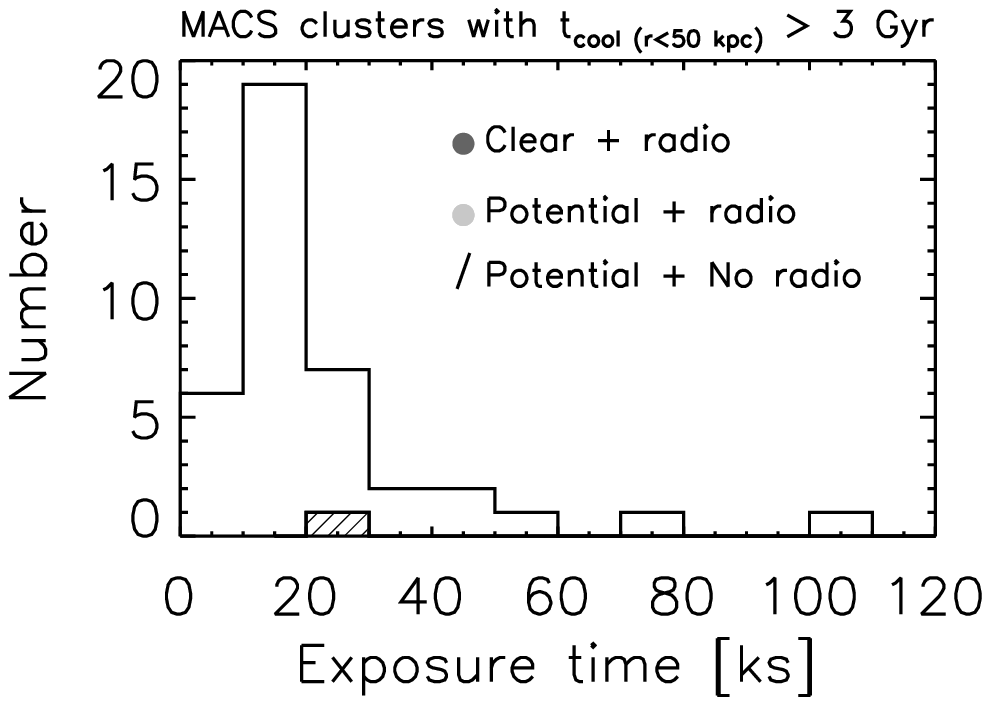}
\end{minipage}
\begin{minipage}[c]{0.99\linewidth}
\includegraphics[width=\linewidth]{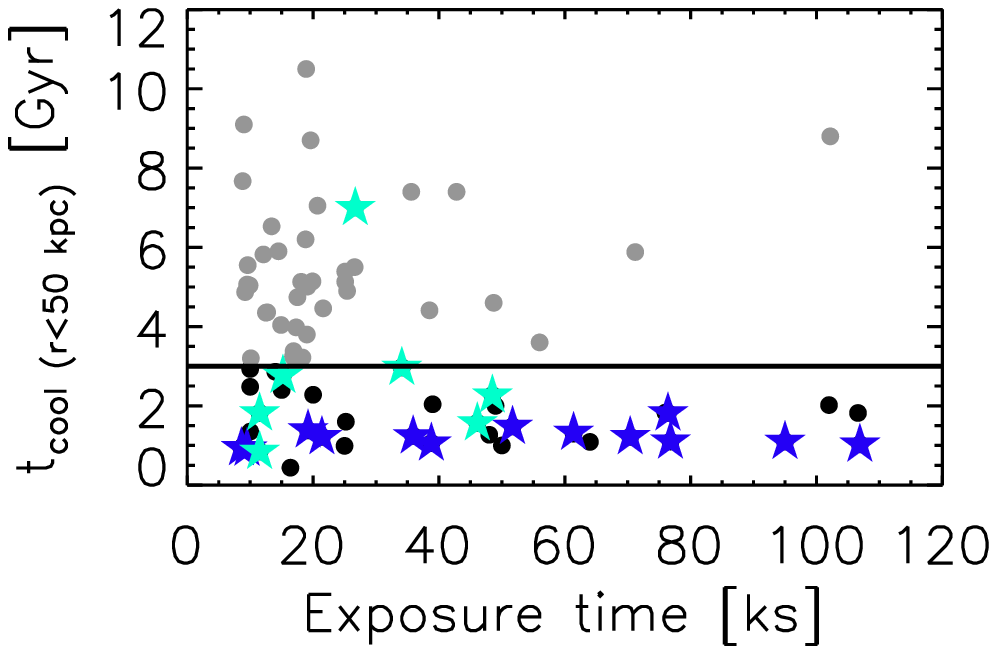}
\end{minipage}
\caption{$Chandra$ exposure time distribution for the 37 MACS clusters with $t_{\rm cool(r<50 kpc)}$ $<$ 3 Gyrs (top panel) and for the 39 MACS clusters with $t_{\rm cool(r<50 kpc)}$ $>$ 3 Gyrs (middle panel). Among these, we outline the 13 with clear bubbles in black (all of these have a radio source associated with the central galaxy) and those with potential cavities and a radio source in grey, as well as those with potential cavities but no radio source in diagonal lines. Note that more than half of the cool cores ($t_{\rm cool(r<50 kpc)}$ $<$ 3 Gyrs) with no cavities have an exposure time below 30~ks, which could explain why we do not detect cavities in these systems. Also shown in the bottom panel is the central cooling time as a function of $Chandra$ exposure time. The MACS clusters with clear bubbles are shown in dark blue, and those with potential bubbles are shown in light blue. The straight line shows the cutoff between those with $t_{\rm cool(r<50 kpc)}$ $<$ 3 Gyrs (light grey circles) and those with $t_{\rm cool(r<50 kpc)}$ $>$ 3 Gyrs (black circles).  }
\label{fig5}
\end{figure}

We find that 37 of the 76 MACS clusters have a strong central cool core ($t_{\rm cool}(r<50\rm{kpc})<3$ Gyrs), corresponding to a cool core fraction of 50 per cent. \citet{Bau2005359} derive detailed cooling time profiles for 38 X-ray luminous clusters from the $ROSAT$ BCS survey ($0.15<z<0.4$) and find that 11 have a central cooling time less than 3 Gyrs (and 55 per cent less than 10 Gyrs). They use the cooling time derived at a radius of 50 kpc as their definition of the central cooling time, whereas we use the cooling time derived $within$ 50 kpc for the central cooling time which gives smaller cooling times. Nonetheless, we find that a significant fraction in our sample are cool core clusters and out of these 37 clusters, we find that 19 have cavities, corresponding to a cavity detection rate in cooling systems of $50$ per cent. Note however, that more than half of the remaining cool cores ($t_{\rm cool(r<50 kpc)}$ $<$ 3 Gyrs) with no cavities have an exposure time below 30~ks (see top panel of Fig. \ref{fig5}), which could explain why we do not detect cavities in these systems. We also plot the central cooling time as a function of $Chandra$ exposure time for all MACS clusters in the bottom panel of Fig. \ref{fig5}, and discuss these results in Section 9.1.

\section{Cavity scaling relations}

\begin{figure}
\centering
\begin{minipage}[c]{0.99\linewidth}
\centering \includegraphics[width=\linewidth]{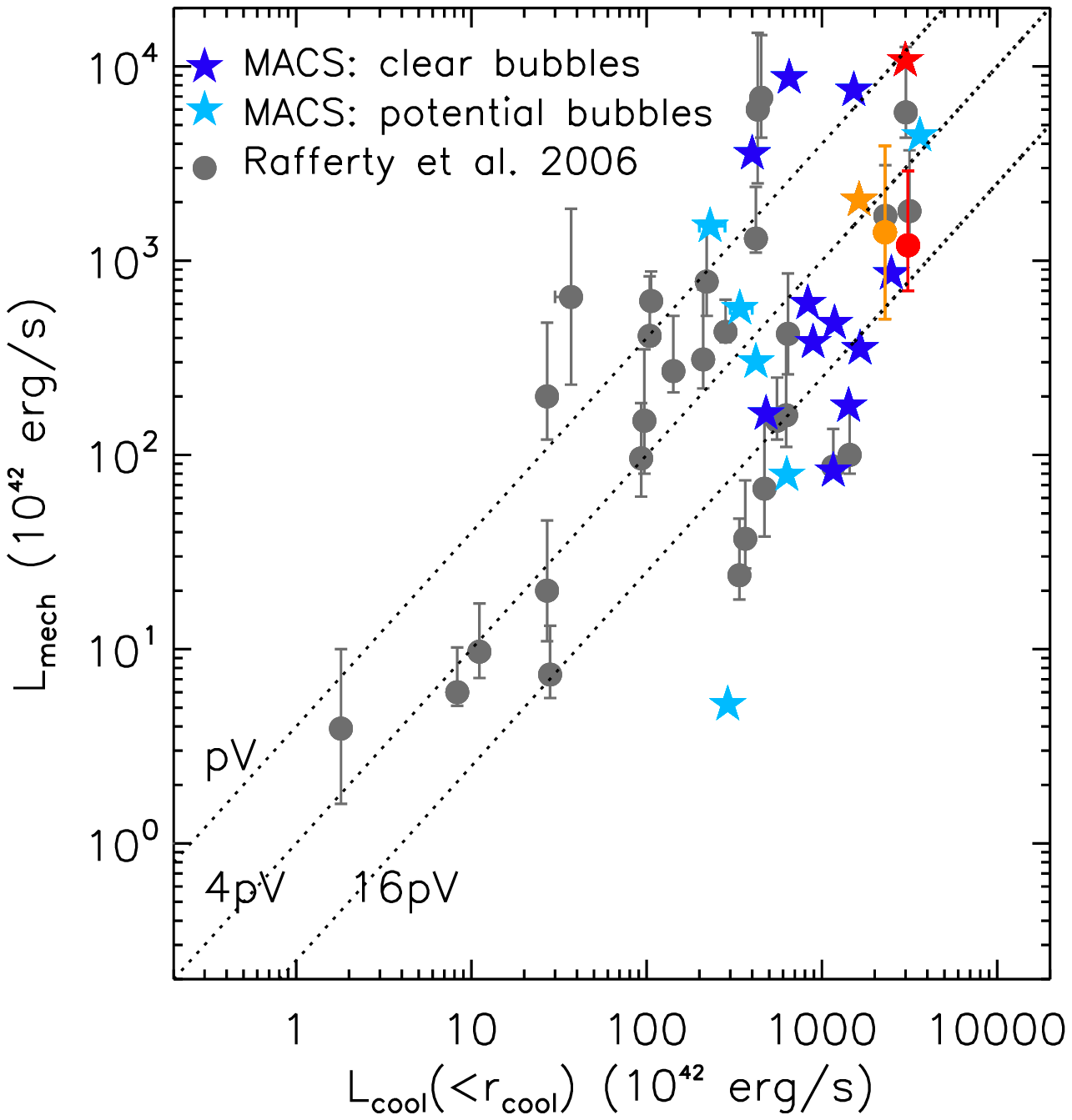}
\end{minipage}
\caption{Mechanical energy being injected into the ICM by the central AGN (estimated from the energetics of the cavities) as a function of energy needed to offset cooling of the ICM by the central galaxy. Those common to both samples are shown in red and yellow.}
\label{fig6}
\end{figure}
\begin{figure}
\centering
\begin{minipage}[c]{0.99\linewidth}
\centering \includegraphics[width=\linewidth]{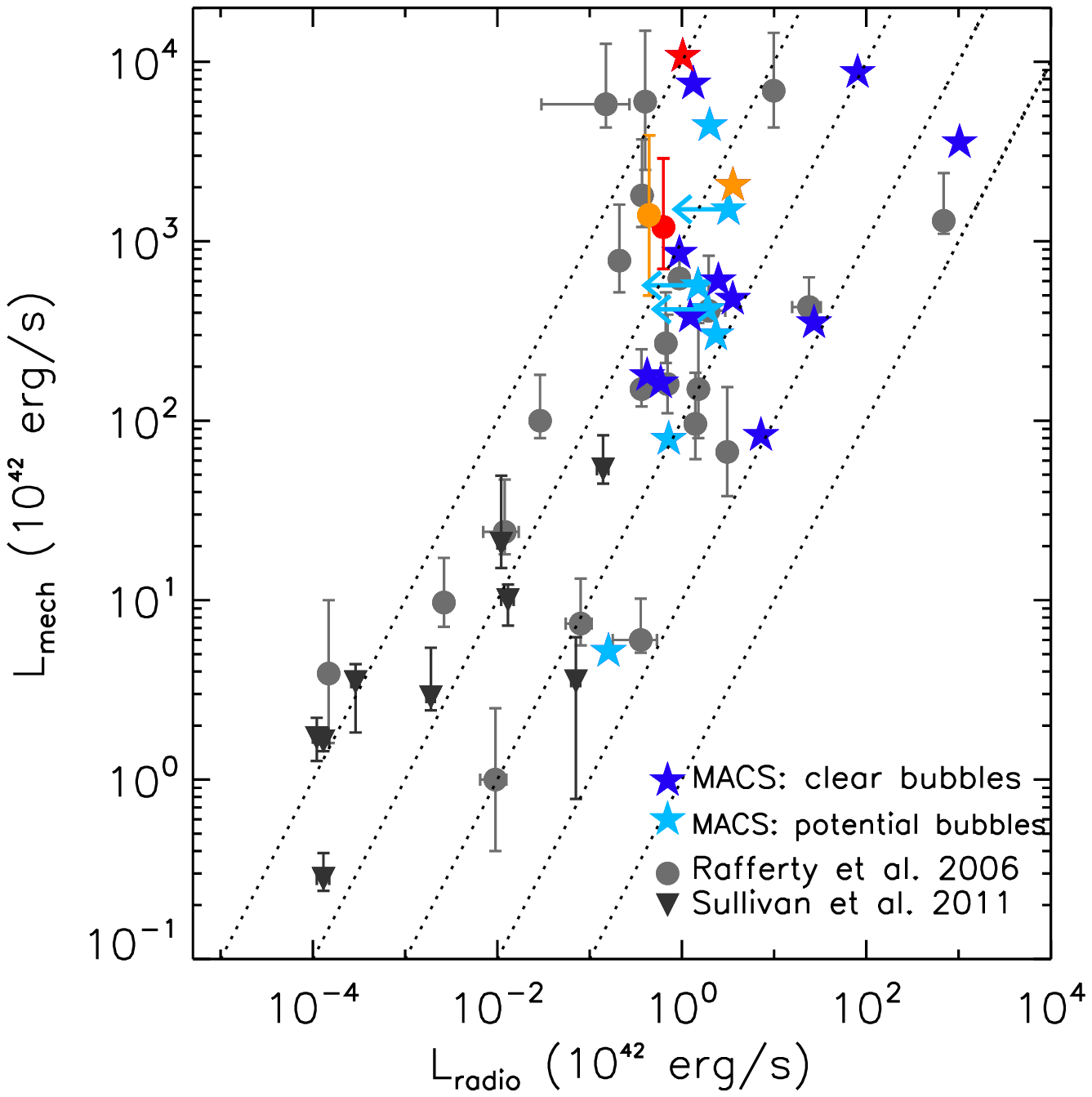}
\end{minipage}
\caption{Mechanical energy being injected into the ICM by the central AGN (estimated from the energetics of the cavities) as a function of total integrated radio luminosity ($10$MHz to $10$GHz) of the central source. Those common to both samples are shown in red and yellow.}
\label{fig7}
\end{figure}

We now address the various scaling relations known for the cavity population in groups and clusters. It has been known for quite some time that AGN feedback is finely tuned to the energy needed to offset cooling of the ICM. If the input of energy were too low, then the core would cool rapidly and cause massive quantities of gas to condense and form stars at rates much higher than what we observe. However, if the input of energy were too high, then this would cause the core to over-heat and we would not observe the low central cooling times in the cores of some clusters.

To illustrate the fine tuning of AGN feedback, we plot the mechanical power arising from the central AGN as a function of the bolometric X-ray luminosity within the cooling radius ($L_{\rm cool}(0.01-100\rm{ keV})$) in Fig. \ref{fig6}. Within this radius, the gas should have had the time to cool if no additional heating sources were present. Although there is some cooling happening in the cores of clusters, studies suggest it is only on the order of 10 per cent \citep{Bir2004607,Raf2006652} or even less than 10 per cent according to the more recent studies with high-resolution spectroscopy \citep{San2008385,San2010402}, and therefore does not affect significantly the relation plotted in Fig. \ref{fig6}. The mechanical power is estimated using the energy stored within the cavities of each system ($4pV$) and the buoyancy rise time ($t_{\rm buoy}$). We also include the cavities of \citet{Raf2006652} and illustrate in red the two systems common to both samples (MACS~J1423.8+2404 and MACS~J0947.2+7623). For MACS~J1423.8+2404, our results agree well with those of \citet{Raf2006652}, but there is a large discrepancy between our estimate of the mechanical power and that of \citet{Raf2006652} in MACS~J0947.2+7623. However, the more recent study on MACS~J0947.2+7623 by \citet[][]{Cav2011732} finds an outburst power consistent with our estimate of $L_{\rm mech}$ ($\sim6\times10^{45}\ergps$). We do not include the data points of \citet{Dun2005364} or \citet{Dun2006373,Dun2008385} as they define the cooling radius as the radius at which $t_{\rm cool}=3$ Gyrs, nor the data points of \citet{Cav2010720} as they only calculate the cooling luminosity within $0.3-2.0$ keV. Fig. \ref{fig6} shows that the MACS sample contains some of the most powerful outbursts known, and that even at these extreme energies, the central AGN is able to offset cooling of the ICM at a fine level. 
      
We also look at the properties of the mechanical power arising from the central AGN as a function of radio luminosity. As these cavities are being created by the central AGN and being filled with relativistic electrons, one would expect a correlation between the mechanical power and radio luminosity, which is a measure of the radio power arising from the AGN. This has been looked at in detail for clusters, ellipticals and more recently groups of galaxies \citep{Bir2004607,Bir2008686,Cav2010720,OSu2011735}, although the scatter remains large and extends over several orders of magnitude ($>3$). We have computed the radio luminosities for our sample (see Table \ref{tab2}) and show in Fig. \ref{fig7} the mechanical power arising from the central AGN ($4pV/t_{\rm buoy}$) as a function of radio luminosity ($L_{\rm radio}$). We include the data points of \citet{OSu2011735} for galaxy groups, as well as those in \citet{Raf2006652}, but use the estimate of $L_{\rm radio}$ given in \citet{Bir2008686}. We do not include the data points of \citet{Cav2010720} as they do not include an estimate of $L_{\rm radio}$ in their paper. We discuss this plot in Section 9.2.

\section{Results}
In this section we review the results of each system. We begin by addressing the systems with ``clear" cavities, and then address those with ``potential" cavities. Many of the MACS clusters have been studied in terms of their mass distribution and cosmological implications \citep[e.g. ][]{All2008383,Ett2009501,Pue2005619}, as well as in terms of how the cluster metallicity content evolves with redshift \citep{Bal2007462,Mau2008174}. These include MACS~J0159.8-0849, MACS~J0242.5-2132, MACS~J0429.6-0253, MACS~J0947.2+7623, MACS~J1423.8+2404, MACS~J1532.8+3021, MACS~J1720.2+3536, MACS~J1931.8-2634, MACS~J2046.0-3430, MACS~J2140.2-2339 and MACS~J1359.8+6231. 

%: MACS~J0159.8-0849, MACS~J0242.5-2132, MACS~J0429.6-0253, MACS~J0947.2+7623, MACS~J1423.8+2404, MACS~J1532.8+3021, MACS~J1720.2+3536, MACS~J1931.8-2634, MACS~J2140.2-2339, MACS~J1359.8+6231, MACS~J2245.0+2637
\subsection{Clusters with ``clear" cavities}
\subsubsection{MACS~J0159.8-0849}
We find two clear X-ray cavities in this system. They are located symmetrically opposite with respect to the central AGN, and at a distance of $\sim25$ kpc from the core. The power stored within them ($P_{\rm buoy}\sim5\times10^{44}\ergps$) falls short by a factor of 2 from the cooling luminosity. The $HST$ optical image shows interesting features, including a potential double or triple nucleus with streams of material leading back to the central nucleus.

\subsubsection{MACS~J0242.5-2132}
We find two cavities in this cluster, with the south-west one being clearer and somewhat larger than its northern counterpart. The optical image shows that there is a bright central isolated galaxy at the centre, which coincides with a very strong radio source of flat spectrum ($\sim1$ Jy bright, the second strongest in our sample). The power stored within the cavities ($P_{\rm buoy}\sim4\times10^{44}\ergps$) falls short by a factor of 4 from the cooling luminosity. This cluster is part of the 34 most luminous MACS clusters \citep{Ebe2010407}.

\subsubsection{MACS~J0429.6-0253}
This cluster has two clear cavities being inflated by the central AGN. The cavities are quite small ($r\sim5$ kpc), and the power stored within them falls short by almost an order of magnitude when compared to the cooling luminosity ($P_{\rm buoy}\sim10^{44}\ergps$). There is a hint of another potential cavity to the north-east, but deeper X-ray observations are needed to confirm this. The optical image shows a dominant central galaxy harbouring a bright nucleus. This cluster is also part of the 34 most luminous MACS clusters \citep{Ebe2010407}.    

\subsubsection{MACS~J0547.0-3904}
We find one cavity in this cluster of radial dimensions $8.5\times11.5$ kpc$^2$, surrounded by bright rims. The power stored within it alone ($P_{\rm buoy}\sim2\times10^{44}\ergps$) can prevent the gas from cooling to almost half of the cooling luminosity. The dominant central galaxy harbours a bright point-like nucleus which shows hints of a jet in the optical, aligned with the jet axis along the X-ray cavity. The X-ray image also shows a bright nucleus dominating in emission. This cluster is part of the 34 most luminous MACS clusters \citep{Ebe2010407}.  

\subsubsection{MACS~J0913.7+4056}
The central galaxy in this cluster, more commonly known in the literature as IRAS 09104+4109, is a hyper-luminous infrared cD galaxy \citep[$L_{\rm IR}\sim12\times10^{12}L_\odot$;][]{Kle1988328} hosting a heavily obscured Type 2 AGN \citep[$\sim1-5\times10^{23}$ cm$^{-2}$;][]{Vig2011}. The $Chandra$ X-ray spectrum of the nucleus shows that there is a prominent Fe K$\alpha$ line at the redshift of the source \citep[see also][]{Iwa2001321}.  

We find two cavities in this system. The cavity to the north-west was first noticed by \citet[][]{Iwa2001321}, but the 9.1~ks $Chandra$ observations were too faint to confirm whether this cavity was real. By using a much deeper observation of 70.4~ks, we confirm the existence of the first cavity and find evidence of a second located at a similar radius to the south-east. Although larger, the cavities carved out in the X-ray image coincide with the location of the radio hot spots seen at 20 cm \citep[][ also see Fig. \ref{fig2} where we show the radio jet axis]{Hin1993415}, and the optical image shows a spectacular set of filaments spreading towards the north-east and south. These filaments run along both cavities that we have identified. However, one of these filaments extends further out ($r_{\rm max(filament)}=51$ kpc and $r_{\rm max(cavity)}=55$ kpc) and is directed more eastward than the northern cavity. As seen in many nearby clusters, including M87 in the Virgo cluster \citep{Boh1995274,Chu2001554} and NGC 1275 in the Perseus Cluster \citep{Fab2003344}, filaments usually trail behind cavities, and could indicate that there is another, older cavity to the north-east, although it is difficult to tell with the current X-ray observations. The power stored within the cavities is substantial ($P_{\rm buoy}\sim8\times10^{45}\ergps$) and is more than sufficient to prevent the ICM from cooling. This cluster harbours the third most powerful outburst in our sample (in terms of total power), and has the cavity with the largest enthalpy in those classified as ``clear" cavities ($pV\sim10^{60}\ergps$). MACS~J0913.7+4056 has one of the most powerful AGN outbursts known in clusters.

\subsubsection{MACS~J0947.2+7623}
MACS~J0947.2+7623 (RBS 0797) is another cluster known to have one of the most powerful AGN outbursts \citep[see][ for a detailed analysis on the AGN outburst energetics]{Cav2011732}. It has a very strong cool core, requiring extreme mechanical feedback from its central AGN ($L_{\rm cool}\sim3\times10^{45}\ergps$), and the X-ray image shows two large X-ray cavities surrounded by bright rims, as well as a point-like nucleus. This cluster is also part of the 34 most luminous MACS clusters \citep{Ebe2010407}. In the optical, we see a large central galaxy hosting a complex and disturbed core. The power stored within each cavity ($P_{\rm buoy}\sim4-7\times10^{45}\ergps$) is more than sufficient to prevent the surrounding gas from cooling. This cluster has the most powerful outburst in our sample and the energetics we derive are consistent with \citet{Cav2011732}, except that we estimate the size of the north-east cavity to be slightly larger. 

Interestingly, \citet{Git2006448} found that the 4.8 GHz radio jet axis is perpendicular to the jet axis running through the cavities (see Fig. \ref{fig2} for the direction of the 4.8 GHz radio jet axis). This was later confirmed by \citet{Bir2008686}. \citet{Cav2011732} also suggested that there is a hint of a small, newly formed cavity at the tip of the southern 4.8 GHz jet. These authors furthermore find that the bright rims surrounding the cavities could be driving weak shocks into the ICM and suggest that the unusually deep surface brightness depressions of the cavities, combined with the ambiguous correlation between the X-ray and radio morphologies, is more consistent with these cavities being elongated along the line of sight. These cavities could therefore be larger than measured in the plane of the sky. 

\subsubsection{MACS~J1411.3+5212}

MACS~J1411.3+5212 (3C295) hosts a very powerful central AGN embedded in a giant cD galaxy. This galaxy has a bright radio core and is one of the most powerful FRII sources known. The cluster shows evidence of non-thermal emission filling the inner radio lobes, consistent with ICCMB photons \citep[e.g.][]{Bru2001372,All2001324,Har2000530}. The radio maps show GHz radio emission filling the inner $\sim25$ kpc of the core, with a jet axis aligned in the north-west to south-east direction \citep[][ see also Fig. \ref{fig2} for the direction of the GHz radio jet axis]{Tay1991101,Tay1992262,Bru2001372,All2001324,Har2000530}. \citet[][]{Bru2001372} also find a prominent Fe K line at the redshift of the source associated with the nucleus of the central galaxy, consistent with this object being a powerful hidden and absorbed AGN. 

During the spectral analysis of this source (Sections 4, 5 and 6), we omitted the central regions containing the ICCMB X-ray emission since our aim was to extract the properties of the thermal gas. The cavity that we find is located at the edge of south-east ICCMB X-ray lobe, and could be an older cavity carved in the past by the radio source but which has now risen buoyantly. The radio maps at GHz frequencies show that the emission is limited to the inner $r<25$ kpc \citep[][]{Tay1991101,Tay1992262,Bru2001372,All2001324,Har2000530}. The cavity we find is beyond this radius ($r_{\rm cavity}\sim35$ kpc), and is consistent with being older since the GHz radio observations do not seem to show emission extending out to this radius. This cavity is large and the power stored within it ($P_{\rm buoy}\sim4\times10^{45}\ergps$) is more than sufficient to prevent the gas from cooling.

\subsubsection{MACS~J1423.8+2404}
This cluster hosts a very large central galaxy harbouring a bright nucleus seen both at optical and X-ray wavelengths. In the optical, the $HST$ image shows that the central galaxy is heavily disturbed: the bright point-like nucleus is surrounded by a thick envelope in the inner regions ($r<2.6$ kpc) and a more diffuse and disrupted envelope out to $r\sim20$ kpc. The optical image also shows two depressions, most likely due to dust: one to the west which coincides with the location of the cavity seen at X-ray wavelengths and a hint of a second to the north-east of the nucleus which coincides with a very small depression seen in the X-rays (although we have not included this inner smaller cavity in our sample). The power stored in the cavities ($P_{\rm buoy}\sim2\times10^{45}\ergps$) is sufficient to prevent the gas from cooling within the cooling radius. This cluster is also the most distant object in the MACS sample in which we have identified cavities ($z=0.5449$), and is part of the 12 most distant clusters detected by the MACS survey at $z>0.5$ \citep{Ebe2007661}. The bright nucleus does not show any evidence of a hidden quasar in the form of Fe K lines at the redshift of the source, as opposed to other MACS clusters mentioned previously with a bright X-ray nucleus. \citet{Kar2008389} looked at the large scale structure of the 12 most distant clusters detected by the MACS survey at $z>0.5$ by tracing the surface brightness density of galaxies near the red sequence, and singled out that MACS~J1423.8+2404 was the most relaxed cluster in the sample. \citet{Lim2010405} also confirms that the cluster seems to be nearly fully virialized as suggested by gravitational lensing analysis. Having such a virialized cluster only $\sim7$ Gyrs after the Big Bang puts constraints on structure formation and evolution in a cosmological context.

\subsubsection{MACS~J1532.8+3021}
MACS~J1532.8+3021 (RX J1532.9+3021) has the fourth most infrared luminous BCG in the \citet{Qui2008176} sample of 62 X-ray luminous clusters, enough that it can be classified as a luminous infrared galaxy (LIRG) with $L_{\rm IR}=22.6\times10^{44}\ergps$. The central galaxy also has one of the most massive molecular gas detections known \citep[$2.5\times10^{11}M_{\odot}$, the most massive detection in][]{Edg2001328} and is the second most optically line-luminous central cluster galaxy in the sample of \citet{Cra1999306} with an optical line emission of $4.2\times10^{42}\ergps$. The galaxy is also particularly blue, suggesting star formation, and has the largest star formation region in the \citet{Ode2008681} cluster sample, with a scale length of $\sim70$ kpc. The cluster is also particularly massive $M>10^{15}M_{\odot}$, as found by the weak lensing study of \citet{Dah2002139}. 

We find one clear cavity associated with the central galaxy, located to the west. It is quite large ($r\sim20$ kpc) and situated almost 40 kpc from the nucleus. This cluster has the third strongest cooling luminosity in our MACS sample with cavities, requiring around $3\times10^{45}\ergps$ to prevent the gas from cooling. We find that the power stored within the cavity falls short by a factor of $\sim2-3$ of the cooling luminosity. The $HST$ image suggests that there is some filamentary structure surrounding the central galaxy (although it is not as clear as in MACS0913.7+4056), and one of the northern filaments seems to sweep around the bottom portion of the cavity.

\subsubsection{MACS~J1720.2+3536}
We find two cavities associated with the central regions of MACS~J1720.2+3536 (Z8201), one to the north and another, though very small and not as clear, to the south-east. If we consider the power stored within the cavities using the buoyancy time-scale, then the power falls short by a factor of 2 of the cooling luminosity. If we consider the sound-crossing time, then the power is more than sufficient to offset cooling of the ICM. The $HST$ image shows that the central galaxy has a bright core surrounded by a large diffuse envelope.   
\citet{Raf2008687} also find that the central galaxy has blue emission extending out to 2.5$''$ ($\sim14$ kpc). This cluster is part of the 34 most luminous MACS clusters \citep{Ebe2010407}.

\subsubsection{MACS~J1931.8-2634}
MACS~J1931.8-2634 is a strong cool core cluster harbouring one of the most powerful AGN outbursts known. The central nucleus has carved two large X-ray cavities to the west and east, greatly disrupting the core \citep{Ehl2011411}. Diffuse 1.4 GHz radio emission fills the central regions and extends in the west-east direction, but no clear radio lobes or jets are seen. The X-ray image (Fig. \ref{fig2}) shows two bright ridges formed to the north and south, some 25 kpc from the nucleus. The north ridge is also made of cool, metal-rich gas rich in H$\alpha$ filaments and young stars. \citet{Ehl2011411} propose that these ridges were formed by the AGN outburst, as well as sloshing of the core along the north to south direction induced by a past merger. \citet{Ehl2011411} found that the temperature profile decreased with radius, but that in the inner-most regions ($r<30$ kpc), the temperature increased once more. This could indicate shock heating. In Fig. \ref{figA1}, we show the spectral profiles of MACS~J1931.8 we obtained during our analysis. Our temperature profile also increases in the inner-most regions ($r<30$ kpc). 

The cavity edges are not well defined, and the radio emission shows no clear radio lobes. Determining a precise measurement of the cavity sizes is therefore difficult. Taking this into account, we estimate that the $4pV$ energy stored within the cavities is roughly $\sim3\times10^{60}\erg$ and the power input is $P_{\rm buoy}\sim8-9\times10^{45}\ergps$ or using the time-crossing time, $P_{\rm c_{\rm s}}\sim4-5\times10^{45}\ergps$. \citet{Ehl2011411} find a total energy of $\sim1-8\times10^{60}\erg$ ($4pV$) and $P_{\rm c_{\rm s}}\sim4-14\times10^{45}\ergps$, both consistent with our results. MACS~J1931.8-2634 therefore has one of the most powerful AGN outbursts  known. This cluster is also part of the 34 most luminous MACS clusters \citep{Ebe2010407}.

\subsubsection{MACS~J2046.0-3430}
We find two cavities associated with this cluster. The first is located to the south, and has a flat shape with bright rims surrounding it. The second cavity is located to the north but is less well defined. We were not able to find any $HST$ archival data and show in Fig. \ref{fig2} the R passband UH 2.2-m image. The optical image shows that the central dominant galaxy is offset from the X-ray point source by $\sim2.7''$ (15 kpc). The power stored in the cavities ($P_{\rm buoy}\sim6\times10^{44}\ergps$) is comparable to the amount of energy needed to prevent the gas from cooling.

\subsubsection{MACS~J2140.2-2339}
MACS~J2140.2-2339 (MS2137.3-2353) is a luminous cluster, dominated by a large central galaxy. The most interesting feature seen in the optical image is a large radial lensing arc some 70 kpc from the nucleus, which was the first of its kind to be discovered \citep{For1992399} and has been looked at in detail by many authors \citep[e.g.][]{Gav2003403,Gav2005443,Com2006642,San2008674,Mer2009500}. This cluster is part of the 34 most luminous MACS clusters \citep{Ebe2010407}. We find a small cavity to the south of the nucleus, only capable of preventing the gas from cooling to 10 per cent of the cooling luminosity. Although we classify this cavity as being ``clear", it is very small and does not have any bright rims surrounding it.

\begin{figure*}
\centering
\begin{minipage}[c]{0.325\linewidth}
\centering \includegraphics[width=\linewidth]{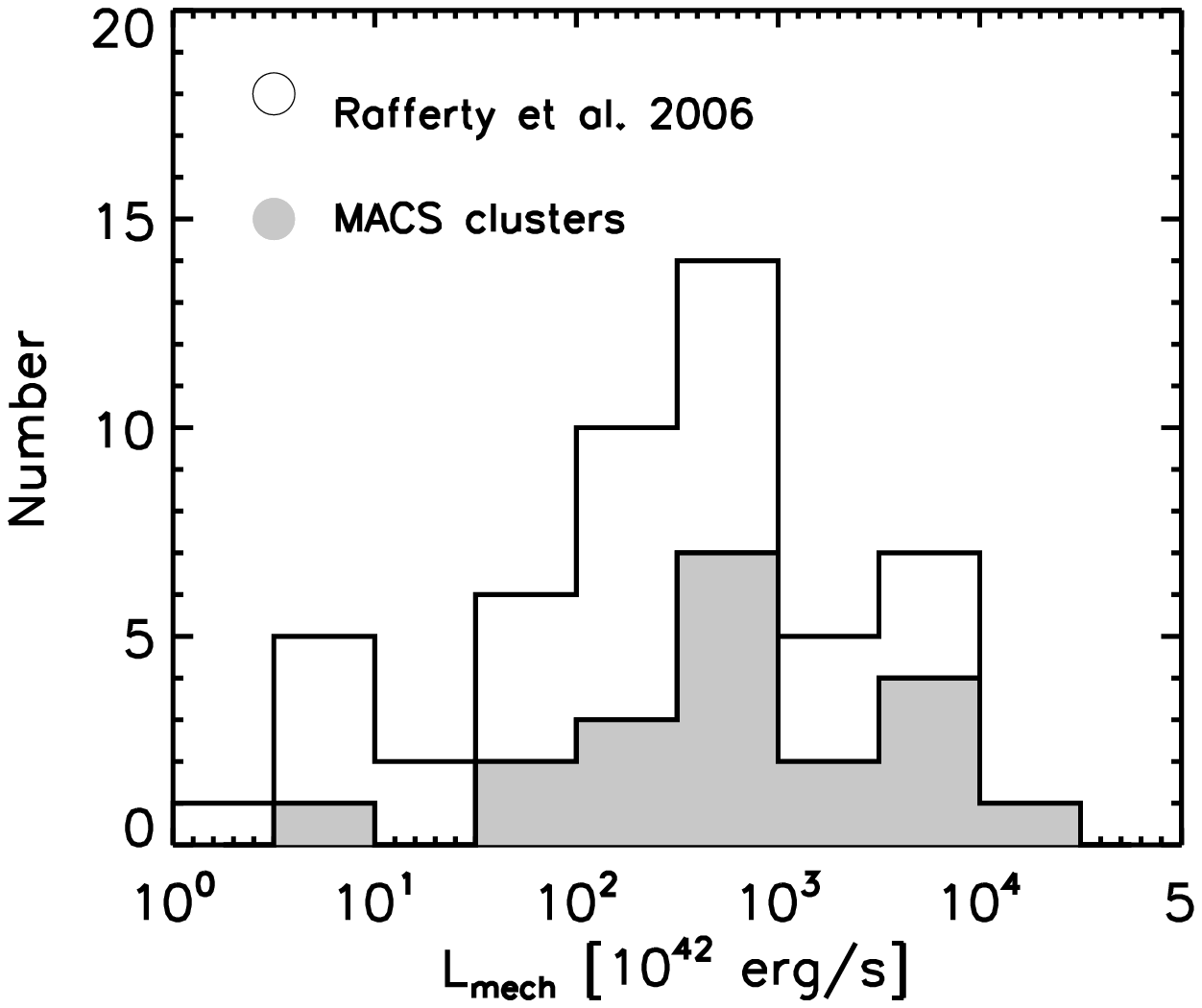}
\end{minipage}
\begin{minipage}[c]{0.325\linewidth}
\centering \includegraphics[width=\linewidth]{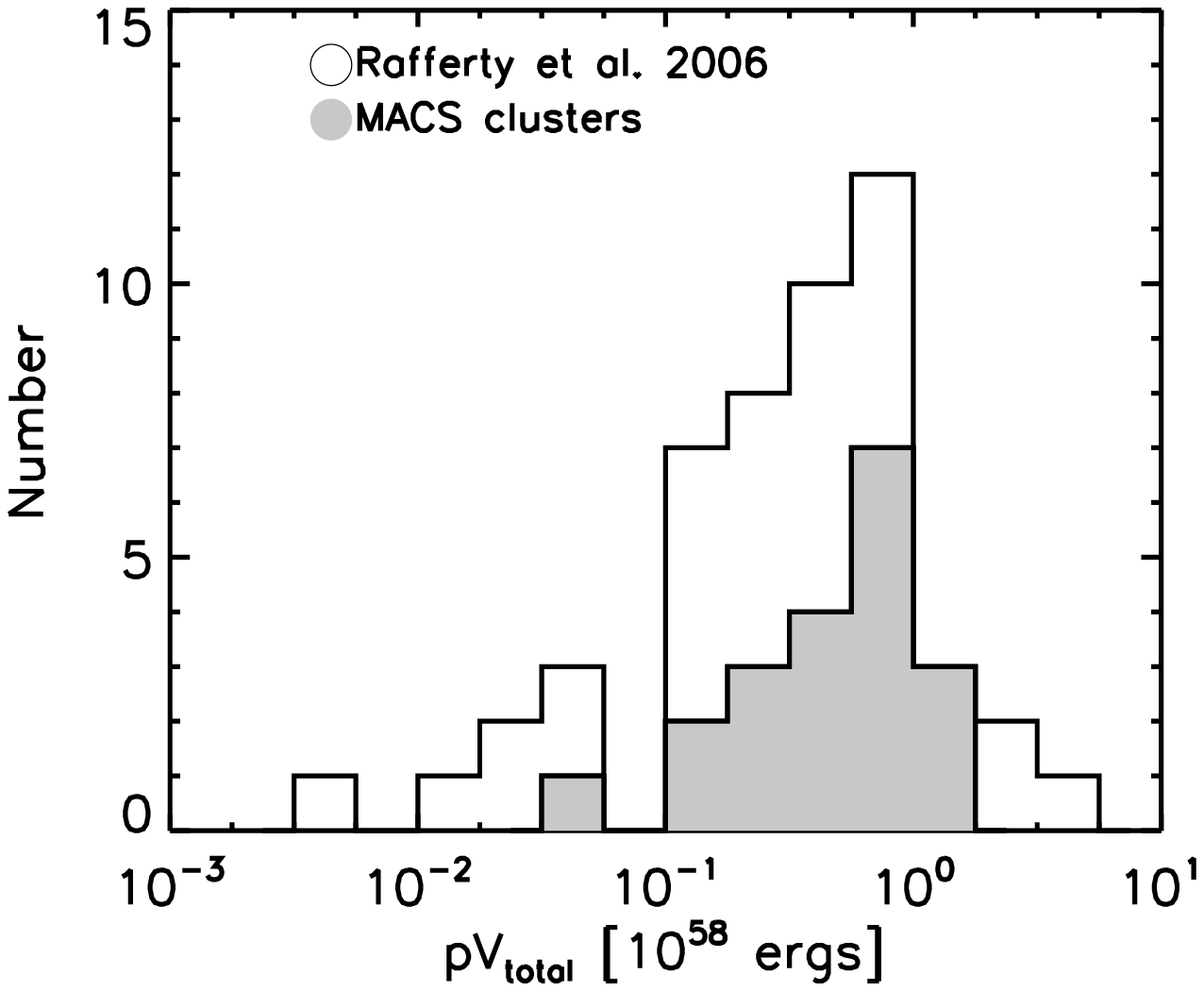}
\end{minipage}
\begin{minipage}[c]{0.325\linewidth}
\centering \includegraphics[width=\linewidth]{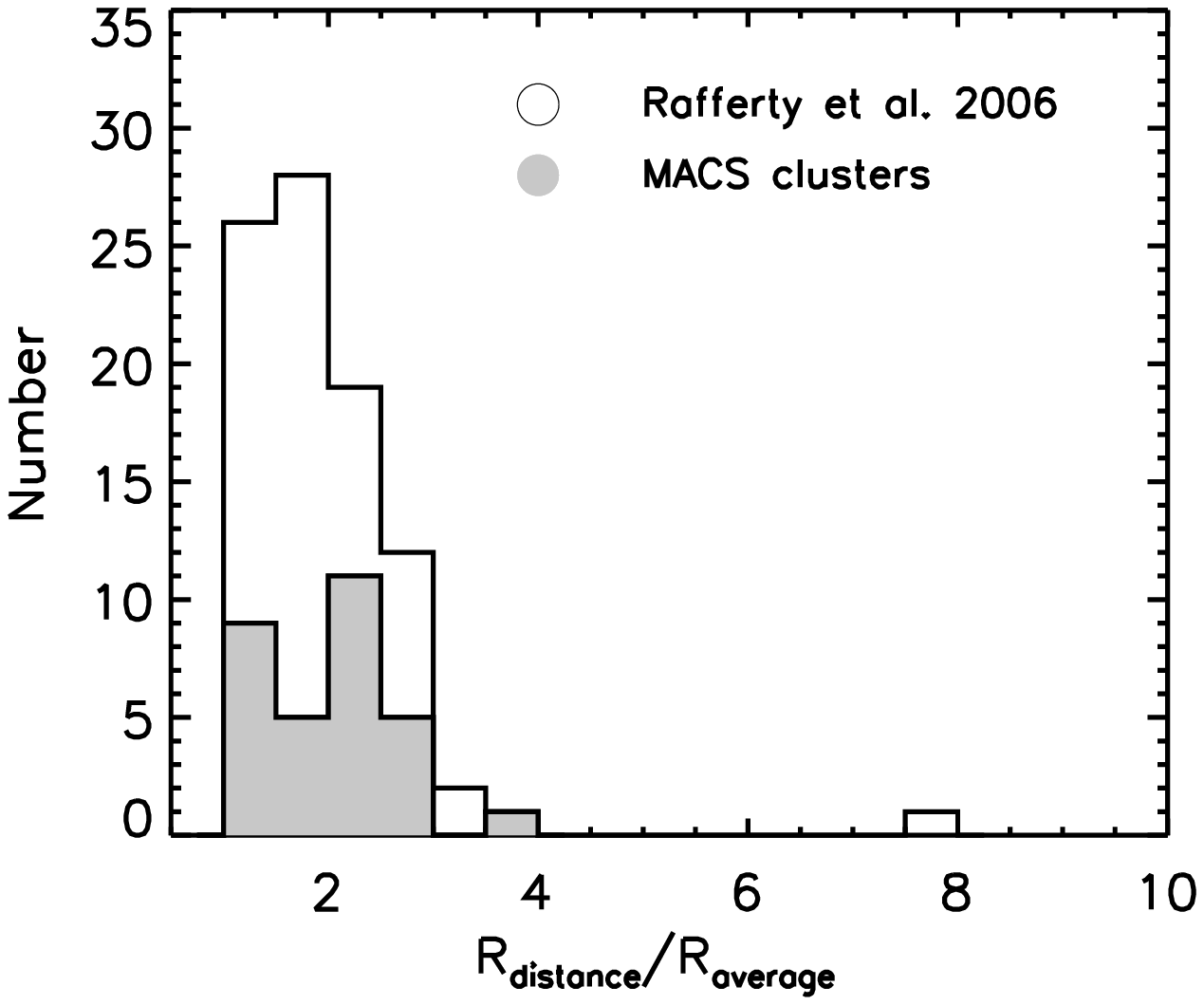}
\end{minipage}
\begin{minipage}[c]{0.325\linewidth}
\centering \includegraphics[width=\linewidth]{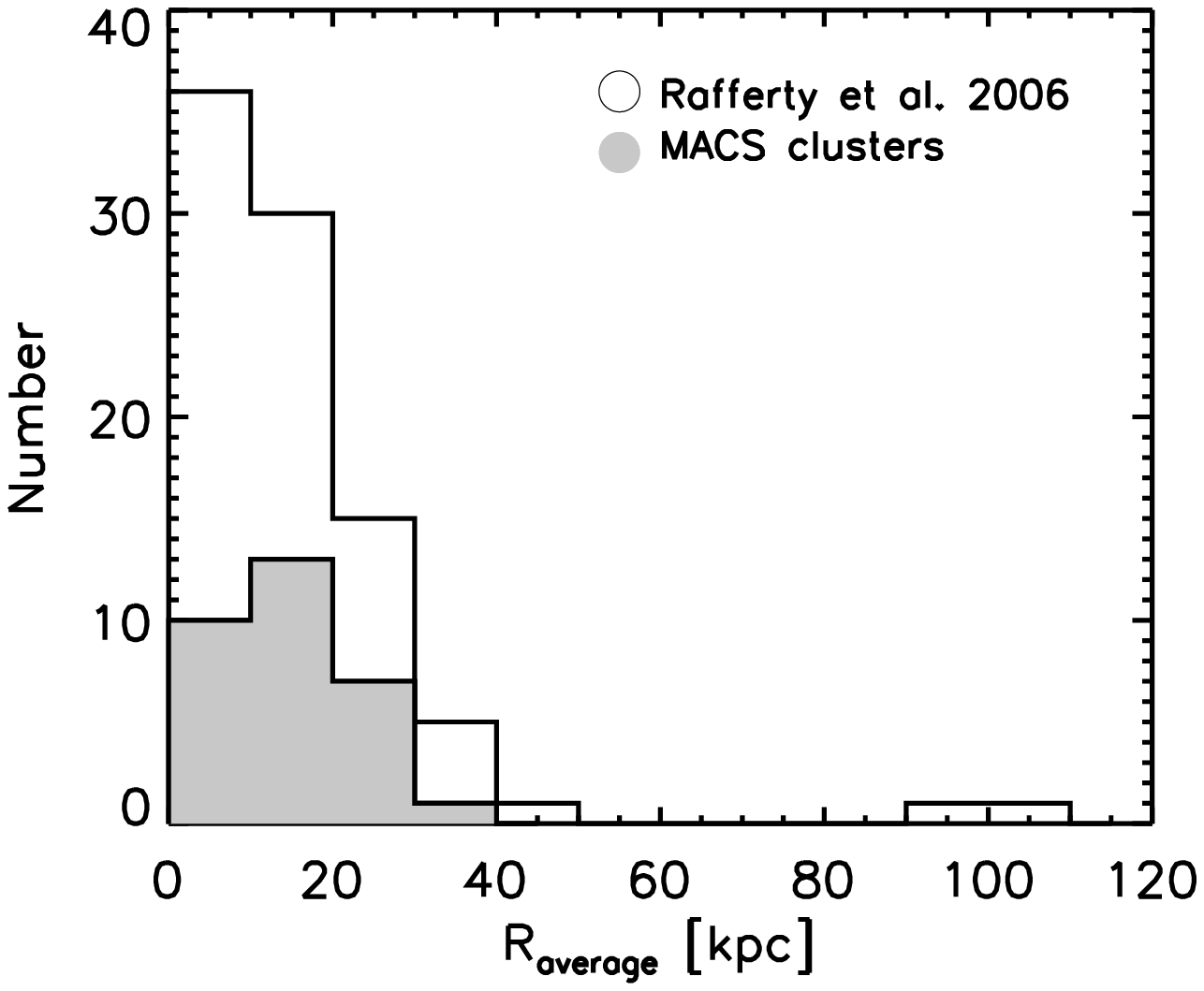}
\end{minipage}
\begin{minipage}[c]{0.325\linewidth}
\centering \includegraphics[width=\linewidth]{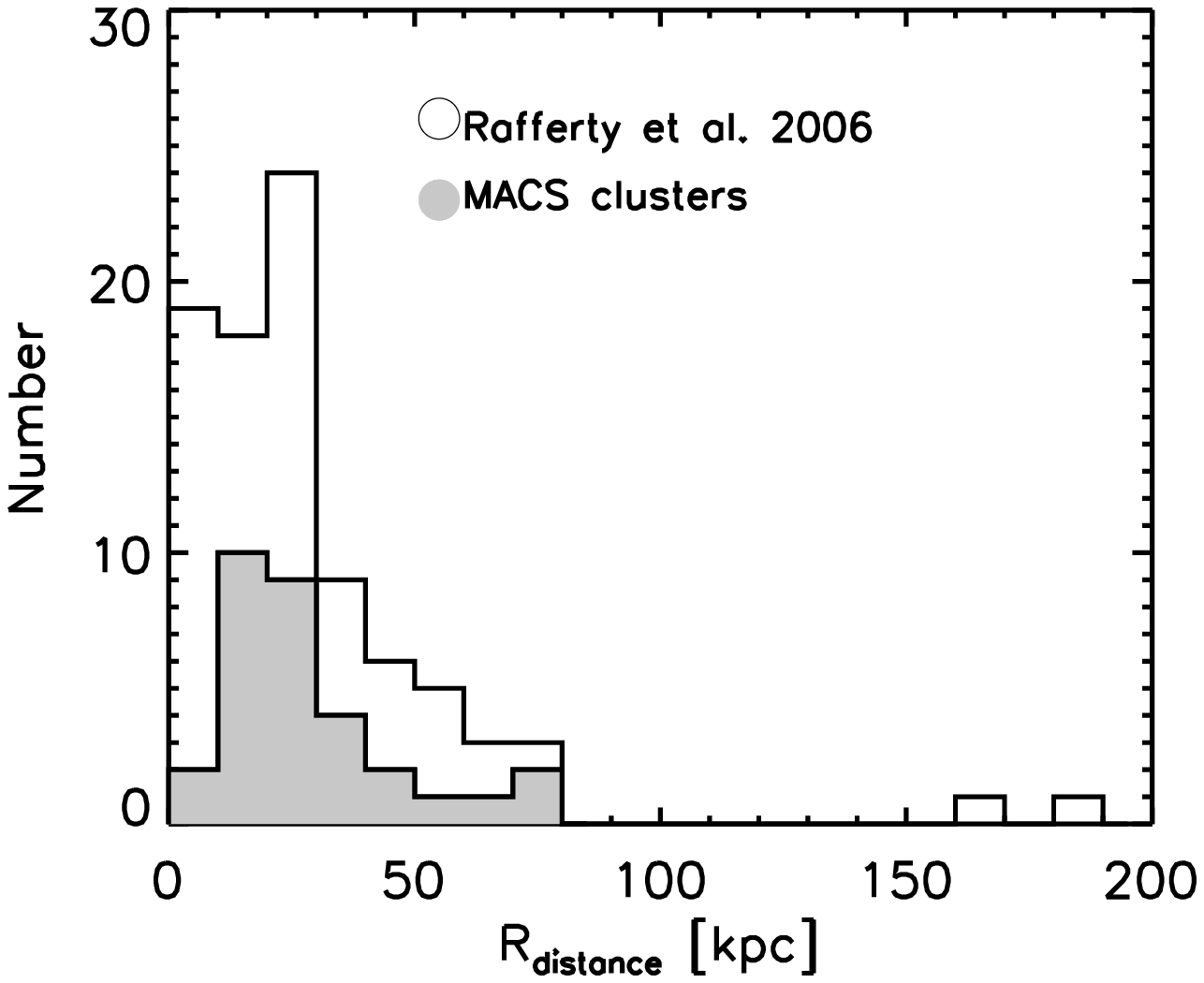}
\end{minipage}
\begin{minipage}[c]{0.325\linewidth}
\centering \includegraphics[width=\linewidth]{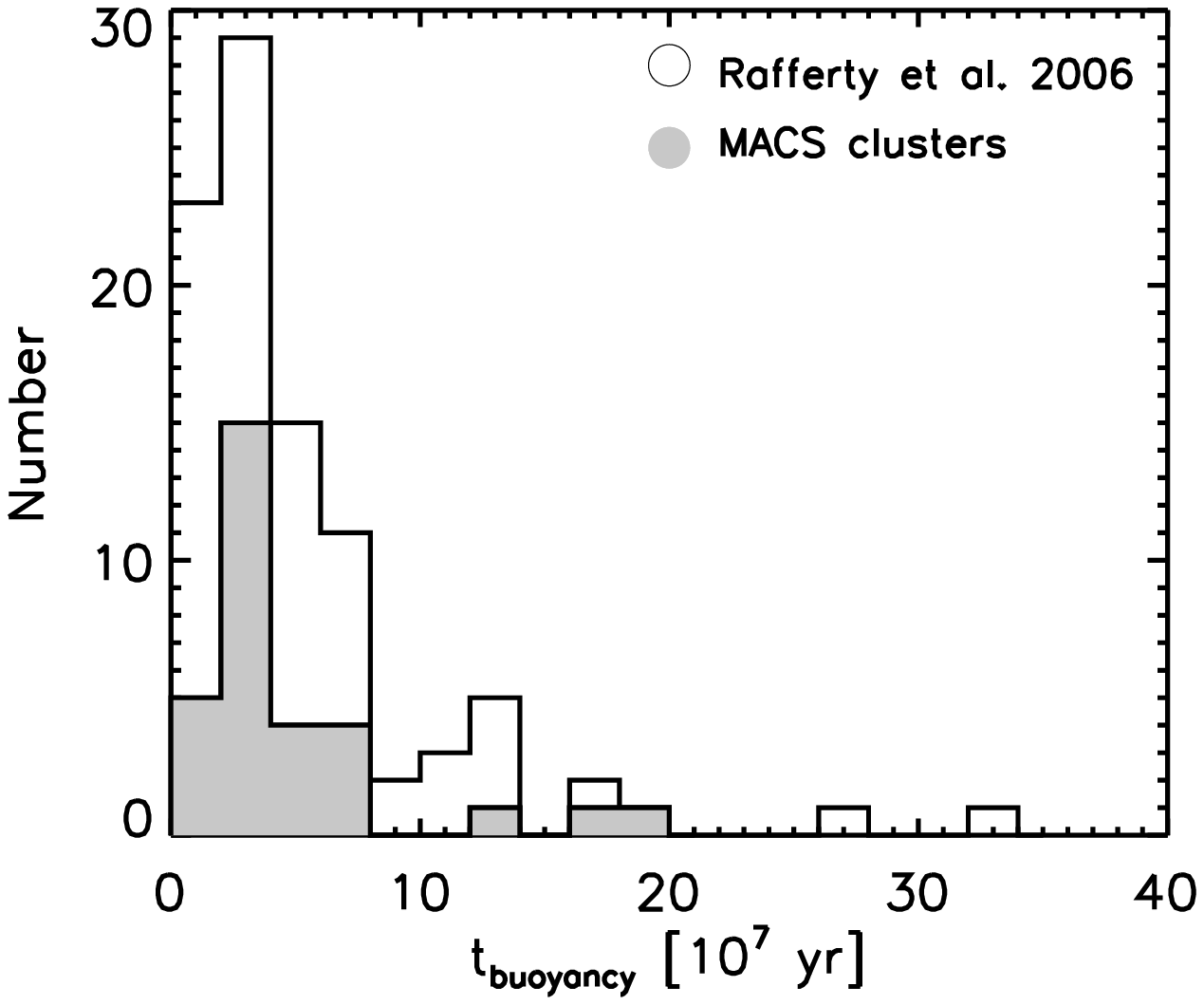}
\end{minipage}
\caption{From left to right and top to bottom: Distribution of the mechanical power being injected by the central AGN, the total energy output stored in the cavities, the ratio of the distance from the core and the average radius of the bubble, the average radius, the distance from the core and the buoyancy rise time. }
\label{fig8}
\end{figure*}

\subsection{Clusters with ``potential" cavities}

We now address the systems with ``potential" cavities. Although we found evidence for cavities in these clusters, many of the X-ray observations have a low number of counts, and deeper observations are needed to confirm if these cavities are real.

\subsubsection{MACS~J0111.5+0855}
We find two cavities associated with the dominant central galaxy in this cluster. They are both quite large (30 kpc in length) and located some 60-70 kpc from the nucleus. However, the data quality for this system is quite poor, and we were not able to deproject the spectra. Furthermore, we were only able to extract two annuli to derive the variations of the thermal properties with radius (see Fig. \ref{figA1}). Although the central cooling time we derive at $r=50$ kpc is below 10 Gyrs, it is not clear whether this cluster has a cool core because the data quality does not allow us to obtain a detailed temperature profile in the inner regions. This cluster also has no central radio source associated with it. If we assume that this cluster has a small cool core, then the cooling luminosity within the cooling radius ($r_{\rm cool}=78$ kpc) is $\sim2\times10^{44}\ergps$. The cavity power is on the order of $P_{\rm buoy}\sim15\times10^{44}\ergps$, but since these cavities are not very well-defined (and only classified as potential cavities), this value could be quite uncertain.

\subsubsection{MACS~J0257.1-2325}
This cluster is located at a redshift of $z=0.5039$ and is part of the 12 most distant clusters detected by the MACS survey at $z>0.5$ \citep{Ebe2007661}. The study by \citet{Kar2008389} looked at the large scale structure in the 12 most distant MACS clusters (including MACS~J1423.8+2404), and found that MACS~J0257.1-2325 has the most complex system in terms of large-scale structure (on Mpc scales), with infalling systems along cosmic filaments. The $HST$ image in Fig. \ref{fig2} shows that the cluster is dominated by a large central galaxy, hosting a bright core and diffuse envelope. We also find many lensed arcs surrounding the central regions ($r<70$ kpc), including a very bright one to the south-west. We find no radio source associated with the central galaxy, and only a mild cooling flow with a central cooling time of $\sim8$ Gyrs. Although, we were able to deproject the spectra, it is not clear if the temperature profile is decreasing in the inner regions because the error bars are still quite large. The cavity we find is located to the north-east some 40 kpc from the nucleus and the power stored within ($P_{\rm buoy}\sim6\times10^{44}\ergps$) is sufficient to prevent the gas from cooling.

\subsubsection{MACS~J1359.1-1929}
This cluster hosts a large central galaxy, surrounded by a diffuse envelope as seen in the $HST$ image, and has a radio source associated with the central regions. We also find a central cooling time of 4 Gyrs and a temperature profile steadily declining in the inner regions, with a cooling luminosity on the order of $4\times10^{44}\ergps$. We were able to identify two cavities in this system, although not well defined and deeper data is needed to confirm their existence. If these cavities are real, then the power stored within them is $P_{\rm buoy}\sim3\times10^{44}\ergps$. The X-ray image also shows 3 small depressions ($\sim6$ kpc in size) immediately surrounding the central X-ray peak, which could also contribute to the heating. 
\begin{figure*}
\centering
\begin{minipage}[c]{0.325\linewidth}
\centering \includegraphics[width=\linewidth]{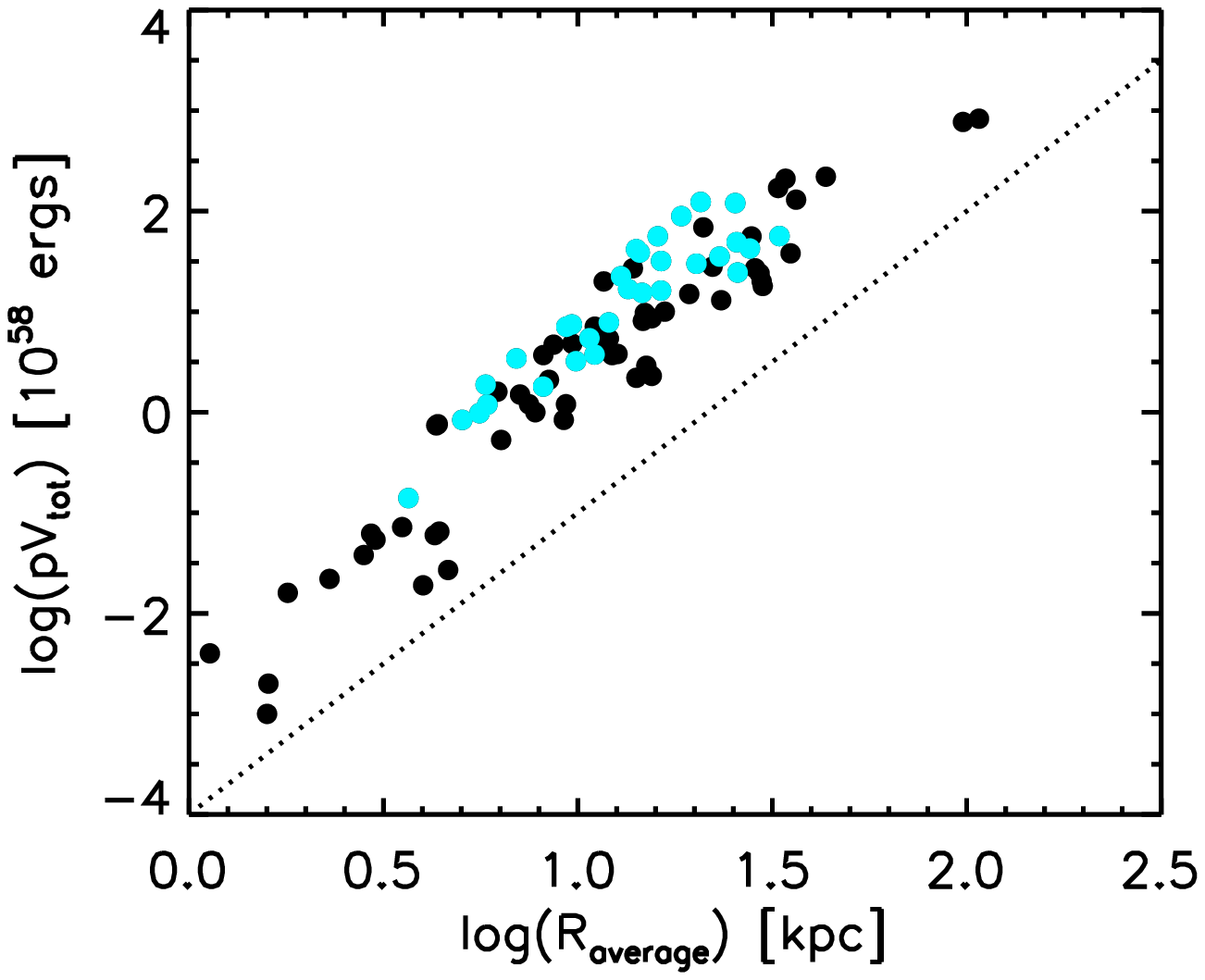}
\end{minipage}
\begin{minipage}[c]{0.325\linewidth}
\centering \includegraphics[width=\linewidth]{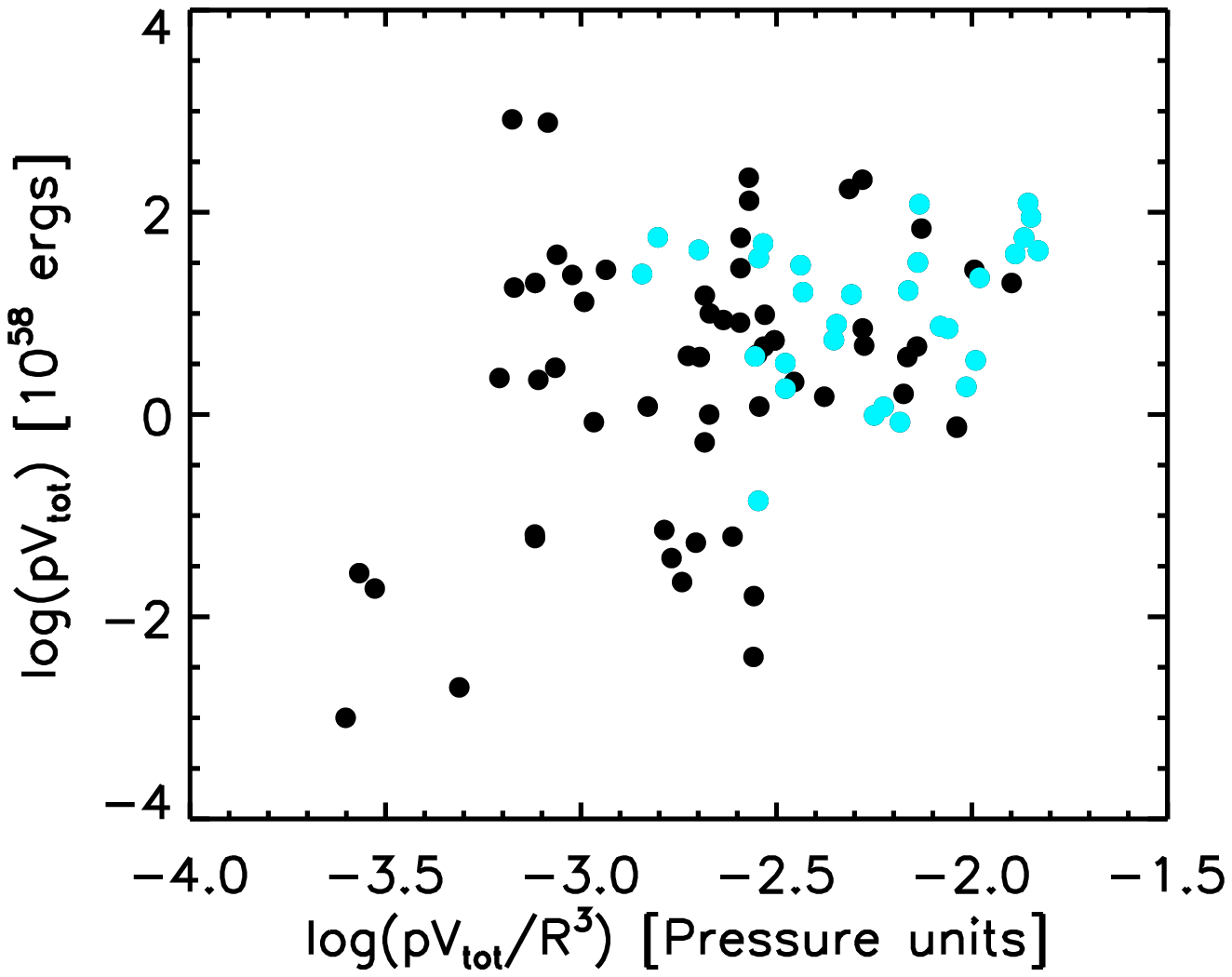}
\end{minipage}
\begin{minipage}[c]{0.325\linewidth}
\centering \includegraphics[width=\linewidth]{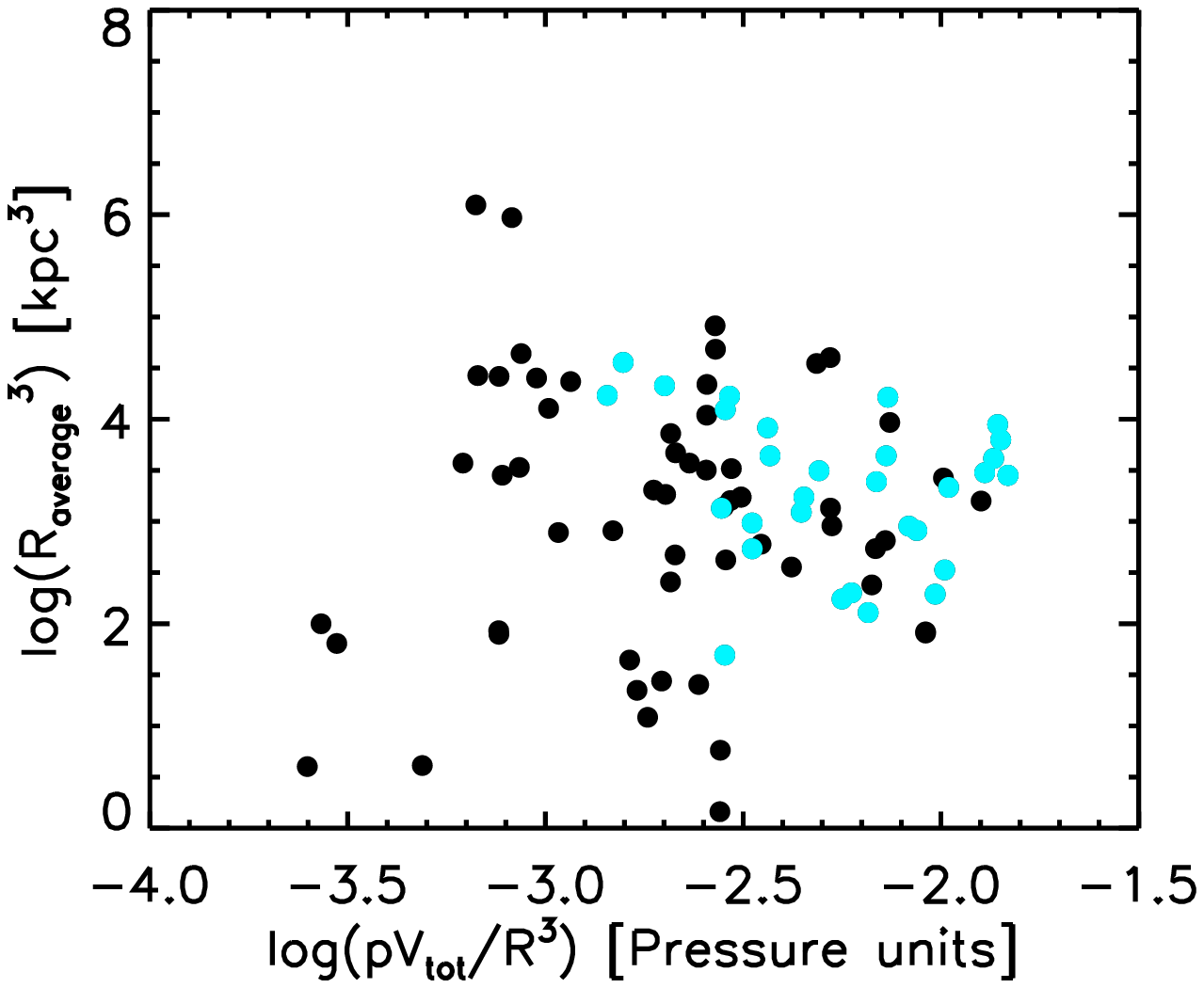}
\end{minipage}
\caption{Left - Energy of \textit{each cavity} as a function its average radius ($R_{\rm average}$). The dotted line shows the slope of 3 one would expect between $pV_{\rm tot}$ and ${R_{\rm average}}$ since $pV_{\rm tot}\propto{R_{\rm average}^3}$. Middle - Energy of \textit{each cavity} as a function of the pseudo-pressure, the latter determined by dividing $pV_{\rm tot}$ by the average radius cubed (${R_{\rm average}}^3$). Right - Average radius cubed (${R_{\rm average}}^3$) of \textit{each cavity} as a function of the pseudo-pressure. The light blue points are for the MACS clusters and the black points are for the \citet{Raf2006652} objects.}
\label{fig9}
\end{figure*}

\subsubsection{MACS~J1359.8+6231}
MACS~J1359.8+6231 (MS 1358.4+6245 or EMSS 1358+6245) is a relaxed cool core cluster known for its lensing properties \citep{Hoe1998504,Fra1997486,All1998296,Zit2011413} and detailed mass estimates \citep{Ara2002572}.

We have identified one small cavity to the north-east located near the central galaxy. The optical image shows that the central regions are dominated by a large galaxy with a bright core and diffuse envelope. The power capabilities of the cavity are quite small with $P_{\rm buoy}\sim5\times10^{42}\ergps$ and are not capable of preventing the cool core from cooling by more than an order of magnitude. This cluster stands out in the $L_{\rm mech}-L_{\rm cool}$ plot shown in Fig. \ref{fig6} since the mechanical energy within the cavity fails to meet the power needed to offset cooling.    

\subsubsection{MACS~J1447.4+0827}
MACS~J1447.4+0827 (RBS 1429) has been recently-added to the MACS cluster sample. It is an ultra-luminous cluster (the most X-ray luminous in our sample with cavities; $L_{\rm X}\sim5\times10^{45}\ergps$) and has a very strong cool core, therefore requiring extreme mechanical feedback from its central AGN ($L_{\rm cool}\sim3-4\times10^{45}\ergps$). With just $\sim10$~ks on the source, we find evidence of a large cavity, although not well defined. The energetics of the cavity in itself are substantial, and the power stored within this outburst is enough to prevent the gas from cooling. This cluster has the most energetic outburst in our sample in terms of enthalpy ($\sim10^{60}$ erg), making it one of the most powerful outbursts known.

\subsubsection{MACS~J2135.2-0102}
MACS~J2135.2-0102 (1RXS J213515.7-010208) is a luminous cluster known for its lensing properties. An earlier study found a series of arcs in the shape of a ``Cosmic Eye" and ``Cosmic Eyelash" associated with this cluster \citep[see][]{Sma2007654,Sia2009698,Swi2010NAT}. However, the data quality of the $Chandra$ X-ray image is very poor, and we were not able to deproject the spectra. We find no radio source associated with the central regions and that the central cooling time is more than 10 Gyrs. There is however at least one depression in the X-ray image within the inner regions (see Fig. \ref{fig2}). Initially, we identified two cavities, one to the north-west and another to the south-east. However, as mentioned in Section 3, the BCG is located almost within the south-eastern cavity. This could be due to projection effects, but we choose to discard this cavity and only consider the north-western one. If the cavity to the north-west is real, then the power capability is quite large ($P_{\rm buoy}\sim4\times10^{44}\ergps$). The optical image also shows an interesting structure with a central galaxy hosting 2 or 3 potential nuclei.

\subsubsection{MACS~J2245.0+2637}
Finally, we present the results for MACS~J2245.0+2637. The $HST$ image shows a central dominant galaxy, with a very bright core. This cluster has a radio source associated with the central regions, as well as a short central cooling time of $\sim4$ Gyrs ($L_{\rm cool}\sim6\times10^{44}\ergps$). We find one cavity in this system, but the data quality remains poor. The cavity is also quite small ($\sim$8 kpc in size) and the power stored within it is only $P_{\rm buoy}\sim1\times10^{44}\ergps$ or $P_{\rm c_{\rm s}}\sim2\times10^{44}\ergps$. This cluster is also part of the 34 most luminous MACS clusters \citep{Ebe2010407}. 

\begin{figure}
\centering
\begin{minipage}[c]{0.9\linewidth}
\centering \includegraphics[width=\linewidth]{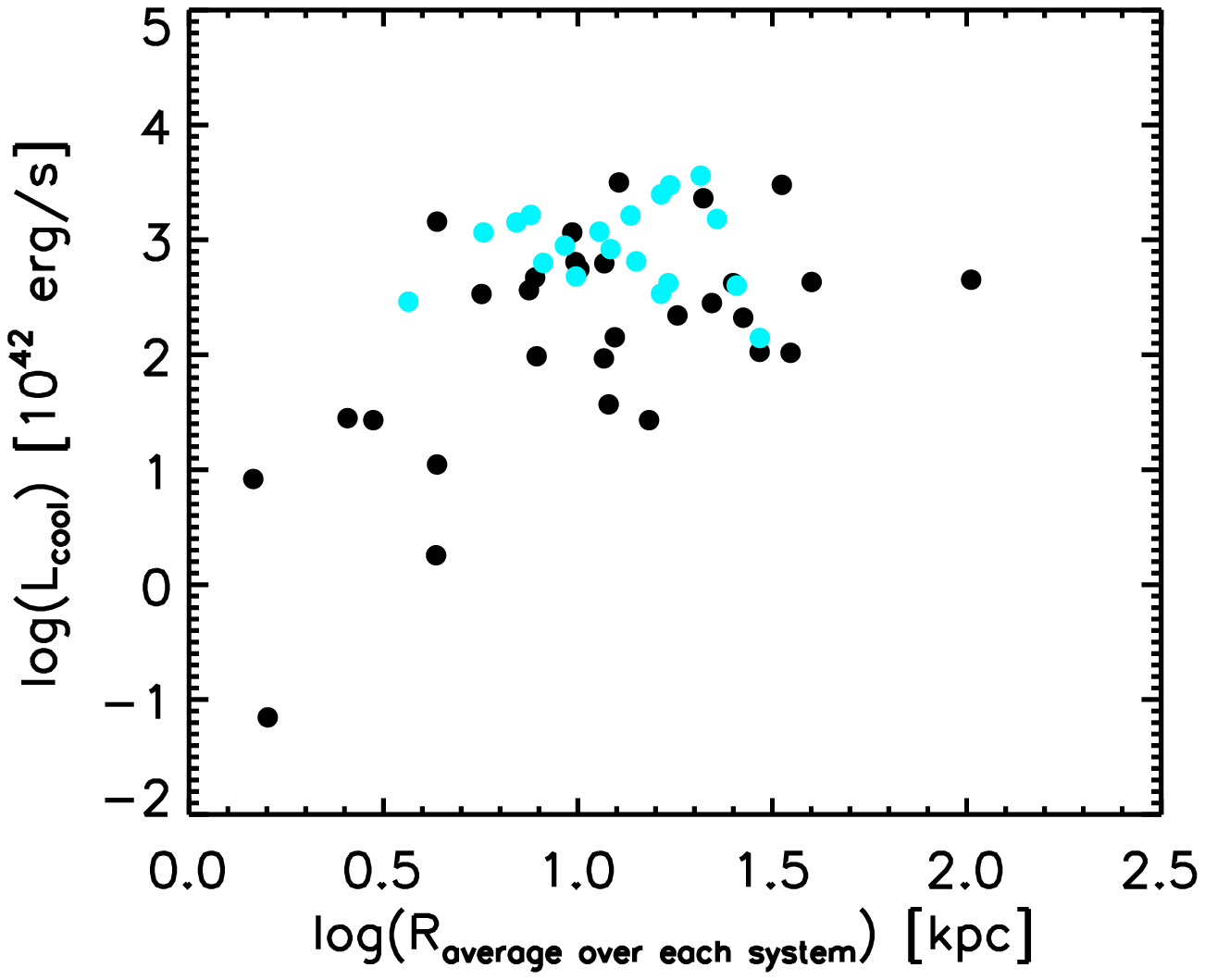}
\end{minipage}
\caption{Shown is the cooling luminosity of the system as a function of the average cavity radius in each system (if there are two cavities in one system, then we calculate and plot the average radius between the two cavities). The light blue points are for the MACS clusters and the black points are for the \citet{Raf2006652} objects.}
\label{fig10}
\end{figure}

\section{Discussion}

\subsection{Selection effects}
We have analysed the $Chandra$ X-ray observations of the MACS cluster sample and searched surface-brightness depressions in these systems. The MACS cluster sample consists of luminous, massive clusters all above a redshift of $z>0.3$. Out of the 76 clusters, we find that 13 have ``clear" surface-brightness depressions associated with their central AGN. We interpret these depressions as cavities being inflated by the central SMBH, which then rise buoyantly into the outskirts. If we include the 7 systems in which we find ``potential" surface-brightness depressions, then the detection rate goes up to 26 per cent. We stress that many of the clusters in which we identified cavities were those that had the ``deepest" $Chandra$ observations (more than 30~ks), and that many of the remaining MACS clusters only have $10-20$~ks observations (or as little as 500-1000 counts within a radius of 200 kpc). Our detection rates should therefore be regarded as lower limits. 

The 76 clusters that have been observed with $Chandra$ were mostly selected on a flux limited basis \citep{Ebe2007661,Ebe2010407}, and should therefore not be more biased towards cool cores than non cool cores. As shown in Fig. \ref{fig4}, we find the same percentage of cool core clusters in the MACS sample (with $Chandra$ data) than non cool core clusters. However, the bottom panel of Fig. \ref{fig5} shows that most of the clusters with deep observations are cool core clusters. This is because many of those that were initially observed for short exposure times and that showed interesting features, such as a peaked X-ray distribution, were followed-up with longer observations. We could therefore be missing many cavities in the non cool core systems which do not have long observations. However, as shown by earlier studies \citep{Bir2004607,Dun2005364,Raf2006652,Dun2006373,Dun2008385}, X-ray cavities seem to lie predominantly in systems that require some form of heating, i.e. cool core clusters. Note also, the $Chandra$ observations targeted the brightest clusters in the sample. We might therefore be missing many of the less energetic outflows found in the less luminous clusters. This does not affect our results, since we draw our conclusions only on the powerful AGN outbursts and compare these to the powerful outbursts at lower redshift, both of which have $L_{\rm mech}\apgt10^{44-45}\ergps$. 

Finally, we mention that although the MACS sample lies within $0.3\leq{z}\leq{0.7}$, the number of clusters with $Chandra$ data decreases steadily with redshift and there are only 4 MACS clusters with $Chandra$ data beyond $z>0.55$. The probability of detecting cavities in systems beyond $z=0.55$ is therefore low considering the limited sample. According to our cosmology, $1''$ corresponds to 4.45 kpc at $z=0.3$ and 7.15 kpc at $z=0.7$. Cavities have typical sizes of 20 kpc in radius (see Fig. \ref{fig8}), whereas $Chandra$ has a pixel size of $\sim0.49''$ and a point spread function of $\sim1''$. A cavity of 20 kpc in radius located at $z=0.3$ would therefore spread across $4.5''$ (or 9 pixels), and if the same cavity were located at $z=0.7$, it would spread across $3''$ (or 6 pixels). Although deeper observations would be needed, and we would be pushing the unique resolution of $Chandra$ to its limit, we could in principle identify cavities in $z=0.6-0.7$ clusters, and might therefore be missing cavities in these further systems.  

\subsection{Detection rates}
In more recent years, there has been considerable interest in studying the correlation between jet power, as determined from cavity energetics, and radio luminosities \citep[e.g.][ and see Fig. \ref{fig7} here]{Bir2004607,Bir2008686,Cav2010720,OSu2011735}. This relation extends from individual elliptical galaxies to groups and even clusters of galaxies, and is interesting for two reason. First, it provides insight into the nature of jets \citep[e.g.][]{Wil1999309}, and secondly, it provides a means to determine jet powers from the radio properties alone \citep[e.g.][]{Bes2007379,Mag2007379}. Although the scatter remains large and more effort is needed to fully understand its nature, this method could in principle be used to determine jet powers in more distant clusters where the X-ray image quality remains too poor to identify surface-brightness depressions \citep[see][ who looked at AGN heating in the 400SD survey using this relation]{Ma2011}. However, deriving jet powers directly from X-ray cavities remains the most direct and reliable method. 

\citet{Bir2004607} initially compiled a study of 80 systems with $Chandra$ X-ray archival data, in which the authors visually searched for surface-brightness depressions associated with the central regions. They found 18 systems with well-defined cavities (16 clusters of galaxies; one galaxy group, HCG 62; and one giant elliptical, M84), corresponding to a detection rate on the order of 20 per cent. \citet{Raf2006652} then followed-up this study, by adding 13 more objects where cavities had been recently discovered. Although many targets overlap between the various samples, \citet{Dun2006373} also provided an in depth study of X-ray cavities in the Brightest 55 sample of galaxy clusters \citep{Edg1990245}, which was then followed by a more extensive study including the Brightest Cluster Sample \citep{Ebe1998301} in \citet{Dun2008385}. The total sample of \citet{Dun2006373,Dun2008385} consists of 71 clusters in the redshift range $0\leq{z}\leq{0.4}$. The authors find 22 systems with clear cavities, similar to our detection rate if we include the ``potential" cavities ($\sim25$ per cent). However, \citet{Dun2006373} concluded that cavities sit mostly in clusters which require some form of heating, i.e. cool core clusters. They find that at least 14 of 20 (70 per cent) cool core clusters have cavities, making the detection rate in these systems very high. Since the majority of their clusters have an X-ray luminosity above $10^{44}\ergps$, similar to the MACS clusters, we can directly compare our rates with theirs. We find that 20 (26 per cent), or at least 13 (17 per cent) of the 76 MACS clusters have cavities. We also find that the majority of our systems with cavities have a short central cooling time (see Fig. \ref{fig3} and Fig. \ref{fig4}). If we consider only the 37 MACS clusters that require some form of heating with $t_{\rm cool(r<50kpc)}<3$ Gyrs, then we find that 19 have cavities. The detection rate is therefore 50 per cent in cooling systems, but could be higher considering that more than half of the remaining 18 MACS clusters with $t_{\rm cool(r<50kpc)}<3$ Gyrs have less than 30~ks observations (Fig. \ref{fig5}). Note also, cool core clusters are more likely to be the ones with bright X-ray cores, since the central densities are high. When a cavity is present, the contrast it creates in these peaked cores is higher, which makes the identification of cavities easier. We might therefore be missing cavities in non cool core systems.

More recently, studies have focussed on extending our knowledge of AGN feedback to the general population of giant ellipticals \citep[e.g.][]{Fin2001547,Jon2002567,Mac2006648,Nul2007,Dun2010404}. \citet{Nul2007} searched through an extensive sample of 160 nearby giant ellipticals, and found 109 with significant emission from surrounding hot gas. From these, they find that 27 have X-ray cavities. This corresponds to a detection rate of only 25 per cent in these cooling systems (compared to 70 per cent in cool core clusters), suggesting that duty cycles are greater in larger systems (see Jones et al., in preparation).

\begin{figure}
%\centering
\begin{minipage}[c]{0.99\linewidth}
\includegraphics[width=\linewidth]{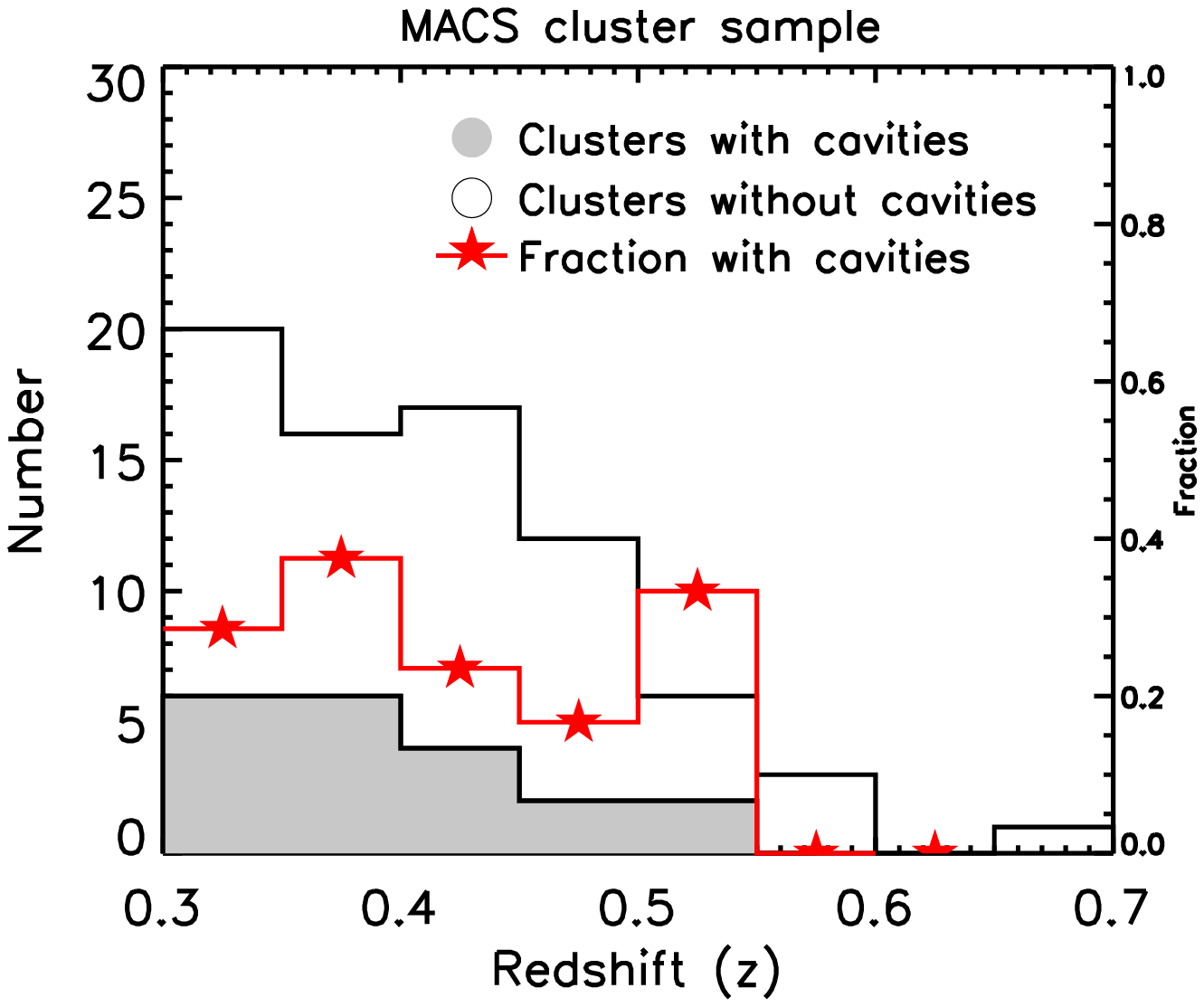}
\end{minipage}
\begin{minipage}[c]{0.99\linewidth}
\includegraphics[width=\linewidth]{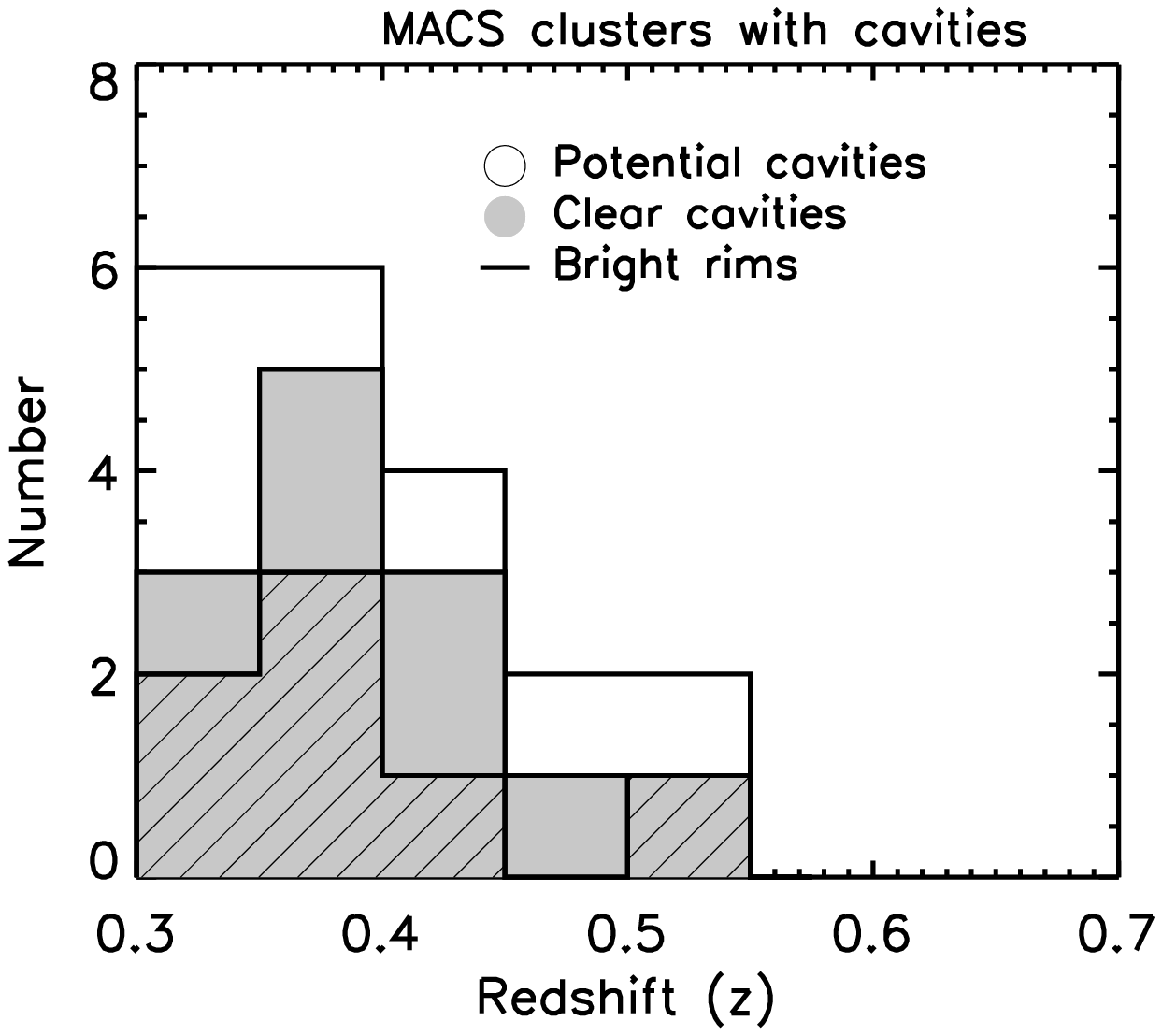}
\end{minipage}
\caption{In the top panel, we show the distribution of clusters in the MACS sample as a function of redshift. We outline those in which we identified cavities (also shown in more detail in the bottom panel). The fraction of clusters with cavities for each redshift bin is shown in red. The fraction remains around $25-30$ per cent until a redshift of 0.55. }
\label{fig11}
\end{figure} 

\subsection{Cavity properties}

In the top panel of Fig. \ref{fig4} we show that most of the systems where we find cavities have a strong cool core with $t_{\rm cool{\rm (r=50kpc)}}<3-5$ Gyrs, therefore requiring extreme feedback from their central AGN to prevent the gas from cooling ($L_{\rm mech}\sim10^{44-45}\ergps$). Although we expect to see some of the most extreme AGN outbursts in this sample, the cavities we find are not larger or more powerful than the counterparts found in nearby ($z<0.3$) clusters with similar $L_{\rm cool}$. 

Fig. \ref{fig6} shows that all but one\footnote[2]{The outlier refers to MACS~J1359.8+6231, where we found a cavity to the north-east, although very small (some $3-4$ kpc in size) and therefore not energetically capable of preventing the gas from cooling.} of the MACS clusters in which we find cavities are consistent with the $L_{\rm mech}-L_{\rm cool}$ correlation, implying that highly-luminous clusters are able to prevent the gas from cooling at the same level than the less-luminous clusters and groups. By contrast, \citet{Nul2007} provided evidence of a turnover at the lower luminosity end, with outbursts being more powerful for a given cooling luminosity. This is consistent with the idea that in smaller systems, the duty cycles are smaller (only $\sim25$ per cent of the time, compared to $\sim70$ per cent in cool core clusters), requiring that the outbursts be more energetic for a given cycle. \citet{Dun2008385} looked at the cavity properties of 6 distant clusters ($0.1\leq{z}\leq{0.4}$) and found that they all sit above the line of exact balance in the $L_{\rm mech}-L_{\rm cool}$ correlation. Our more complete study of distant clusters finds differently. Essentially, our clusters agree well with the $L_{\rm mech}-L_{\rm cool}$ correlation with some above and some below the line of exact balance. There is therefore no turnover at the higher luminosity end.

In Fig. \ref{fig8}, we show the distribution of mechanical power ($L_{\rm mech}$), cavity energetics ($pV_{\rm tot}$) and buoyancy rise time ($t_{\rm buoy}$), as well as the cavity radius ($R_{\rm average}$) and distance from the nucleus ($R_{\rm distance}$). According to these figures, we find that the MACS cluster outbursts follow the same trends as the clusters in \citet{Raf2006652}. The \citet{Raf2006652} sample consists of clusters that are $z\aplt0.3$, some of which require the same heating powers as the MACS clusters to prevent the gas from cooling ($L_{\rm mech}\sim10^{44-45}\ergps$). In other words, the MACS outbursts are powerful, but not more powerful than the lower redshift counterparts. 

It is also important to emphasize that we can only interpret the results concerning the largest and most powerful outbursts. We are probably missing a number of smaller cavities which could not be detected since our clusters are at a relatively high redshift and the resolution of $Chandra$ can only go so far. We will have to wait for future missions with better spatial resolution than $Chandra$ to study the small cavities at high redshift.  

Fig. \ref{fig8} also shows that the $pV_{\rm tot}$ distribution rises smoothly, peaks at $10^{59-60}$ erg and then decreases steeply. By contrast, $L_{\rm mech}$ seems to rise and decrease more smoothly, although this distribution would be altered if we included the giant ellipticals for which $L_{\rm mech}$ peaks at $10^{41-42}\ergps$. Since $pV_{\rm tot}$ depends only on the thermal pressure and cavity size, the sudden drop at $10^{59-60}\erg$ might be telling us something about how cavities form. As they are being inflated by the central AGN, it could be that they reach a maximum size in which they no longer can expand no matter the physical conditions. There could therefore be a maximum $pV_{\rm tot}$ value that can be reached in a single outburst. To provide further insight into this idea, we plot in the left panel of Fig. \ref{fig9} the energy in each cavity ($pV_{\rm tot}$) as a function of the average radius (${R_{\rm average}}=(R_{\rm w}\times{R_{\rm l})^{0.5}}$). Since $pV_{\rm tot}\propto{R_{\rm average}}^3$, we expect a correlation between these two quantities, and in a log-log plot, the slope should be 3 (this is shown with the dotted-line). However, this plot also shows that the most powerful outflows can only be created by the largest cavities. If a small cavity was being created in a high-pressure environment, the resulting $pV_{\rm tot}$ could be large. However, Fig. \ref{fig9} shows that we do not observe such cavities. Furthermore, the middle and right panels of Fig. \ref{fig9} show that powerful outbursts are created in a variety of pressure environments, and that the size of each cavity has no dependence on the surrounding pressure. In other words, large cavities are created both in high and low pressure environments and the maximum allowed cavity radius in clusters does not depend on the surrounding pressure. By contrast, one could argue that large cavities can only be created if the surrounding thermal pressure is not large. Fig. \ref{fig9} suggests otherwise. Finally, in Fig. \ref{fig10}, we plot the cooling luminosity as a function of the average cavity radii in each system. \citet{Dun2008385} provided some evidence for a trend between these two quantities, and although the scatter remains large, our data support this idea (the larger the X-ray cooling, the larger the cavities). Again, this emphasised that somehow the central AGN knows about the X-ray cooling and provides the cavities needed to offset cooling of the ICM down to a fine level.

As shown by Fig. \ref{fig8}, the distribution of cavity radii and distances from the nucleus also suggest that extremely large and distant cavities, such as those found in MS 0735.6+7421 \citep{McN2005433,Git2007660}, are rare (these cavities are the outliers seen in the $R_{\rm average}$ and $R_{\rm distance}$ distributions of Fig. \ref{fig8}). Furthermore, $R_{\rm distance}/R_{\rm average}$ peaks strongly at a value of 2 and then falls off quickly. In other words, cavities travel up to a distance of $2{\times}R_{\rm average}$ (i.e. their diameter) and then either disintegrate or, more likely, become too difficult to detect \citep{Mcn200745}. More distant cavities have been detected in the Perseus cluster, but only through very deep X-ray observations \citep[$\sim500$~ks;][]{San2007381,Fab2011}.   

Finally, we mention that the shapes of the cavities we find vary from almost perfectly circular to elongated and even flattened (e.g. MACS~J2046.0-3430). Two clusters also seem to have a misalignment between the X-ray peak and central galaxy (MACS~J2046.0-3430, MACS~J2135.2-0102), which may simply be due to projection effects. If a cavity is viewed along a different line of sight, projection effects can alter its shape and make it difficult to detect. The cavities we find may therefore be much larger and even more numerous than what we find \citep[see also][]{San2009393}. There are also two MACS clusters with long observations ($>100$~ks) and short central cooling times ($t_{\rm cool}(r<50\rm{kpc})<$ 3 Gyrs), but no cavities (MACS~J1311.0-0311 at $z=0.49$ and MACS~J1621.3+3810 at $z=0.465$; see top panel in Fig. \ref{fig5}). Statistically, we expect that some systems will have a jet axis aligned with the line of sight, rendering the cavities almost impossible to detect. This could explain why we do not see cavities in these two systems.

\begin{figure*}
\centering
\begin{minipage}[c]{0.45\linewidth}
\centering \includegraphics[width=\linewidth]{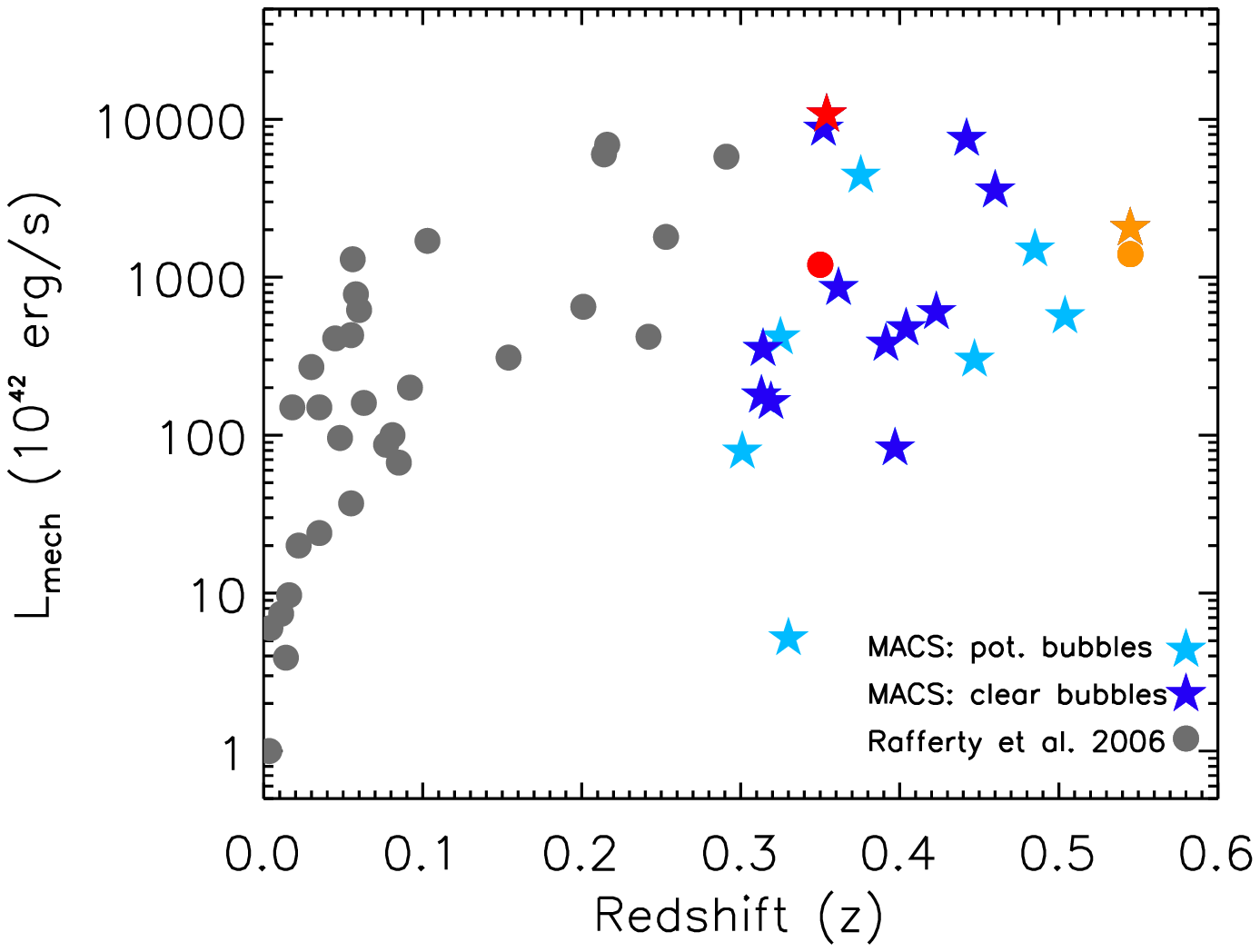}
\end{minipage}
\begin{minipage}[c]{0.45\linewidth}
\centering \includegraphics[width=\linewidth]{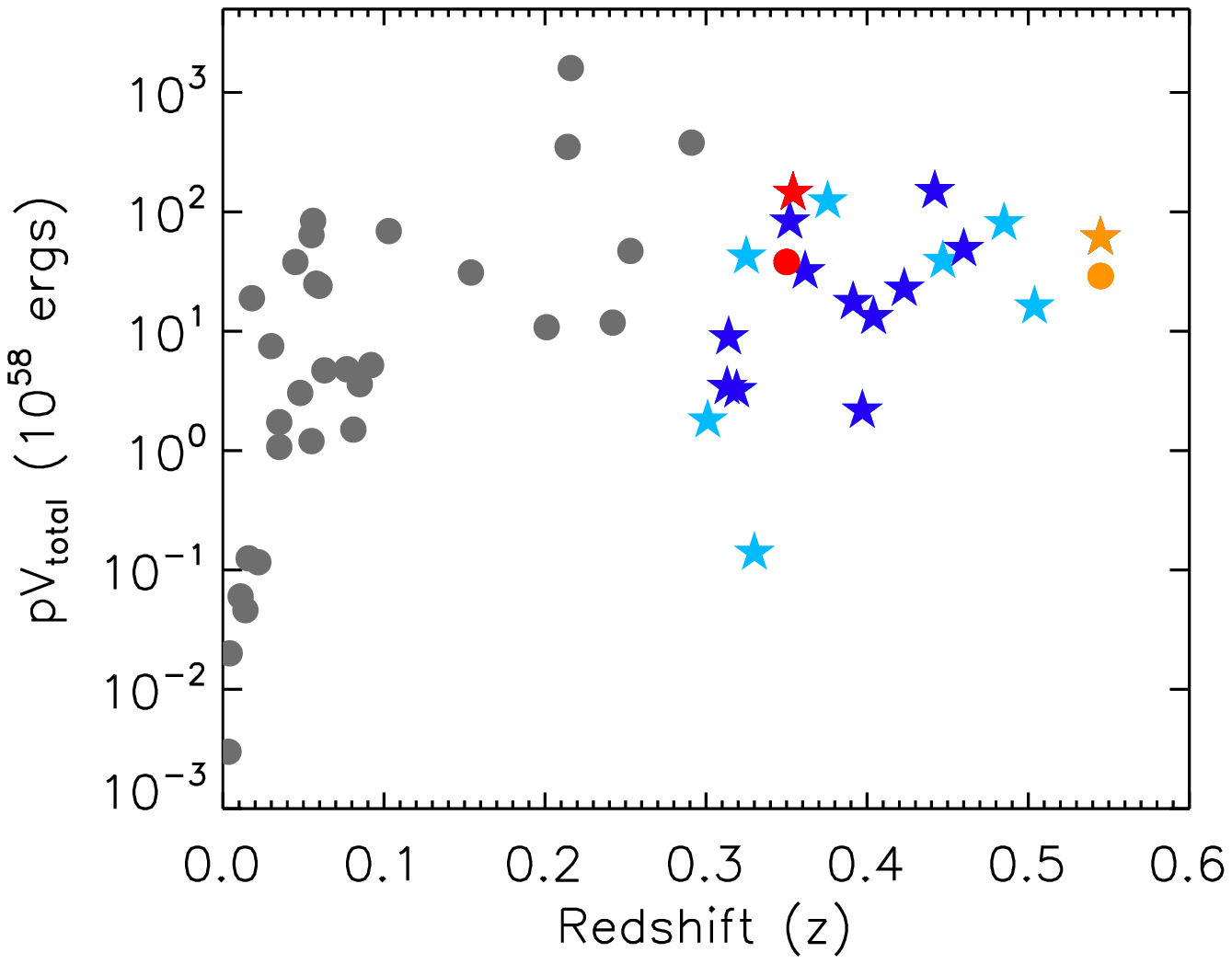}
\end{minipage}
\begin{minipage}[c]{0.45\linewidth}
\centering \includegraphics[width=\linewidth]{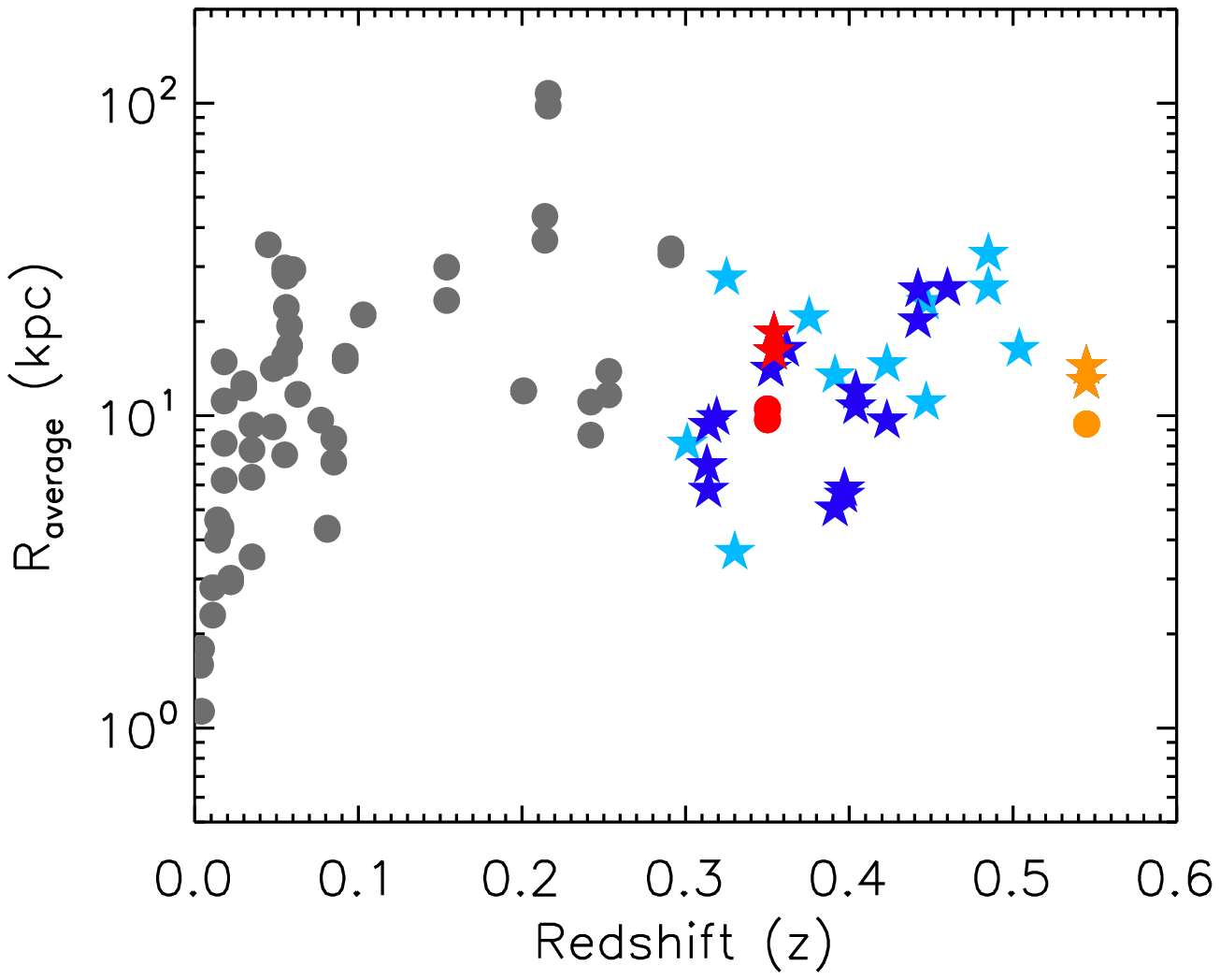}
\end{minipage}
\begin{minipage}[c]{0.45\linewidth}
\centering \includegraphics[width=\linewidth]{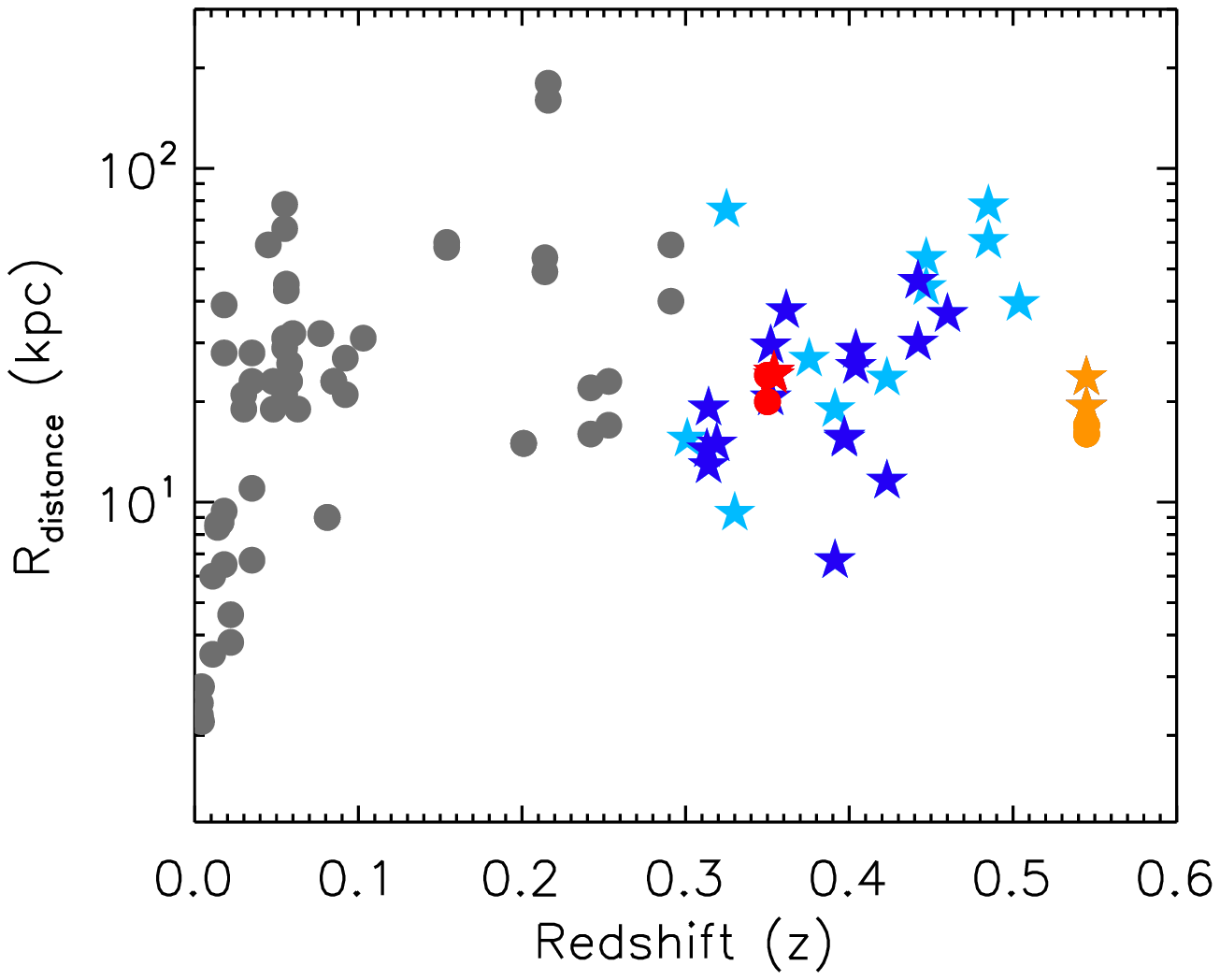}
\end{minipage}
\caption{Profiles of the mechanical energy being injected by the central AGN (top-left), the total energy output stored in the cavities (top-right), the average radius of a bubble (bottom-left) and distance from the core to the cavity centre (bottom-right) as a function of redshift. The MACS clusters are shown with the star symbols (dark blue for ``clear" cavities and light blue for ``potential" cavities), and those of \citet{Raf2006652} are shown with the grey circles. Objects common to both samples are shown in red and yellow.}
\label{fig12}
\end{figure*}

\subsection{Evolution of cavity properties}

The MACS sample consists of clusters within $0.3\leq{z}\leq{0.7}$, for which we were able to find surface-brightness depressions in objects up to $z\sim0.6$. In the top panel of Fig. \ref{fig11} we show the fraction of MACS clusters with cavities as a function of redshift, and in the bottom panel we show the redshift distribution of those with cavities. This fraction remains constant around $0.25-0.30$ up to $z\sim0.6$, and is consistent with the fraction found in previous cavity studies by \citet[][ $\sim30$ per cent]{Bir2004607} and \citet[][ $\sim25$ per cent]{Raf2006652}. Therefore, we find no significant evolution in the fraction of clusters with cavities as a function of redshift. 

An interesting finding in this study has been the lack of any significant evolution in the cavity properties. This is shown in Fig. \ref{fig12}, where we plot the mechanical and energy outputs ($L_{\rm mech}$ and $pV_{\rm tot}$), as well as average cavity radius ($R_{\rm average}$) and distance from the nucleus ($R_{\rm distance}$) for each cavity, as a function of redshift. This figure shows that at least for the largest and most powerful outbursts, there is no evidence for evolution in the cavity properties. These cavities are not larger (or smaller) at higher redshift, and the distance to the nucleus remains on average the same, i.e. cavities at higher redshifts are not able to rise more (or less) than their lower redshift counterparts. The cavity energetics of the most powerful outbursts also remain the same. 

Therefore, AGNs lying at the centres of clusters have not only been able to power extreme mechanical outbursts as early as $\sim7-8$ Gyrs after the Big Bang (corresponding to a redshift of $z\sim0.6$), but they have also been able to maintain these extreme outbursts up to present day ($L_{\rm mech(max)}\sim10^{45}\ergps$). In a cosmological context, this means that ``radio mode" feedback has been in place for at least the past $5$ Gyrs and shows no sign of evolution since then. Based on the CDM cosmological simulations of \citet{Cro2006365}, ``radio mode" feedback occurs at late times and involves a phase where the AGN is accreting at sub-Eddington rates and driving powerful outflows that suppress cooling flows from $z\sim1$ to present. In this phase, it is thought that black holes do not grow significantly in mass. However, in order to maintain the powerful MACS outbursts for the past 5 Gyrs, the black hole mass must have been initially very high. If we equate the energy required to power the outflows over the last 5 Gyrs (assuming an average output of $\sim10^{45}\ergps$), to the energy released by a black hole ($\eta{M_{\rm BH}}c^2$, assuming an efficiency of $\eta=0.1$) then it takes at least a $10^9M_{\odot}$ black hole to power such outflows. Here we assume that the duty cycles of dominant cluster galaxies are high ($>70-90$ per cent), as supported by the observation that most or even all of the clusters that require some form of heating to counterbalance cooling have jetted outflows \citep[][]{Dun2006373}. Note also, at least 50 per cent of our clusters requiring feedback from their central AGN seem to have jetted outflows, therefore supporting the idea that duty cycles are high (see Section 9.2). Even with a duty cycle of 50 per cent, the black hole masses required to power the outflows are on the order of $10^9M_{\odot}$ for an efficiency of $\eta\sim0.1$. This rivals the largest known black holes masses, and suggests that some, especially those lying at the centres of the most extreme cooling flows, may be very massive and even ultramassive with $>10^{10}M_{\odot}$. 

Likewise, we could argue that to power outflows of $10^{45}\ergps$ for the past 5 Gyrs, this requires the accreted mass ($M_{\rm acc}=E_{\rm cav}/\eta{c^2}$) to be at least on the order of $10^9M_{\odot}$. If the initial mass of the black hole was $\sim10^8M_{\odot}$, then by accreting some $10^9M_{\odot}$ in mass, we should have seen some variation in the cavity properties with redshift. Indeed, the mass of the black hole is one of the fundamental properties that regulates the power capabilities (another would be the spin). Since we do not see any evidence for evolution in the outflow properties, this strongly suggests that the $10^9M_{\odot}$ increase in mass does not significantly affect the AGN, and therefore that the black holes are initially very massive with $M_{\rm BH}>10^{9-10}M_{\odot}$. 

\citet{Sam2011731} looked at nebular emission in 77 BCGs belonging to the 160SD X-ray cluster survey \citep{Vik1998502,Mul2003594} and found no [O\thinspace{\sc ii}]$\lambda3727$ or H$\alpha$ emission stronger than $15$\AA\ and $5$\AA\ respectively in any of the BCGs. The authors suggest that this may be due to a significant decrease in the number of \textit{strong} cooling flow clusters from $z\sim0.5$ to today (which does not mean that there are no \textit{strong} cool cores, just less). They also suggest that this decline could be due to over-heating of the core by AGN \citep[as supported by][]{Ma2011} or cluster mergers, which happen more frequently at higher redshift. By using the relation between jet power and radio power \citep{Bir2004607,Bir2008686,Cav2010720,OSu2011735}, \citet{Ma2011} find that the power injected by radio sources within $r<250$ kpc is significant compared to the power radiated by the cluster atmosphere. However, as shown in Fig. \ref{fig6}, at least for the central AGN in the MACS clusters, the outbursts we find are able to balance cooling down to a fine level up to $z\sim0.6$. The central AGN in clusters therefore does not seem to overheat the core at higher redshift and can not account for the decrease in strong cool cores with redshift. Other studies in the literature also find a decrease in the number of strong cooling flow clusters with redshift \citep[see][]{Vik2007,San2008483}, but suggest that it is most likely due to the higher major merger rate of the past.          

As a last note, we mention that many of the clusters in which we find cavities have a bright X-ray nucleus associated with the central AGN. These include MACS~J0547.0-3904, MACS~J0913.7+4056, MACS~J0947.2+7623, MACS~J1411.3+5212, MACS~J1423.8+2404, MACS~J1931.8-2634 and MACS~J2046.0-3430. Most of these either have an Fe K line in the spectrum of the nucleus or a spectrum that can best be fitted by a power-law, indicating the presence of an active AGN. BCGs with bright X-ray cores are rare at low redshift, and many that have similar $L_{\rm mech}$ as those in the MACS clusters do not have any evidence of an X-ray nucleus associated with their central AGN \citep{Hla2011}. This may be indicating that we are seeing some evolution, not in terms of the mechanical properties of AGN feedback, but in terms of the radiative properties of the central AGN, thus suggesting that we are starting to see the transition between ``quasar mode" and ``radio mode" feedback. This will be analysed in detail in a forthcoming paper (Hlavacek-Larrondo et al., in preparation). 

\section{Concluding remarks}
We have analysed the $Chandra$ X-ray observations of the MACS cluster sample and searched for surface-brightness depressions associated with the central AGN in each cluster. The MACS sample consists of very X-ray luminous clusters within $0.3\leq{z}\leq{0.7}$ (median $L_{\rm X, RASS}=7\times10^{44}\ergps$). Out of the 76 clusters, we find 13 with ``clear" cavities and 7 with ``potential" cavities, bringing the detection rate to $\sim25$ per cent. Most of the clusters in which we find cavities have a short central cooling time below $3-5$ Gyrs, consistent with the idea that cavities sit predominantly in cool core clusters. By combining our results with those of previous cavity studies, the latter focussing on systems at $z<0.3$, we find no evidence for evolution in any of the cavity properties. Although we expect to see some of the most extreme outbursts ($L_{\rm mech}\sim10^{44-45}\ergps$), the cavities are not larger (or smaller) at higher redshift, and they are not able to rise to further (or lesser) radii. The energy and power capabilities of the most powerful outbursts also remain the same. Extreme ``radio mode" AGN feedback therefore starts to operate as early as $7-8$ Gyrs after the Big Bang and shows no sign of evolution since then. In other words, AGNs lying at the centre of clusters are able to operate at early times and with extreme powers.

\section*{Acknowledgments}
JHL thanks the Cambridge Trusts, Natural Sciences and Engineering Research Council of Canada (NSERC), as well as the Fonds Quebecois de la Recherche sur la Nature et les Technologies (FQRNT). ACF thanks the Royal Society for support. GBT acknowledges support for this provided by the National Aeronautics and Space Administration through Chandra Award Numbers GO0-11139X and GO1-12156X issued by the Chandra X-ray Observatory Center, which is operated by the Smithsonian Astrophysical Observatory for and on behalf of the National Aeronautics Space Administration under contract NAS8-03060. HE gratefully acknowledges financial support from SAO grants GO2-3168X, GO5-6133X, GO9-0146X, and GO0-11137X, as well as STScI grants GO-10491, GO-10875, and GO-12166. This research work has used the TIFR GMRT Sky Survey (http://tgss.ncra.tifr.res.in) data products.

\label{lastpage}
\bibliographystyle{mn2e}
\bibliography{bibli}

\begin{thebibliography}{}

\bibitem[\protect\citeauthoryear{{Allen}}{{Allen}}{1998}]{All1998296}
{Allen} S.~W.,  1998, \mnras, 296, 392

\bibitem[\protect\citeauthoryear{{Allen}, {Rapetti}, {Schmidt}, {Ebeling},
  {Morris} \& {Fabian}}{{Allen} et~al.}{2008}]{All2008383}
{Allen} S.~W.,  {Rapetti} D.~A.,  {Schmidt} R.~W.,  {Ebeling} H.,  {Morris}
  R.~G.,    {Fabian} A.~C.,  2008, \mnras, 383, 879

\bibitem[\protect\citeauthoryear{{Allen}, {Taylor}, {Nulsen}, {Johnstone},
  {David}, {Ettori}, {Fabian}, {Forman}, {Jones} \& {McNamara}}{{Allen}
  et~al.}{2001}]{All2001324}
{Allen} S.~W.,  {Taylor} G.~B.,  {Nulsen} P.~E.~J.,  {Johnstone} R.~M.,
  {David} L.~P.,  {Ettori} S.,  {Fabian} A.~C.,  {Forman} W.,  {Jones} C.,
  {McNamara} B.,  2001, \mnras, 324, 842

\bibitem[\protect\citeauthoryear{{Anders} \& {Grevesse}}{{Anders} \&
  {Grevesse}}{1989}]{And198953}
{Anders} E.,  {Grevesse} N.,  1989, \gca, 53, 197

\bibitem[\protect\citeauthoryear{{Arabadjis}, {Bautz} \& {Garmire}}{{Arabadjis}
  et~al.}{2002}]{Ara2002572}
{Arabadjis} J.~S.,  {Bautz} M.~W.,    {Garmire} G.~P.,  2002, \apj, 572, 66

\bibitem[\protect\citeauthoryear{{Arnaud}}{{Arnaud}}{1996}]{Arn1996101}
{Arnaud} K.~A.,  1996, in {G.~H.~Jacoby \& J.~Barnes} ed., Astronomical Data
  Analysis Software and Systems V Vol.~101 of Astronomical Society of the
  Pacific Conference Series, {XSPEC: The First Ten Years}.
pp 17--+

\bibitem[\protect\citeauthoryear{{Balestra}, {Tozzi}, {Ettori}, {Rosati},
  {Borgani}, {Mainieri}, {Norman} \& {Viola}}{{Balestra}
  et~al.}{2007}]{Bal2007462}
{Balestra} I.,  {Tozzi} P.,  {Ettori} S.,  {Rosati} P.,  {Borgani} S.,
  {Mainieri} V.,  {Norman} C.,    {Viola} M.,  2007, \aap, 462, 429

\bibitem[\protect\citeauthoryear{{Bauer}, {Fabian}, {Sanders}, {Allen} \&
  {Johnstone}}{{Bauer} et~al.}{2005}]{Bau2005359}
{Bauer} F.~E.,  {Fabian} A.~C.,  {Sanders} J.~S.,  {Allen} S.~W.,
  {Johnstone} R.~M.,  2005, \mnras, 359, 1481

\bibitem[\protect\citeauthoryear{{Becker}, {White} \& {Helfand}}{{Becker}
  et~al.}{1994}]{Bec199461}
{Becker} R.~H.,  {White} R.~L.,    {Helfand} D.~J.,  1994, in {D.~R.~Crabtree,
  R.~J.~Hanisch, \& J.~Barnes} ed., Astronomical Data Analysis Software and
  Systems III Vol.~61 of Astronomical Society of the Pacific Conference Series,
  {The VLA's FIRST Survey}.
pp 165--+

\bibitem[\protect\citeauthoryear{{Best}, {von der Linden}, {Kauffmann},
  {Heckman} \& {Kaiser}}{{Best} et~al.}{2007}]{Bes2007379}
{Best} P.~N.,  {von der Linden} A.,  {Kauffmann} G.,  {Heckman} T.~M.,
  {Kaiser} C.~R.,  2007, \mnras, 379, 894

\bibitem[\protect\citeauthoryear{{B{\^i}rzan}, {McNamara}, {Nulsen}, {Carilli}
  \& {Wise}}{{B{\^i}rzan} et~al.}{2008}]{Bir2008686}
{B{\^i}rzan} L.,  {McNamara} B.~R.,  {Nulsen} P.~E.~J.,  {Carilli} C.~L.,
  {Wise} M.~W.,  2008, \apj, 686, 859

\bibitem[\protect\citeauthoryear{{B{\^i}rzan}, {Rafferty}, {McNamara}, {Wise}
  \& {Nulsen}}{{B{\^i}rzan} et~al.}{2004}]{Bir2004607}
{B{\^i}rzan} L.,  {Rafferty} D.~A.,  {McNamara} B.~R.,  {Wise} M.~W.,
  {Nulsen} P.~E.~J.,  2004, \apj, 607, 800

\bibitem[\protect\citeauthoryear{{Bock}, {Large} \& {Sadler}}{{Bock}
  et~al.}{1999}]{Boc1999117}
{Bock} D.~C.-J.,  {Large} M.~I.,    {Sadler} E.~M.,  1999, \aj, 117, 1578

\bibitem[\protect\citeauthoryear{{Bohringer}, {Nulsen}, {Braun} \&
  {Fabian}}{{Bohringer} et~al.}{1995}]{Boh1995274}
{Bohringer} H.,  {Nulsen} P.~E.~J.,  {Braun} R.,    {Fabian} A.~C.,  1995,
  \mnras, 274, L67

\bibitem[\protect\citeauthoryear{{Brunetti}, {Cappi}, {Setti}, {Feretti} \&
  {Harris}}{{Brunetti} et~al.}{2001}]{Bru2001372}
{Brunetti} G.,  {Cappi} M.,  {Setti} G.,  {Feretti} L.,    {Harris} D.~E.,
  2001, \aap, 372, 755

\bibitem[\protect\citeauthoryear{{Burenin}, {Vikhlinin}, {Hornstrup},
  {Ebeling}, {Quintana} \& {Mescheryakov}}{{Burenin} et~al.}{2007}]{Bur2007172}
{Burenin} R.~A.,  {Vikhlinin} A.,  {Hornstrup} A.,  {Ebeling} H.,  {Quintana}
  H.,    {Mescheryakov} A.,  2007, \apjs, 172, 561

\bibitem[\protect\citeauthoryear{{Carilli}, {Perley}, {Dreher} \&
  {Leahy}}{{Carilli} et~al.}{1991}]{Car1991383}
{Carilli} C.~L.,  {Perley} R.~A.,  {Dreher} J.~W.,    {Leahy} J.~P.,  1991,
  \apj, 383, 554

\bibitem[\protect\citeauthoryear{{Cavagnolo}, {McNamara}, {Nulsen}, {Carilli},
  {Jones} \& {B{\^i}rzan}}{{Cavagnolo} et~al.}{2010}]{Cav2010720}
{Cavagnolo} K.~W.,  {McNamara} B.~R.,  {Nulsen} P.~E.~J.,  {Carilli} C.~L.,
  {Jones} C.,    {B{\^i}rzan} L.,  2010, \apj, 720, 1066

\bibitem[\protect\citeauthoryear{{Cavagnolo}, {McNamara}, {Wise}, {Nulsen},
  {Br{\"u}ggen}, {Gitti} \& {Rafferty}}{{Cavagnolo} et~al.}{2011}]{Cav2011732}
{Cavagnolo} K.~W.,  {McNamara} B.~R.,  {Wise} M.~W.,  {Nulsen} P.~E.~J.,
  {Br{\"u}ggen} M.,  {Gitti} M.,    {Rafferty} D.~A.,  2011, \apj, 732, 71

\bibitem[\protect\citeauthoryear{{Churazov}, {Br{\"u}ggen}, {Kaiser},
  {B{\"o}hringer} \& {Forman}}{{Churazov} et~al.}{2001}]{Chu2001554}
{Churazov} E.,  {Br{\"u}ggen} M.,  {Kaiser} C.~R.,  {B{\"o}hringer} H.,
  {Forman} W.,  2001, \apj, 554, 261

\bibitem[\protect\citeauthoryear{{Coble}, {Bonamente}, {Carlstrom}, {Dawson},
  {Hasler}, {Holzapfel}, {Joy}, {La Roque}, {Marrone} \& {Reese}}{{Coble}
  et~al.}{2007}]{Cob2008134}
{Coble} K.,  {Bonamente} M.,  {Carlstrom} J.~E.,  {Dawson} K.,  {Hasler} N.,
  {Holzapfel} W.,  {Joy} M.,  {La Roque} S.,  {Marrone} D.~P.,    {Reese}
  E.~D.,  2007, \aj, 134, 897

\bibitem[\protect\citeauthoryear{{Cohen}, {Lane}, {Cotton}, {Kassim}, {Lazio},
  {Perley}, {Condon} \& {Erickson}}{{Cohen} et~al.}{2007}]{Coh2007134}
{Cohen} A.~S.,  {Lane} W.~M.,  {Cotton} W.~D.,  {Kassim} N.~E.,  {Lazio}
  T.~J.~W.,  {Perley} R.~A.,  {Condon} J.~J.,    {Erickson} W.~C.,  2007, \aj,
  134, 1245

\bibitem[\protect\citeauthoryear{{Comerford}, {Meneghetti}, {Bartelmann} \&
  {Schirmer}}{{Comerford} et~al.}{2006}]{Com2006642}
{Comerford} J.~M.,  {Meneghetti} M.,  {Bartelmann} M.,    {Schirmer} M.,  2006,
  \apj, 642, 39

\bibitem[\protect\citeauthoryear{{Condon}, {Cotton}, {Greisen}, {Yin},
  {Perley}, {Taylor} \& {Broderick}}{{Condon} et~al.}{1998}]{Con1998115}
{Condon} J.~J.,  {Cotton} W.~D.,  {Greisen} E.~W.,  {Yin} Q.~F.,  {Perley}
  R.~A.,  {Taylor} G.~B.,    {Broderick} J.~J.,  1998, \aj, 115, 1693

\bibitem[\protect\citeauthoryear{{Crawford}, {Allen}, {Ebeling}, {Edge} \&
  {Fabian}}{{Crawford} et~al.}{1999}]{Cra1999306}
{Crawford} C.~S.,  {Allen} S.~W.,  {Ebeling} H.,  {Edge} A.~C.,    {Fabian}
  A.~C.,  1999, \mnras, 306, 857

\bibitem[\protect\citeauthoryear{{Croton}, {Springel}, {White}, {De Lucia},
  {Frenk}, {Gao}, {Jenkins}, {Kauffmann}, {Navarro} \& {Yoshida}}{{Croton}
  et~al.}{2006}]{Cro2006365}
{Croton} D.~J.,  {Springel} V.,  {White} S.~D.~M.,  {De Lucia} G.,  {Frenk}
  C.~S.,  {Gao} L.,  {Jenkins} A.,  {Kauffmann} G.,  {Navarro} J.~F.,
  {Yoshida} N.,  2006, \mnras, 365, 11

\bibitem[\protect\citeauthoryear{{Dahle}, {Kaiser}, {Irgens}, {Lilje} \&
  {Maddox}}{{Dahle} et~al.}{2002}]{Dah2002139}
{Dahle} H.,  {Kaiser} N.,  {Irgens} R.~J.,  {Lilje} P.~B.,    {Maddox} S.~J.,
  2002, \apjs, 139, 313

\bibitem[\protect\citeauthoryear{{Davis}}{{Davis}}{2001}]{Dav2001562}
{Davis} J.~E.,  2001, \apj, 562, 575

\bibitem[\protect\citeauthoryear{{Dong}, {Rasmussen} \& {Mulchaey}}{{Dong}
  et~al.}{2010}]{Don2010712}
{Dong} R.,  {Rasmussen} J.,    {Mulchaey} J.~S.,  2010, \apj, 712, 883

\bibitem[\protect\citeauthoryear{{Dressler} \& {Gunn}}{{Dressler} \&
  {Gunn}}{1992}]{Dre199278}
{Dressler} A.,  {Gunn} J.~E.,  1992, \apjs, 78, 1

\bibitem[\protect\citeauthoryear{{Dunn}, {Allen}, {Taylor}, {Shurkin},
  {Gentile}, {Fabian} \& {Reynolds}}{{Dunn} et~al.}{2010}]{Dun2010404}
{Dunn} R.~J.~H.,  {Allen} S.~W.,  {Taylor} G.~B.,  {Shurkin} K.~F.,  {Gentile}
  G.,  {Fabian} A.~C.,    {Reynolds} C.~S.,  2010, \mnras, 404, 180

\bibitem[\protect\citeauthoryear{{Dunn} \& {Fabian}}{{Dunn} \&
  {Fabian}}{2006}]{Dun2006373}
{Dunn} R.~J.~H.,  {Fabian} A.~C.,  2006, \mnras, 373, 959

\bibitem[\protect\citeauthoryear{{Dunn} \& {Fabian}}{{Dunn} \&
  {Fabian}}{2008}]{Dun2008385}
{Dunn} R.~J.~H.,  {Fabian} A.~C.,  2008, \mnras, 385, 757

\bibitem[\protect\citeauthoryear{{Dunn}, {Fabian} \& {Taylor}}{{Dunn}
  et~al.}{2005}]{Dun2005364}
{Dunn} R.~J.~H.,  {Fabian} A.~C.,    {Taylor} G.~B.,  2005, \mnras, 364, 1343

\bibitem[\protect\citeauthoryear{{Ebeling}}{{Ebeling}}{2009}]{Ebe2009}
{Ebeling} H.,  2009, in HST Proposal {A Snapshot Survey of The Most Massive
  Clusters of Galaxies}.
pp 12166--+

\bibitem[\protect\citeauthoryear{{Ebeling}, {Barrett}, {Donovan}, {Ma}, {Edge}
  \& {van Speybroeck}}{{Ebeling} et~al.}{2007}]{Ebe2007661}
{Ebeling} H.,  {Barrett} E.,  {Donovan} D.,  {Ma} C.-J.,  {Edge} A.~C.,    {van
  Speybroeck} L.,  2007, \apjl, 661, L33

\bibitem[\protect\citeauthoryear{{Ebeling}, {Edge}, {Allen}, {Crawford},
  {Fabian} \& {Huchra}}{{Ebeling} et~al.}{2000}]{Ebe2000318}
{Ebeling} H.,  {Edge} A.~C.,  {Allen} S.~W.,  {Crawford} C.~S.,  {Fabian}
  A.~C.,    {Huchra} J.~P.,  2000, \mnras, 318, 333

\bibitem[\protect\citeauthoryear{{Ebeling}, {Edge}, {Bohringer}, {Allen},
  {Crawford}, {Fabian}, {Voges} \& {Huchra}}{{Ebeling}
  et~al.}{1998}]{Ebe1998301}
{Ebeling} H.,  {Edge} A.~C.,  {Bohringer} H.,  {Allen} S.~W.,  {Crawford}
  C.~S.,  {Fabian} A.~C.,  {Voges} W.,    {Huchra} J.~P.,  1998, \mnras, 301,
  881

\bibitem[\protect\citeauthoryear{{Ebeling}, {Edge} \& {Henry}}{{Ebeling}
  et~al.}{2001}]{Ebe2001553}
{Ebeling} H.,  {Edge} A.~C.,    {Henry} J.~P.,  2001, \apj, 553, 668

\bibitem[\protect\citeauthoryear{{Ebeling}, {Edge}, {Mantz}, {Barrett},
  {Henry}, {Ma} \& {van Speybroeck}}{{Ebeling} et~al.}{2010}]{Ebe2010407}
{Ebeling} H.,  {Edge} A.~C.,  {Mantz} A.,  {Barrett} E.,  {Henry} J.~P.,  {Ma}
  C.~J.,    {van Speybroeck} L.,  2010, \mnras, 407, 83

\bibitem[\protect\citeauthoryear{{Edge}}{{Edge}}{2001}]{Edg2001328}
{Edge} A.~C.,  2001, \mnras, 328, 762

\bibitem[\protect\citeauthoryear{{Edge}, {Stewart}, {Fabian} \&
  {Arnaud}}{{Edge} et~al.}{1990}]{Edg1990245}
{Edge} A.~C.,  {Stewart} G.~C.,  {Fabian} A.~C.,    {Arnaud} K.~A.,  1990,
  \mnras, 245, 559

\bibitem[\protect\citeauthoryear{{Ehlert}, {Allen}, {von der Linden},
  {Simionescu}, {Werner}, {Taylor}, {Gentile}, {Ebeling}, {Allen}, {Applegate},
  {Dunn}, {Fabian}, {Kelly}, {Million}, {Morris}, {Sanders} \&
  {Schmidt}}{{Ehlert} et~al.}{2011}]{Ehl2011411}
{Ehlert} S.,  {Allen} S.~W.,  {von der Linden} A.,  {Simionescu} A.,  {Werner}
  N.,  {Taylor} G.~B.,  {Gentile} G.,  {Ebeling} H.,  {Allen} M.~T.,
  {Applegate} D.,  {Dunn} R.~J.~H.,  {Fabian} A.~C.,  {Kelly} P.,  {Million}
  E.~T.,  {Morris} R.~G.,  {Sanders} J.~S.,    {Schmidt} R.~W.,  2011, \mnras,
  411, 1641

\bibitem[\protect\citeauthoryear{{Ettori}, {Morandi}, {Tozzi}, {Balestra},
  {Borgani}, {Rosati}, {Lovisari} \& {Terenziani}}{{Ettori}
  et~al.}{2009}]{Ett2009501}
{Ettori} S.,  {Morandi} A.,  {Tozzi} P.,  {Balestra} I.,  {Borgani} S.,
  {Rosati} P.,  {Lovisari} L.,    {Terenziani} F.,  2009, \aap, 501, 61

\bibitem[\protect\citeauthoryear{{Fabian}, {Sanders}, {Allen}, {Canning},
  {Churazov}, {Crawford}, {Forman}, {GaBany}, {Hlavacek-Larrondo}, {Johnstone},
  {Russell}, {Reynolds}, {Salome}, {Taylor} \& {Young}}{{Fabian}
  et~al.}{2011}]{Fab2011}
{Fabian} A.~C.,  {Sanders} J.~S.,  {Allen} S.~W.,  {Canning} R.~E.~A.,
  {Churazov} E.,  {Crawford} C.~S.,  {Forman} W.,  {GaBany} J.,
  {Hlavacek-Larrondo} J.,  {Johnstone} R.~M.,  {Russell} H.~R.,  {Reynolds}
  C.~S.,  {Salome} P.,  {Taylor} G.~B.,    {Young} A.~J.,  2011, ArXiv e-prints

\bibitem[\protect\citeauthoryear{{Fabian}, {Sanders}, {Allen}, {Crawford},
  {Iwasawa}, {Johnstone}, {Schmidt} \& {Taylor}}{{Fabian}
  et~al.}{2003}]{Fab2003344}
{Fabian} A.~C.,  {Sanders} J.~S.,  {Allen} S.~W.,  {Crawford} C.~S.,  {Iwasawa}
  K.,  {Johnstone} R.~M.,  {Schmidt} R.~W.,    {Taylor} G.~B.,  2003, \mnras,
  344, L43

\bibitem[\protect\citeauthoryear{{Fabian}, {Sanders}, {Taylor}, {Allen},
  {Crawford}, {Johnstone} \& {Iwasawa}}{{Fabian} et~al.}{2006}]{Fab2006366}
{Fabian} A.~C.,  {Sanders} J.~S.,  {Taylor} G.~B.,  {Allen} S.~W.,  {Crawford}
  C.~S.,  {Johnstone} R.~M.,    {Iwasawa} K.,  2006, \mnras, 366, 417

\bibitem[\protect\citeauthoryear{{Finoguenov} \& {Jones}}{{Finoguenov} \&
  {Jones}}{2001}]{Fin2001547}
{Finoguenov} A.,  {Jones} C.,  2001, \apjl, 547, L107

\bibitem[\protect\citeauthoryear{{Forman}, {Nulsen}, {Heinz}, {Owen}, {Eilek},
  {Vikhlinin}, {Markevitch}, {Kraft}, {Churazov} \& {Jones}}{{Forman}
  et~al.}{2005}]{For2005635}
{Forman} W.,  {Nulsen} P.,  {Heinz} S.,  {Owen} F.,  {Eilek} J.,  {Vikhlinin}
  A.,  {Markevitch} M.,  {Kraft} R.,  {Churazov} E.,    {Jones} C.,  2005,
  \apj, 635, 894

\bibitem[\protect\citeauthoryear{{Fort}, {Le Fevre}, {Hammer} \&
  {Cailloux}}{{Fort} et~al.}{1992}]{For1992399}
{Fort} B.,  {Le Fevre} O.,  {Hammer} F.,    {Cailloux} M.,  1992, \apjl, 399,
  L125

\bibitem[\protect\citeauthoryear{{Franx}, {Illingworth}, {Kelson}, {van Dokkum}
  \& {Tran}}{{Franx} et~al.}{1997}]{Fra1997486}
{Franx} M.,  {Illingworth} G.~D.,  {Kelson} D.~D.,  {van Dokkum} P.~G.,
  {Tran} K.-V.,  1997, \apjl, 486, L75+

\bibitem[\protect\citeauthoryear{{Gavazzi}}{{Gavazzi}}{2005}]{Gav2005443}
{Gavazzi} R.,  2005, \aap, 443, 793

\bibitem[\protect\citeauthoryear{{Gavazzi}, {Fort}, {Mellier}, {Pell{\'o}} \&
  {Dantel-Fort}}{{Gavazzi} et~al.}{2003}]{Gav2003403}
{Gavazzi} R.,  {Fort} B.,  {Mellier} Y.,  {Pell{\'o}} R.,    {Dantel-Fort} M.,
  2003, \aap, 403, 11

\bibitem[\protect\citeauthoryear{{Gioia} \& {Luppino}}{{Gioia} \&
  {Luppino}}{1994}]{Gio199494}
{Gioia} I.~M.,  {Luppino} G.~A.,  1994, \apjs, 94, 583

\bibitem[\protect\citeauthoryear{{Gioia}, {Maccacaro}, {Schild}, {Wolter},
  {Stocke}, {Morris} \& {Henry}}{{Gioia} et~al.}{1990}]{Gio199072}
{Gioia} I.~M.,  {Maccacaro} T.,  {Schild} R.~E.,  {Wolter} A.,  {Stocke} J.~T.,
   {Morris} S.~L.,    {Henry} J.~P.,  1990, \apjs, 72, 567

\bibitem[\protect\citeauthoryear{{Gitti}, {Feretti} \& {Schindler}}{{Gitti}
  et~al.}{2006}]{Git2006448}
{Gitti} M.,  {Feretti} L.,    {Schindler} S.,  2006, \aap, 448, 853

\bibitem[\protect\citeauthoryear{{Gitti}, {McNamara}, {Nulsen} \&
  {Wise}}{{Gitti} et~al.}{2007}]{Git2007660}
{Gitti} M.,  {McNamara} B.~R.,  {Nulsen} P.~E.~J.,    {Wise} M.~W.,  2007,
  \apj, 660, 1118

\bibitem[\protect\citeauthoryear{{Graham}, {Fabian} \& {Sanders}}{{Graham}
  et~al.}{2008}]{Gra2009386}
{Graham} J.,  {Fabian} A.~C.,    {Sanders} J.~S.,  2008, \mnras, 386, 278

\bibitem[\protect\citeauthoryear{{Gregory} \& {Condon}}{{Gregory} \&
  {Condon}}{1991}]{Gre199175}
{Gregory} P.~C.,  {Condon} J.~J.,  1991, \apjs, 75, 1011

\bibitem[\protect\citeauthoryear{{Griffith} \& {Wright}}{{Griffith} \&
  {Wright}}{1993}]{Gri1993105}
{Griffith} M.~R.,  {Wright} A.~E.,  1993, \aj, 105, 1666

\bibitem[\protect\citeauthoryear{{Harris}, {Nulsen}, {Ponman}, {Bautz},
  {Cameron}, {David}, {Donnelly} \& {Forman}}{{Harris}
  et~al.}{2000}]{Har2000530}
{Harris} D.~E.,  {Nulsen} P.~E.~J.,  {Ponman} T.~J.,  {Bautz} M.,  {Cameron}
  R.~A.,  {David} L.~P.,  {Donnelly} R.~H.,    {Forman} W.~R. e.~a.,  2000,
  \apjl, 530, L81

\bibitem[\protect\citeauthoryear{{Hines} \& {Wills}}{{Hines} \&
  {Wills}}{1993}]{Hin1993415}
{Hines} D.~C.,  {Wills} B.~J.,  1993, \apj, 415, 82

\bibitem[\protect\citeauthoryear{{Hlavacek-Larrondo} \&
  {Fabian}}{{Hlavacek-Larrondo} \& {Fabian}}{2011}]{Hla2011}
{Hlavacek-Larrondo} J.,  {Fabian} A.~C.,  2011, \mnras, pp 138--+

\bibitem[\protect\citeauthoryear{{Hoekstra}, {Franx}, {Kuijken} \&
  {Squires}}{{Hoekstra} et~al.}{1998}]{Hoe1998504}
{Hoekstra} H.,  {Franx} M.,  {Kuijken} K.,    {Squires} G.,  1998, \apj, 504,
  636

\bibitem[\protect\citeauthoryear{{Iwasawa}, {Fabian} \& {Ettori}}{{Iwasawa}
  et~al.}{2001}]{Iwa2001321}
{Iwasawa} K.,  {Fabian} A.~C.,    {Ettori} S.,  2001, \mnras, 321, L15

\bibitem[\protect\citeauthoryear{{Jones}, {Forman}, {Vikhlinin}, {Markevitch},
  {David}, {Warmflash}, {Murray} \& {Nulsen}}{{Jones}
  et~al.}{2002}]{Jon2002567}
{Jones} C.,  {Forman} W.,  {Vikhlinin} A.,  {Markevitch} M.,  {David} L.,
  {Warmflash} A.,  {Murray} S.,    {Nulsen} P.~E.~J.,  2002, \apjl, 567, L115

\bibitem[\protect\citeauthoryear{{Kalberla}, {Burton}, {Hartmann}, {Arnal},
  {Bajaja}, {Morras} \& {P{\"o}ppel}}{{Kalberla} et~al.}{2005}]{Kal2005440}
{Kalberla} P.~M.~W.,  {Burton} W.~B.,  {Hartmann} D.,  {Arnal} E.~M.,  {Bajaja}
  E.,  {Morras} R.,    {P{\"o}ppel} W.~G.~L.,  2005, \aap, 440, 775

\bibitem[\protect\citeauthoryear{{Kartaltepe}, {Ebeling}, {Ma} \&
  {Donovan}}{{Kartaltepe} et~al.}{2008}]{Kar2008389}
{Kartaltepe} J.~S.,  {Ebeling} H.,  {Ma} C.~J.,    {Donovan} D.,  2008, \mnras,
  389, 1240

\bibitem[\protect\citeauthoryear{{Kleinmann}, {Hamilton}, {Keel},
  {Wynn-Williams}, {Eales}, {Becklin} \& {Kuntz}}{{Kleinmann}
  et~al.}{1988}]{Kle1988328}
{Kleinmann} S.~G.,  {Hamilton} D.,  {Keel} W.~C.,  {Wynn-Williams} C.~G.,
  {Eales} S.~A.,  {Becklin} E.~E.,    {Kuntz} K.~D.,  1988, \apj, 328, 161

\bibitem[\protect\citeauthoryear{{Limousin}, {Ebeling}, {Ma}, {Swinbank},
  {Smith}, {Richard}, {Edge}, {Jauzac}, {Kneib}, {Marshall} \&
  {Schrabback}}{{Limousin} et~al.}{2010}]{Lim2010405}
{Limousin} M.,  {Ebeling} H.,  {Ma} C.-J.,  {Swinbank} A.~M.,  {Smith} G.~P.,
  {Richard} J.,  {Edge} A.~C.,  {Jauzac} M.,  {Kneib} J.-P.,  {Marshall} P.,
  {Schrabback} T.,  2010, \mnras, 405, 777

\bibitem[\protect\citeauthoryear{{Ma}, {McNamara}, {Nulsen}, {Schaffer} \&
  {Vikhlinin}}{{Ma} et~al.}{2011}]{Ma2011}
{Ma} C.~.,  {McNamara} B.~R.,  {Nulsen} P.~E.~J.,  {Schaffer} R.,
  {Vikhlinin} A.,  2011, ArXiv e-prints

\bibitem[\protect\citeauthoryear{{Machacek}, {Nulsen}, {Jones} \&
  {Forman}}{{Machacek} et~al.}{2006}]{Mac2006648}
{Machacek} M.,  {Nulsen} P.~E.~J.,  {Jones} C.,    {Forman} W.~R.,  2006, \apj,
  648, 947

\bibitem[\protect\citeauthoryear{{Magliocchetti} \&
  {Br{\"u}ggen}}{{Magliocchetti} \& {Br{\"u}ggen}}{2007}]{Mag2007379}
{Magliocchetti} M.,  {Br{\"u}ggen} M.,  2007, \mnras, 379, 260

\bibitem[\protect\citeauthoryear{{Mann} \& {Ebeling}}{{Mann} \&
  {Ebeling}}{2011}]{Man2011}
{Mann} A.~W.,  {Ebeling} E.,  2011

\bibitem[\protect\citeauthoryear{{Mauch}, {Murphy}, {Buttery}, {Curran},
  {Hunstead}, {Piestrzynski}, {Robertson} \& {Sadler}}{{Mauch}
  et~al.}{2003}]{Mau2003342}
{Mauch} T.,  {Murphy} T.,  {Buttery} H.~J.,  {Curran} J.,  {Hunstead} R.~W.,
  {Piestrzynski} B.,  {Robertson} J.~G.,    {Sadler} E.~M.,  2003, \mnras, 342,
  1117

\bibitem[\protect\citeauthoryear{{Maughan}, {Jones}, {Forman} \& {Van
  Speybroeck}}{{Maughan} et~al.}{2008}]{Mau2008174}
{Maughan} B.~J.,  {Jones} C.,  {Forman} W.,    {Van Speybroeck} L.,  2008,
  \apjs, 174, 117

\bibitem[\protect\citeauthoryear{{McNamara} \& {Nulsen}}{{McNamara} \&
  {Nulsen}}{2007}]{Mcn200745}
{McNamara} B.~R.,  {Nulsen} P.~E.~J.,  2007, \araa, 45, 117

\bibitem[\protect\citeauthoryear{{McNamara}, {Nulsen}, {Wise}, {Rafferty},
  {Carilli}, {Sarazin} \& {Blanton}}{{McNamara} et~al.}{2005}]{McN2005433}
{McNamara} B.~R.,  {Nulsen} P.~E.~J.,  {Wise} M.~W.,  {Rafferty} D.~A.,
  {Carilli} C.,  {Sarazin} C.~L.,    {Blanton} E.~L.,  2005, \nat, 433, 45

\bibitem[\protect\citeauthoryear{{McNamara}, {Wise}, {Nulsen}, {David},
  {Sarazin}, {Bautz}, {Markevitch}, {Vikhlinin}, {Forman}, {Jones} \&
  {Harris}}{{McNamara} et~al.}{2000}]{McN2000534}
{McNamara} B.~R.,  {Wise} M.,  {Nulsen} P.~E.~J.,  {David} L.~P.,  {Sarazin}
  C.~L.,  {Bautz} M.,  {Markevitch} M.,  {Vikhlinin} A.,  {Forman} W.~R.,
  {Jones} C.,    {Harris} D.~E.,  2000, \apjl, 534, L135

\bibitem[\protect\citeauthoryear{{Merten}, {Cacciato}, {Meneghetti}, {Mignone}
  \& {Bartelmann}}{{Merten} et~al.}{2009}]{Mer2009500}
{Merten} J.,  {Cacciato} M.,  {Meneghetti} M.,  {Mignone} C.,    {Bartelmann}
  M.,  2009, \aap, 500, 681

\bibitem[\protect\citeauthoryear{{Mullis}, {McNamara}, {Quintana}, {Vikhlinin},
  {Henry}, {Gioia}, {Hornstrup}, {Forman} \& {Jones}}{{Mullis}
  et~al.}{2003}]{Mul2003594}
{Mullis} C.~R.,  {McNamara} B.~R.,  {Quintana} H.,  {Vikhlinin} A.,  {Henry}
  J.~P.,  {Gioia} I.~M.,  {Hornstrup} A.,  {Forman} W.,    {Jones} C.,  2003,
  \apj, 594, 154

\bibitem[\protect\citeauthoryear{{Nulsen}, {Jones}, {Forman}, {David},
  {McNamara}, {Rafferty}, {B{\^i}rzan} \& {Wise}}{{Nulsen}
  et~al.}{2007}]{Nul2007}
{Nulsen} P.~E.~J.,  {Jones} C.,  {Forman} W.~R.,  {David} L.~P.,  {McNamara}
  B.~R.,  {Rafferty} D.~A.,  {B{\^i}rzan} L.,    {Wise} M.~W.,  2007, in
  {H.~B{\"o}hringer, G.~W.~Pratt, A.~Finoguenov, \& P.~Schuecker } ed., Heating
  versus Cooling in Galaxies and Clusters of Galaxies {AGN Heating Through
  Cavities and Shocks}.
pp 210--+

\bibitem[\protect\citeauthoryear{{O'Dea}, {Baum}, {Privon}, {Noel-Storr},
  {Quillen}, {Zufelt}, {Park}, {Edge}, {Russell}, {Fabian}, {Donahue},
  {Sarazin}, {McNamara}, {Bregman} \& {Egami}}{{O'Dea}
  et~al.}{2008}]{Ode2008681}
{O'Dea} C.~P.,  {Baum} S.~A.,  {Privon} G.,  {Noel-Storr} J.,  {Quillen} A.~C.,
   {Zufelt} N.,  {Park} J.,  {Edge} A.,  {Russell} H.,  {Fabian} A.~C.,
  {Donahue} M.,  {Sarazin} C.~L.,  {McNamara} B.,  {Bregman} J.~N.,    {Egami}
  E.,  2008, \apj, 681, 1035

\bibitem[\protect\citeauthoryear{{O'Sullivan}, {Giacintucci}, {David}, {Gitti},
  {Vrtilek}, {Raychaudhury} \& {Ponman}}{{O'Sullivan}
  et~al.}{2011}]{OSu2011735}
{O'Sullivan} E.,  {Giacintucci} S.,  {David} L.~P.,  {Gitti} M.,  {Vrtilek}
  J.~M.,  {Raychaudhury} S.,    {Ponman} T.~J.,  2011, \apj, 735, 11

\bibitem[\protect\citeauthoryear{{Perley} \& {Taylor}}{{Perley} \&
  {Taylor}}{1991}]{Tay1991101}
{Perley} R.~A.,  {Taylor} G.~B.,  1991, \aj, 101, 1623

\bibitem[\protect\citeauthoryear{{Perlman}, {Horner}, {Jones}, {Scharf},
  {Ebeling}, {Wegner} \& {Malkan}}{{Perlman} et~al.}{2002}]{Per2002140}
{Perlman} E.~S.,  {Horner} D.~J.,  {Jones} L.~R.,  {Scharf} C.~A.,  {Ebeling}
  H.,  {Wegner} G.,    {Malkan} M.,  2002, \apjs, 140, 265

\bibitem[\protect\citeauthoryear{{Peterson} \& {Fabian}}{{Peterson} \&
  {Fabian}}{2006}]{Pet2006427}
{Peterson} J.~R.,  {Fabian} A.~C.,  2006, \physrep, 427, 1

\bibitem[\protect\citeauthoryear{{Puetzfeld}, {Pohl} \& {Zhu}}{{Puetzfeld}
  et~al.}{2005}]{Pue2005619}
{Puetzfeld} D.,  {Pohl} M.,    {Zhu} Z.-H.,  2005, \apj, 619, 657

\bibitem[\protect\citeauthoryear{{Quillen}, {Zufelt}, {Park}, {O'Dea}, {Baum},
  {Privon}, {Noel-Storr}, {Edge}, {Russell}, {Fabian}, {Donahue}, {Bregman},
  {McNamara} \& {Sarazin}}{{Quillen} et~al.}{2008}]{Qui2008176}
{Quillen} A.~C.,  {Zufelt} N.,  {Park} J.,  {O'Dea} C.~P.,  {Baum} S.~A.,
  {Privon} G.,  {Noel-Storr} J.,  {Edge} A.,  {Russell} H.,  {Fabian} A.,
  {Donahue} M.,  {Bregman} J.~N.,  {McNamara} B.~R.,    {Sarazin} C.~L.,  2008,
  \apjs, 176, 39

\bibitem[\protect\citeauthoryear{{Rafferty}, {McNamara} \& {Nulsen}}{{Rafferty}
  et~al.}{2008}]{Raf2008687}
{Rafferty} D.~A.,  {McNamara} B.~R.,    {Nulsen} P.~E.~J.,  2008, \apj, 687,
  899

\bibitem[\protect\citeauthoryear{{Rafferty}, {McNamara}, {Nulsen} \&
  {Wise}}{{Rafferty} et~al.}{2006}]{Raf2006652}
{Rafferty} D.~A.,  {McNamara} B.~R.,  {Nulsen} P.~E.~J.,    {Wise} M.~W.,
  2006, \apj, 652, 216

\bibitem[\protect\citeauthoryear{{Reid}, {Brewer}, {Brucato}, {McKinley},
  {Maury}, {Mendenhall}, {Mould}, {Mueller}, {Neugebauer}, {Phinney},
  {Sargent}, {Schombert} \& {Thicksten}}{{Reid} et~al.}{1991}]{Rei1991103}
{Reid} I.~N.,  {Brewer} C.,  {Brucato} R.~J.,  {McKinley} W.~R.,  {Maury} A.,
  {Mendenhall} D.,  {Mould} J.~R.,  {Mueller} J.,  {Neugebauer} G.,  {Phinney}
  J.,  {Sargent} W.~L.~W.,  {Schombert} J.,    {Thicksten} R.,  1991, \pasp,
  103, 661

\bibitem[\protect\citeauthoryear{{Rengelink}, {Tang}, {de Bruyn}, {Miley},
  {Bremer}, {Roettgering} \& {Bremer}}{{Rengelink} et~al.}{1997}]{Ren1997124}
{Rengelink} R.~B.,  {Tang} Y.,  {de Bruyn} A.~G.,  {Miley} G.~K.,  {Bremer}
  M.~N.,  {Roettgering} H.~J.~A.,    {Bremer} M.~A.~R.,  1997, \aaps, 124, 259

\bibitem[\protect\citeauthoryear{{Russell}, {Fabian}, {Sanders}, {Johnstone},
  {Blundell}, {Brandt} \& {Crawford}}{{Russell} et~al.}{2010}]{Rus2009402}
{Russell} H.~R.,  {Fabian} A.~C.,  {Sanders} J.~S.,  {Johnstone} R.~M.,
  {Blundell} K.~M.,  {Brandt} W.~N.,    {Crawford} C.~S.,  2010, \mnras, 402,
  1561

\bibitem[\protect\citeauthoryear{{Samuele}, {McNamara}, {Vikhlinin} \&
  {Mullis}}{{Samuele} et~al.}{2011}]{Sam2011731}
{Samuele} R.,  {McNamara} B.~R.,  {Vikhlinin} A.,    {Mullis} C.~R.,  2011,
  \apj, 731, 31

\bibitem[\protect\citeauthoryear{{Sand}, {Treu}, {Ellis}, {Smith} \&
  {Kneib}}{{Sand} et~al.}{2008}]{San2008674}
{Sand} D.~J.,  {Treu} T.,  {Ellis} R.~S.,  {Smith} G.~P.,    {Kneib} J.-P.,
  2008, \apj, 674, 711

\bibitem[\protect\citeauthoryear{{Sanders} \& {Fabian}}{{Sanders} \&
  {Fabian}}{2007}]{San2007381}
{Sanders} J.~S.,  {Fabian} A.~C.,  2007, \mnras, 381, 1381

\bibitem[\protect\citeauthoryear{{Sanders}, {Fabian}, {Allen}, {Morris},
  {Graham} \& {Johnstone}}{{Sanders} et~al.}{2008}]{San2008385}
{Sanders} J.~S.,  {Fabian} A.~C.,  {Allen} S.~W.,  {Morris} R.~G.,  {Graham}
  J.,    {Johnstone} R.~M.,  2008, \mnras, 385, 1186

\bibitem[\protect\citeauthoryear{{Sanders}, {Fabian}, {Frank}, {Peterson} \&
  {Russell}}{{Sanders} et~al.}{2010}]{San2010402}
{Sanders} J.~S.,  {Fabian} A.~C.,  {Frank} K.~A.,  {Peterson} J.~R.,
  {Russell} H.~R.,  2010, \mnras, 402, 127

\bibitem[\protect\citeauthoryear{{Sanders}, {Fabian} \& {Taylor}}{{Sanders}
  et~al.}{2009}]{San2009393}
{Sanders} J.~S.,  {Fabian} A.~C.,    {Taylor} G.~B.,  2009, \mnras, 393, 71

\bibitem[\protect\citeauthoryear{{Santos}, {Rosati}, {Tozzi}, {B{\"o}hringer},
  {Ettori} \& {Bignamini}}{{Santos} et~al.}{2008}]{San2008483}
{Santos} J.~S.,  {Rosati} P.,  {Tozzi} P.,  {B{\"o}hringer} H.,  {Ettori} S.,
   {Bignamini} A.,  2008, \aap, 483, 35

\bibitem[\protect\citeauthoryear{{Siana}, {Smail}, {Swinbank}, {Richard},
  {Teplitz}, {Coppin}, {Ellis}, {Stark}, {Kneib} \& {Edge}}{{Siana}
  et~al.}{2009}]{Sia2009698}
{Siana} B.,  {Smail} I.,  {Swinbank} A.~M.,  {Richard} J.,  {Teplitz} H.~I.,
  {Coppin} K.~E.~K.,  {Ellis} R.~S.,  {Stark} D.~P.,  {Kneib} J.-P.,    {Edge}
  A.~C.,  2009, \apj, 698, 1273

\bibitem[\protect\citeauthoryear{{Sijacki} \& {Springel}}{{Sijacki} \&
  {Springel}}{2006}]{Sij2006366}
{Sijacki} D.,  {Springel} V.,  2006, \mnras, 366, 397

\bibitem[\protect\citeauthoryear{{Smail}, {Swinbank}, {Richard}, {Ebeling},
  {Kneib}, {Edge}, {Stark}, {Ellis}, {Dye}, {Smith} \& {Mullis}}{{Smail}
  et~al.}{2007}]{Sma2007654}
{Smail} I.,  {Swinbank} A.~M.,  {Richard} J.,  {Ebeling} H.,  {Kneib} J.-P.,
  {Edge} A.~C.,  {Stark} D.,  {Ellis} R.~S.,  {Dye} S.,  {Smith} G.~P.,
  {Mullis} C.,  2007, \apjl, 654, L33

\bibitem[\protect\citeauthoryear{{Stocke}, {Morris}, {Gioia}, {Maccacaro},
  {Schild}, {Wolter}, {Fleming} \& {Henry}}{{Stocke} et~al.}{1991}]{Sto199176}
{Stocke} J.~T.,  {Morris} S.~L.,  {Gioia} I.~M.,  {Maccacaro} T.,  {Schild} R.,
   {Wolter} A.,  {Fleming} T.~A.,    {Henry} J.~P.,  1991, \apjs, 76, 813

\bibitem[\protect\citeauthoryear{{Swinbank}, {Smail}, {Longmore}, {Harris},
  {Baker}, {De Breuck}, {Richard}, {Edge}, {Ivison}, {Blundell} \&
  {Coppin}}{{Swinbank} et~al.}{2010}]{Swi2010NAT}
{Swinbank} A.~M.,  {Smail} I.,  {Longmore} S.,  {Harris} A.~I.,  {Baker} A.~J.,
   {De Breuck} C.,  {Richard} J.,  {Edge} A.~C.,  {Ivison} R.~J.,  {Blundell}
  R.,    {Coppin} K.~E.~K. e.~a.,  2010, \nat, 464, 733

\bibitem[\protect\citeauthoryear{{Taylor} \& {Perley}}{{Taylor} \&
  {Perley}}{1992}]{Tay1992262}
{Taylor} G.~B.,  {Perley} R.~A.,  1992, \aap, 262, 417

\bibitem[\protect\citeauthoryear{{Tr{\"u}mper}}{{Tr{\"u}mper}}{1983}]{Tru1983}
{Tr{\"u}mper} J.,  1983, Adv. Space Res., 27, 1404

\bibitem[\protect\citeauthoryear{{Vignali}, {Piconcelli}, {Lanzuisi}, {Feltre},
  {Feruglio}, {Maiolino}, {Fiore}, {Fritz}, {La Parola}, {Mignoli} \&
  {Pozzi}}{{Vignali} et~al.}{2011}]{Vig2011}
{Vignali} C.,  {Piconcelli} E.,  {Lanzuisi} G.,  {Feltre} A.,  {Feruglio} C.,
  {Maiolino} R.,  {Fiore} F.,  {Fritz} J.,  {La Parola} V.,  {Mignoli} M.,
  {Pozzi} F.,  2011, ArXiv e-prints

\bibitem[\protect\citeauthoryear{{Vikhlinin}, {Burenin}, {Forman}, {Jones},
  {Hornstrup}, {Murray} \& {Quintana}}{{Vikhlinin} et~al.}{2007}]{Vik2007}
{Vikhlinin} A.,  {Burenin} R.,  {Forman} W.~R.,  {Jones} C.,  {Hornstrup} A.,
  {Murray} S.~S.,    {Quintana} H.,  2007, in {H.~B{\"o}hringer, G.~W.~Pratt,
  A.~Finoguenov, \& P.~Schuecker } ed., Heating versus Cooling in Galaxies and
  Clusters of Galaxies {Lack of Cooling Flow Clusters at z > 0.5}.
pp 48--+

\bibitem[\protect\citeauthoryear{{Vikhlinin}, {McNamara}, {Forman}, {Jones},
  {Quintana} \& {Hornstrup}}{{Vikhlinin} et~al.}{1998}]{Vik1998502}
{Vikhlinin} A.,  {McNamara} B.~R.,  {Forman} W.,  {Jones} C.,  {Quintana} H.,
   {Hornstrup} A.,  1998, \apj, 502, 558

\bibitem[\protect\citeauthoryear{{Voges}, {Aschenbach}, {Boller},
  {Br{\"a}uninger}, {Briel}, {Burkert}, {Dennerl}, {Englhauser}, {Gruber},
  {Haberl} \& et al.}{{Voges} et~al.}{1999}]{Vog1999}
{Voges} W.,  {Aschenbach} B.,  {Boller} T.,  {Br{\"a}uninger} H.,  {Briel} U.,
  {Burkert} W.,  {Dennerl} K.,  {Englhauser} J.,  {Gruber} R.,  {Haberl} F.,
  et al. H.,  1999, \aap, 349, 389

\bibitem[\protect\citeauthoryear{{Willott}, {Rawlings}, {Blundell} \&
  {Lacy}}{{Willott} et~al.}{1999}]{Wil1999309}
{Willott} C.~J.,  {Rawlings} S.,  {Blundell} K.~M.,    {Lacy} M.,  1999,
  \mnras, 309, 1017

\bibitem[\protect\citeauthoryear{{Wright}, {Ables} \& {Allen}}{{Wright}
  et~al.}{1983}]{Wri1983205}
{Wright} A.~E.,  {Ables} J.~G.,    {Allen} D.~A.,  1983, \mnras, 205, 793

\bibitem[\protect\citeauthoryear{{Zitrin}, {Broadhurst}, {Coe}, {Liesenborgs},
  {Ben{\'{\i}}tez}, {Rephaeli}, {Ford} \& {Umetsu}}{{Zitrin}
  et~al.}{2011}]{Zit2011413}
{Zitrin} A.,  {Broadhurst} T.,  {Coe} D.,  {Liesenborgs} J.,  {Ben{\'{\i}}tez}
  N.,  {Rephaeli} Y.,  {Ford} H.,    {Umetsu} K.,  2011, \mnras, 413, 1753

\end{thebibliography}

\appendix

\section{Unsharp-masked and ellipse-subtracted X-ray images}
We show the unsharp-masked and ellipse-subtracted images for all clusters in which we identified cavities (Fig. \ref{figA2}). The first consists of subtracting a strongly smoothed image from a lightly smoothed image, and the second consists of subtracting an elliptical model of the cluster emission from the original image. These techniques enhance deviations in the original image, and the images were used as a first step to identify cavities. However, we only considered that a cluster had a cavity if we could also see a faint depression in the original $0.5-7$ keV $Chandra$ image. There are two clusters for each horizontal line in Fig. \ref{figA2}, with the unsharp-masked $0.5-7$ keV X-ray image shown in the left panel and the ellipse-subtracted $0.5-7$ keV X-ray image shown in the right panel. The red cross shows the location of the BCG nucleus. We also indicate in the lower-left corner of each image the binning (${\rm B_{X}}$) and smoothing (${\rm S_{X}}$) scales adopted for creating the images. For ${\rm B_{X}}$, ${\rm X}$ indicates the binning factor where ${\rm X}=1$ corresponds to no binning and ${\rm X}=2$ gives an image where each of pixel corresponds to 4 pixels in the original image. For ${\rm S_{X}}$, ${\rm X}$ indicates the smoothing factor and corresponds to the sigma of a gaussian in units of pixels. For each unsharp-masked image, we show the 2 smoothing scales used to create the image (${\rm S_{X_1}-S_{X_2}}$), and for each ellipse-subtracted image, we show the smoothing scale adopted (${\rm S_{X_3}}$) for the original image before creating and subtracting an elliptical model. The first 13 clusters have ``clear" cavities, while the remaining 7 only have ``potential" cavities. Deeper observations are needed to confirm if the ``potential" cavities are real.  

\begin{figure*}
\centering
\begin{minipage}[c]{0.49\linewidth}
\centering \includegraphics[width=\linewidth]{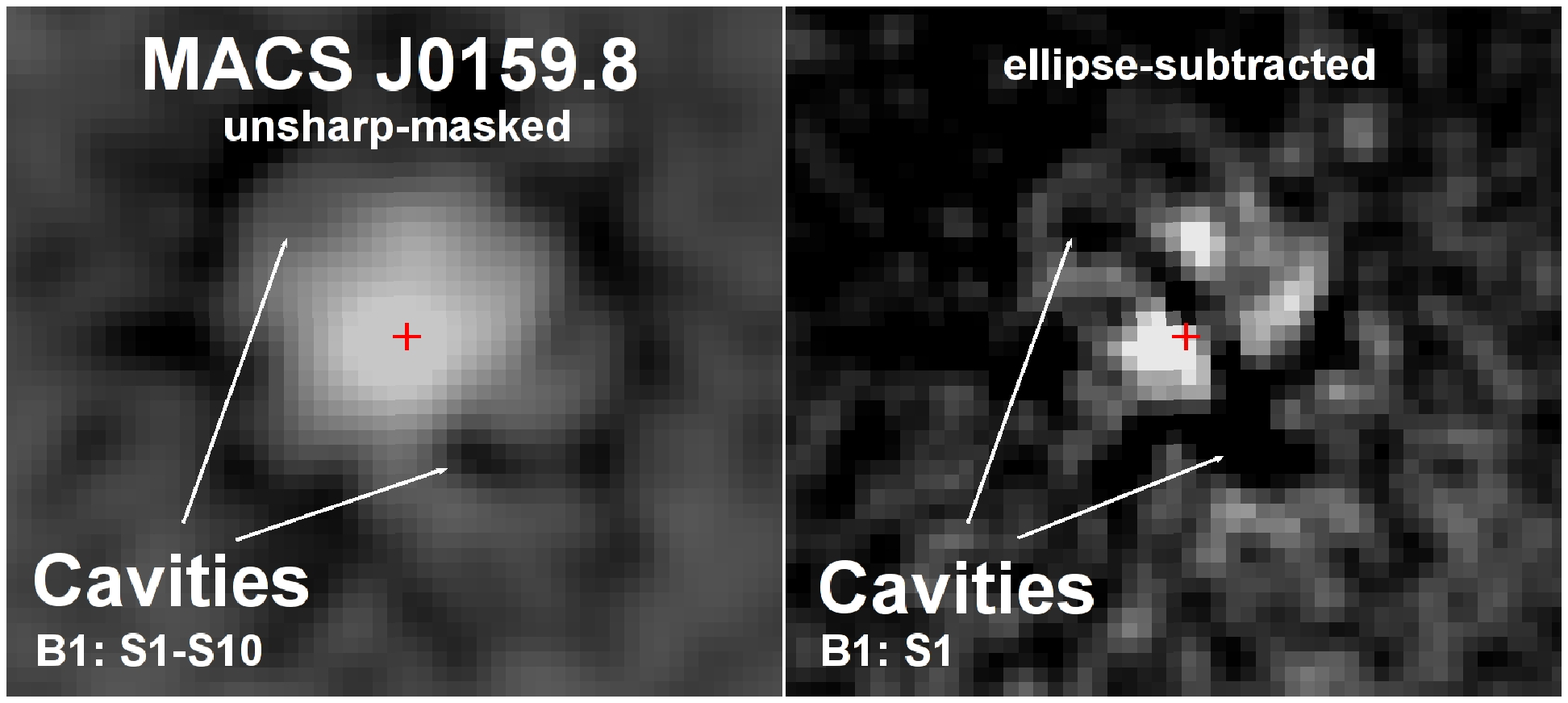}
\end{minipage}
\begin{minipage}[c]{0.49\linewidth}
\centering \includegraphics[width=\linewidth]{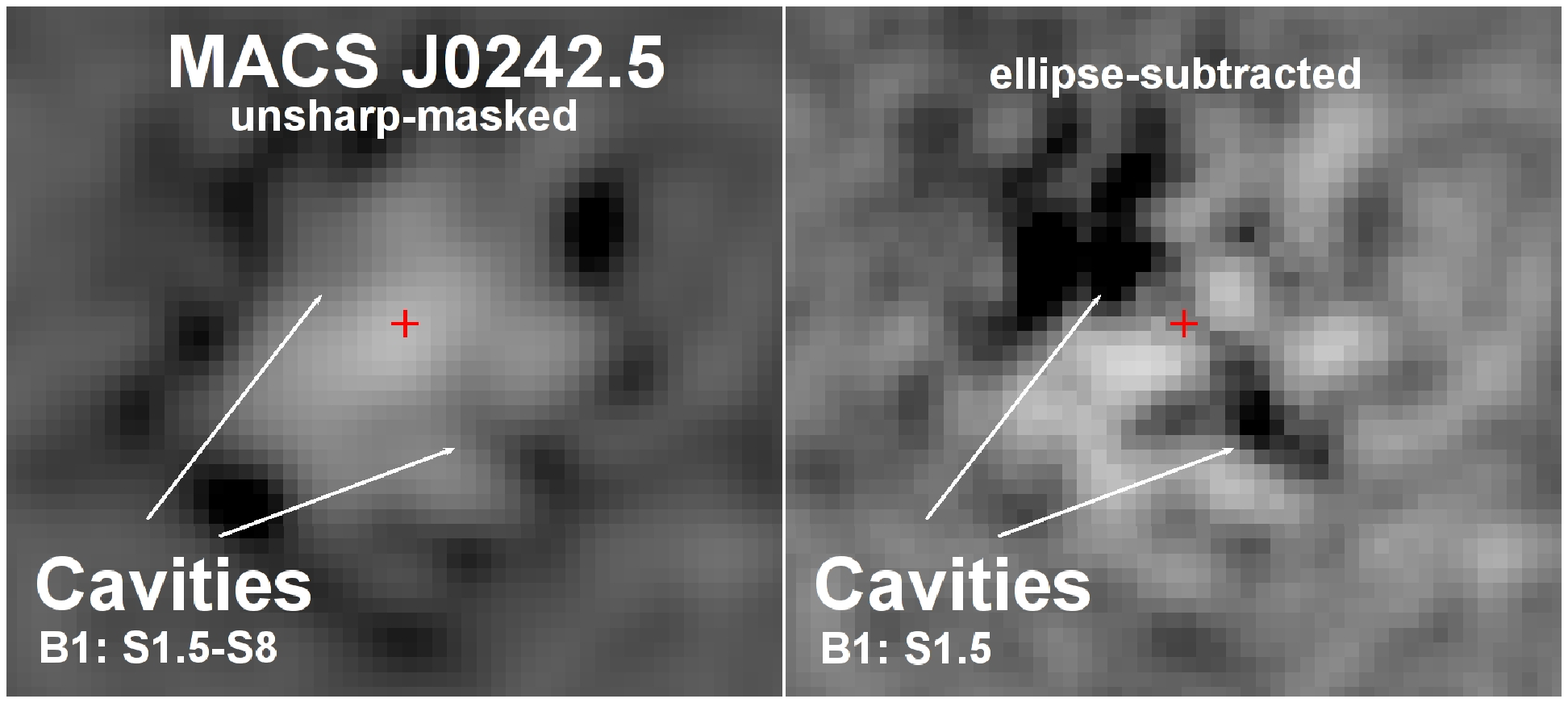}
\end{minipage}
\begin{minipage}[c]{0.49\linewidth}
\centering \includegraphics[width=\linewidth]{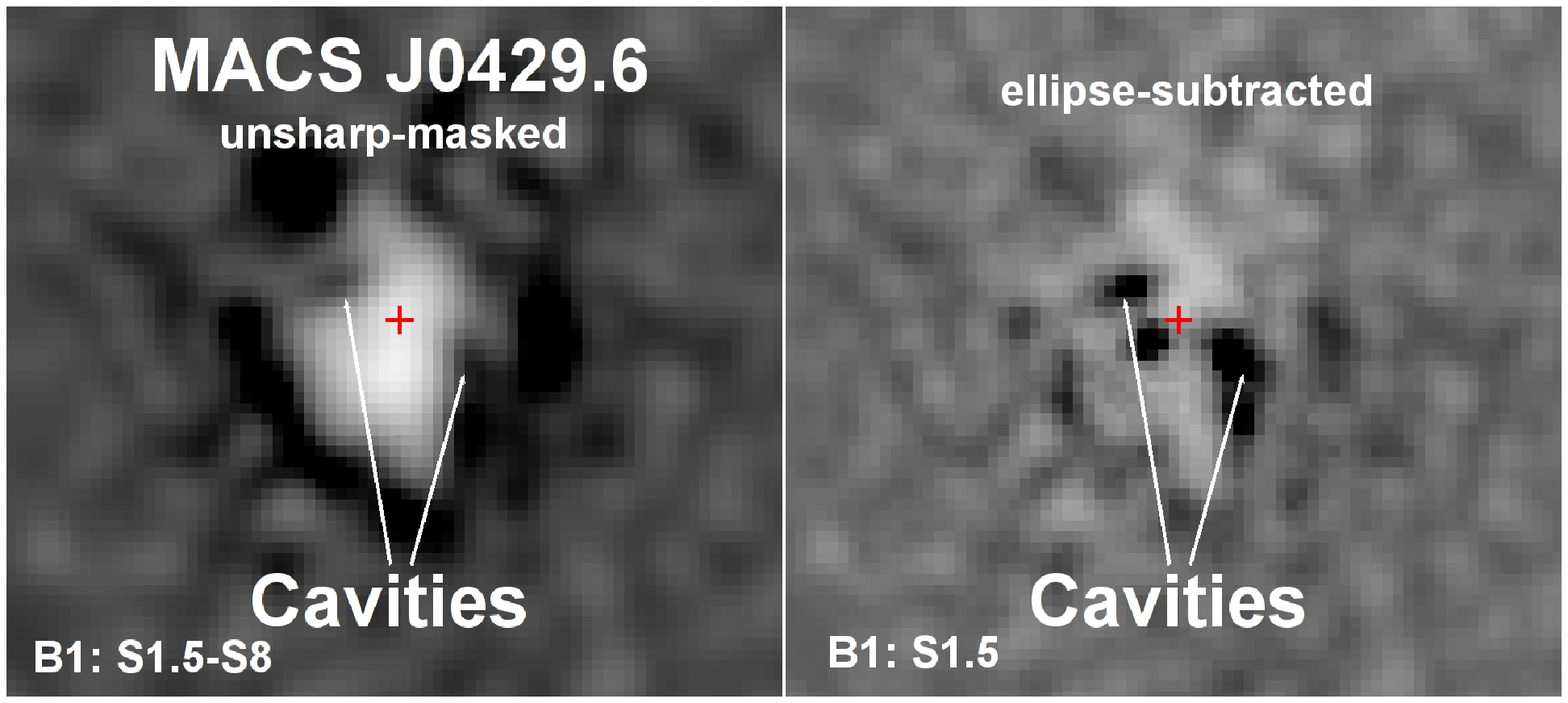}
\end{minipage}
\begin{minipage}[c]{0.49\linewidth}
\centering \includegraphics[width=\linewidth]{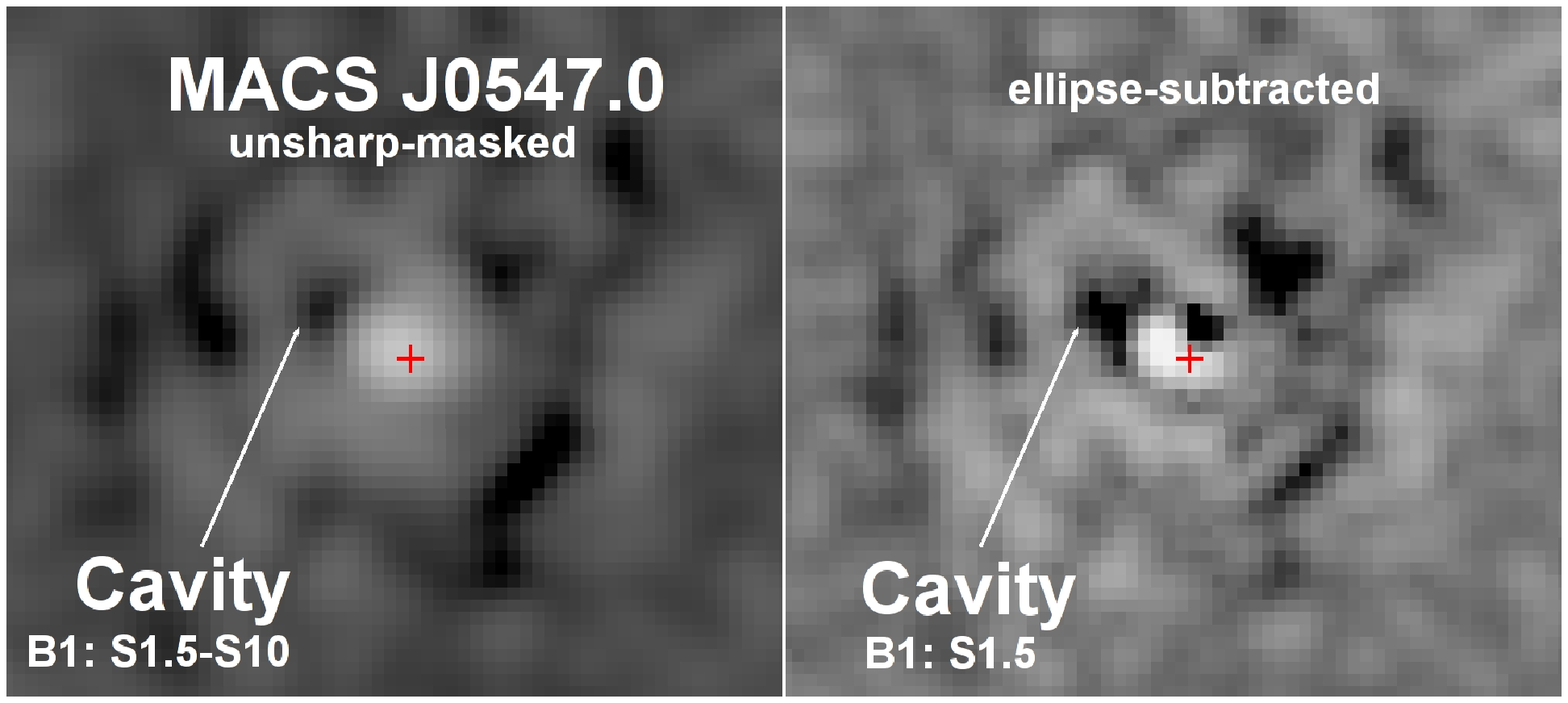}
\end{minipage}
\begin{minipage}[c]{0.49\linewidth}
\centering \includegraphics[width=\linewidth]{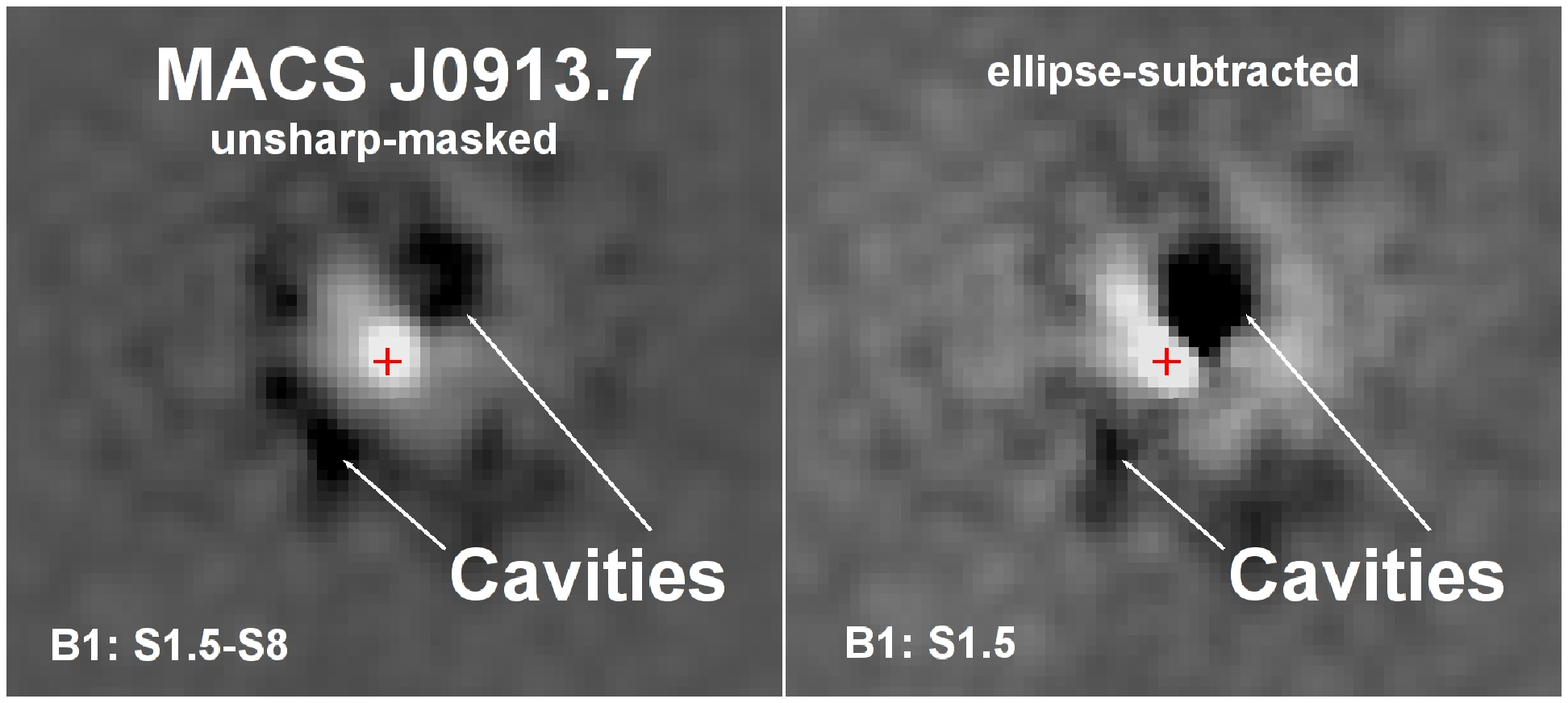}
\end{minipage}
\begin{minipage}[c]{0.49\linewidth}
\centering \includegraphics[width=\linewidth]{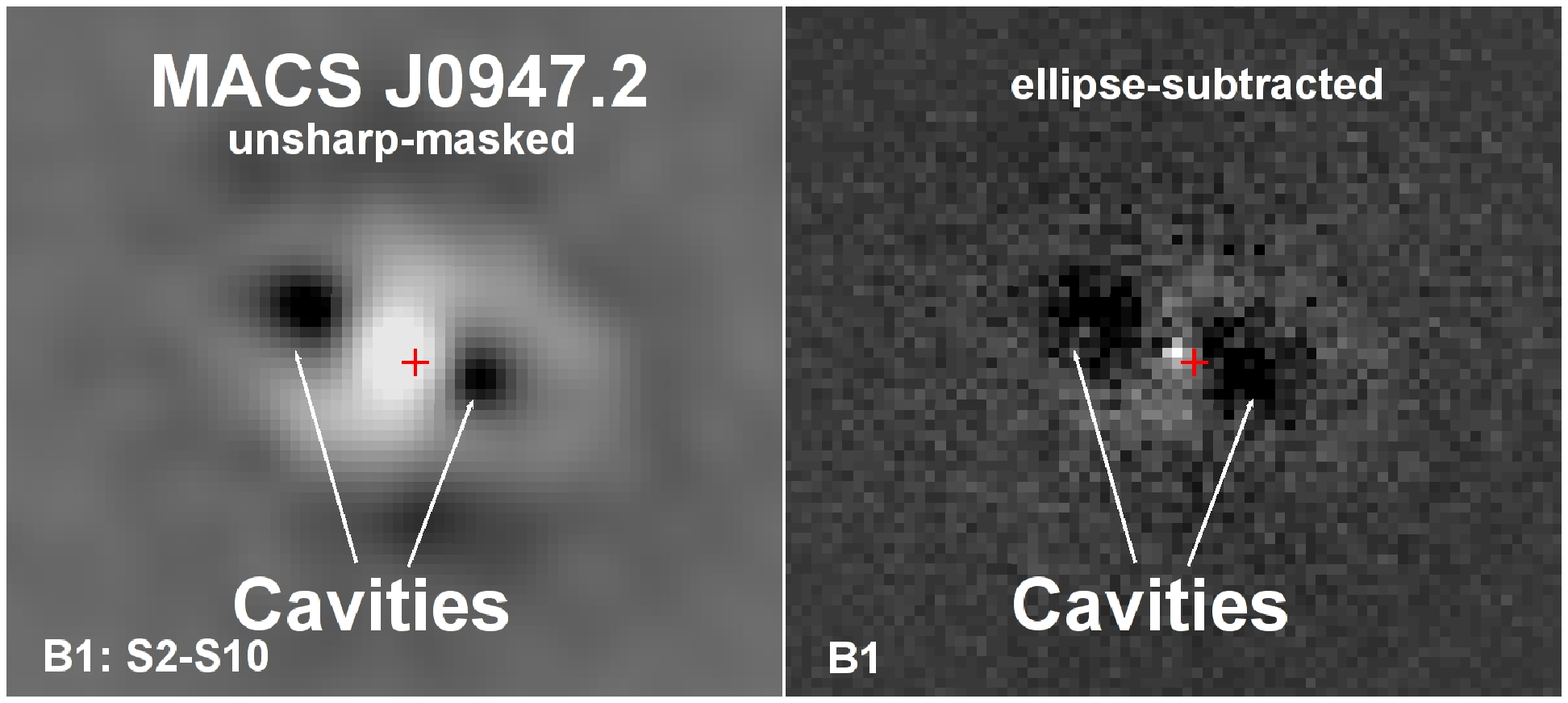}
\end{minipage}
\begin{minipage}[c]{0.49\linewidth}
\centering \includegraphics[width=\linewidth]{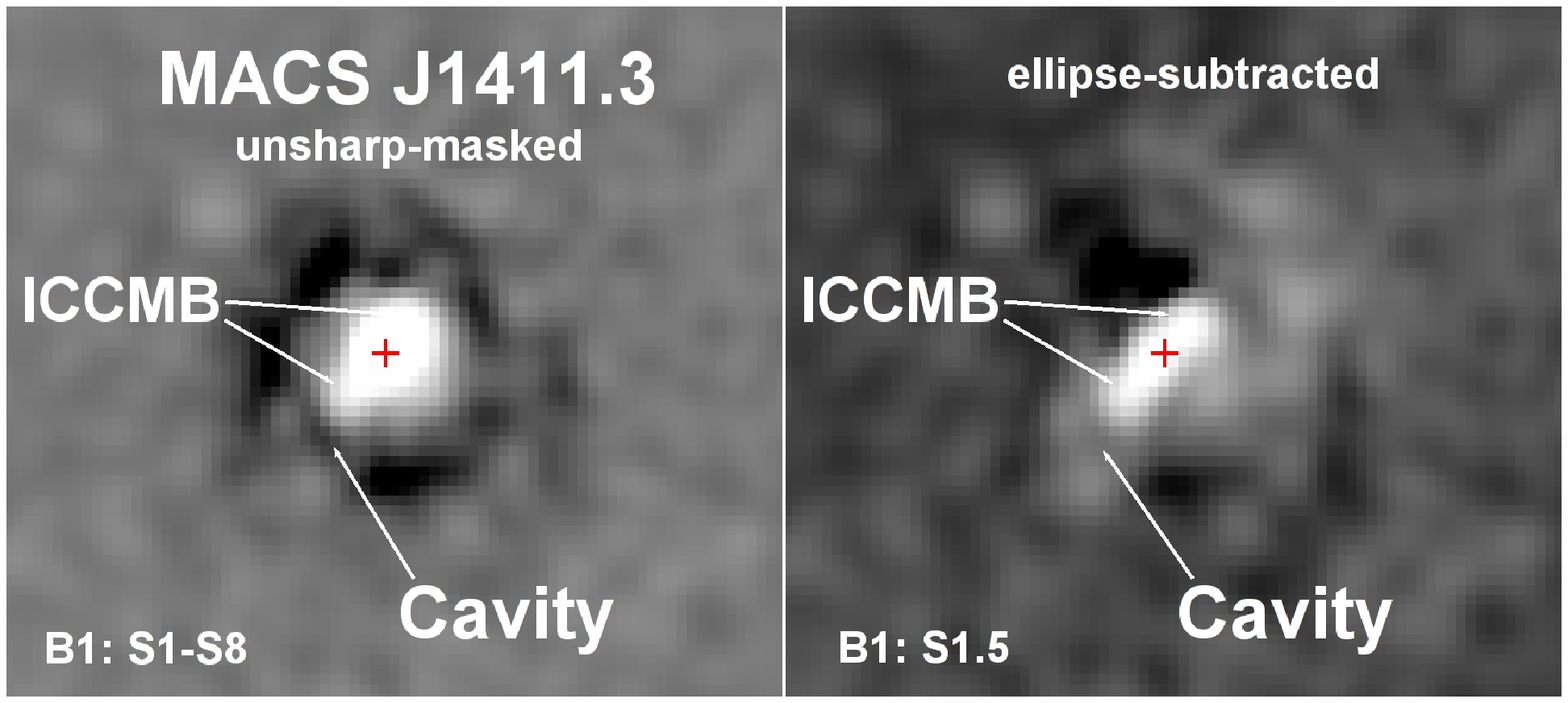}
\end{minipage}
\begin{minipage}[c]{0.49\linewidth}
\centering \includegraphics[width=\linewidth]{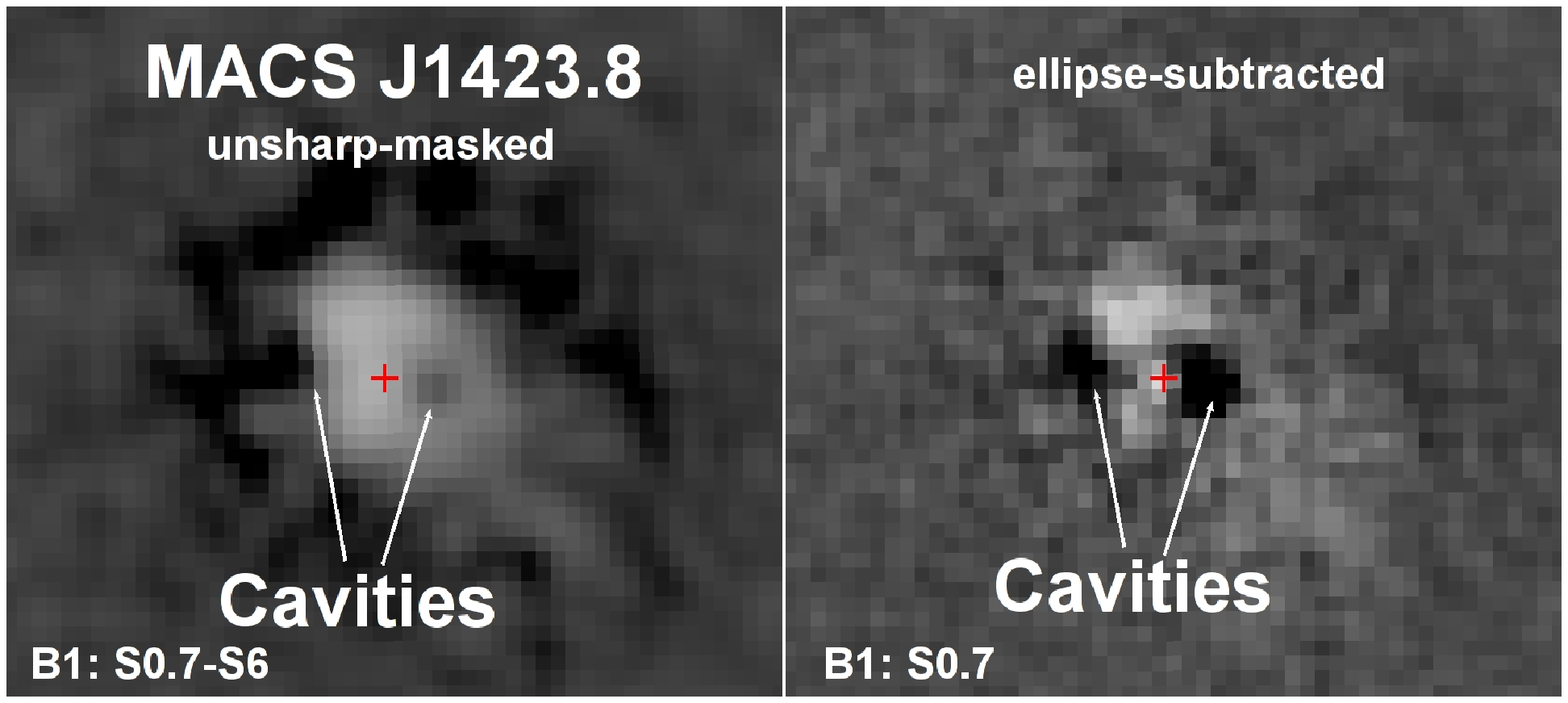}
\end{minipage}
\begin{minipage}[c]{0.49\linewidth}
\centering \includegraphics[width=\linewidth]{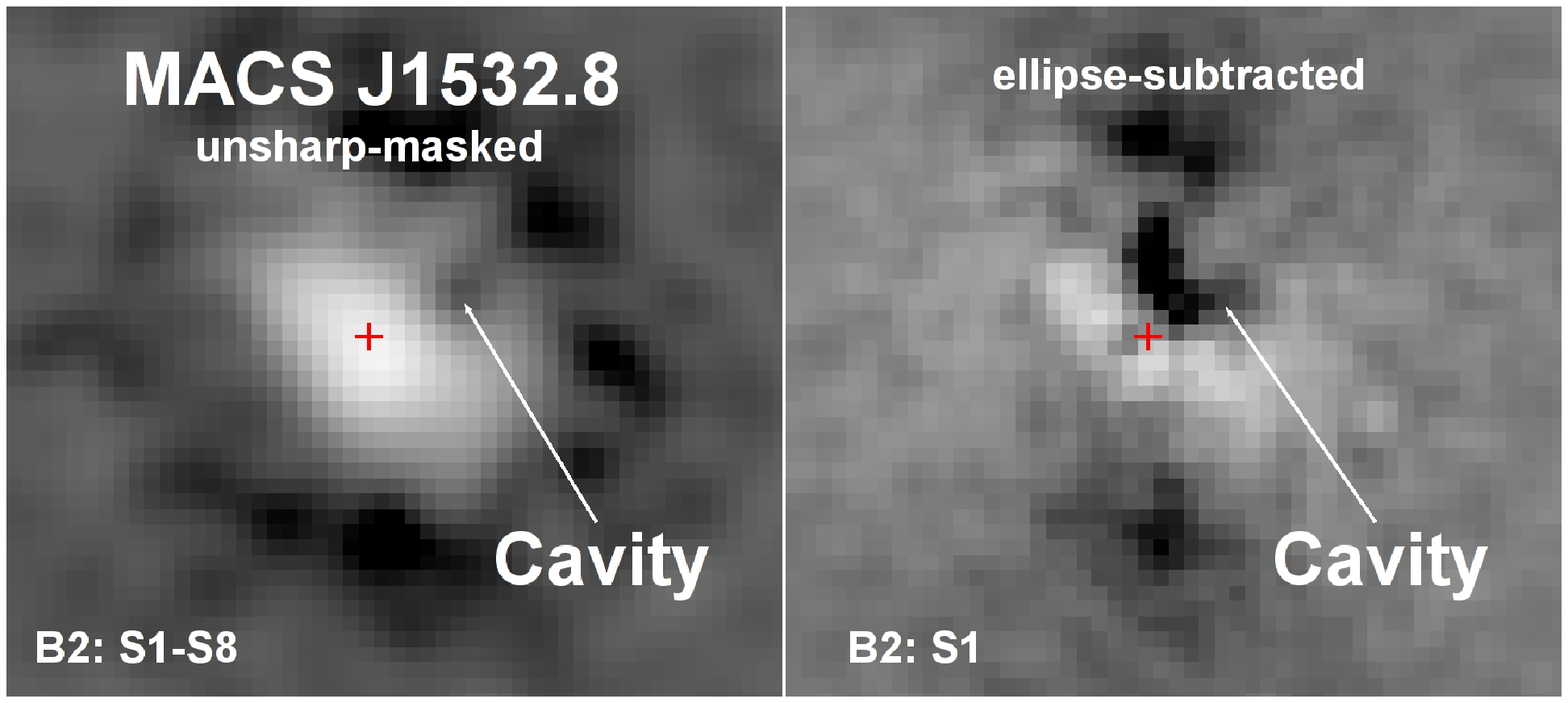}
\end{minipage}
\begin{minipage}[c]{0.49\linewidth}
\centering \includegraphics[width=\linewidth]{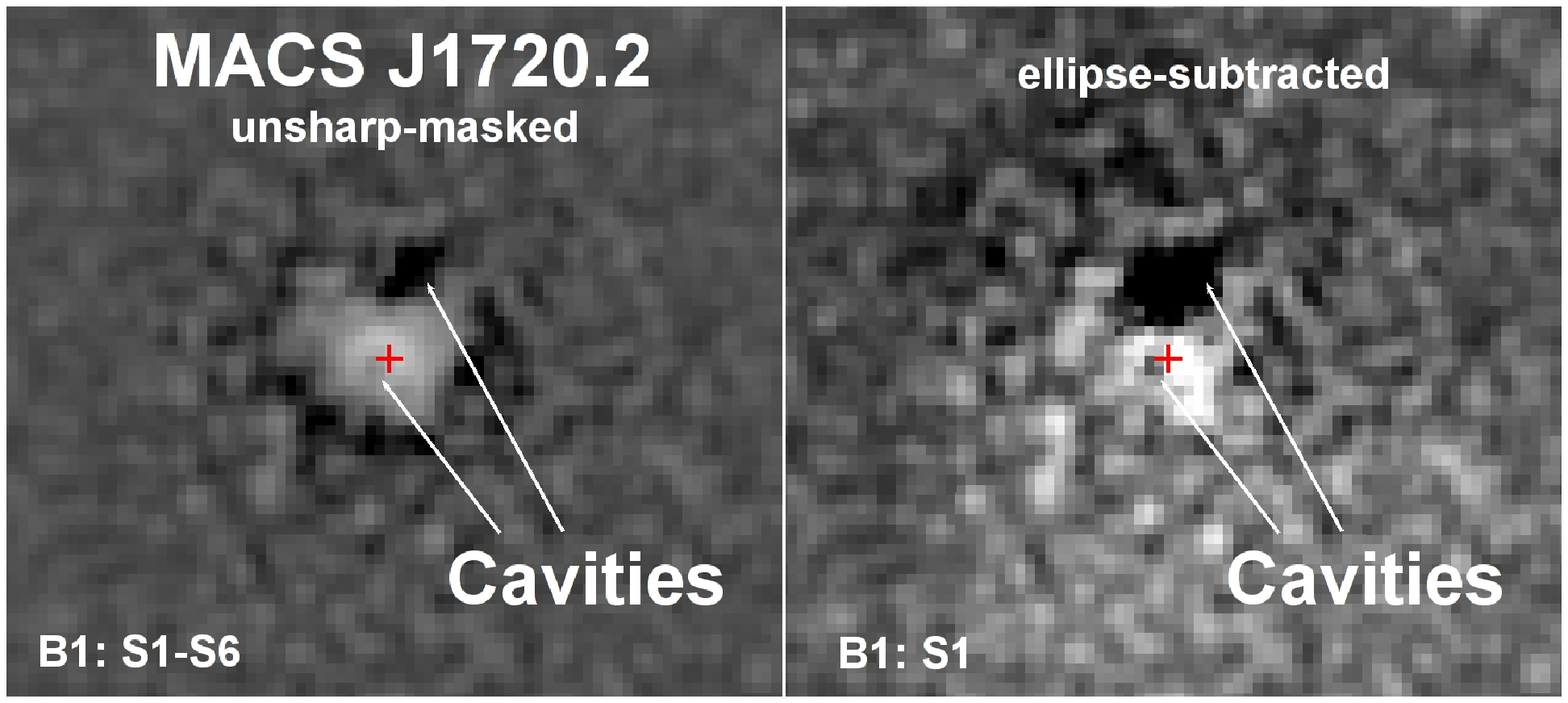}
\end{minipage}
\begin{minipage}[c]{0.49\linewidth}
\centering \includegraphics[width=\linewidth]{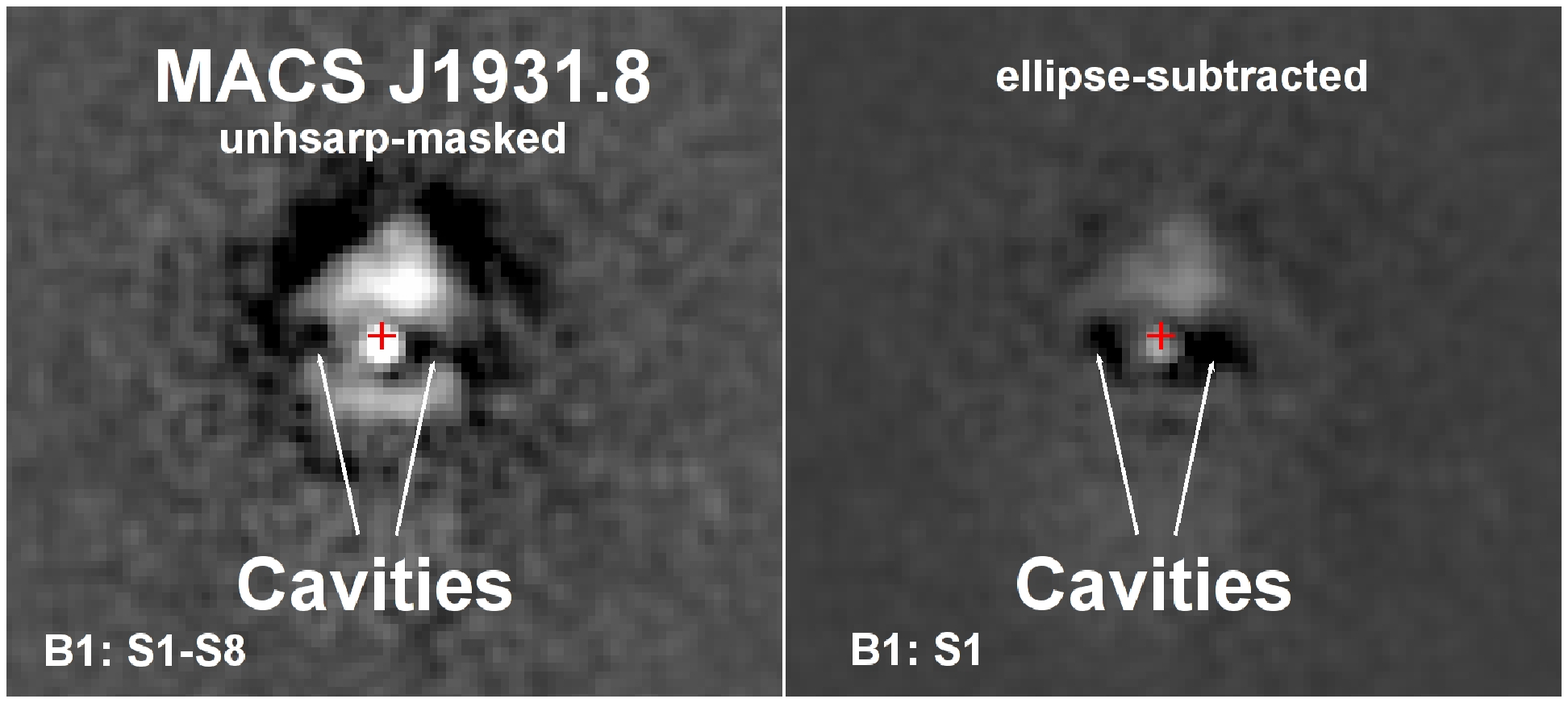}
\end{minipage}
\begin{minipage}[c]{0.49\linewidth}
\centering \includegraphics[width=\linewidth]{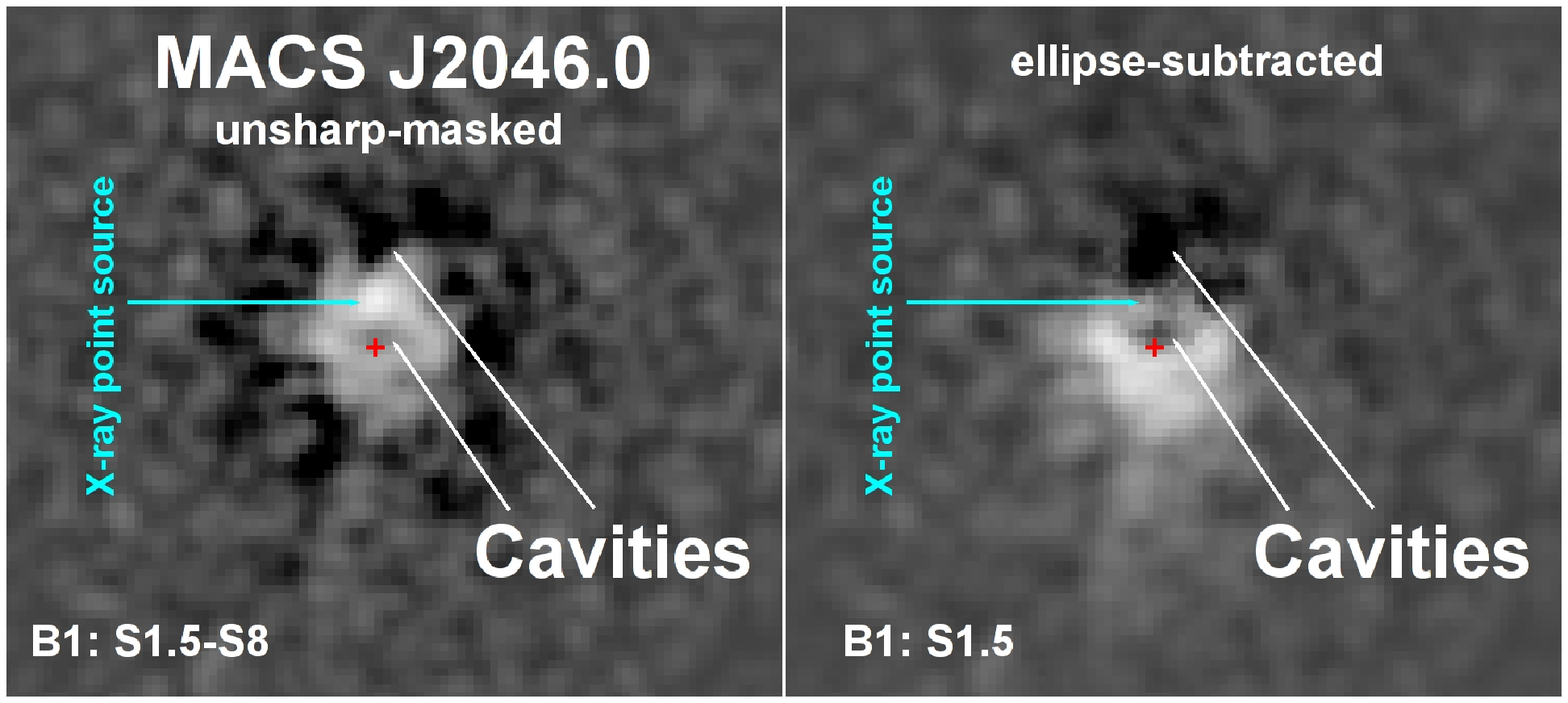}
\end{minipage}
\caption{See caption on next page.}
\end{figure*}
\begin{figure*}
\ContinuedFloat
\centering
\begin{minipage}[c]{0.49\linewidth}
\centering \includegraphics[width=\linewidth]{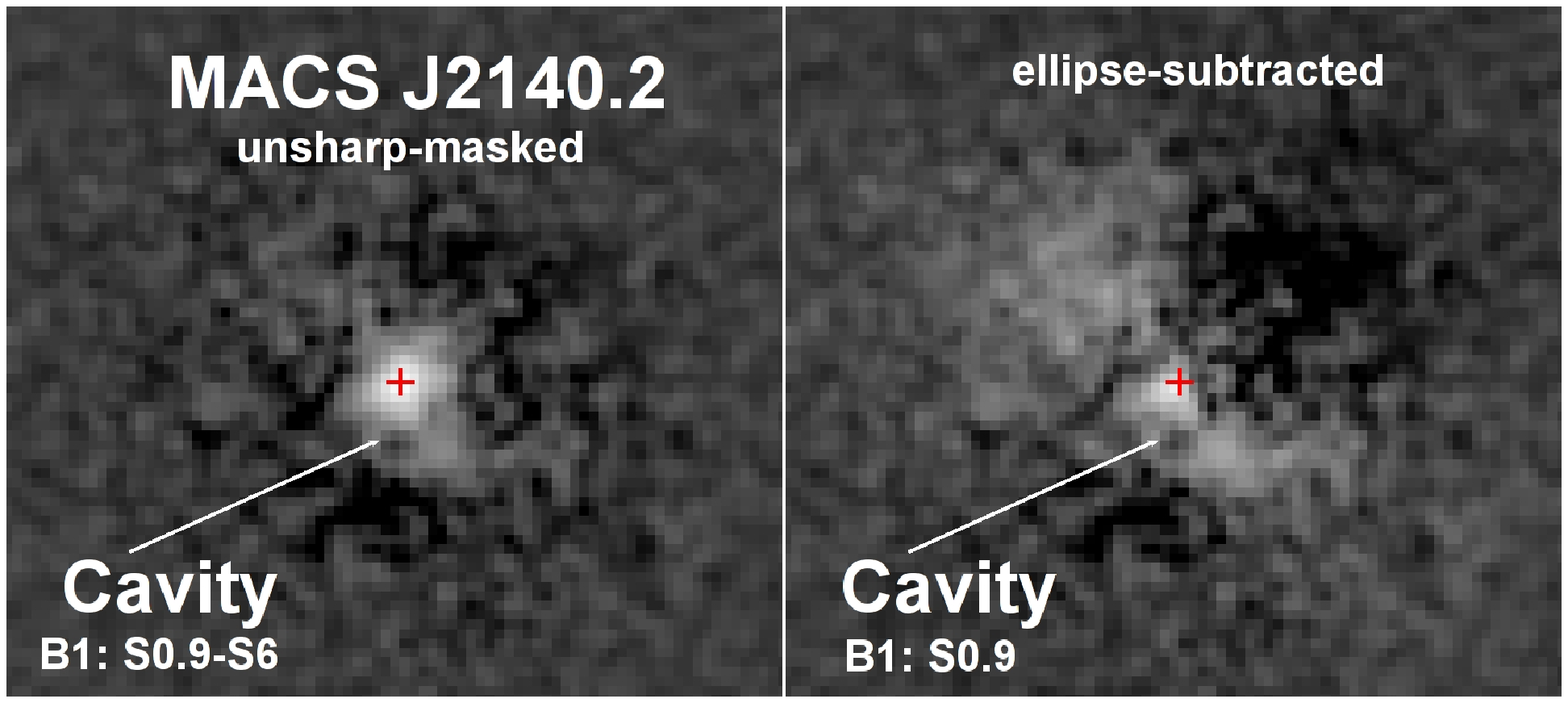}
\end{minipage}
\begin{minipage}[c]{0.49\linewidth}
\centering \includegraphics[width=\linewidth]{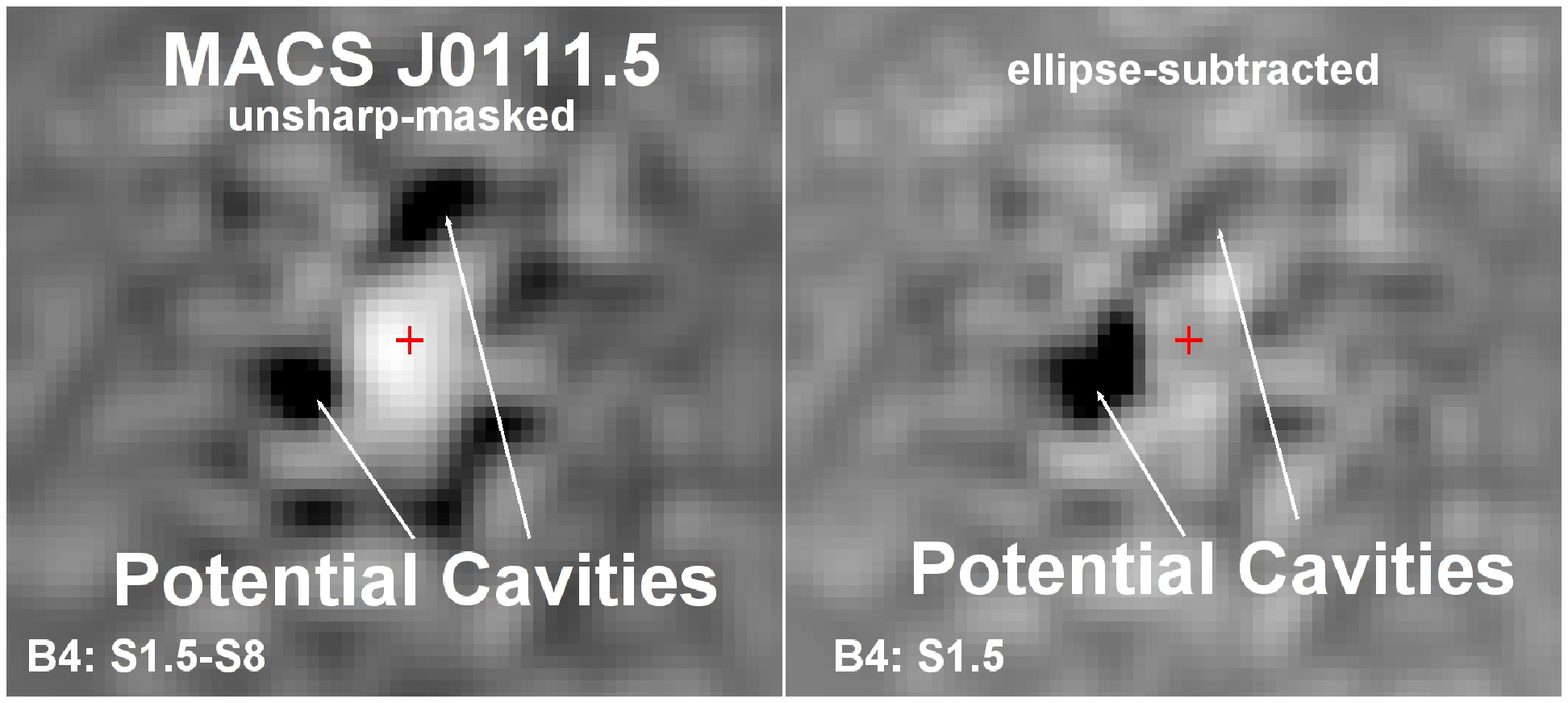}
\end{minipage}
\begin{minipage}[c]{0.49\linewidth}
\centering \includegraphics[width=\linewidth]{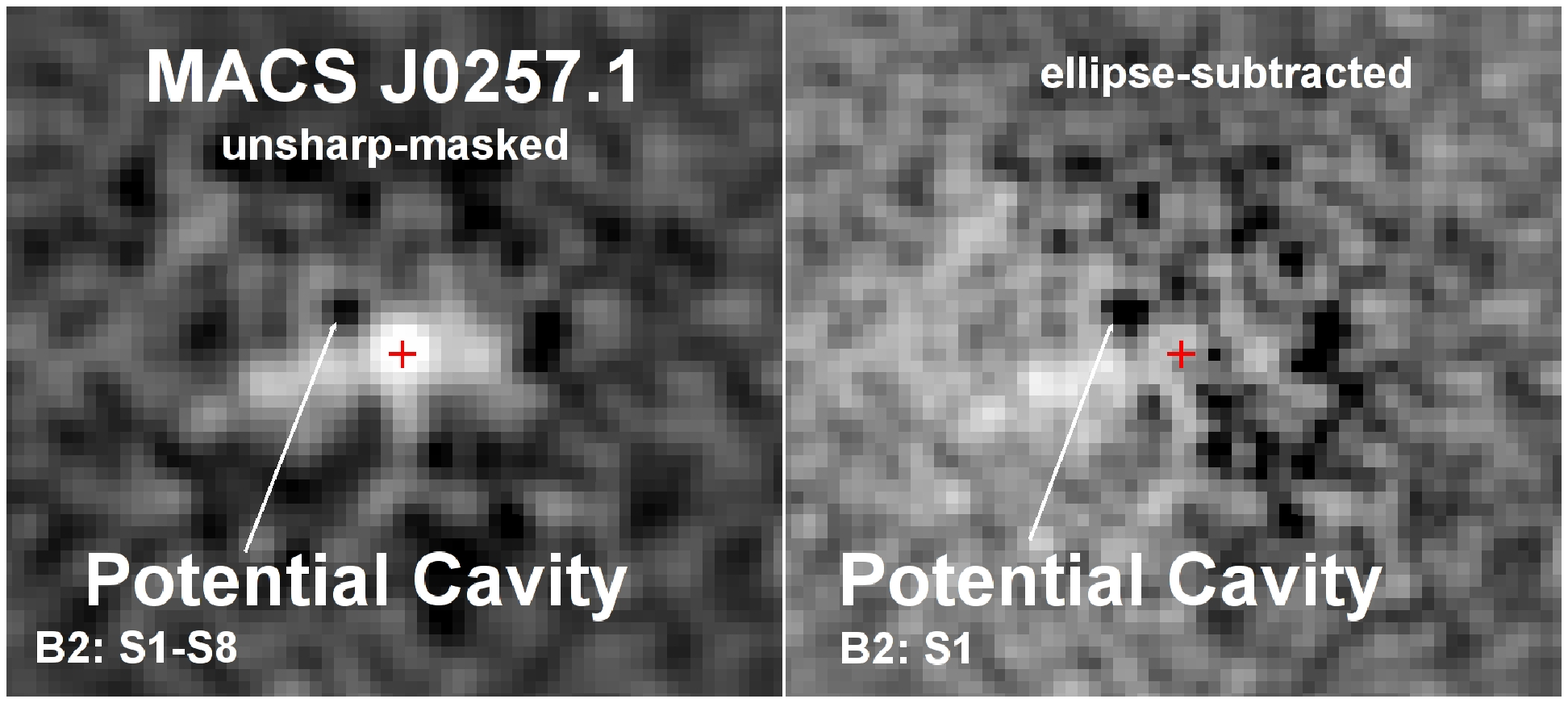}
\end{minipage}
\begin{minipage}[c]{0.49\linewidth}
\centering \includegraphics[width=\linewidth]{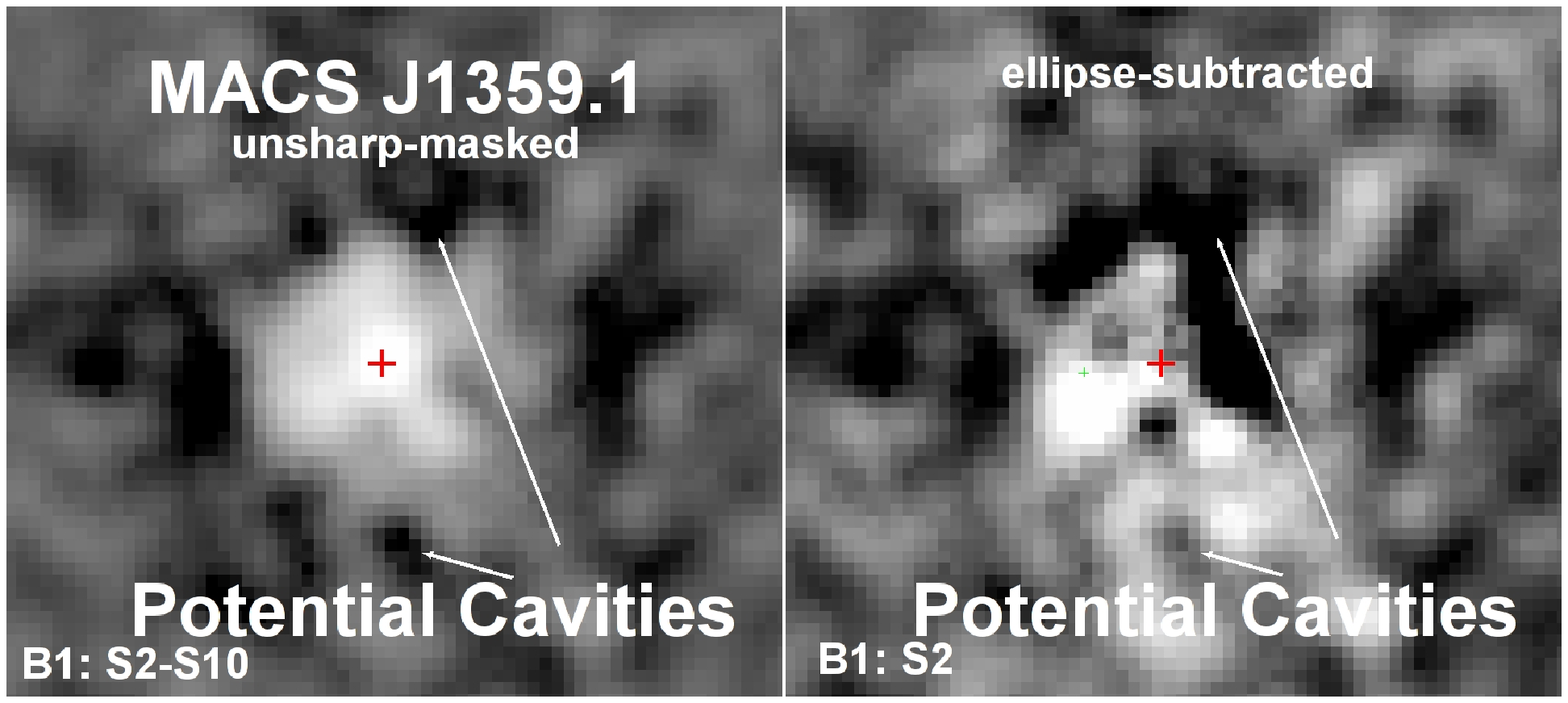}
\end{minipage}
\begin{minipage}[c]{0.49\linewidth}
\centering \includegraphics[width=\linewidth]{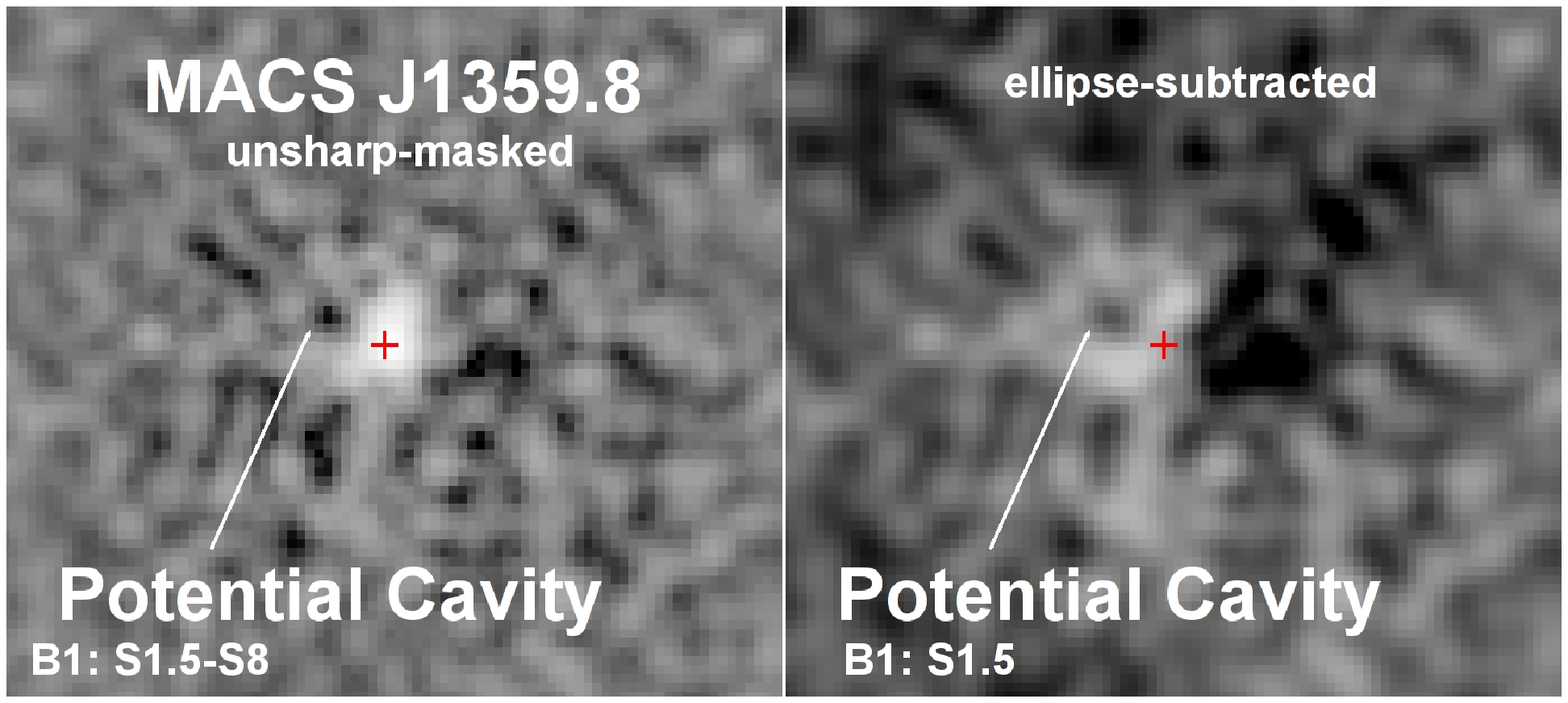}
\end{minipage}
\begin{minipage}[c]{0.49\linewidth}
\centering \includegraphics[width=\linewidth]{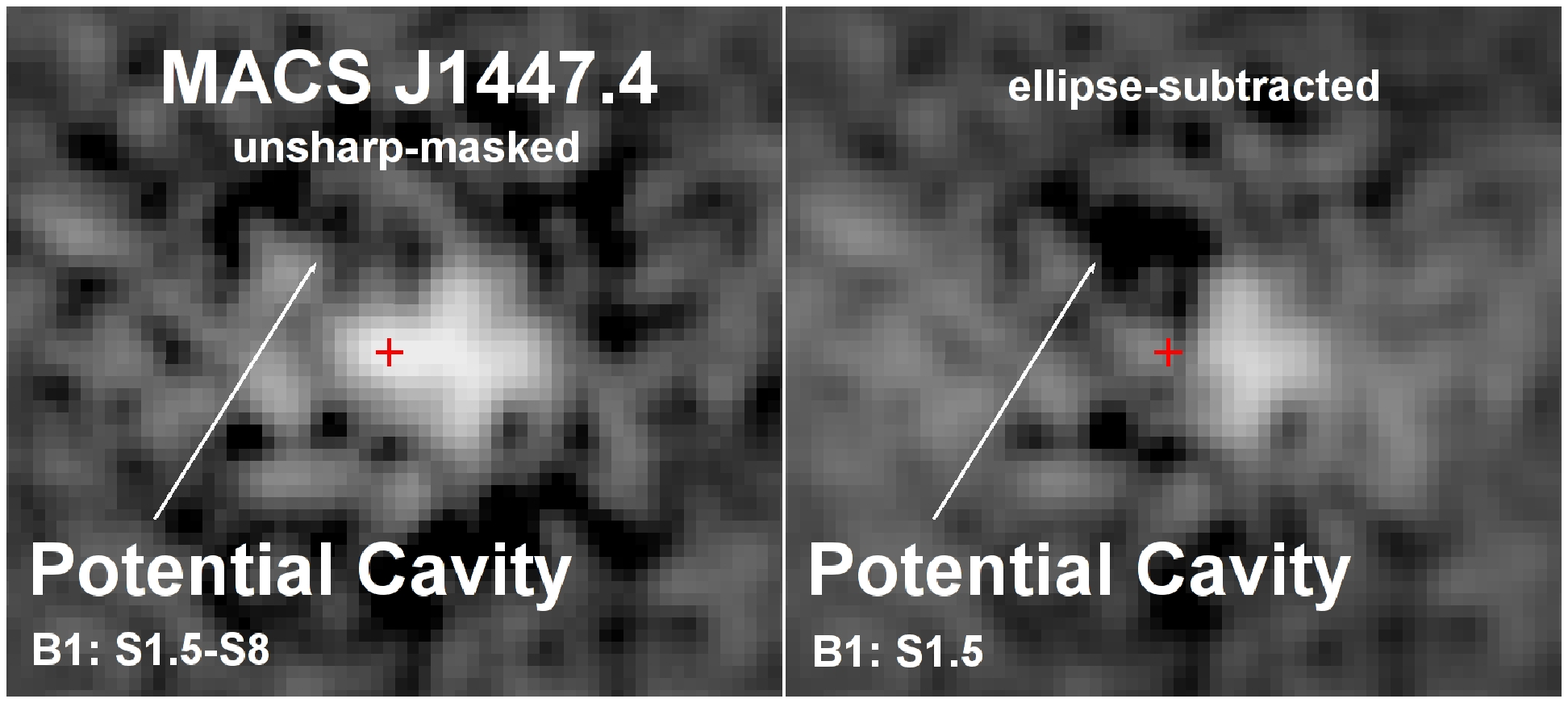}
\end{minipage}
\begin{minipage}[c]{0.49\linewidth}
\centering \includegraphics[width=\linewidth]{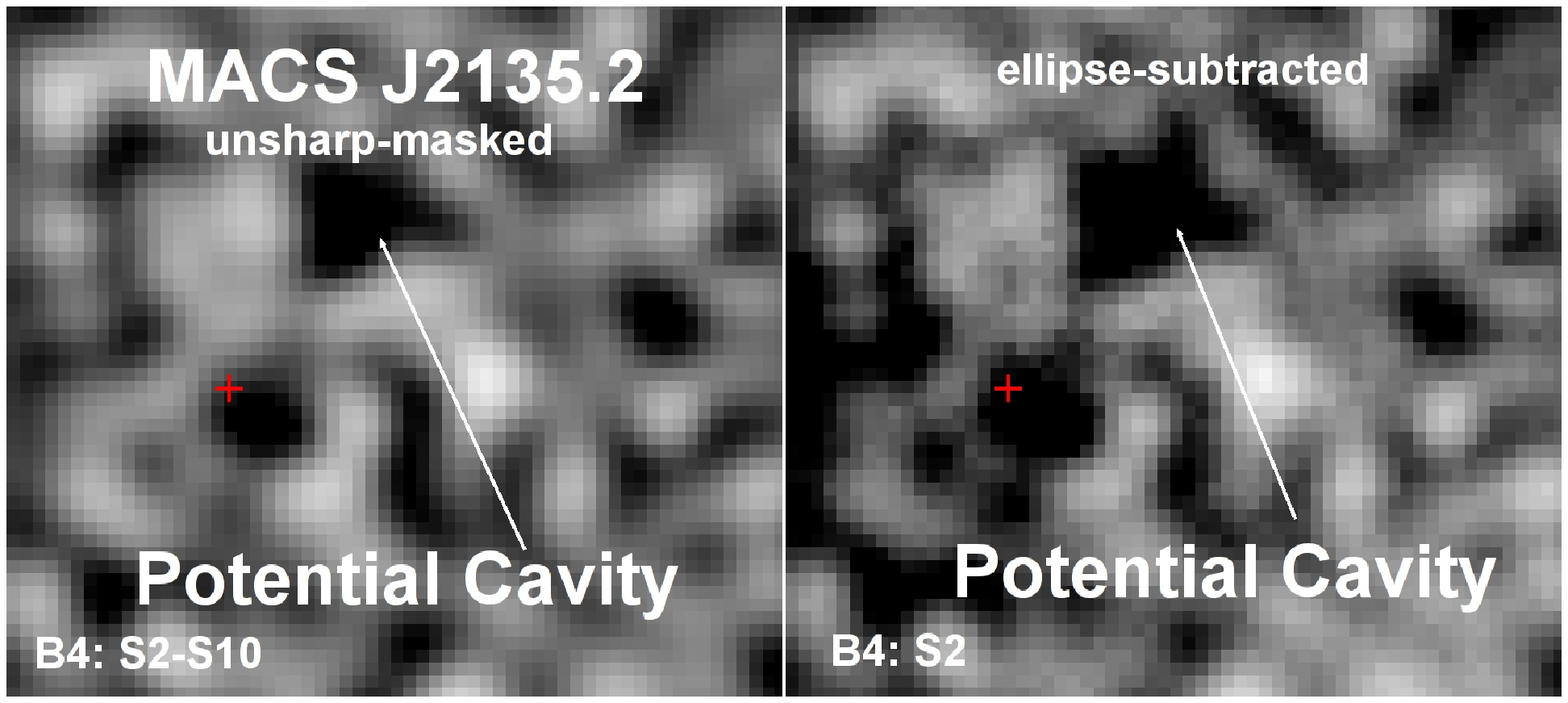}
\end{minipage}
\begin{minipage}[c]{0.49\linewidth}
\centering \includegraphics[width=\linewidth]{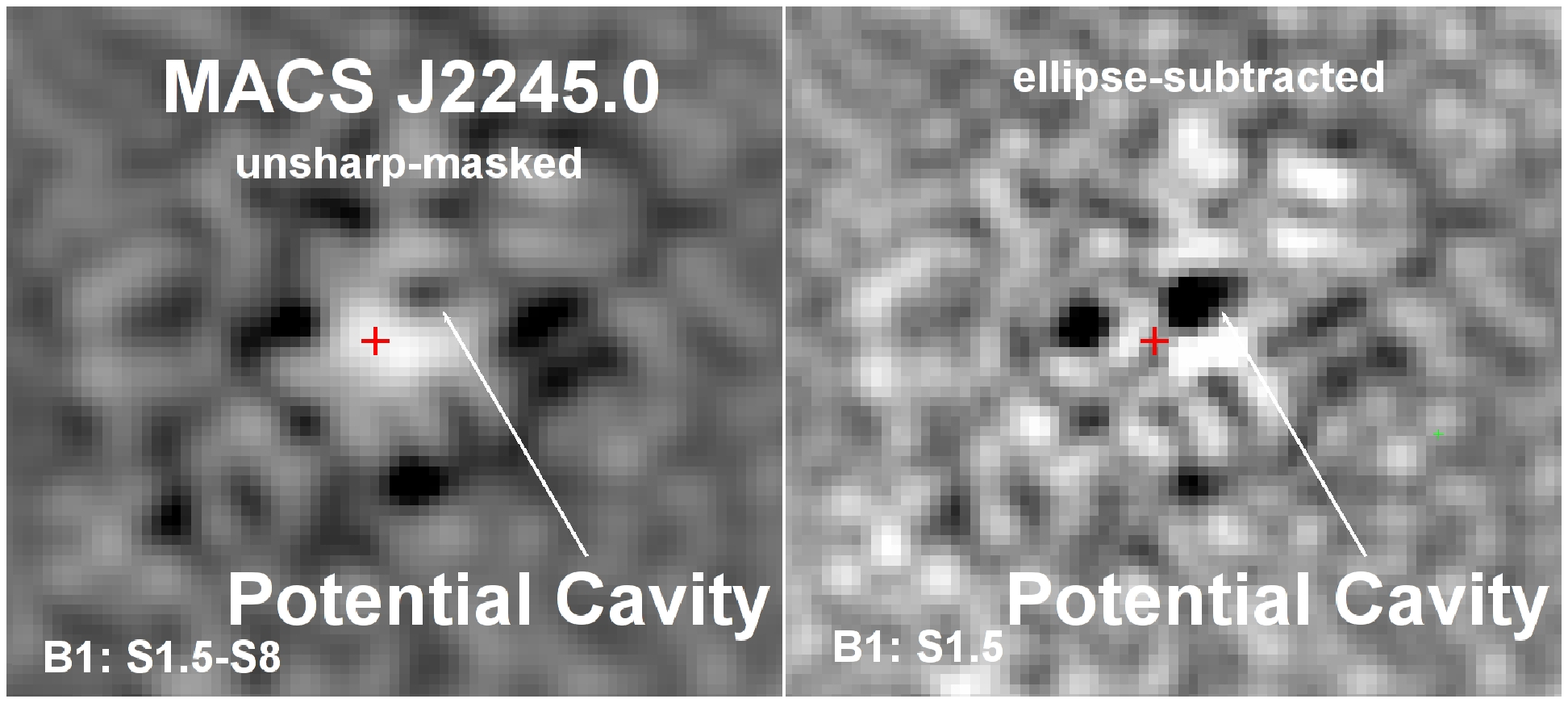}
\end{minipage}
\caption{Unsharp-masked and ellipse-subtracted images for all clusters in which we identified cavities. Same layout and scaling as Fig. \ref{fig2}: two clusters for each horizontal line with the unsharp-masked $0.5-7$ keV X-ray image shown in the left panel and the ellipse-subtracted $0.5-7$ keV X-ray image shown in the right panel. We identify the cavities in each system, and we show with the red cross symbol in the X-ray image, the location of the BCG. We also show in the lower-left corner of each image the binning (${\rm B_{X}}$) and smoothing (${\rm S_{X}}$) scales adopted for creating the images. For each unsharp-masked image, we show the 2 smoothing scales used to create the image (${\rm S_{X_1}-S_{X_2}}$), and for each ellipse-subtracted image, we show the smoothing scale adopted (${\rm S_{X_3}}$) for the original image before creating and subtracting an elliptical model. If no ${\rm S_{X}}$ is shown, then no smoothing was applied to the image. The first 13 clusters have ``clear" cavities, while the remaining 7 only have ``potential" cavities (see ``potential cavities" indicated in the figure). We emphasize that deeper observations are needed to confirm if the ``potential" cavities are real.}
\label{figA2}
\end{figure*}

\section{Spectral profiles}
Fig. \ref{figA1} shows the projected (black symbols) and deprojected (red symbols) temperature, electron density, electron pressure and entropy profiles for each cluster in which a cavity was identified. We used $\chi^2$ statistics and each annulus had a minimum of 900 counts, but for clusters in which the data was of good quality we allowed the annuli to have up to 4000 counts. For MACS~J0111.5+0855 and MACS~J2135.2-0102, we could not deproject the spectra because of the limiting data quality. In this case, we used C-statistics and each annulus only had $\sim400$ counts. 

\begin{figure*}
\centering
\begin{minipage}[c]{0.245\linewidth}
\centering \includegraphics[width=\linewidth]{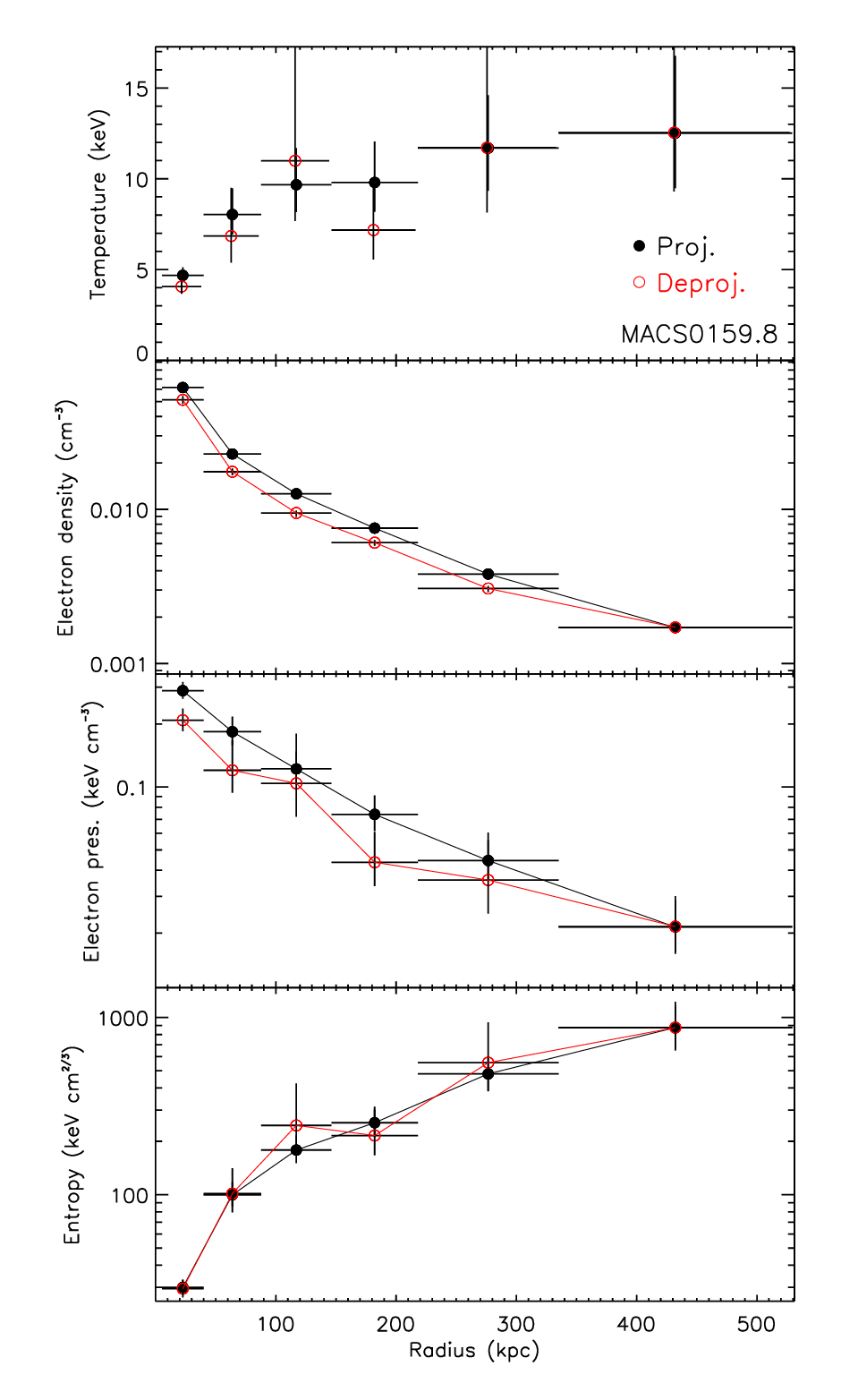}
\end{minipage}
\begin{minipage}[c]{0.245\linewidth}
\centering \includegraphics[width=\linewidth]{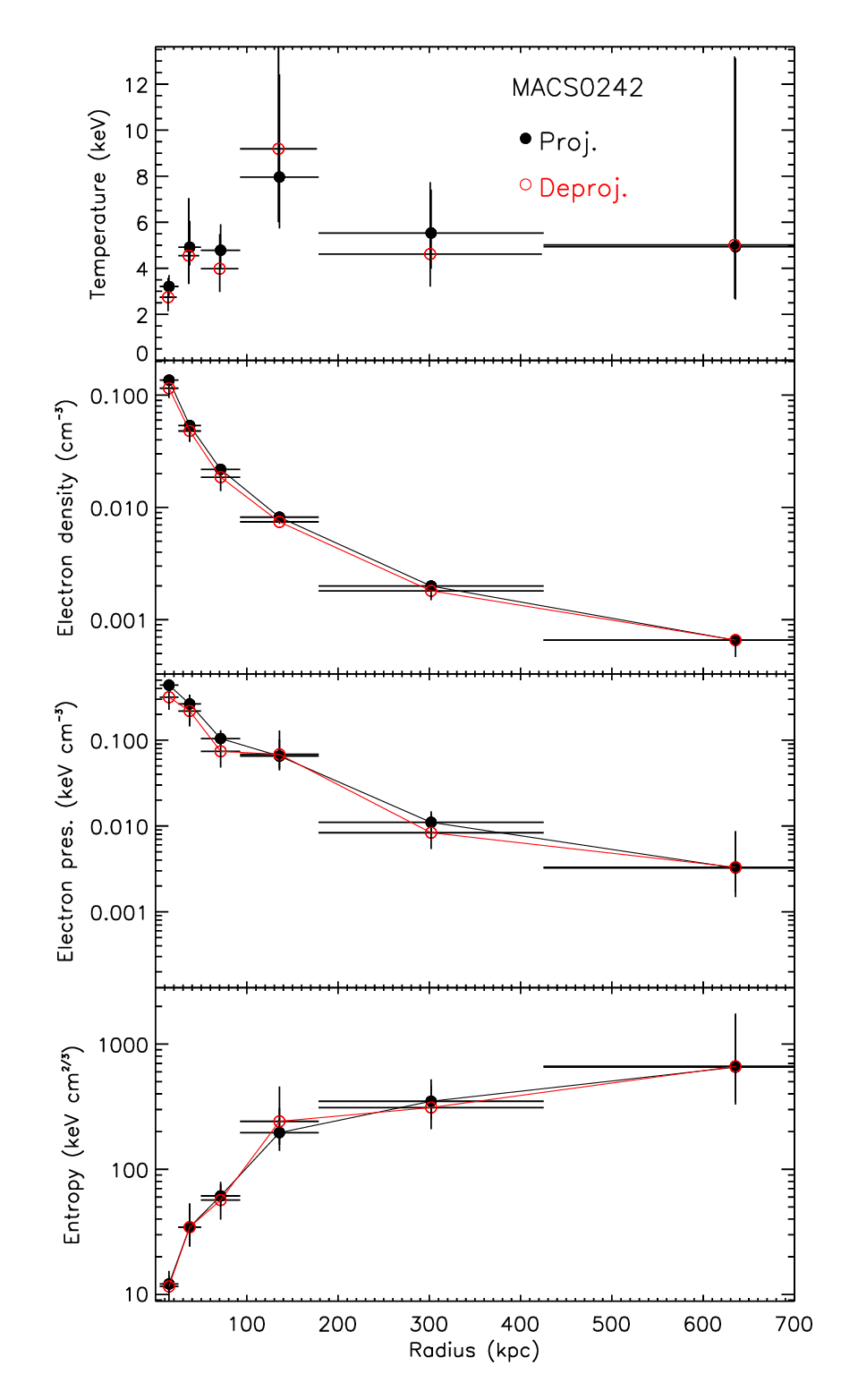}
\end{minipage}
\begin{minipage}[c]{0.245\linewidth}
\centering \includegraphics[width=\linewidth]{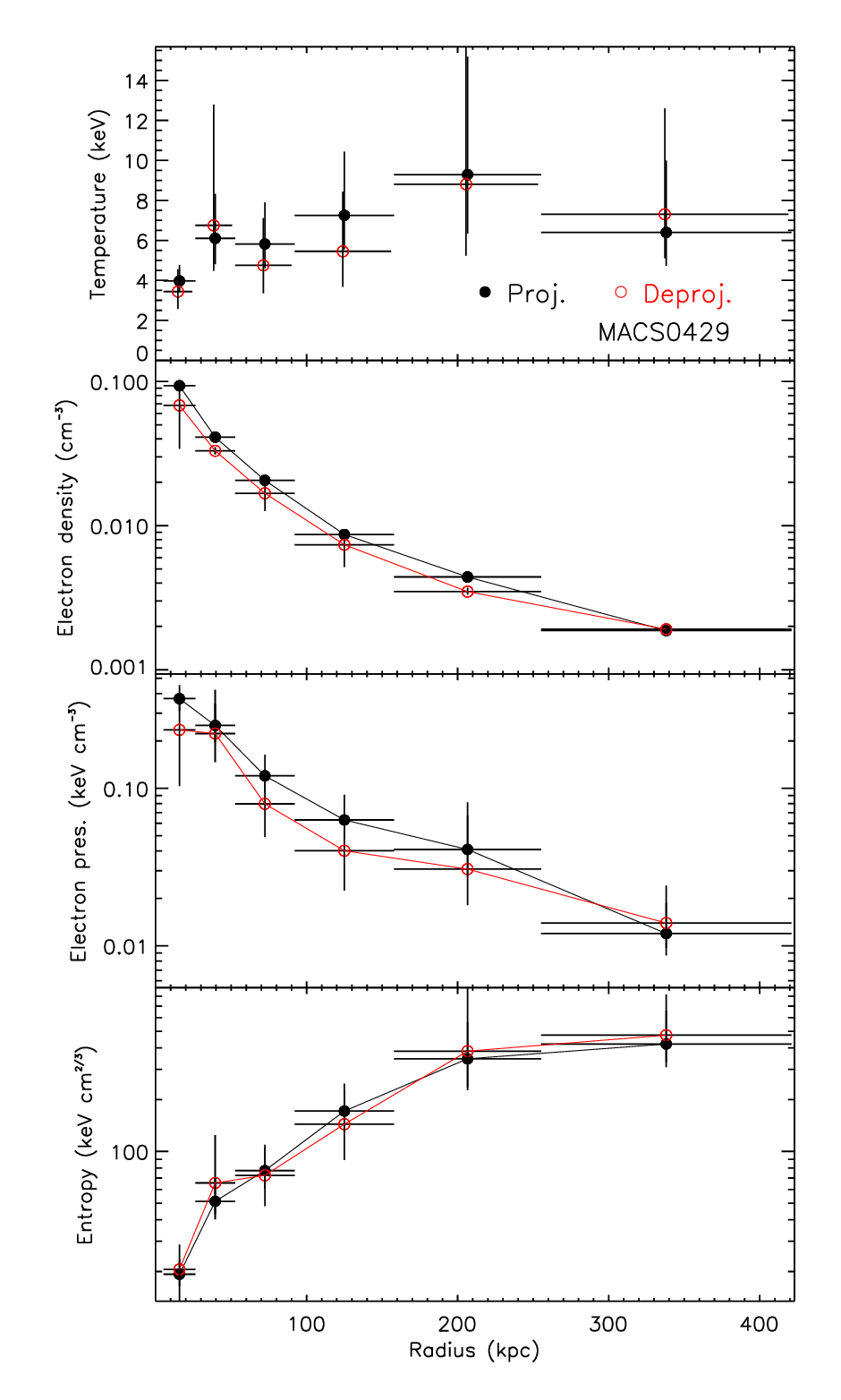}
\end{minipage}
\begin{minipage}[c]{0.245\linewidth}
\centering \includegraphics[width=\linewidth]{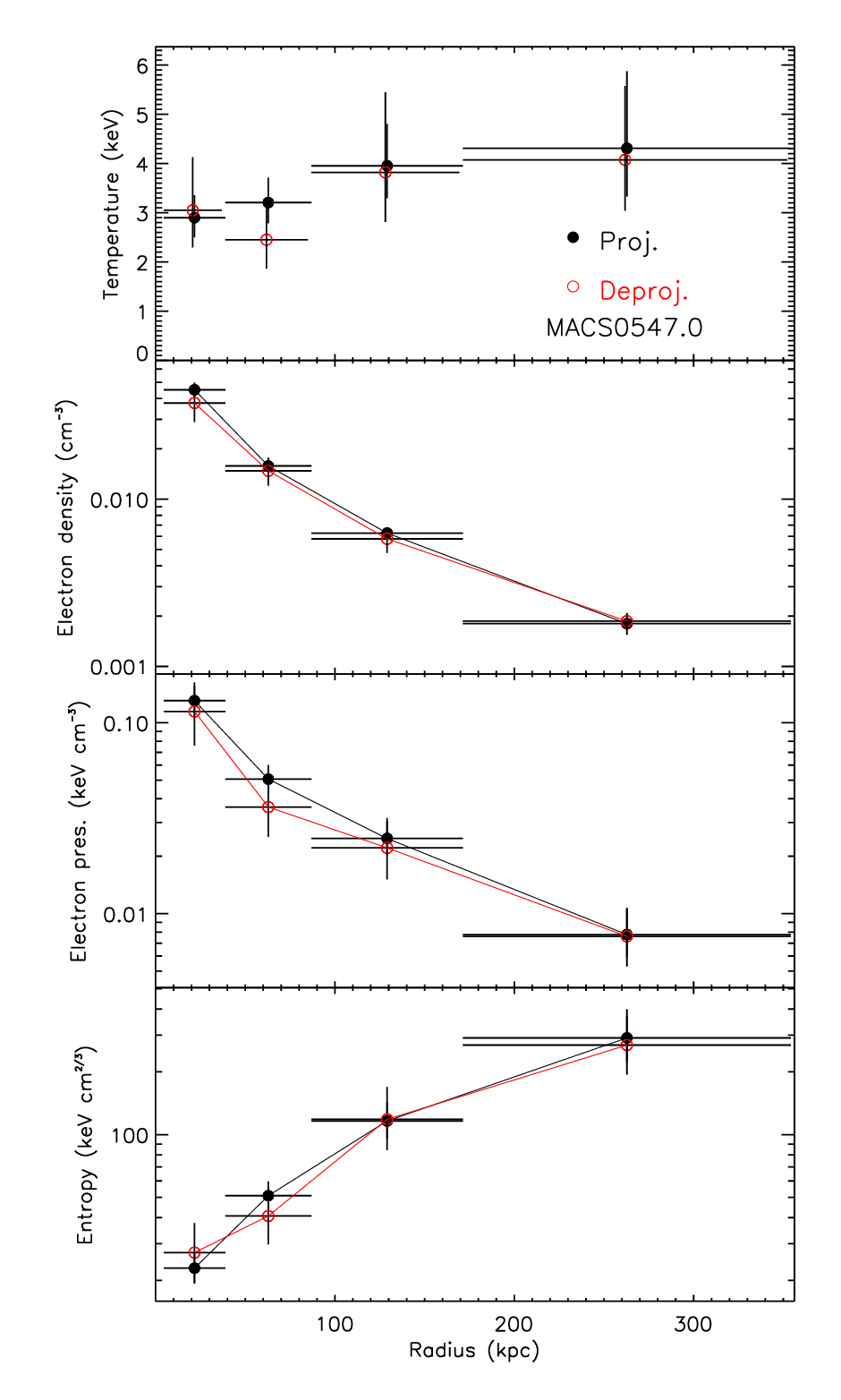}
\end{minipage}
\begin{minipage}[c]{0.245\linewidth}
\centering \includegraphics[width=\linewidth]{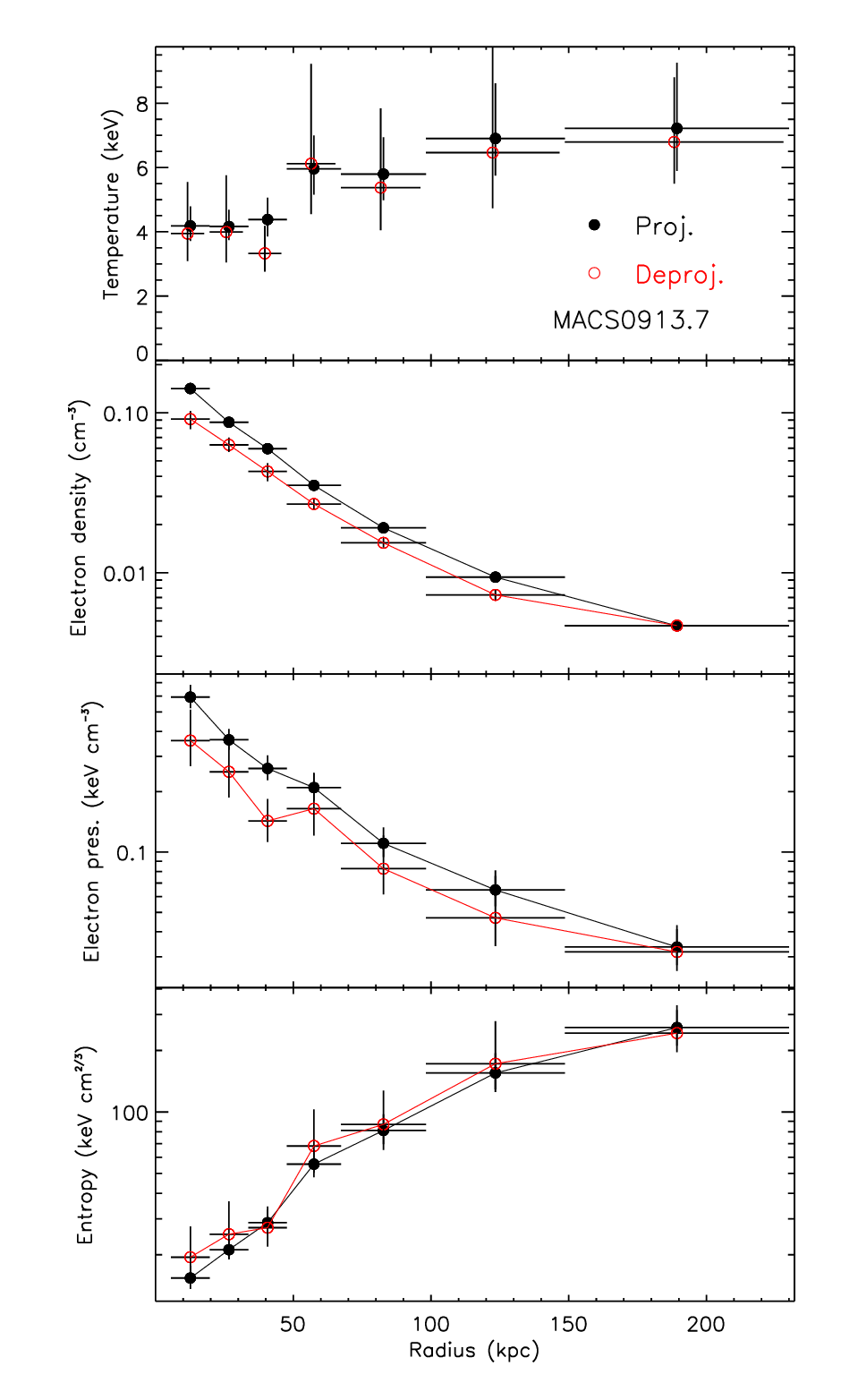}
\end{minipage}
\begin{minipage}[c]{0.245\linewidth}
\centering \includegraphics[width=\linewidth]{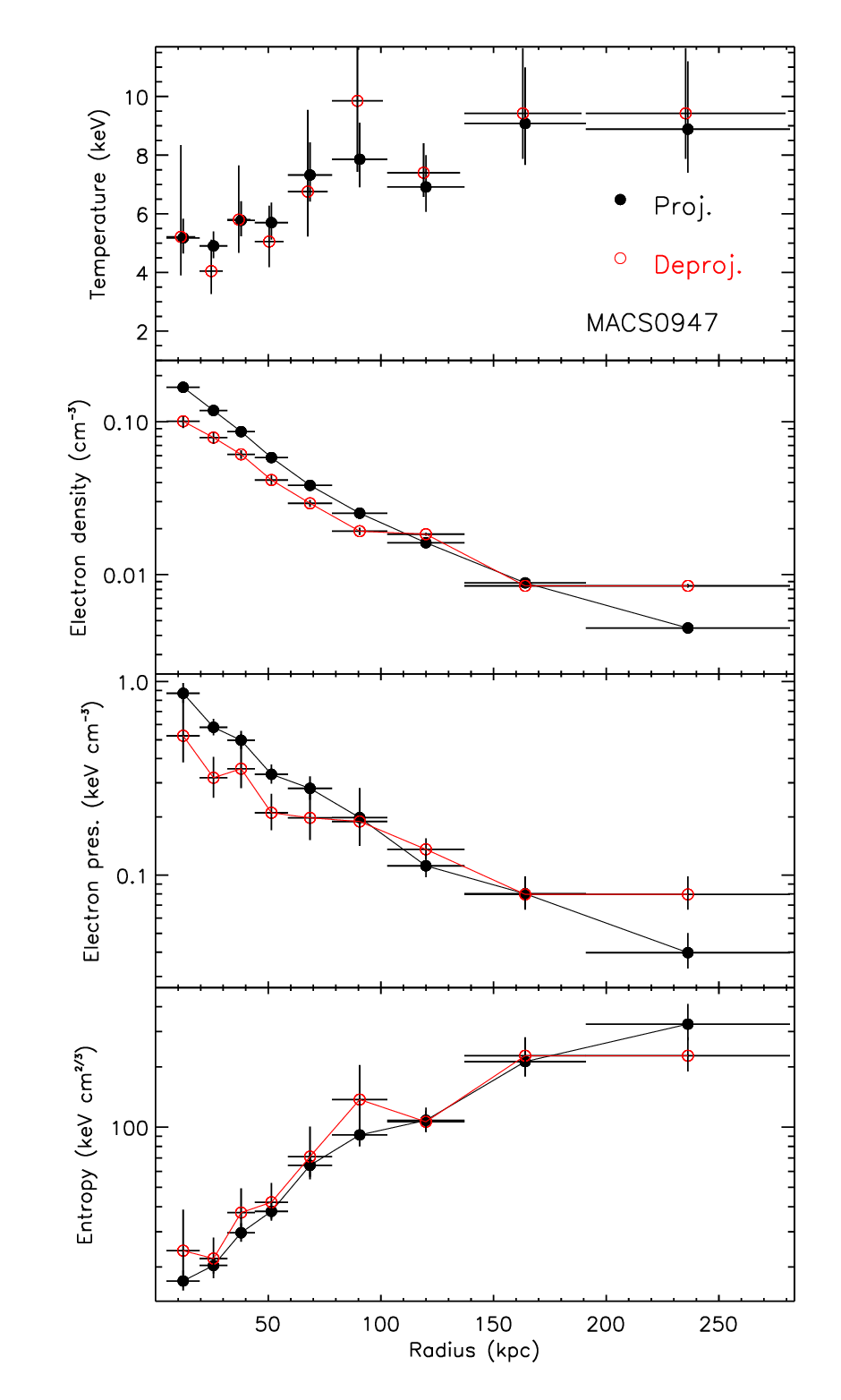}
\end{minipage}
\begin{minipage}[c]{0.245\linewidth}
\centering \includegraphics[width=\linewidth]{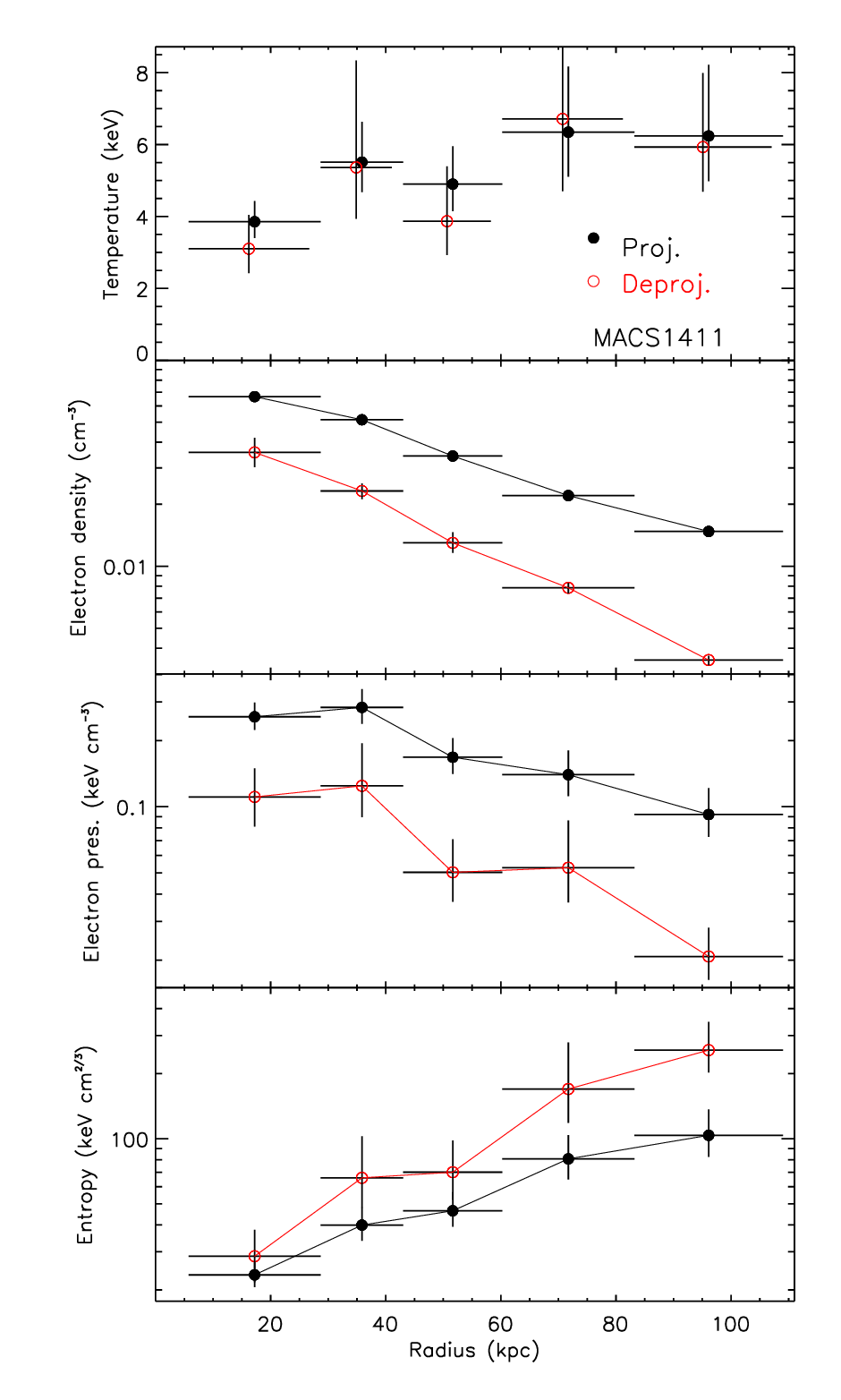}
\end{minipage}
\begin{minipage}[c]{0.245\linewidth}
\centering \includegraphics[width=\linewidth]{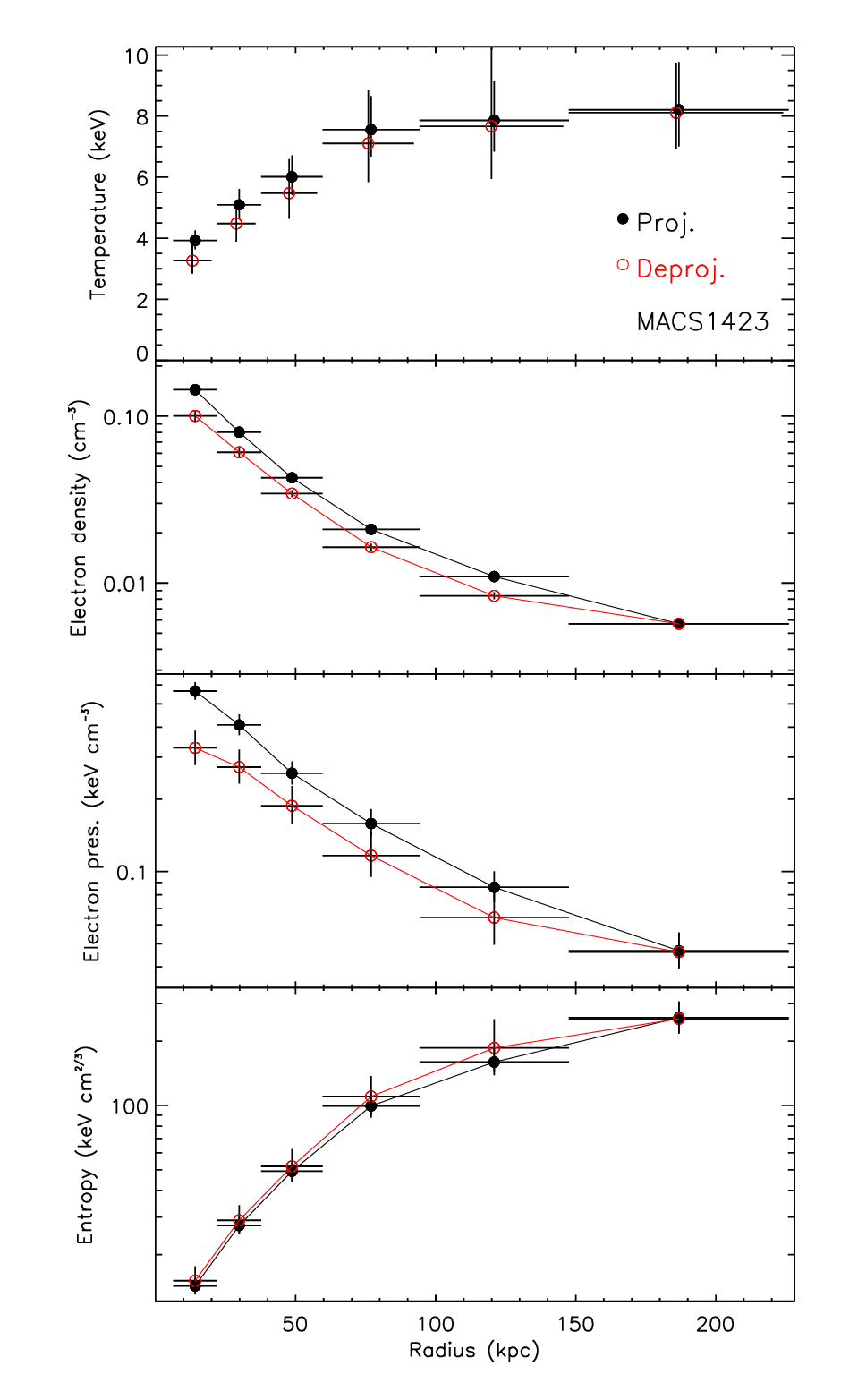}
\end{minipage}
\begin{minipage}[c]{0.245\linewidth}
\centering \includegraphics[width=\linewidth]{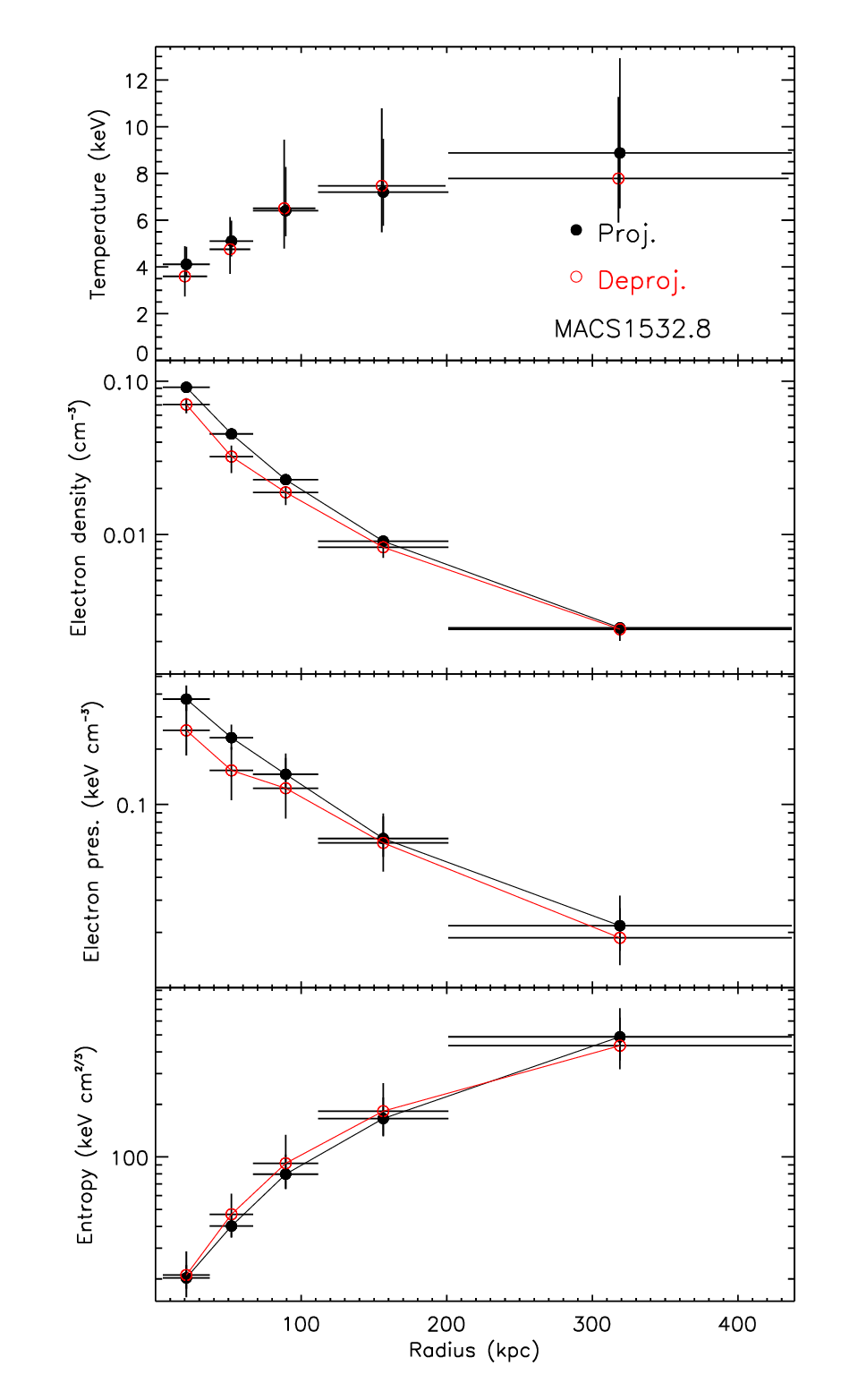}
\end{minipage}
\begin{minipage}[c]{0.245\linewidth}
\centering \includegraphics[width=\linewidth]{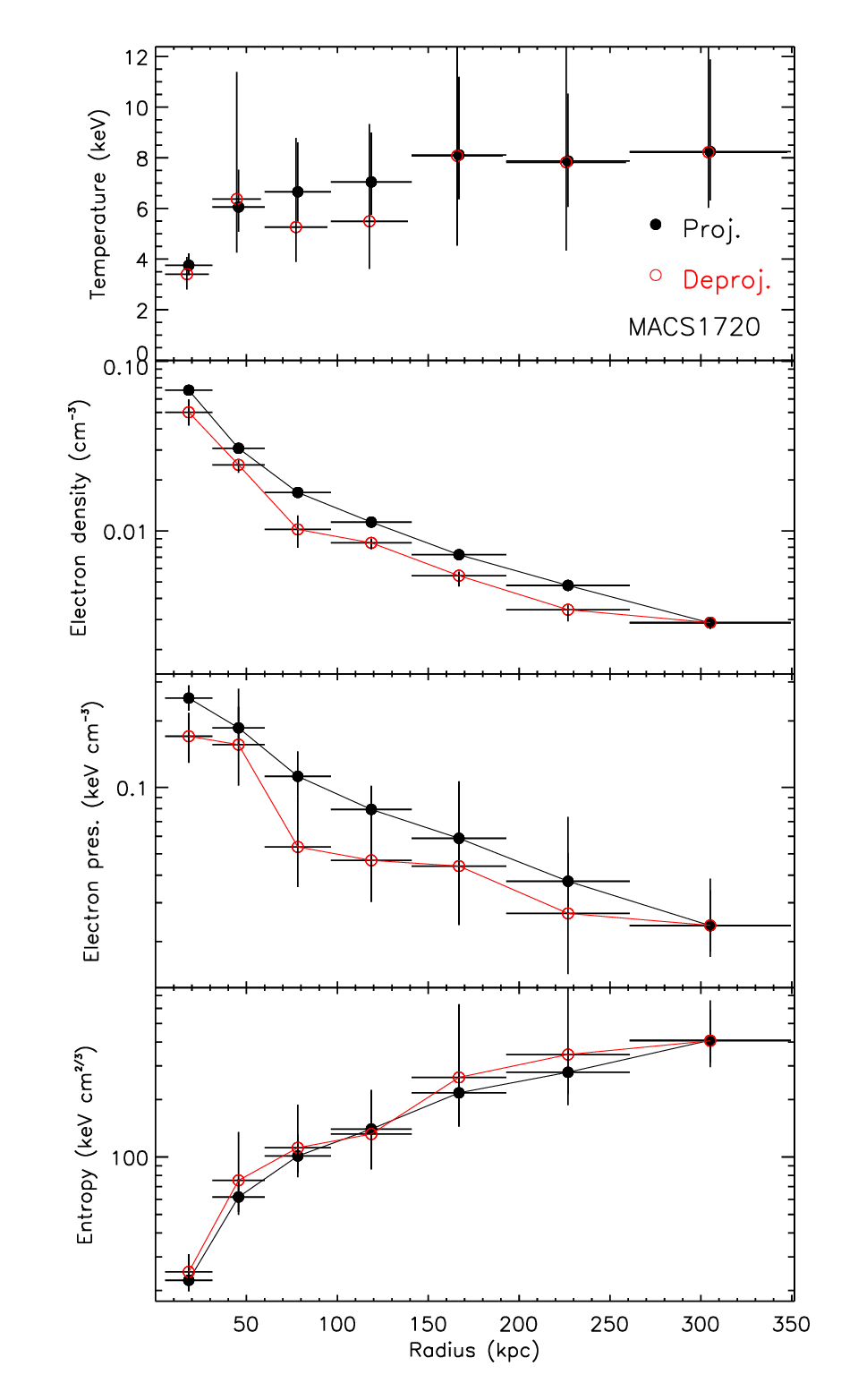}
\end{minipage}
\begin{minipage}[c]{0.245\linewidth}
\centering \includegraphics[width=\linewidth]{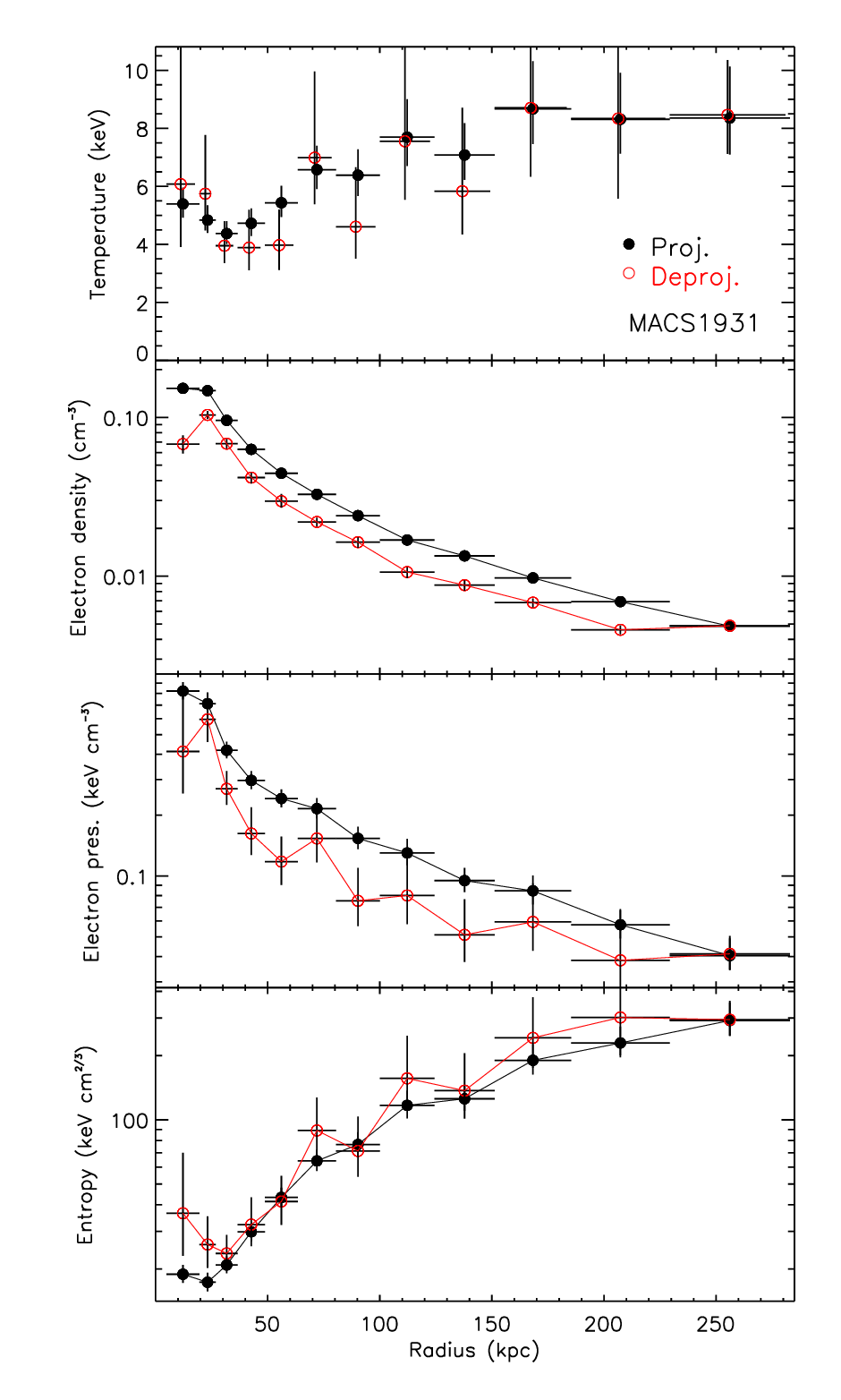}
\end{minipage}
\begin{minipage}[c]{0.245\linewidth}
\centering \includegraphics[width=\linewidth]{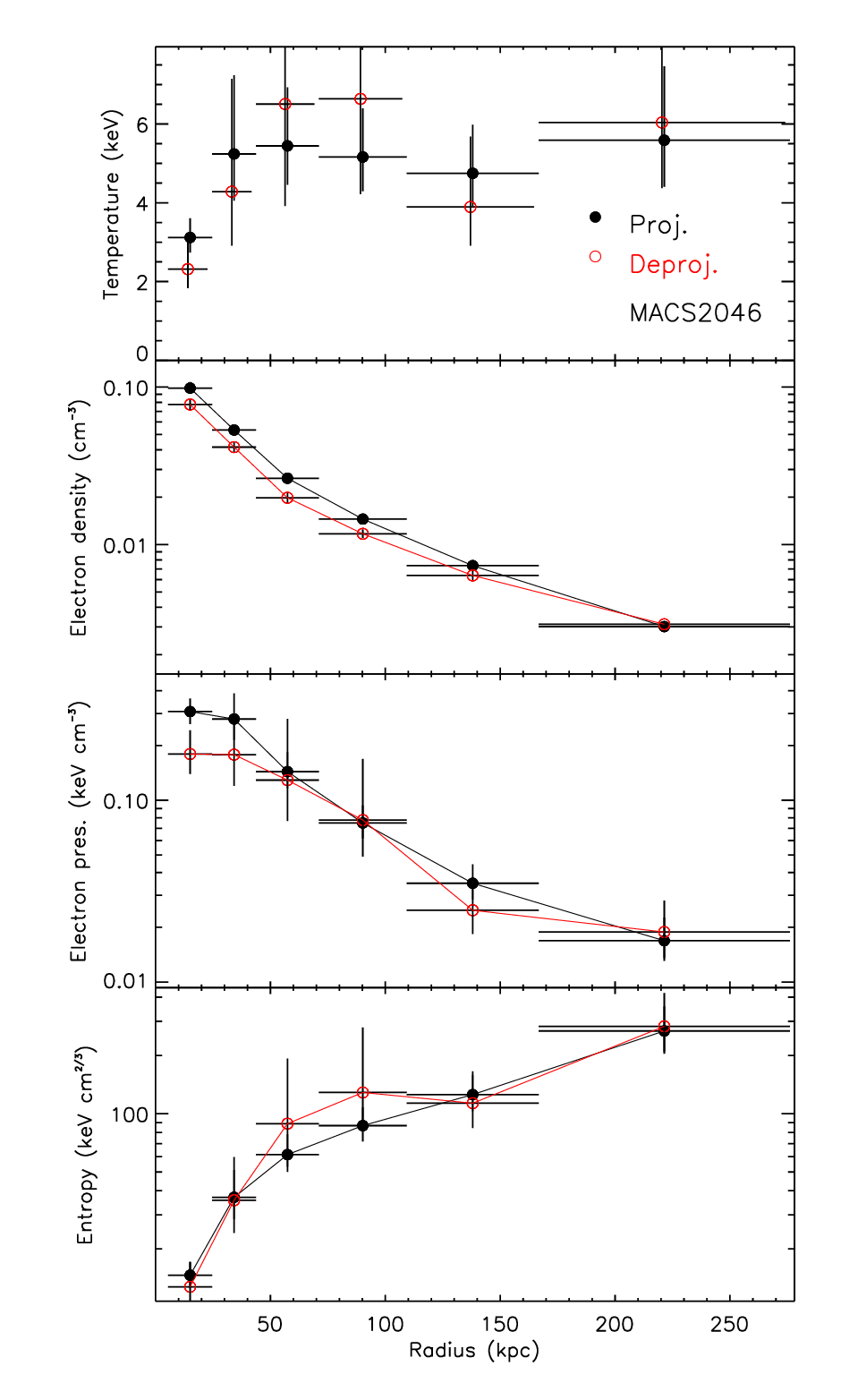}
\end{minipage}
\caption{Projected (black symbols) and deprojected (red symbols) temperature, electron density, electron pressure and entropy profiles for each cluster in which a cavity was identified.}
\end{figure*}
\begin{figure*}
\ContinuedFloat
\begin{minipage}[c]{0.245\linewidth}
\centering \includegraphics[width=\linewidth]{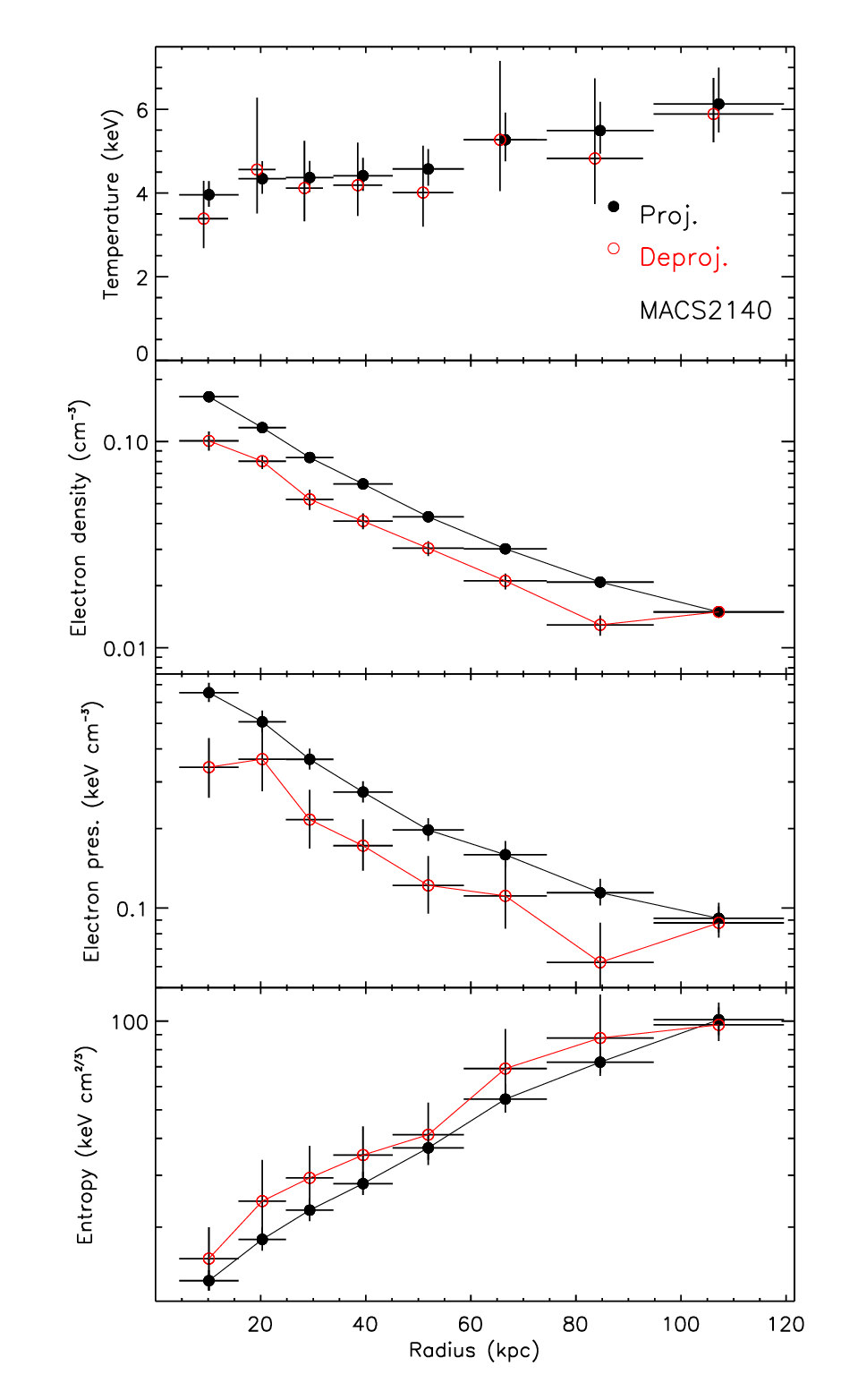}
\end{minipage}
\begin{minipage}[c]{0.245\linewidth}
\centering \includegraphics[width=\linewidth]{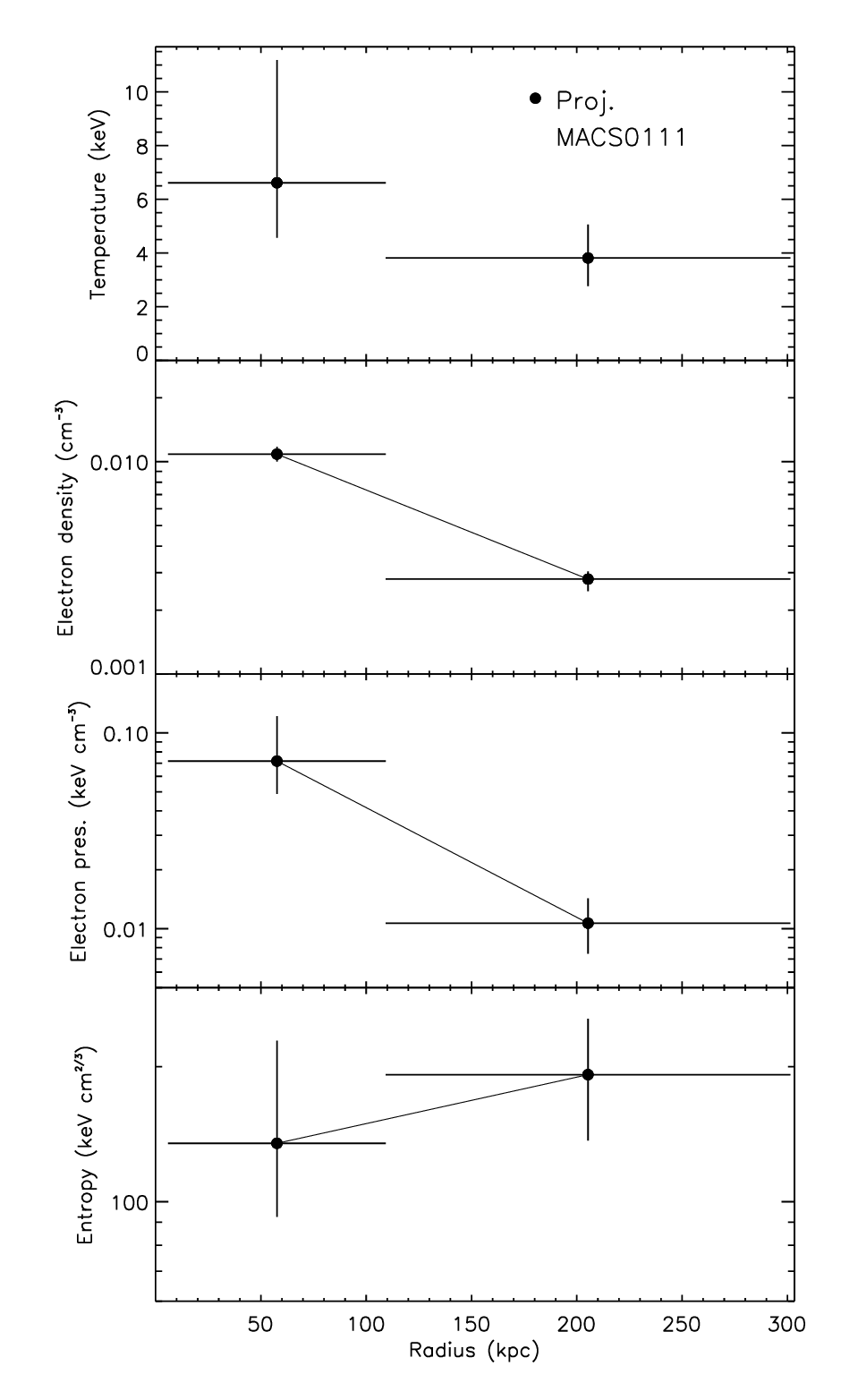}
\end{minipage}
\begin{minipage}[c]{0.245\linewidth}
\centering \includegraphics[width=\linewidth]{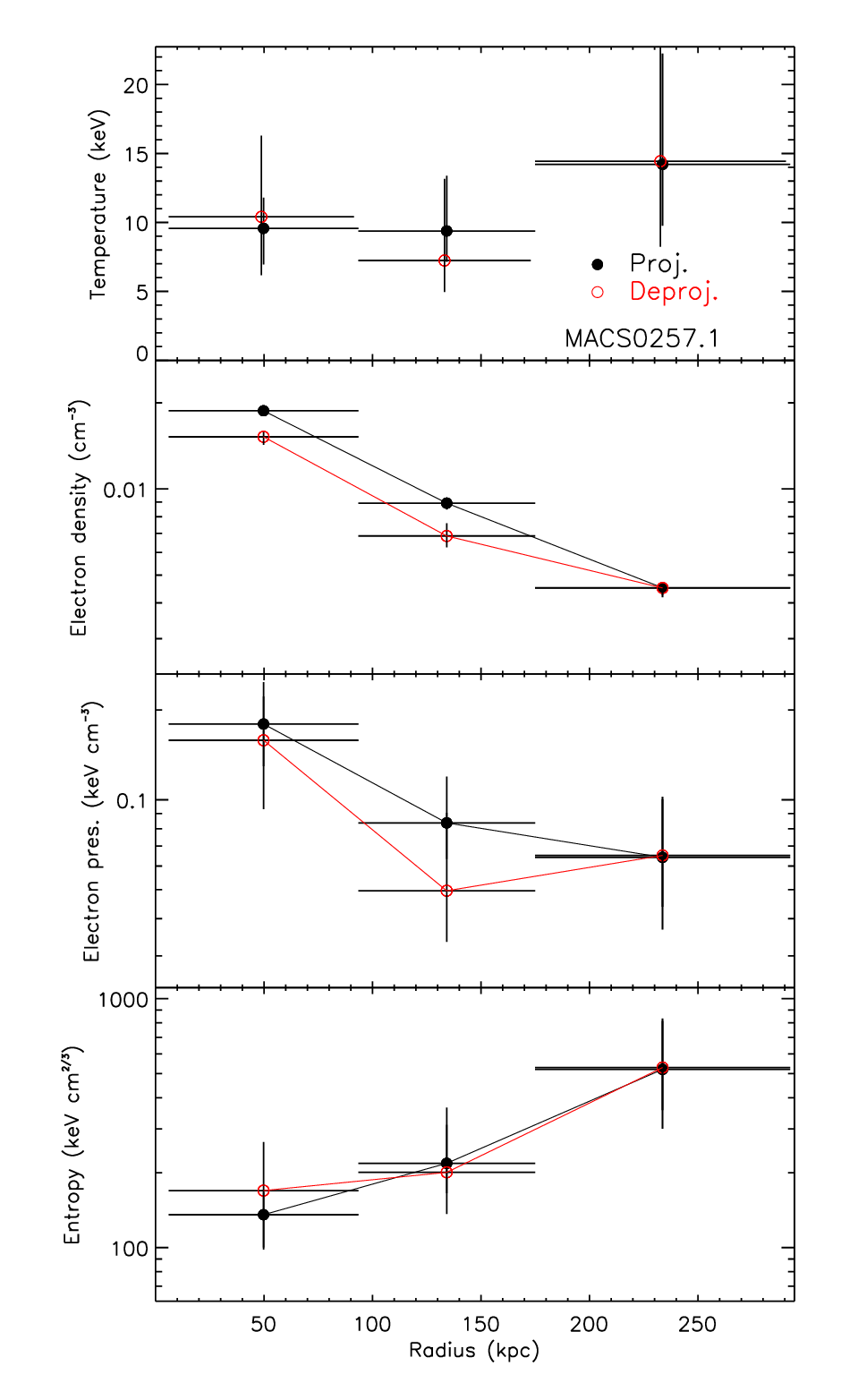}
\end{minipage}
\begin{minipage}[c]{0.245\linewidth}
\centering \includegraphics[width=\linewidth]{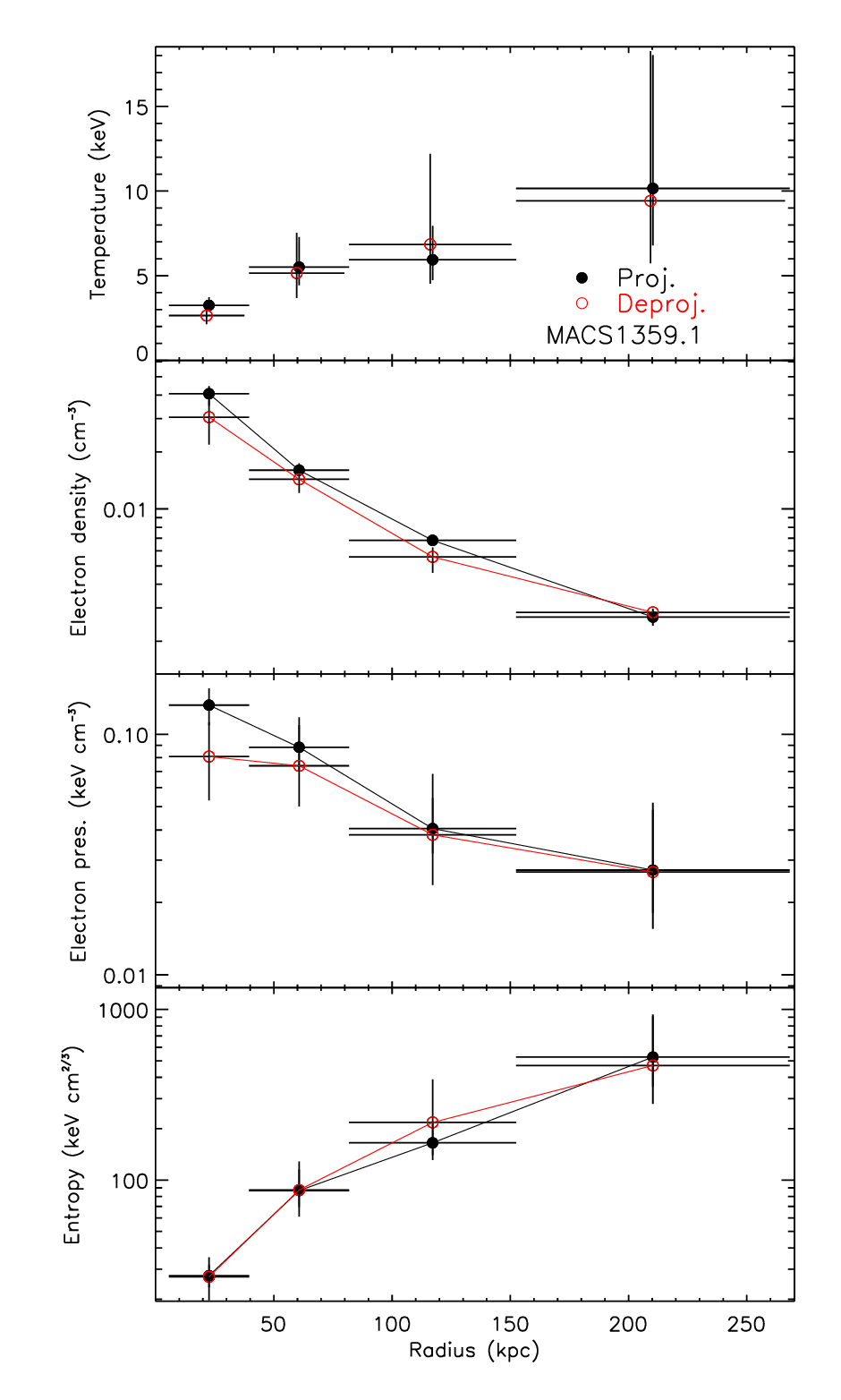}
\end{minipage}
\begin{minipage}[c]{0.245\linewidth}
\centering \includegraphics[width=\linewidth]{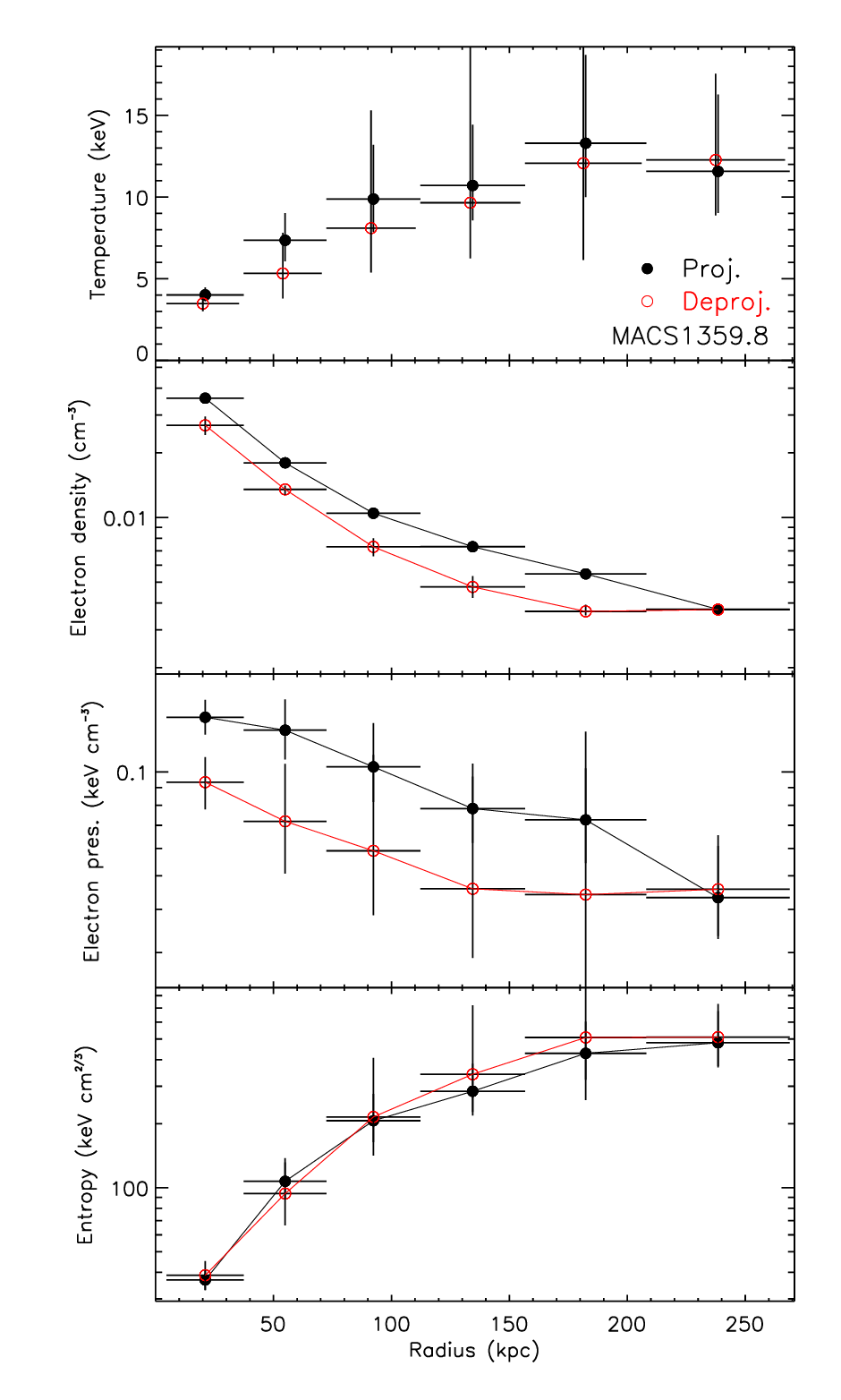}
\end{minipage}
\begin{minipage}[l]{0.245\linewidth}
\centering \includegraphics[width=\linewidth]{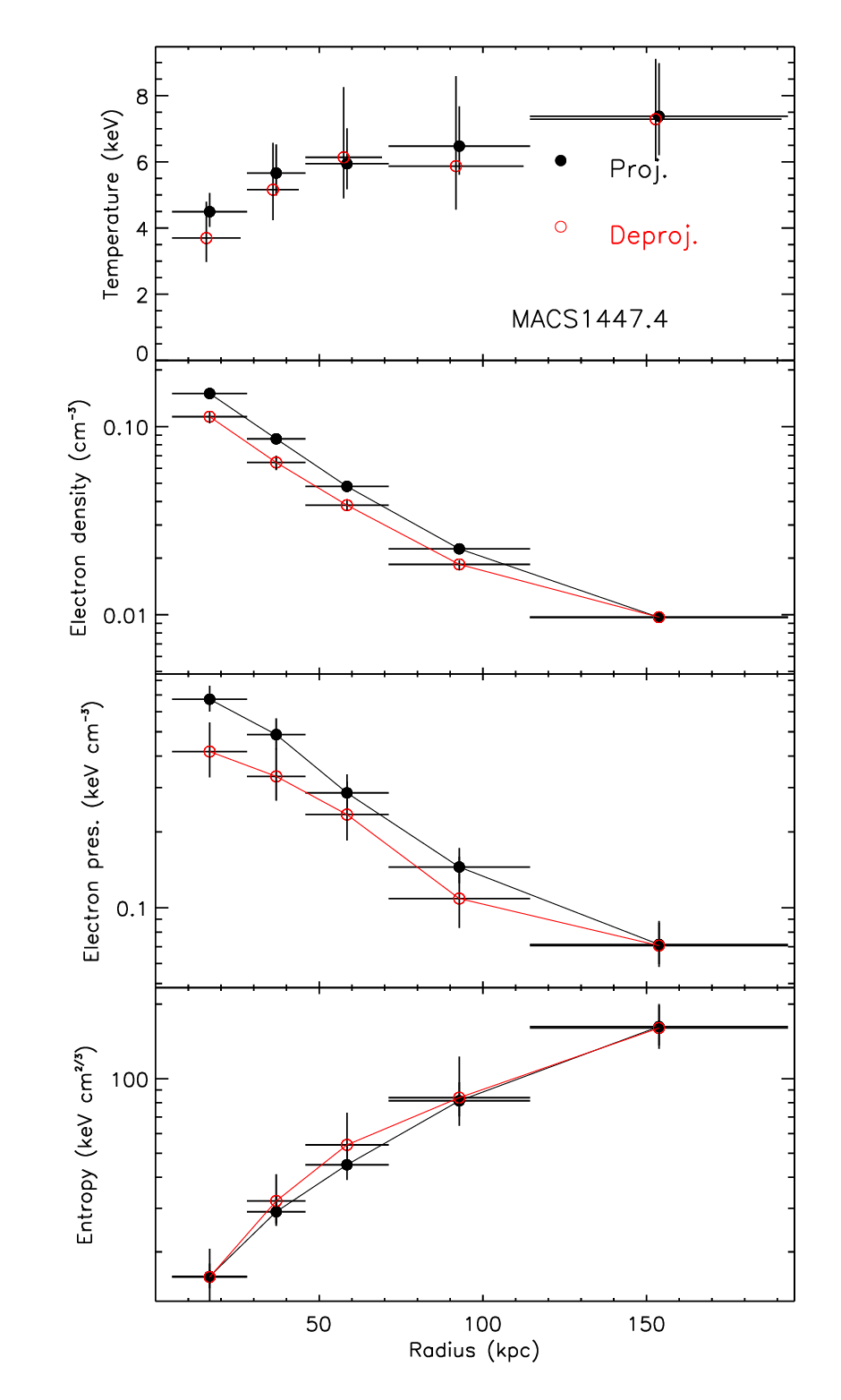}
\end{minipage}
\begin{minipage}[c]{0.245\linewidth}
\centering \includegraphics[width=\linewidth]{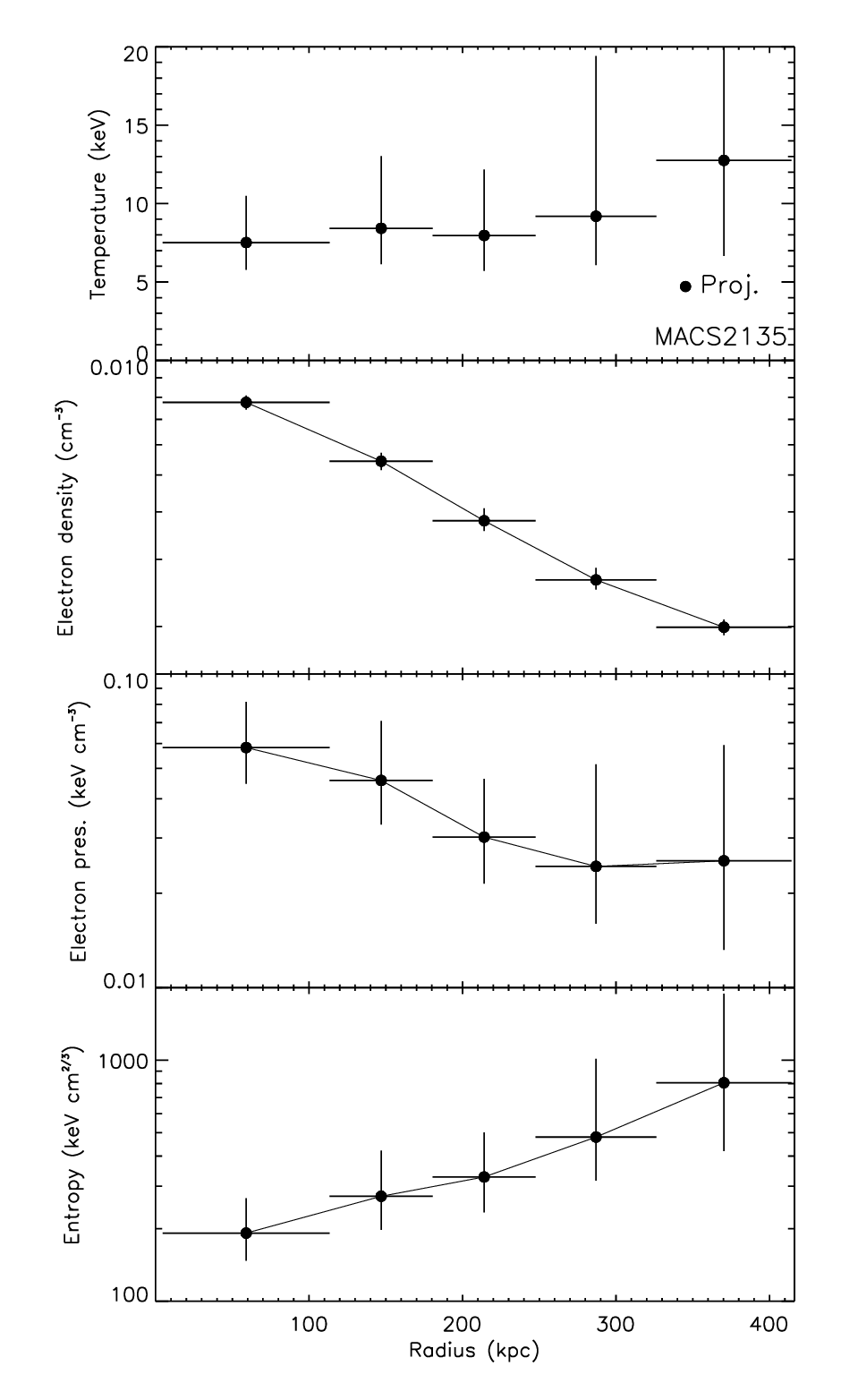}
\end{minipage}
\begin{minipage}[l]{0.245\linewidth}
\centering \includegraphics[width=\linewidth]{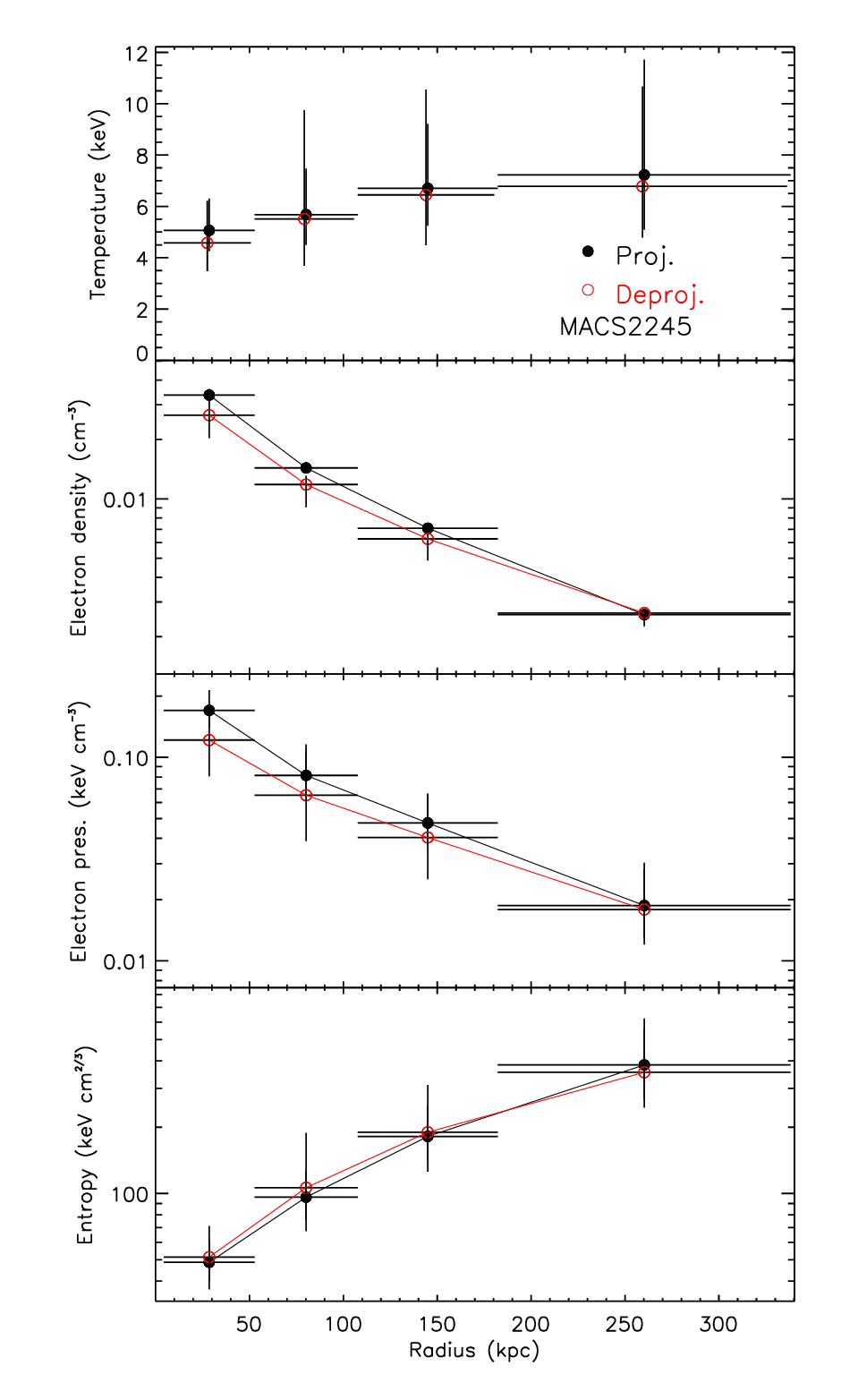}
\end{minipage}
\caption{Continued - Projected (black symbols) and deprojected (red symbols) temperature, electron density, electron pressure and entropy profiles for each cluster in which a cavity was identified.}
\label{figA1}
\end{figure*}

\end{document}